%% file: PaperI_Redline_Corrections.tex
\documentclass[preprint,11pt]{emulateapj}

\shorttitle{X-Shaped Radio Galaxies}
\shortauthors{Roberts et al.}
\usepackage{datetime}
\usepackage{rotate}

\begin{document}

\title{THE ABUNDANCE OF X-SHAPED RADIO SOURCES I. \\ VLA SURVEY OF 52 SOURCES WITH OFF-AXIS DISTORTIONS}

\author{David H.\ Roberts, Jake P. Cohen, \& Jing Lu}
\affil{Department of Physics MS-057, Brandeis University, Waltham, MA 02454-0911 USA}
\email{roberts@brandeis.edu}
\author{Lakshmi Saripalli \& Ravi Subrahmanyan}
\affil{Raman Research Institute, C.\ V.\ Raman Avenue, Sadashivanagar, Bangalore 560080, India}

\begin{abstract}

\citet{C2007} identified a sample of 100 candidate X-shaped radio galaxies using the NRAO FIRST survey; these are small-axial-ratio extended radio sources with off-axis emission. Here we present radio images of 52 of these sources that have been made from archival Very Large Array data with resolution of about 1\arcsec. Fifty-one of the 52 were observed at $1.4$~GHz, seven were observed at $1.4$~GHz and $5$~GHz, and  one was observed only at $5$~GHz. We also present overlays of the SDSS red images for 48 of the sources, and DSS\,II overlays for the remainder. Optical counterparts have been identified for most sources, but there remain a few empty fields.
Our higher resolution VLA images along with FIRST survey images of the sources in the sample reveal that extended extragalactic radio sources with small axial ratios are largely (60\%) cases of double radio sources with twin lobes that have off-axis extensions, usually with inversion-symmetric structure.  The available radio images indicate that at most 20\% sources might be genuine X-shaped radio sources that could have formed by a restarting of beams in a new direction following an interruption and axis flip.  The remaining 20\% are in neither of these categories.  The implications of this result for the gravitational wave background are discussed in \cite{PaperII}.

\end{abstract}


\section{INTRODUCTION} \label{s:intro}

The X-shaped radio galaxy (XRG) population has been the subject of several recent works that have sought to understand the formation mechanism for the non-classical symmetric double lobed radio sources that have peculiar off-axis extended radio structures. The two main contending formation scenarios differ quite starkly: the rapid axis flip and slow axis precession  mechanisms require the AGN beam axis to have undergone a rapid or slow rotation by a large angle (e.g.\ due to accretion disk instabilities; \cite{M2002}; \cite{D2002}, \cite{G2011}) whereas the backflow-origin scenario requires a backflow to either have been deflected by the thermal halo surrounding the host elliptical (\cite{L1984}; \cite{W1995}) or be escaping from a high pressure region along the steepest pressure gradient along the minor axis \citep{C2002}. Besides these models  there are at least two other models that seek to explain the X-structures of
radio galaxies involving either twin radio-loud AGN with axes oriented at large angles with respect to each other \citep{L2007}  or jet-shell interactions \citep{GK2012}. The models are not without drawbacks  (e.g.\ see \cite{S2009}  and \cite{GK2012}). While it is possible that the proposed mechanisms may all give rise to non-classical radio structures in different instances it might also be that one mechanism is more commonly responsible  for off-axis emission than others.

While theoretical modeling studies of the proposed mechanisms is a way forward, progress requires detailed radio-optical-X-ray observations of sufficient quality that would allow much needed
characterization of the population against which the models could be tested. A few studies in this direction (\cite{C2002, S2009, HK2010, M2011}) have been carried out which reported the relatively high ellipticities of the XRG host ellipticals, host minor axis preference of the wings, host major axis proximity of the main lobes, prevalence of X-ray coronae surrounding XRG hosts
and the relatively young ages of the hosts. Besides, Saripalli \& Subrahmanyan have examined the observational data in the context of the formation of a wider class of radio structures and revealed a connection between XRGs and the parent class of sources with lobe distortions suggesting the possibility of a more widespread or common phenomenon at work
in creating peculiar radio galaxy morphologies. 

We present herein our attempt at resolving this issue using archival data on  the sample of XRG candidates compiled by \cite{C2007}, which forms a useful resource for taking the characterization studies further. In this paper we have imaged existing (as yet unanalyzed) archival VLA data of a subsample of his 100 XRG candidates. Here we present new maps of 52 radio galaxies, all of those for which archival L-band A-array and/or C-band B-array data exist.  Each of the sources is discussed briefly. Several have been followed up in the optical and have redshifts available \citep{L2010}. We also provide new optical identifications for several of these sources.

In Section~\ref{s:results} we present our images of 52 sources, with overlays on the corresponding optical fields. Section~\ref{s:discussion} discusses the classification of the sources by the location of their deviant off-axis lobes. The goal is to identify those sources that we believe are {\em candidates} for ``true X-shaped morphology.'' {\em We define a ``true X-shaped morphology'' as one where the deviant off-axis emission is not traced to either of the main lobes of the radio galaxy and instead is seen as an independent transverse feature centered on the host.} The sources in our sample that meet this definition are listed in Section~\ref{s:nature}; particularly good examples are  J1043+3131 and J1327$-$0203. Our results are summarized in Section~\ref{s:summary}.

In \cite{PaperII} we discuss the implications of our results for the gravitational wave background.

\section{OBSERVATIONS AND DATA ANALYSIS}

The NRAO Data Archive was searched for historical VLA data on the 100 sources in Cheung's sample. We used all existing L band observations in the A-array and C band observations in the B-array. Thus the resulting images had resolution of about one arc second. The typical observation was a snapshot of a few minutes duration. The data were calibrated in AIPS using standard techniques and self-calibrated in DIFMAP, with final images made using the AIPS task IMAGR. In a few cases no flux calibrator was available, so the flux scale was bootstrapped from observations of the phase calibration sources at nearby times.

\section{RESULTS} \label{s:results}

For each source we provide contour images at each band for which historical VLA data were available, and overlays of the radio structures on optical images from the Sloan Digital Sky Survey (SDSS), or DSS~II images if there is no SDSS image of the field (four cases). The contours are spaced by factors of $\sqrt{2}$ and run from the lowest contour levels to the peaks given in the individual captions. In the overlays the contours are spaced by factors of two and some may be omitted to make the optical ID clearer. Below we provide notes describing interesting aspects for some of the sources.

In the new analysis we present we detect compact cores in several sources.  We detect 30 new radio cores while for 15 the new higher resolution imaging has failed to detect cores. All detected cores are seen at the IDs suggested by \cite{C2007} except in J1202+4915 where a core is detected but no ID is seen and in J0846+3956 where a likely core is present at a newly identified, faint object. Large-scale radio emission seen in the FIRST images is resolved out in several sources in the new imaging. For 14 sources there is at least 80 percent of the flux captured in our higher resolution maps where as for 12 sources more than 50 percent of the extended flux is missed. The observed properties of the sources are given in Table~\ref{tab:Observed}.

Below we provide notes to each of the sources where we also provide description of the morphology.

\subsection{Notes on Individual Sources}

\begin{figure}[ht] 
\includegraphics[width=0.45\columnwidth]{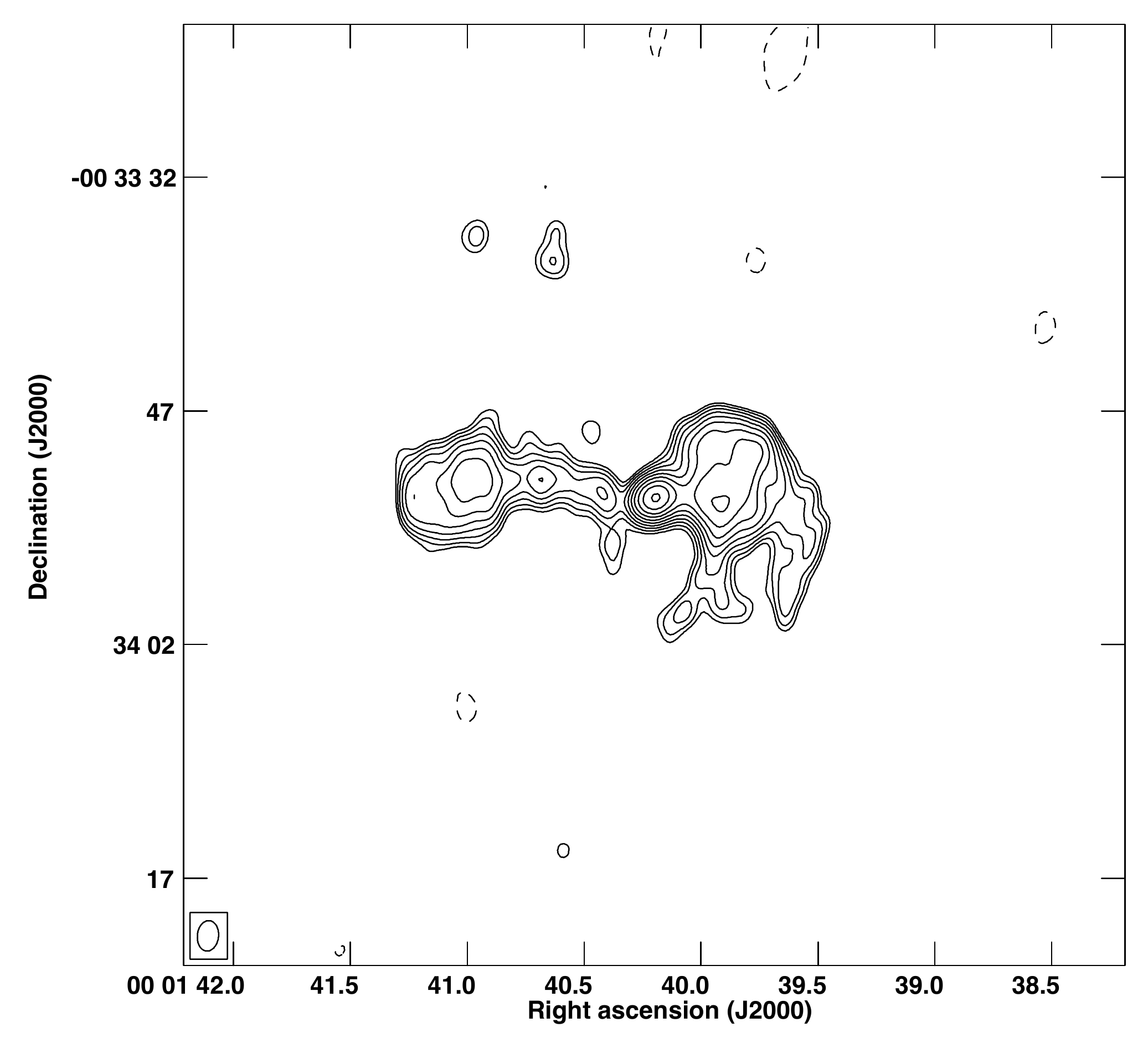}
\includegraphics[width=0.45\columnwidth]{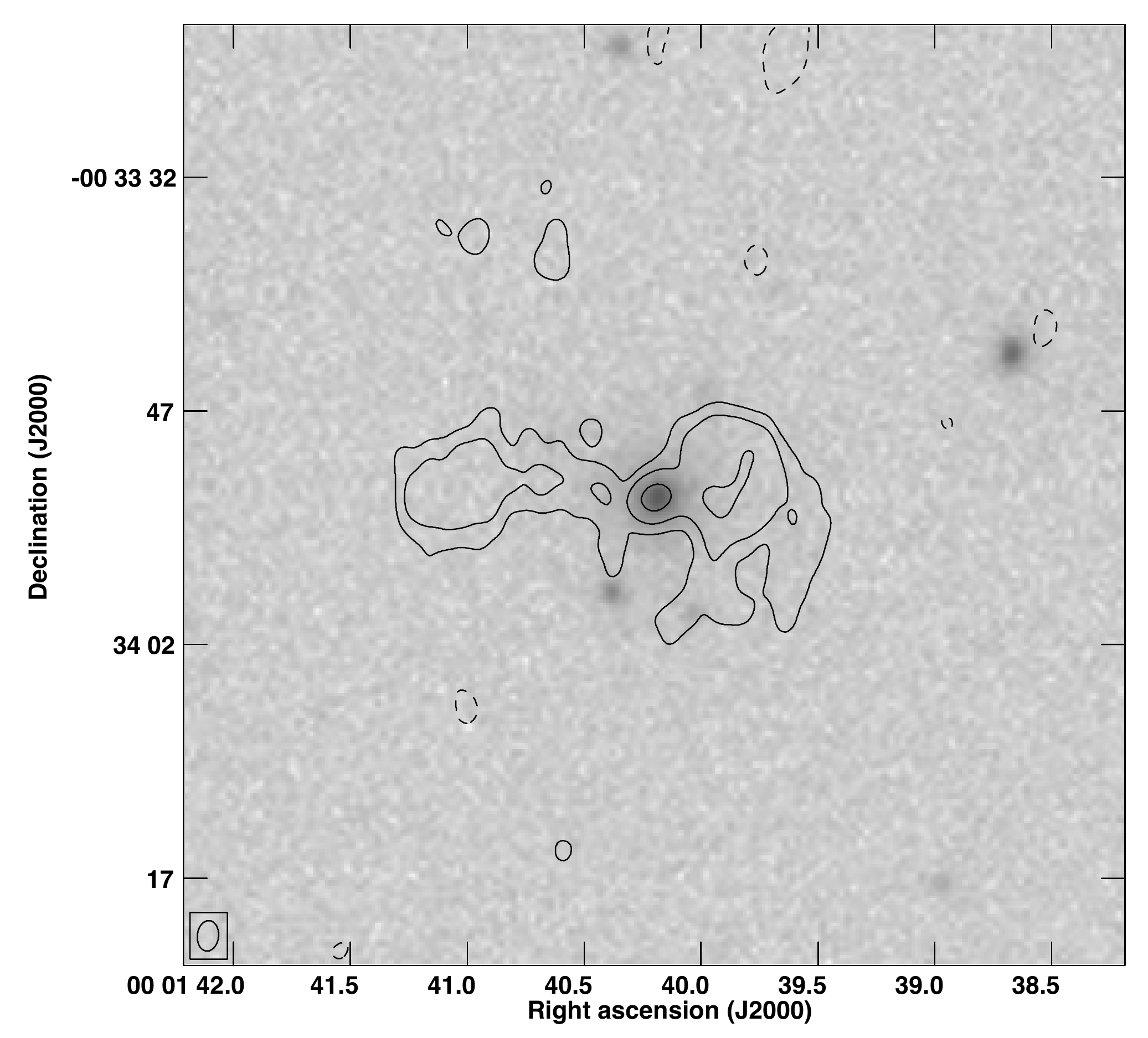}
\caption[J0001$-$0033 (L)]{J0001$-$0033. (left) VLA image at L band, (right) VLA image overlaid on red SDSS image. In all of the figures the contours are spaced by factors of $\sqrt{2}$; in the overly every-other contour is skipped. Here the lowest contour = 0.1~mJy/beam and the peak  = 2.36~mJy/beam.\label{fig:J0001-0033}}
\end{figure}

\noindent J0001$-$0033 (Figure~\ref{fig:J0001-0033}). Our new image shows the core clearly and a narrow, jet-like feature that connects it to the eastern lobe. Both lobes although edge-brightened and well confined lack compact hotspots, showing instead presence of recessed emission peaks. Diffuse extension to the north from the eastern lobe seen in the FIRST image is completely resolved out in our higher resolution map.

\begin{figure}[ht] 
\includegraphics[width=0.45\columnwidth]{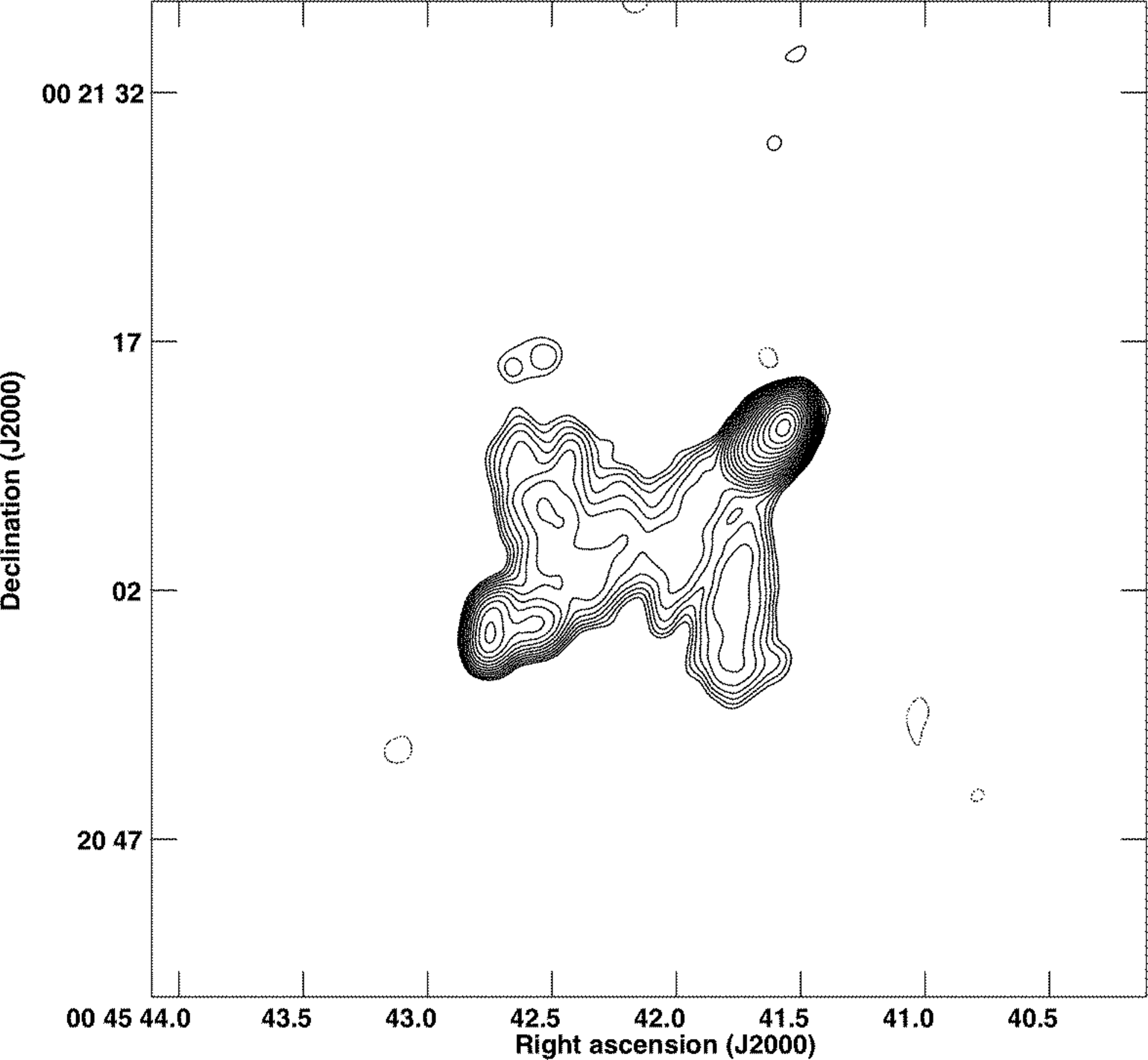}
\includegraphics[width=0.45\columnwidth]{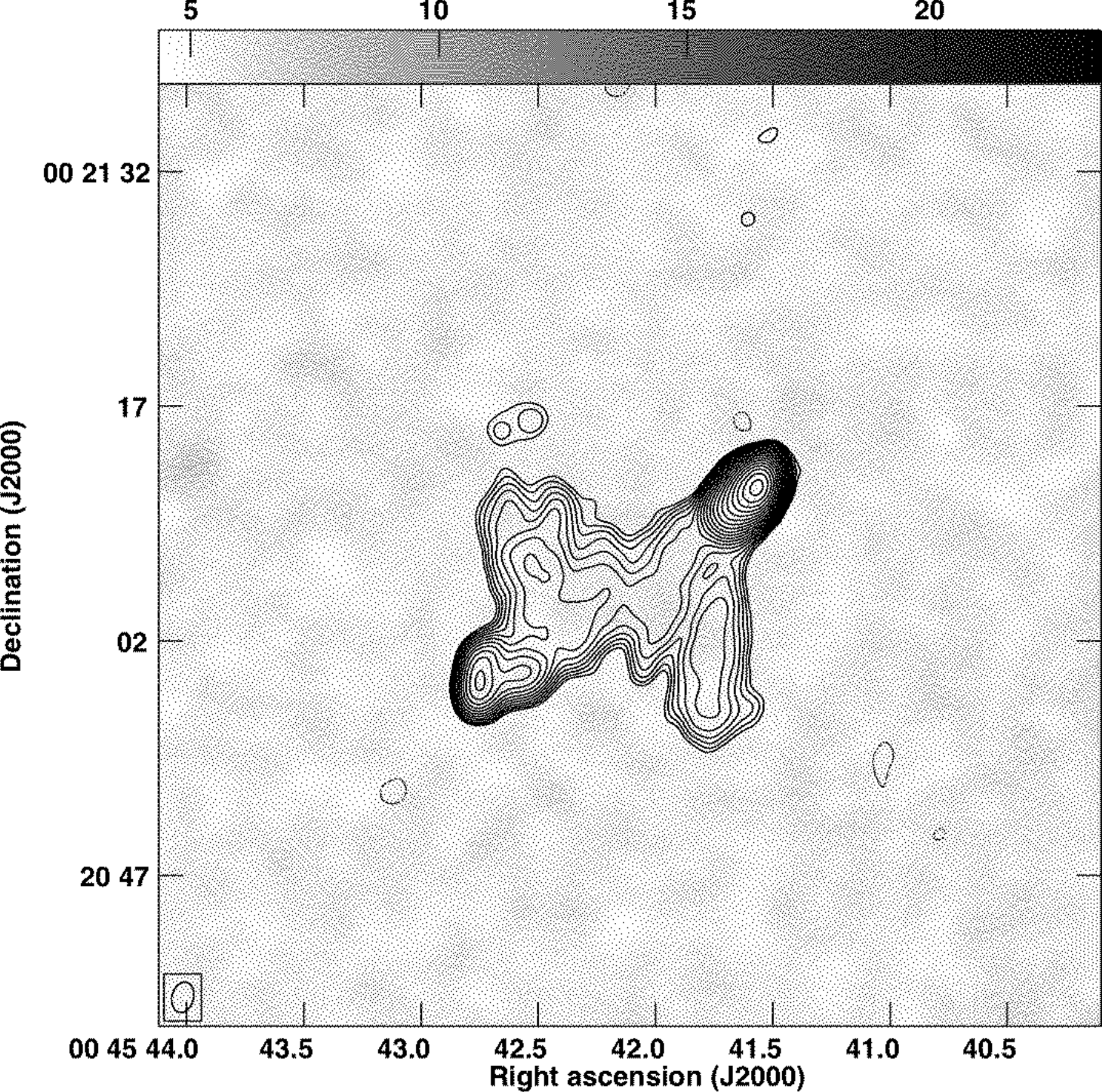}\\
\includegraphics[width=0.45\columnwidth]{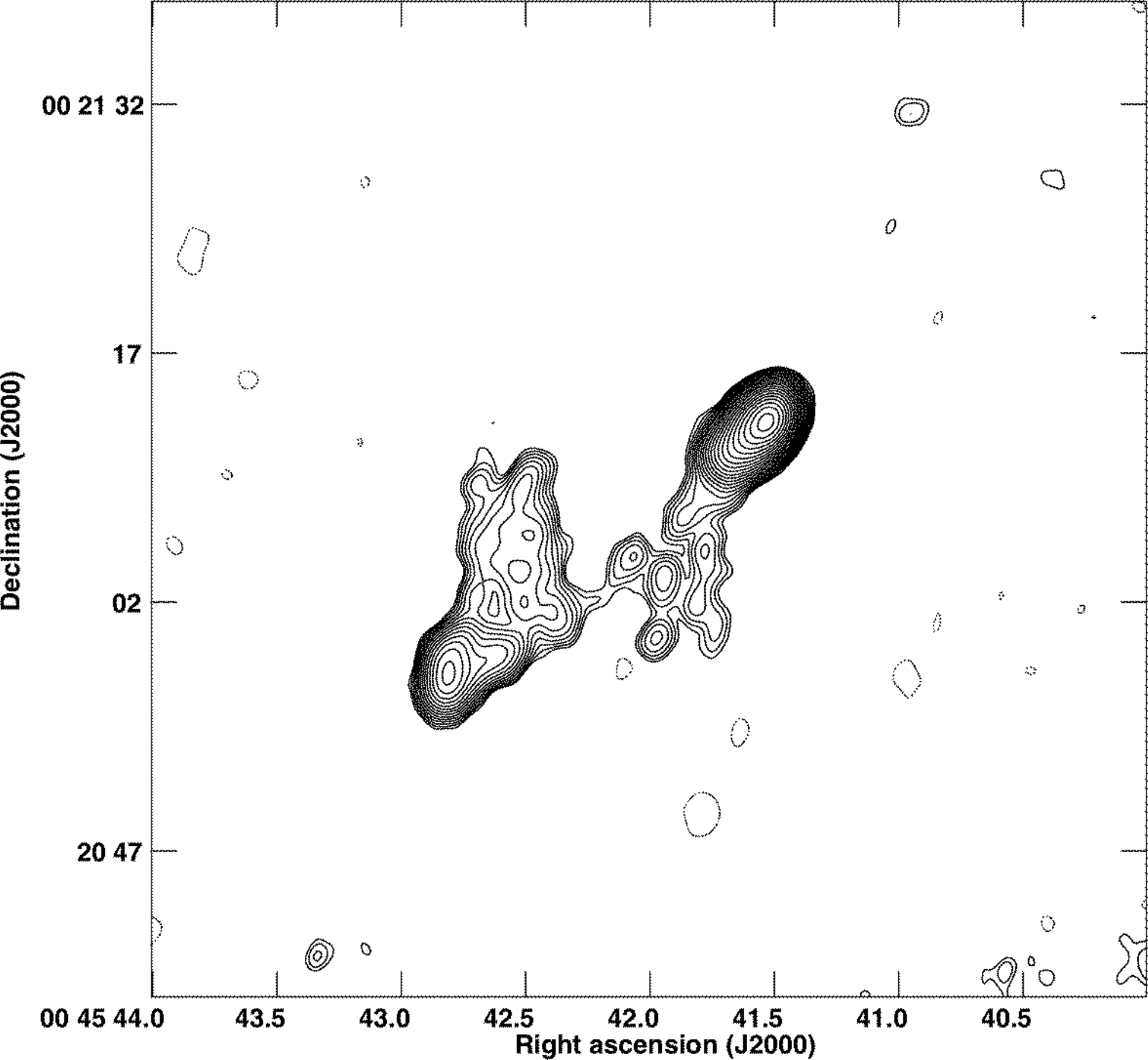}
\includegraphics[width=0.45\columnwidth]{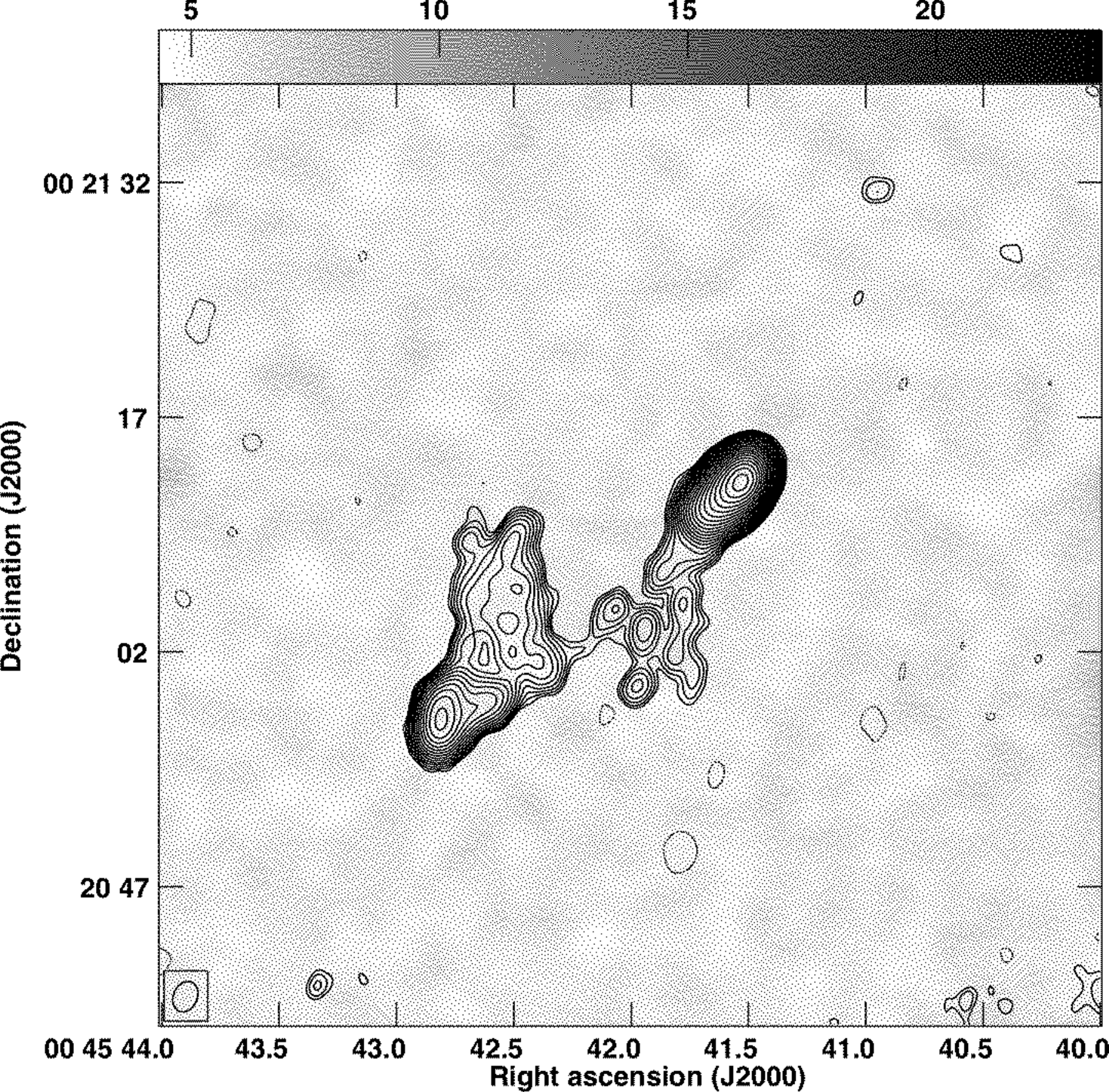}\\
\caption[J0045+0021 (L \& C)]{J0045+0021. (top) (left) VLA image at L band and (right) VLA image overlaid on red SDSS image. Lowest contour= 0.3~mJy/beam, peak =178~mJy/beam. (bottom) (left) VLA image at C band and (right) VLA image overlaid on red DSS II image. Lowest contour = 0.125~mJy/beam, peak  = 60.7~mJy/beam. The red DSS\,II plate does not show any candidate sources.  \label{fig:J0045+0021}}
\end{figure}

\noindent J0045+0021 (Figure~\ref{fig:J0045+0021}). For this source we have presented images at two frequencies. The source, which appears as a classic XRG in the FIRST image continues to exhibit the diffuse orthogonally, oriented extensions in both our maps. The C-band map clearly shows the connection between the transverse extension and the eastern lobe, which also shows a sharp inner edge. A compact core midway between the lobes is detected in our C-band map.

\begin{figure}[ht] 
\includegraphics[width=0.45\columnwidth]{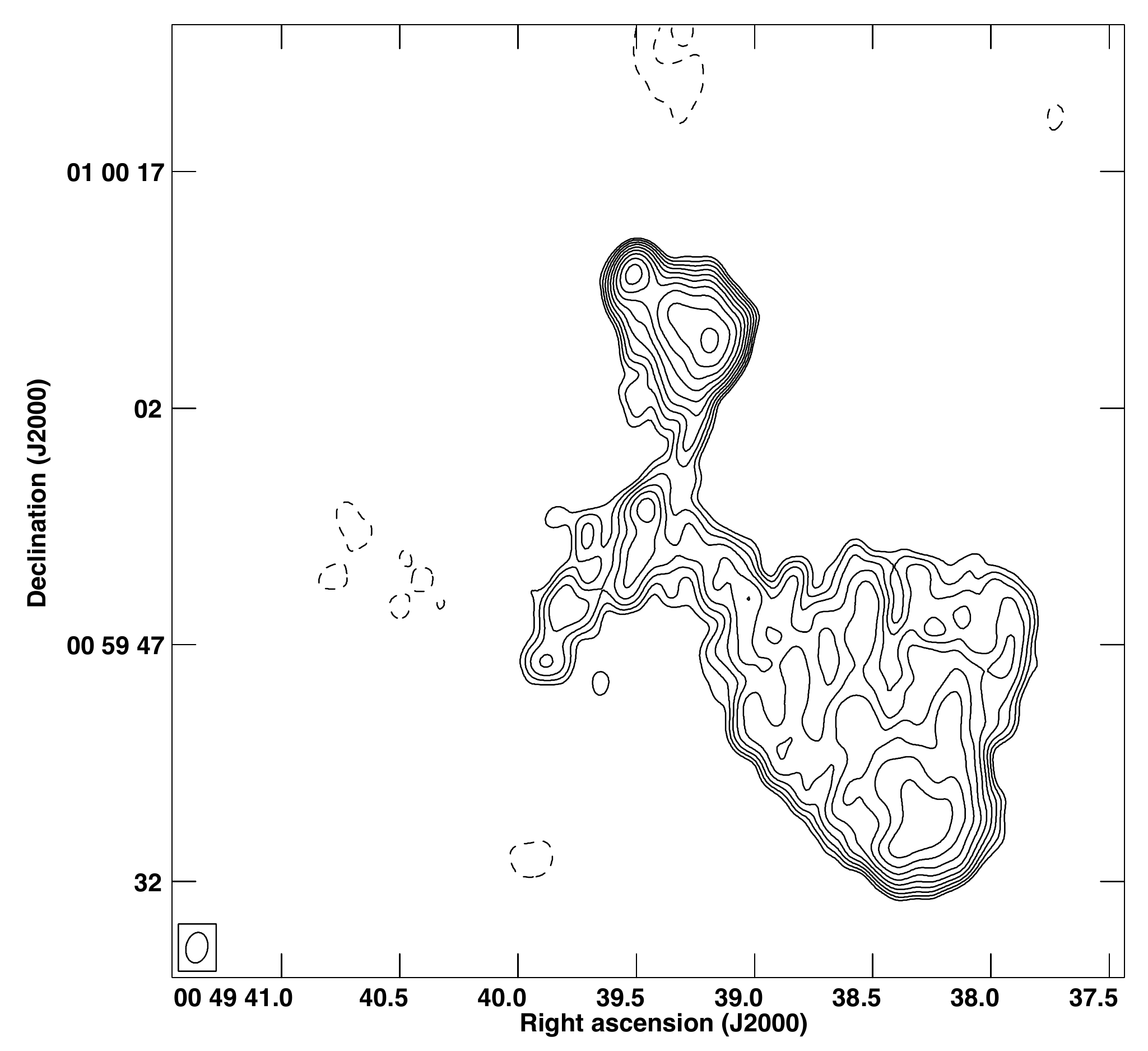}
\includegraphics[width=0.45\columnwidth]{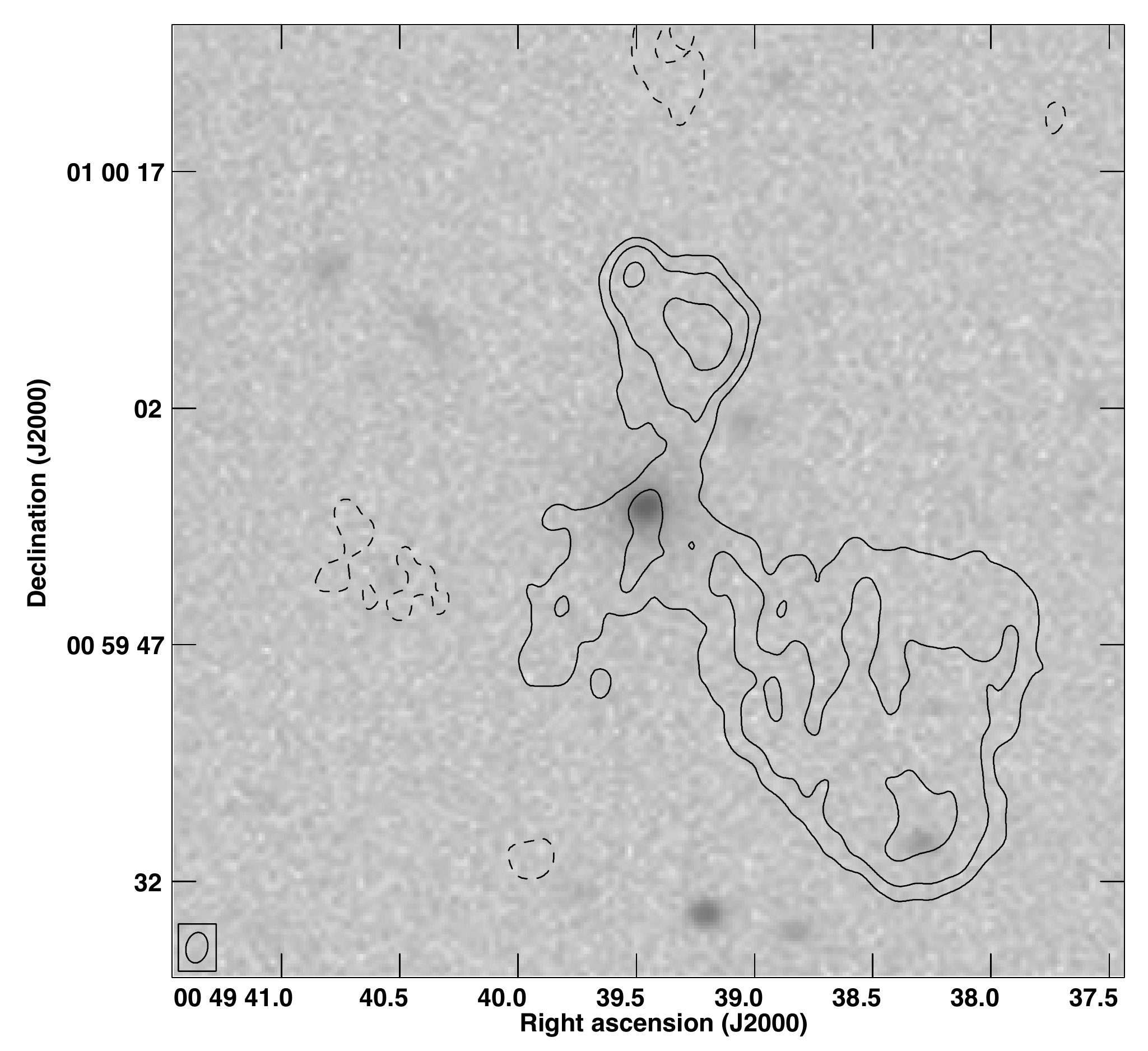}
\caption[J0049+0059 (L)]{J0049+0059.  (left) VLA image at L band, (right) VLA image overlaid on red SDSS image. Lowest contour = 0.1~mJy/beam, peak  = 2.51~mJy/beam.\label{fig:J0049+0059}}
\end{figure}

\noindent J0049+0059 (Figure~\ref{fig:J0049+0059}). Although the FIRST map shows an edge-brightened radio galaxy our new map has revealed a complex structure for this source. Interestingly, the northern lobe is resolved into a source with an edge-brightened double structure. The lobe is extended along a position angle similar to the southern lobe however it is offset by nearly 10\arcsec\ to the west. The offset northern lobe is connected to the core by a weak narrow extension. A narrow, linear feature is also seen extending SE from the core in the opposite direction. The southern lobe has a faint optical ID at the location of the peak at the leading end. We speculate whether the radio source is in fact a collection of three, extended and independent radio sources.

\begin{figure}[ht] 
\includegraphics[width=0.45\columnwidth]{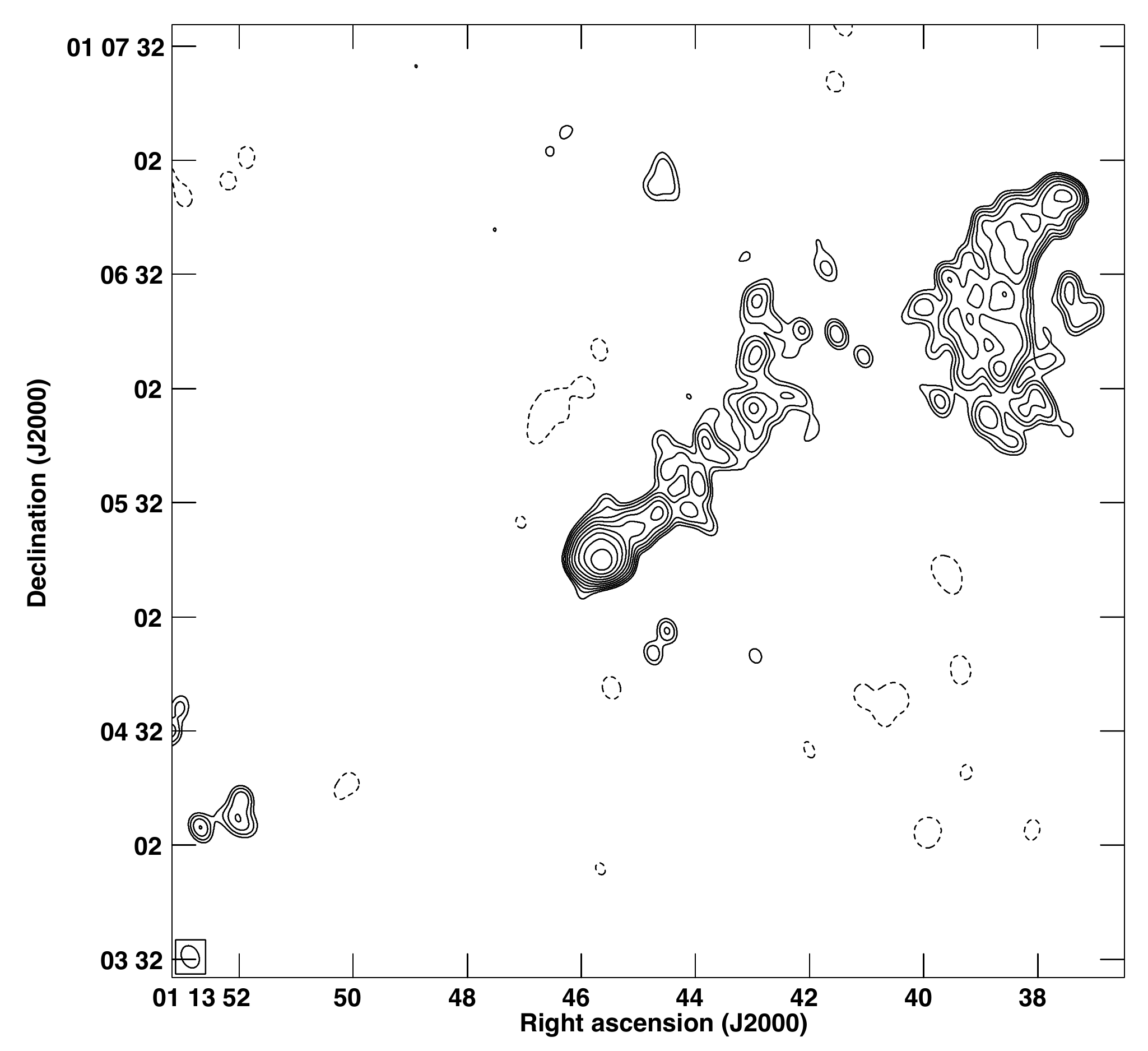}
\includegraphics[width=0.45\columnwidth]{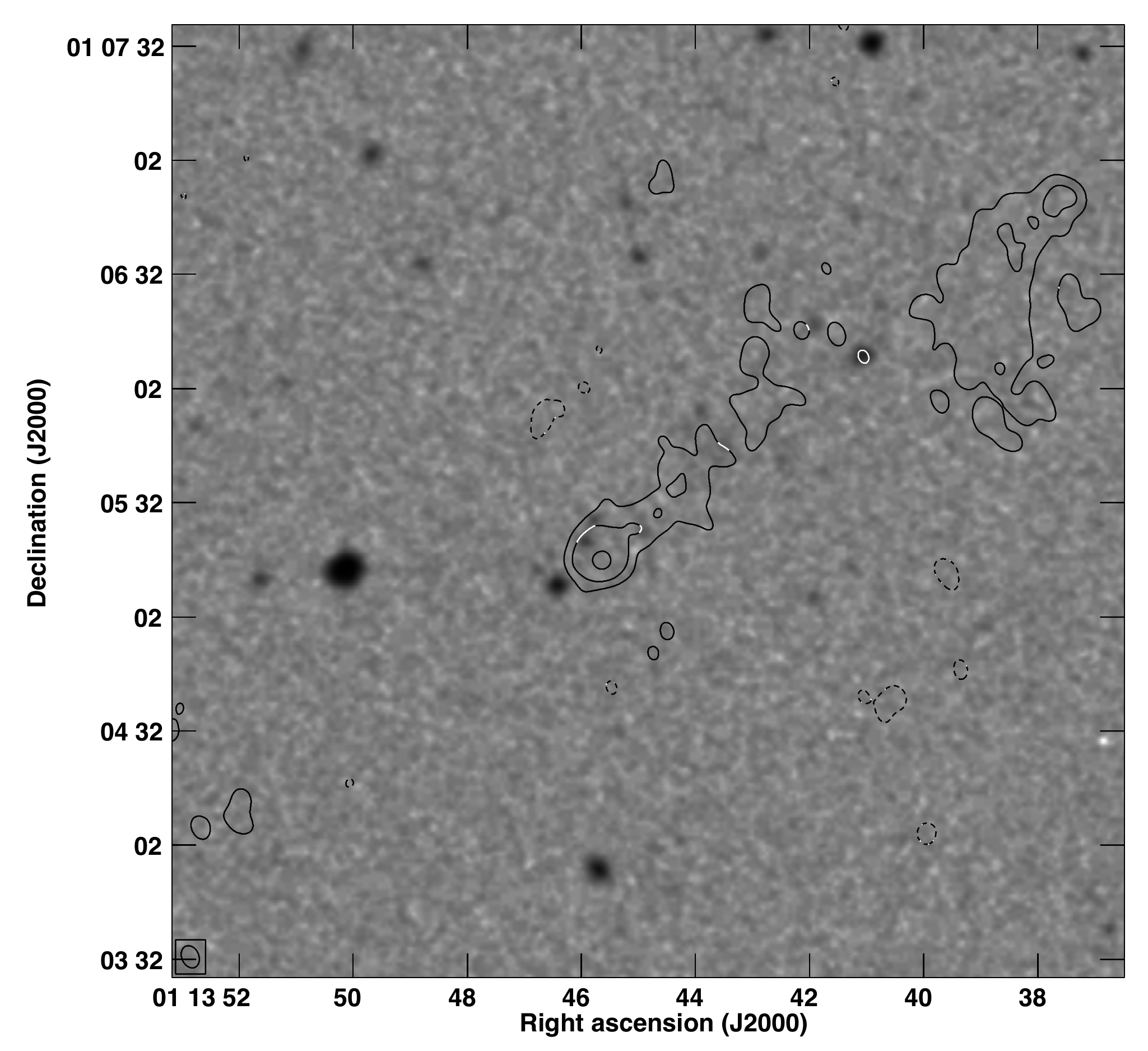}\\
\includegraphics[width=0.45\columnwidth]{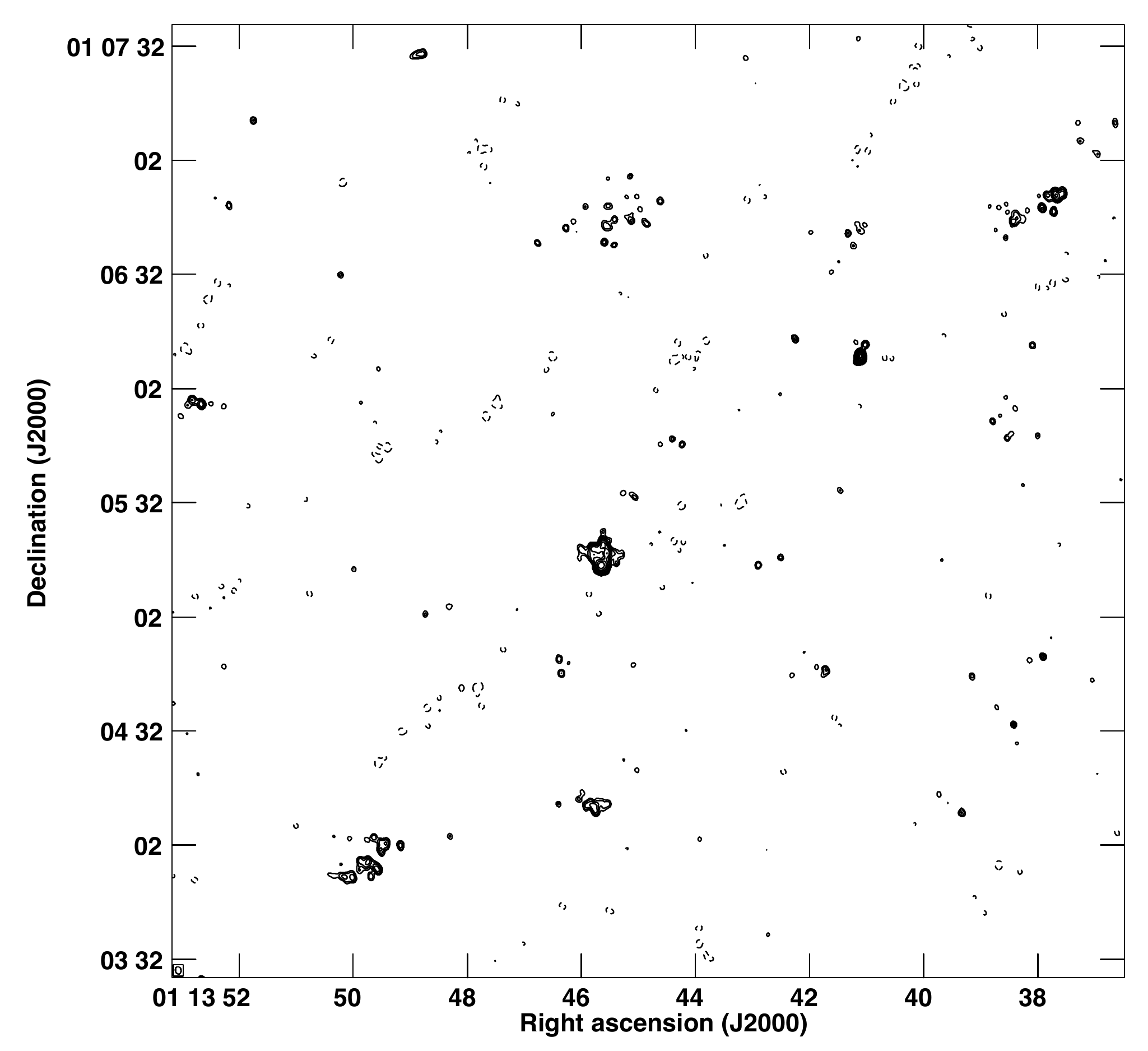}
\includegraphics[width=0.45\columnwidth]{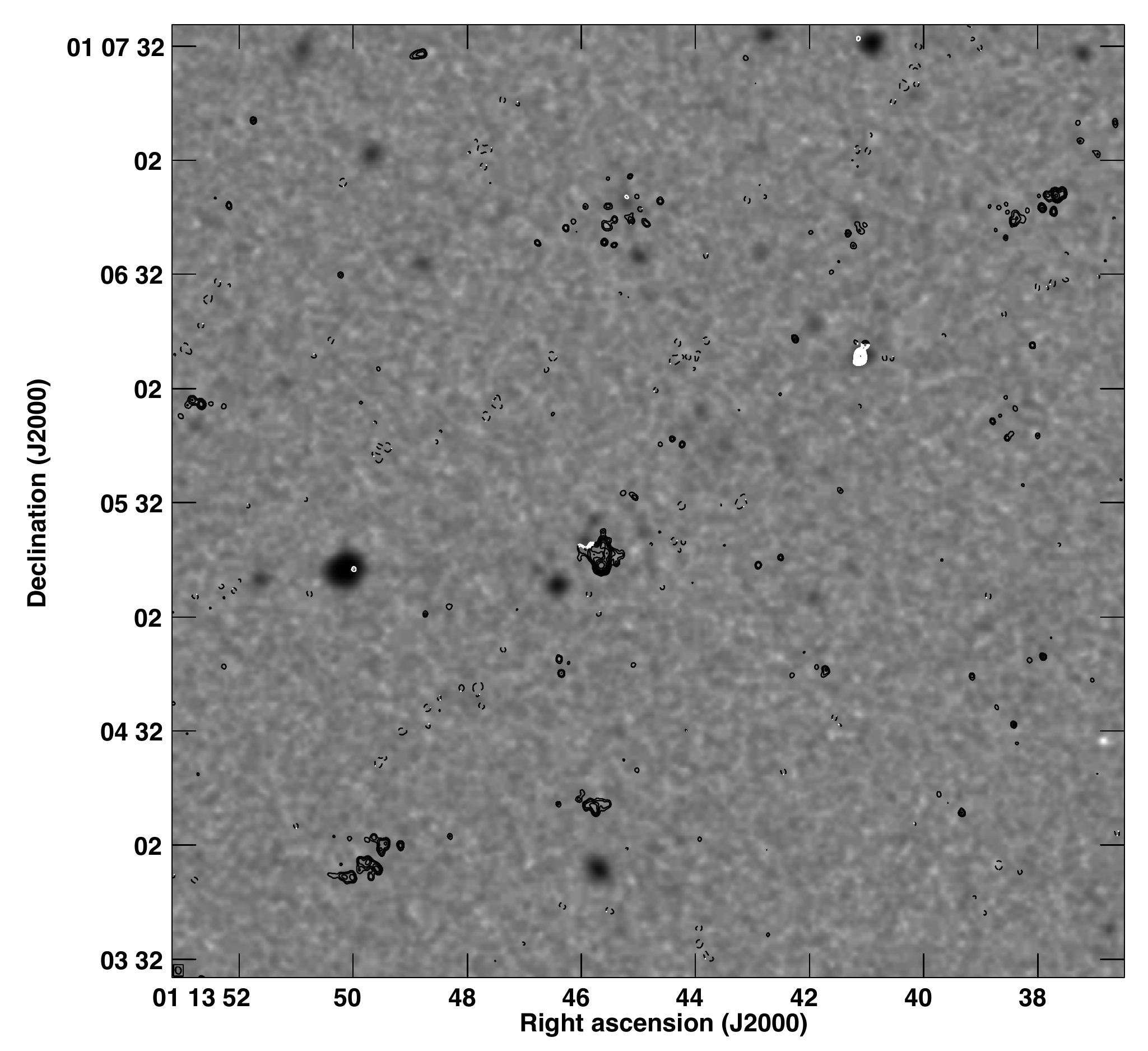}\\
\caption[J0113+0106 (L \& C)]{J0113+0106.  (top) (left) VLA image at L band, (right) VLA image overlaid on red SDSS image. Lowest contour = 0.7~mJy/beam, peak = 18.0~mJy/beam.  (bottom) (left) VLA image at C band, (right) VLA image overlaid on red SDSS image. Lowest contour = 0.1~mJy/beam, peak = 2.56~mJy/beam.\label{fig:J0113+0106} }  \end{figure}

\noindent J0113+0106 (Figure~\ref{fig:J0113+0106}). The inversion symmetric transverse extensions seen in the FIRST image are also seen in the L-band image but are completely resolved out in the C-band image. In both the FIRST and L-band image the transverse extensions show inner edges and connect to respective lobes. 

\begin{figure}[ht] 
\includegraphics[width=0.45\columnwidth]{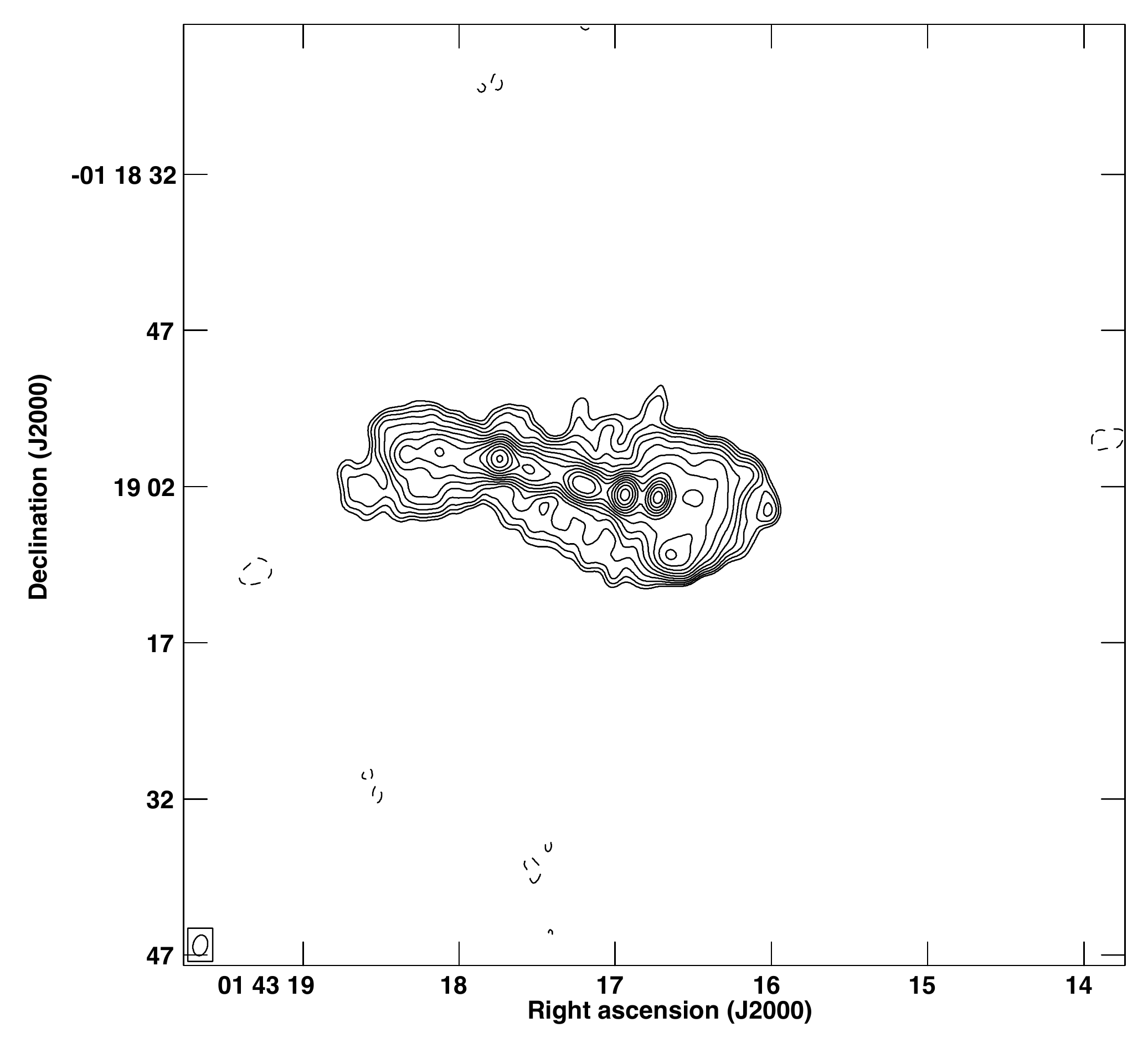}
\includegraphics[width=0.45\columnwidth]{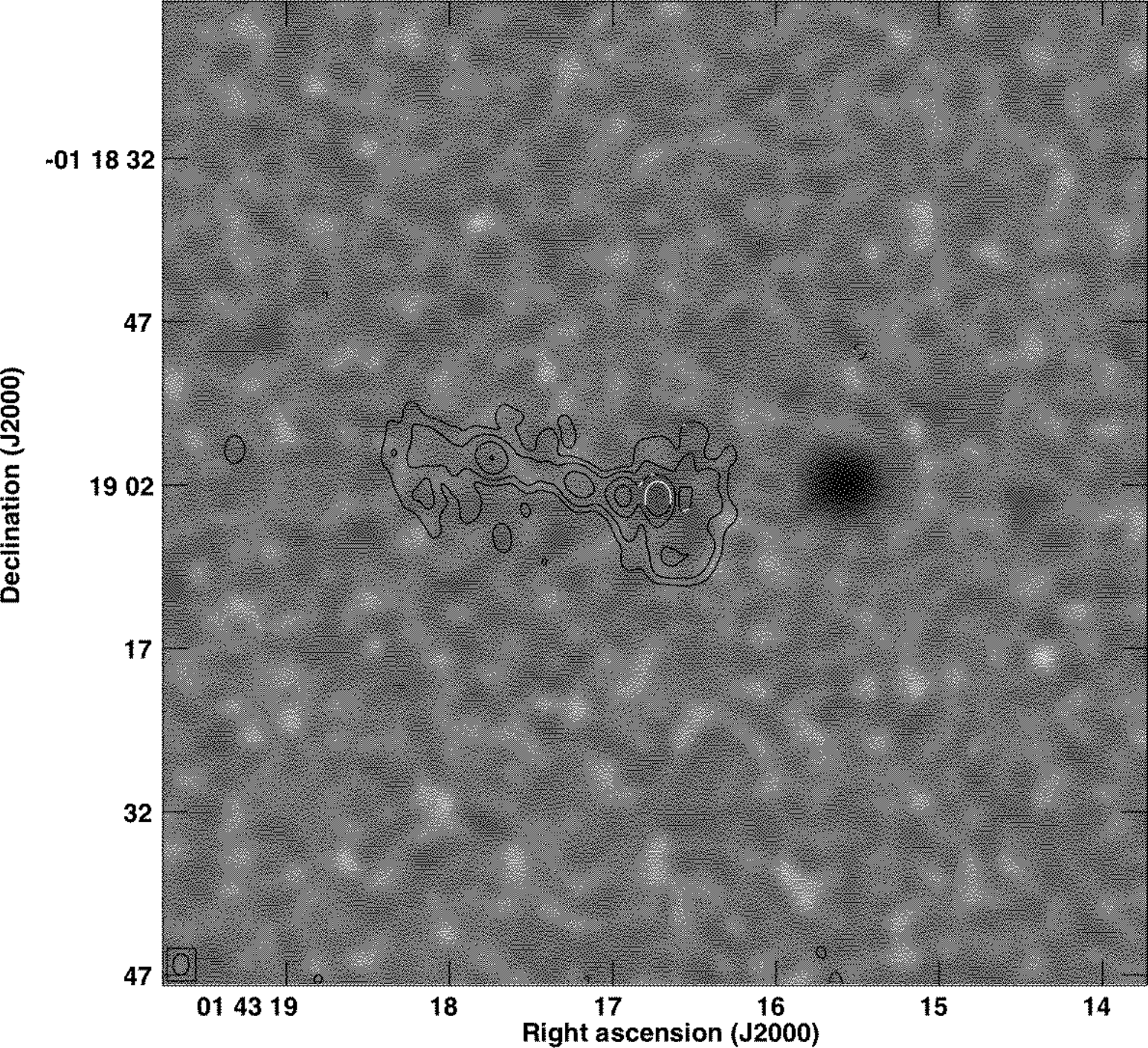}\\
\includegraphics[width=0.45\columnwidth]{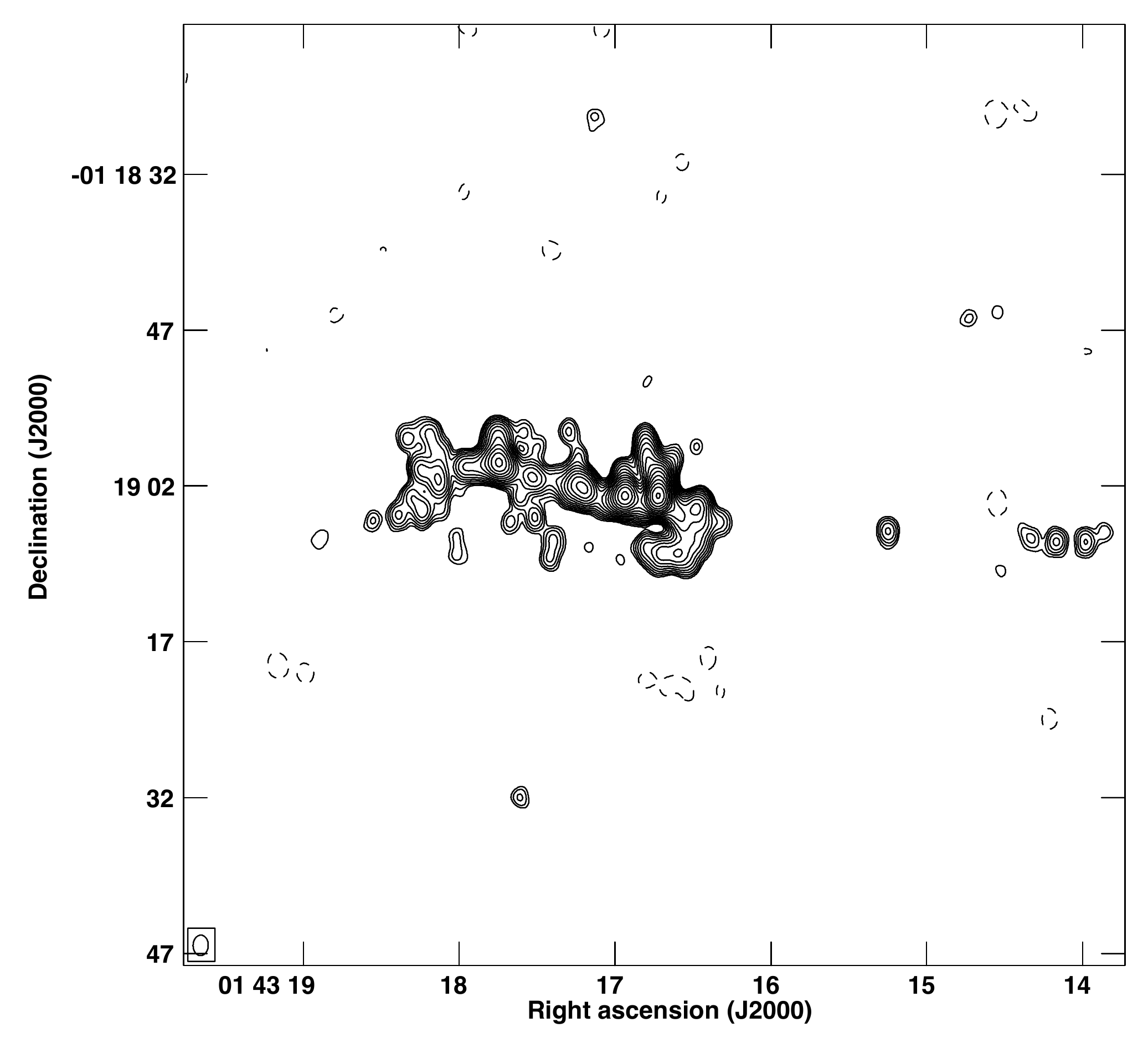}
\includegraphics[width=0.45\columnwidth]{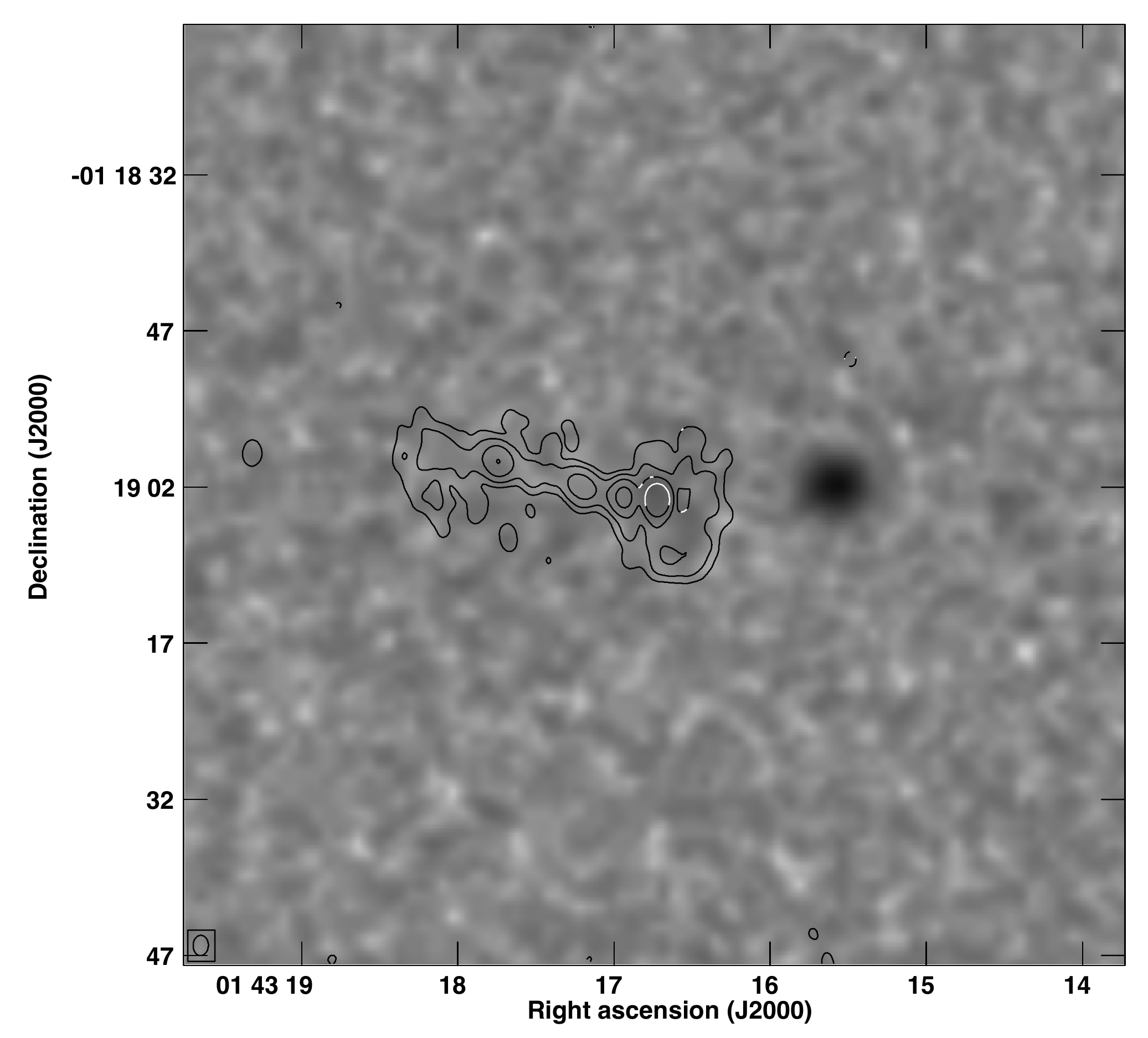}\\
\caption[J0143$-$0119 (L \& C)]{J0143$-$0119.  (top) (left) VLA image at L band and (right) VLA image overlaid on red SDSS image. Lowest contour = 0.2~mJy/beam, peak = 44.4~mJy/beam. (bottom) (left) VLA image at C band and (right) VLA image overlaid on red SDSS image. Lowest contour = 0.1~mJy/beam, peak  = 50.3~mJy/beam. \label{fig:J0143-0119}}
\end{figure}

\noindent J0143$-$0119 (Figure~\ref{fig:J0143-0119}).  There is impressive structure revealed in our map of this source. The extended diffuse region orthogonal to the source axis visible in the FIRST image is completely resolved out. A central bright knotty jet is seen with the peak at extreme west identified with a galaxy, making the source extremely asymmetric. The nature of the source is not clear.

\begin{figure}[ht] 
\includegraphics[width=0.45\columnwidth]{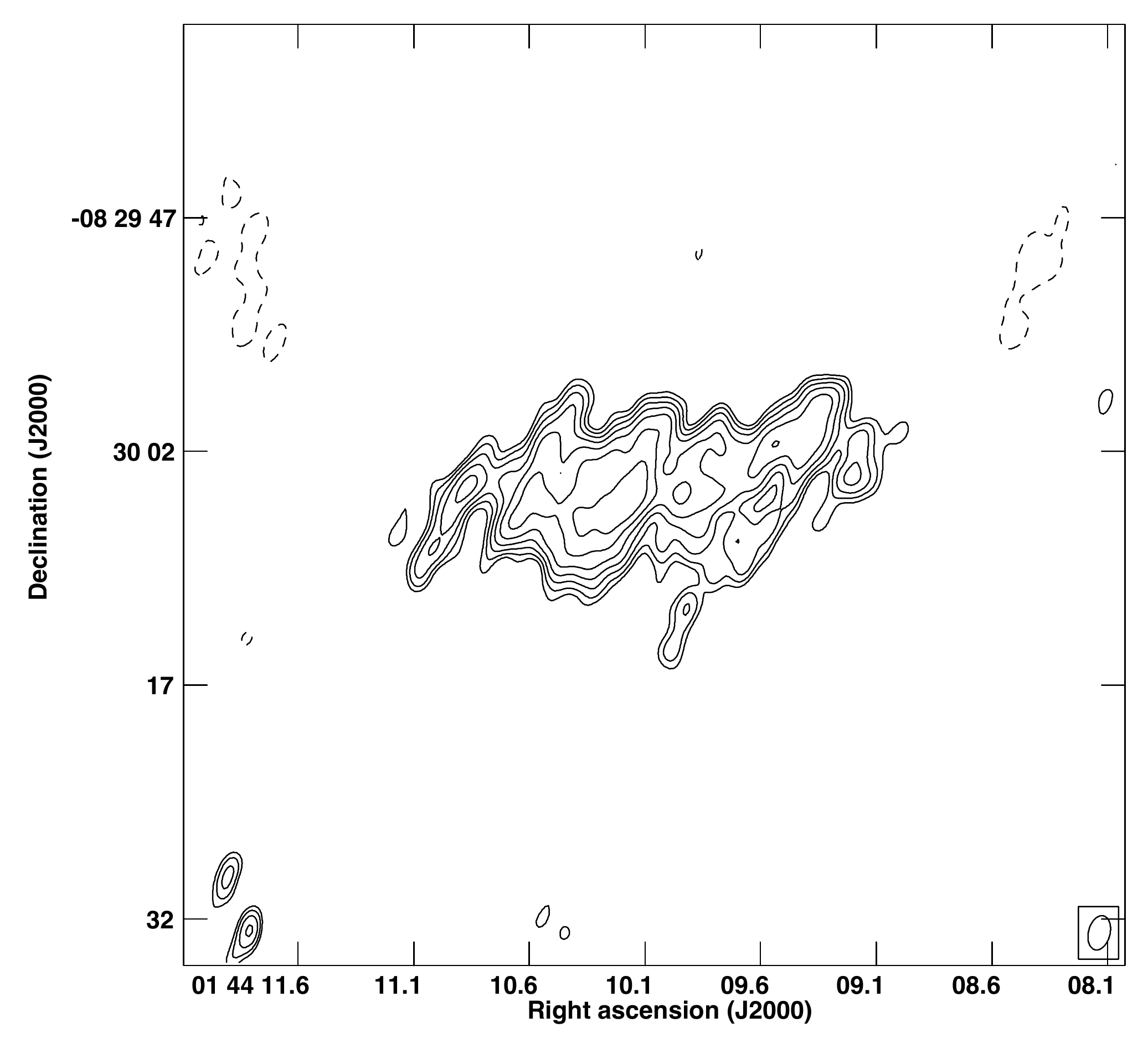}
\includegraphics[width=0.45\columnwidth]{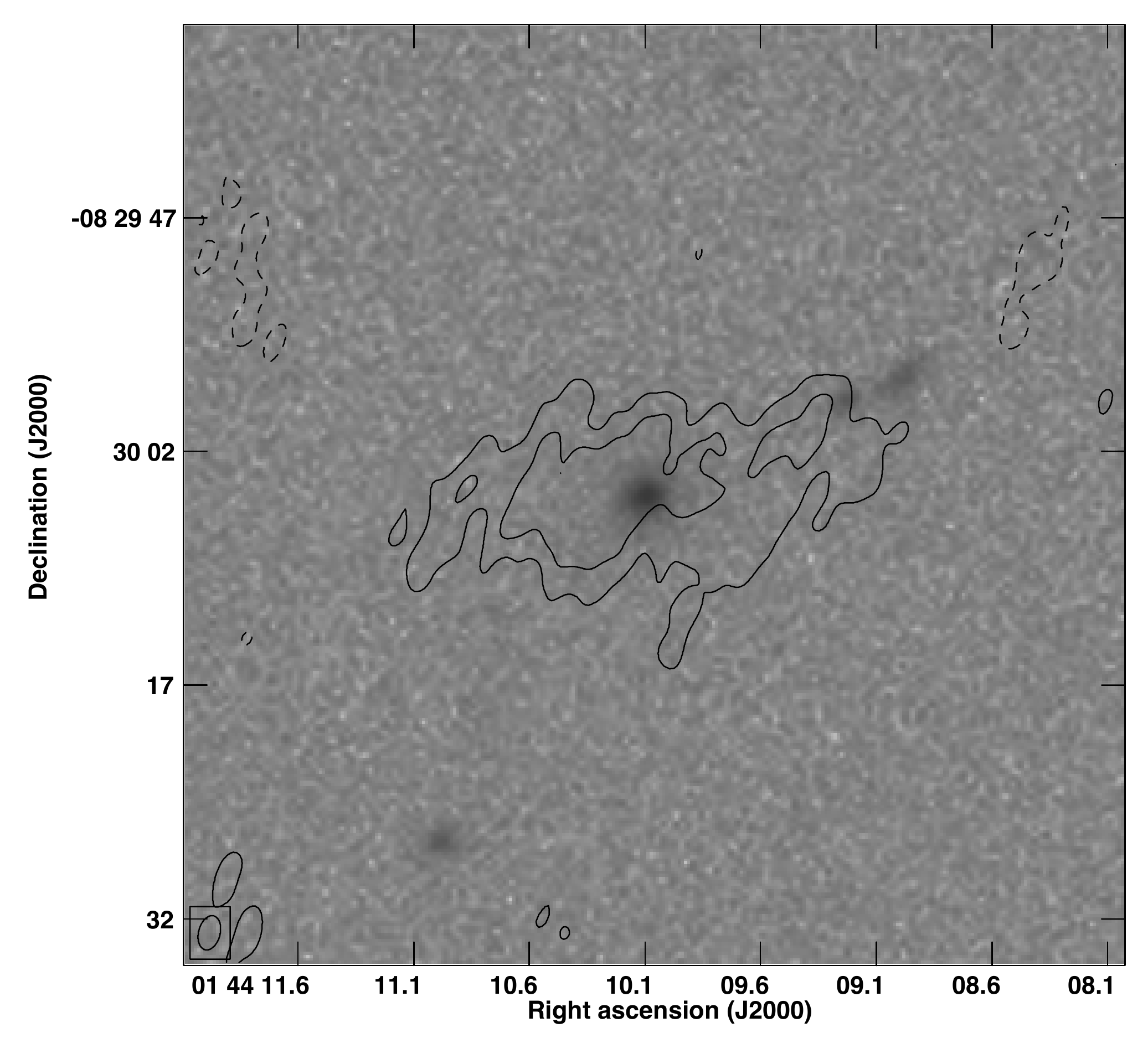}
\caption[J0144$-$0830 (L)]{J0144$-$0830.  (left) VLA image at L band, (right) VLA image overlaid on red SDSS image. Lowest contour  = 0.09~mJy/beam, peak  =1.40~mJy/beam. \label{fig:J0144-0830}}
\end{figure}

\noindent J0144$-$0830 (Figure~\ref{fig:J0144-0830}).  Our map does not reveal any compact features in this source, which is seen as a centrally bright X-shaped source in the FIRST image.

\begin{figure}[ht] 
\includegraphics[width=0.45\columnwidth]{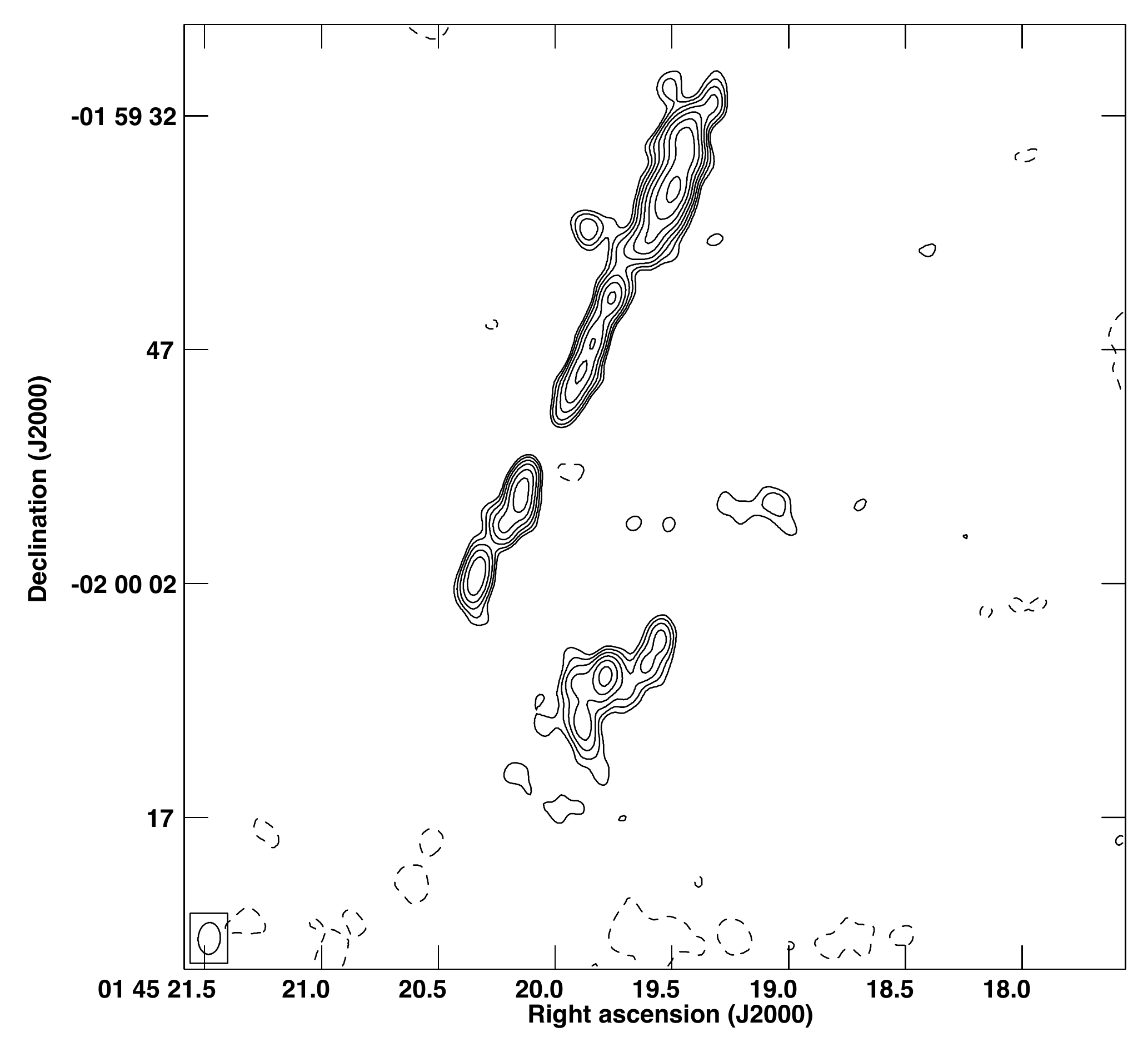}
\includegraphics[width=0.45\columnwidth]{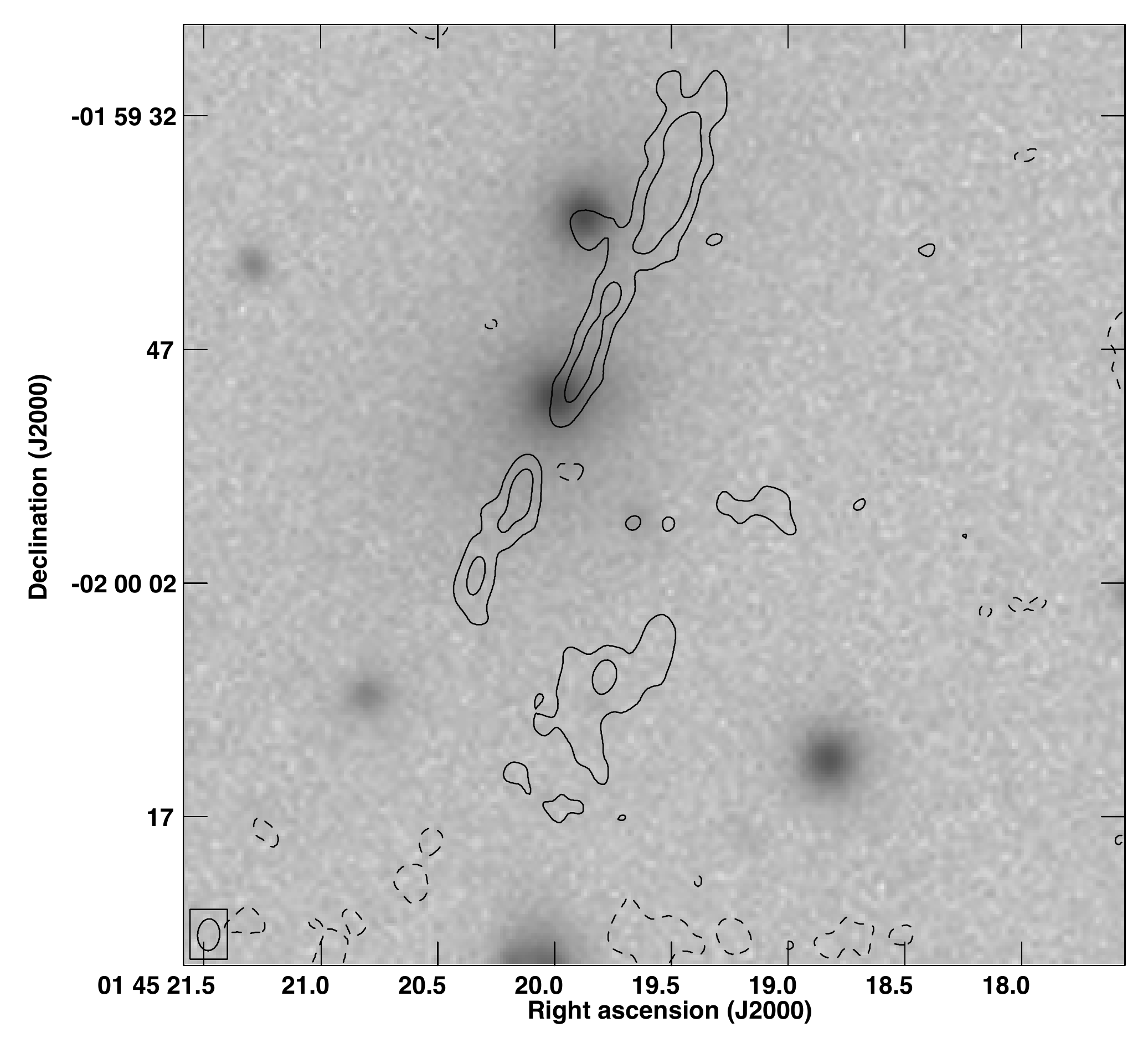}
\caption[J0145$-$0159 (L)]{J0145$-$0159. (left) VLA image at L band, (right) VLA image overlaid on red SDSS image. Lowest contour = 0.2~mJy/beam, peak  = 2.56~mJy/beam. \label{fig:J0145-0159}}
\end{figure}

\noindent J0145$-$0159 (Figure~\ref{fig:J0145-0159}). Although the FIRST map shows an edge-brightened radio galaxy our map shows only a central narrow twin-jet feature with all the extended emission seen in the FIRST map resolved out. The bright galaxy ID is located at the base of the northern jet. The map reveals a feature offset from the southern jet corresponding to the bright extended emission at the end of the southern lobe seen in the FIRST map. Our map reveals no hotspots in this edge-brightened source.

\begin{figure}[ht] 
\includegraphics[width=0.45\columnwidth]{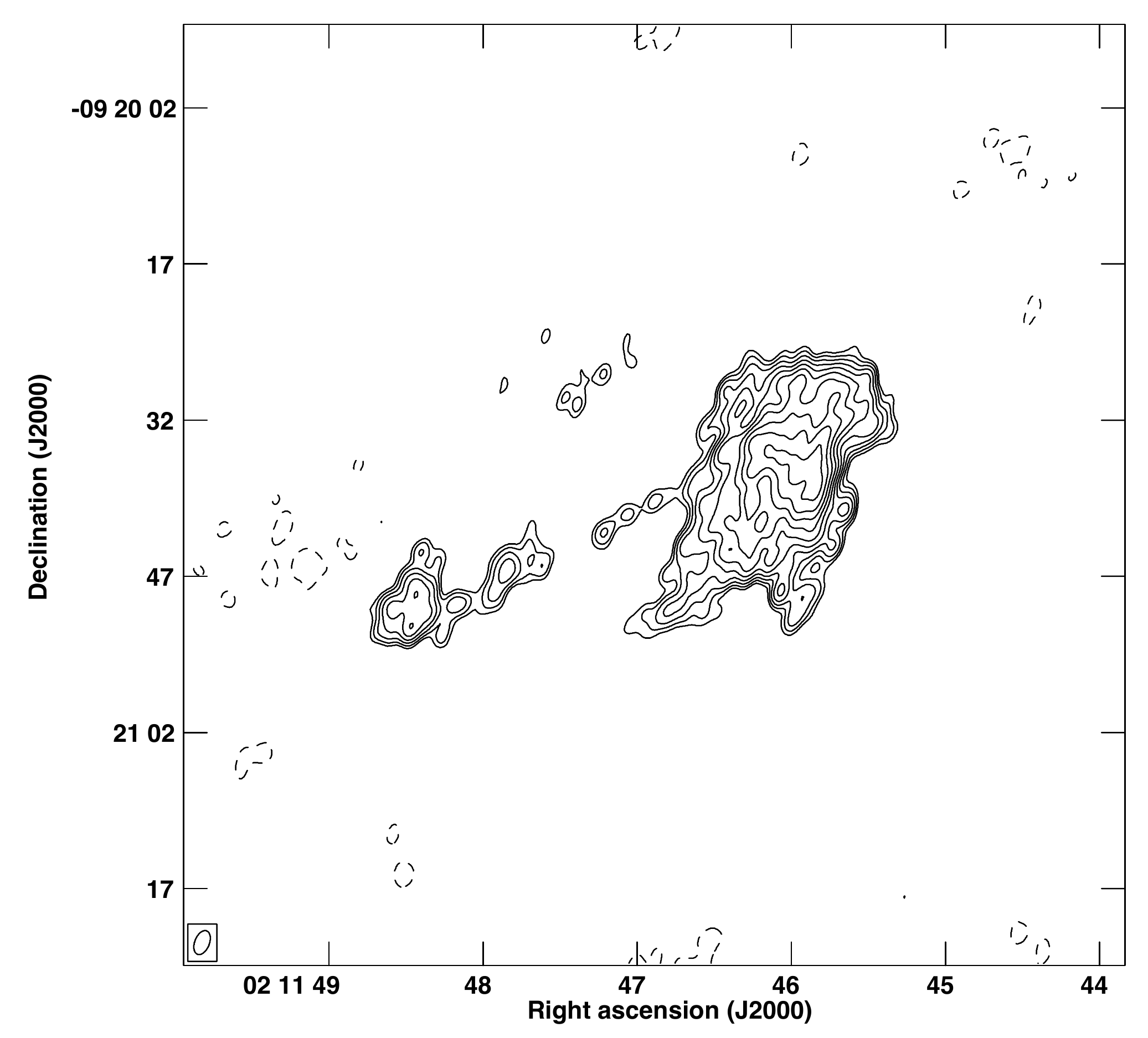}
\includegraphics[width=0.45\columnwidth]{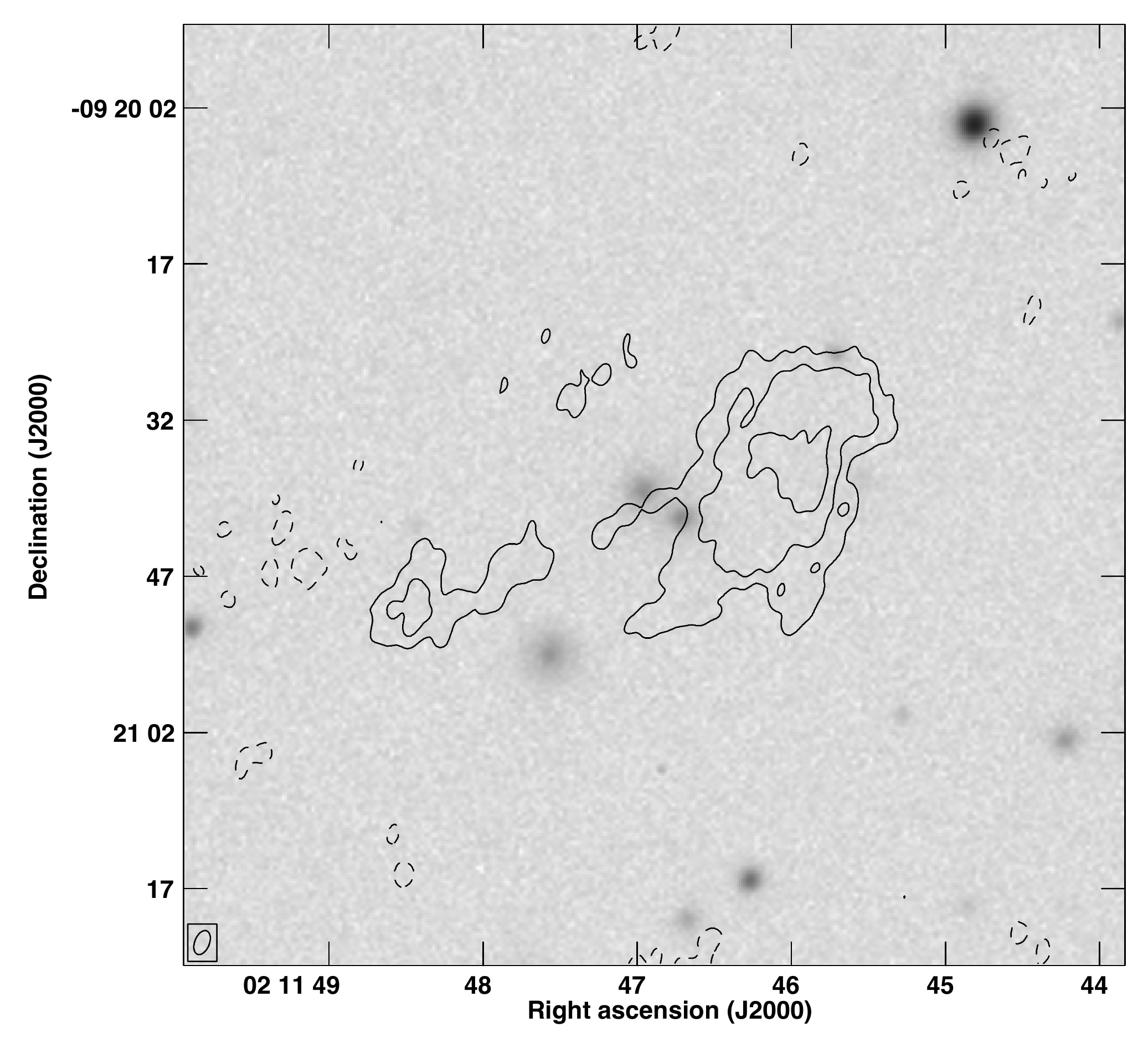}
\caption[J0211$-$0920 (L)]{J0211$-$0920.  (left) VLA image at L band, (right) VLA image overlaid on red SDSS image. Lowest contour = 0.1~mJy/beam, peak  = 2.81~mJy/beam. \label{fig:J0211-0920}}
\end{figure}

\noindent J0211$-$0920 (Figure~\ref{fig:J0211-0920}). The map reveals the NE galaxy constituting the galaxy pair to be the likely host. Two narrow jets are seen leading to the lobes where transverse emission is seen only for the NW lobe. 

\begin{figure}[ht] 
\includegraphics[width=0.45\columnwidth]{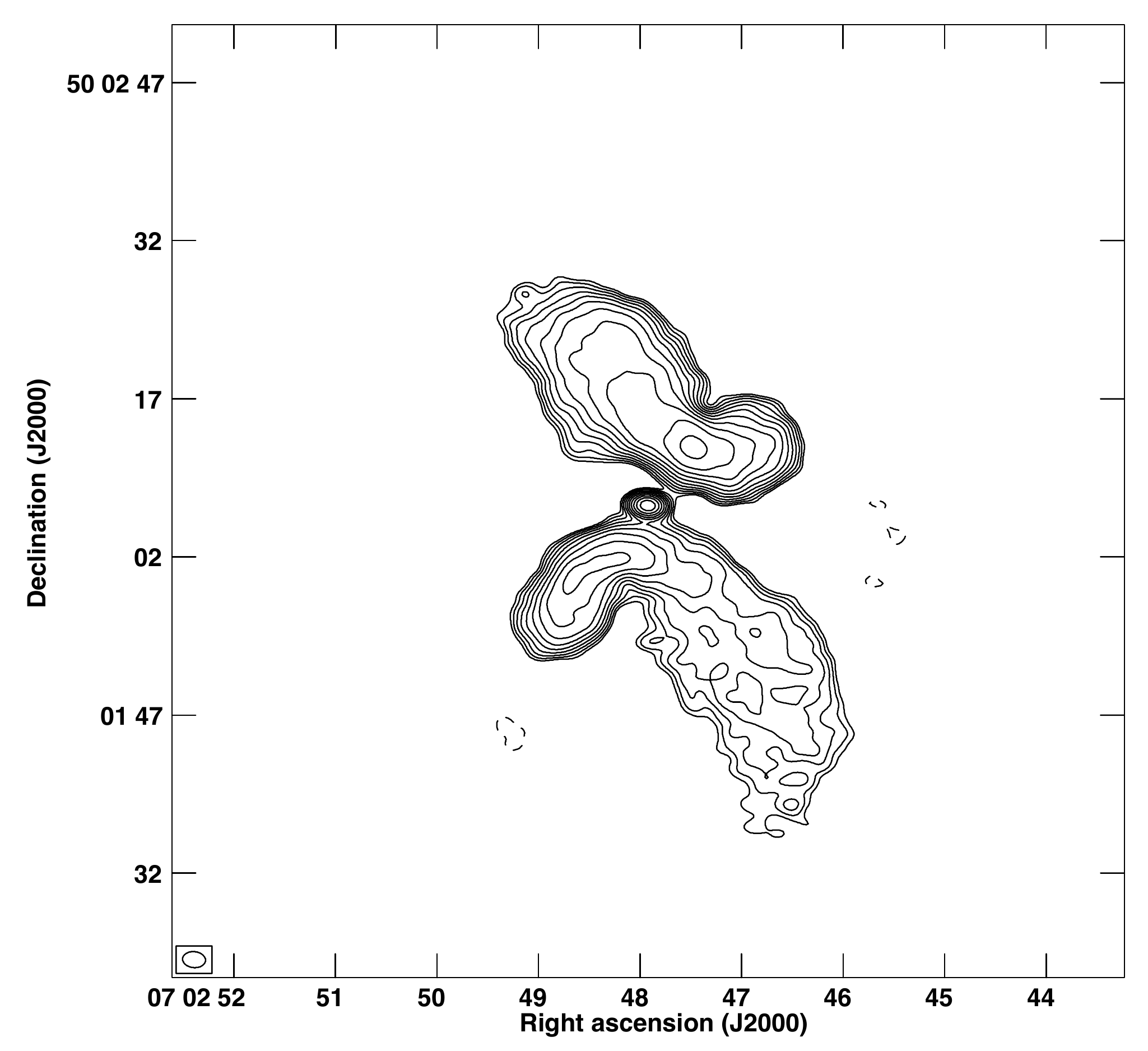}
\includegraphics[width=0.45\columnwidth]{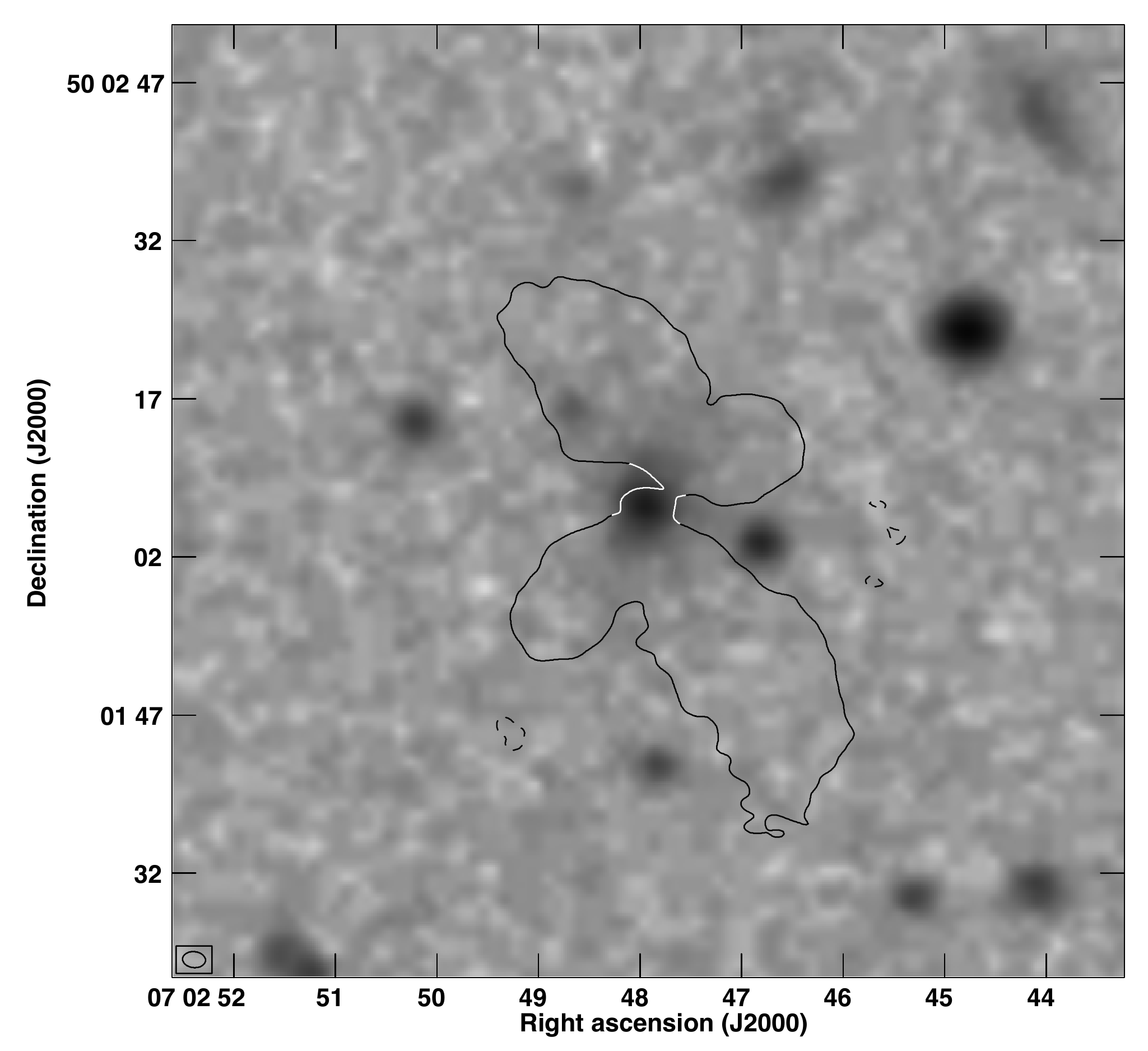}
\caption[J0702+5002 (L)]{J0702+5002. (left) VLA image at L band, (right) VLA image overlaid on red DSS\,II image. Lowest contour = 0.2~mJy/beam, peak = 7.17~mJy/beam.  \label{fig:J0702+5002}}
\end{figure}

\noindent J0702+5002 (Figure~\ref{fig:J0702+5002}).  The new map has revealed two distinct lobes on either side of a previously unseen central core. The lobes are at least as long as the source extent. Each of the lobes extends orthogonally in opposite directions. Neither of the lobes is seen to have compact hotspots.

\begin{figure}[ht] 
\includegraphics[width=0.45\columnwidth]{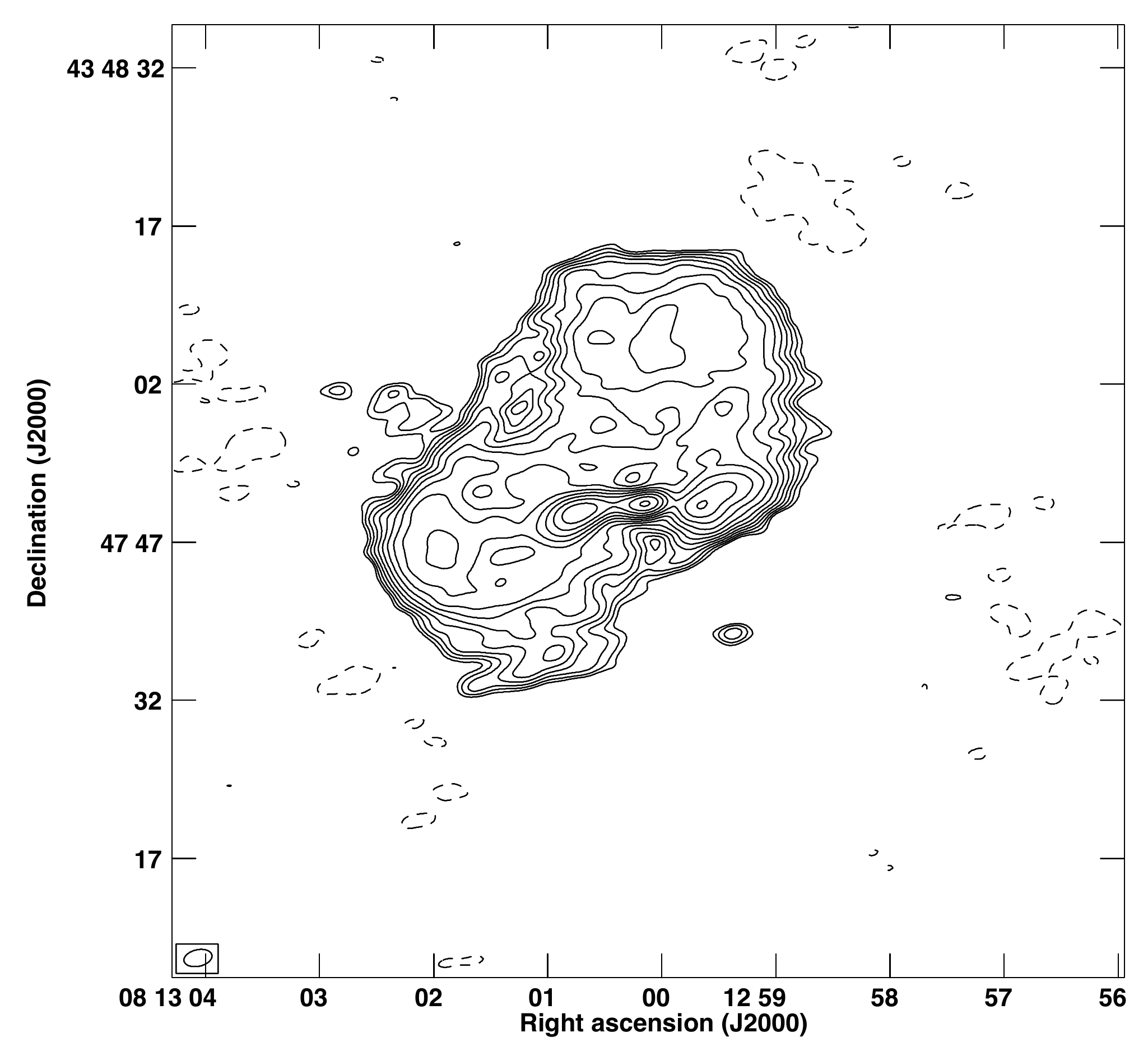}
\includegraphics[width=0.45\columnwidth]{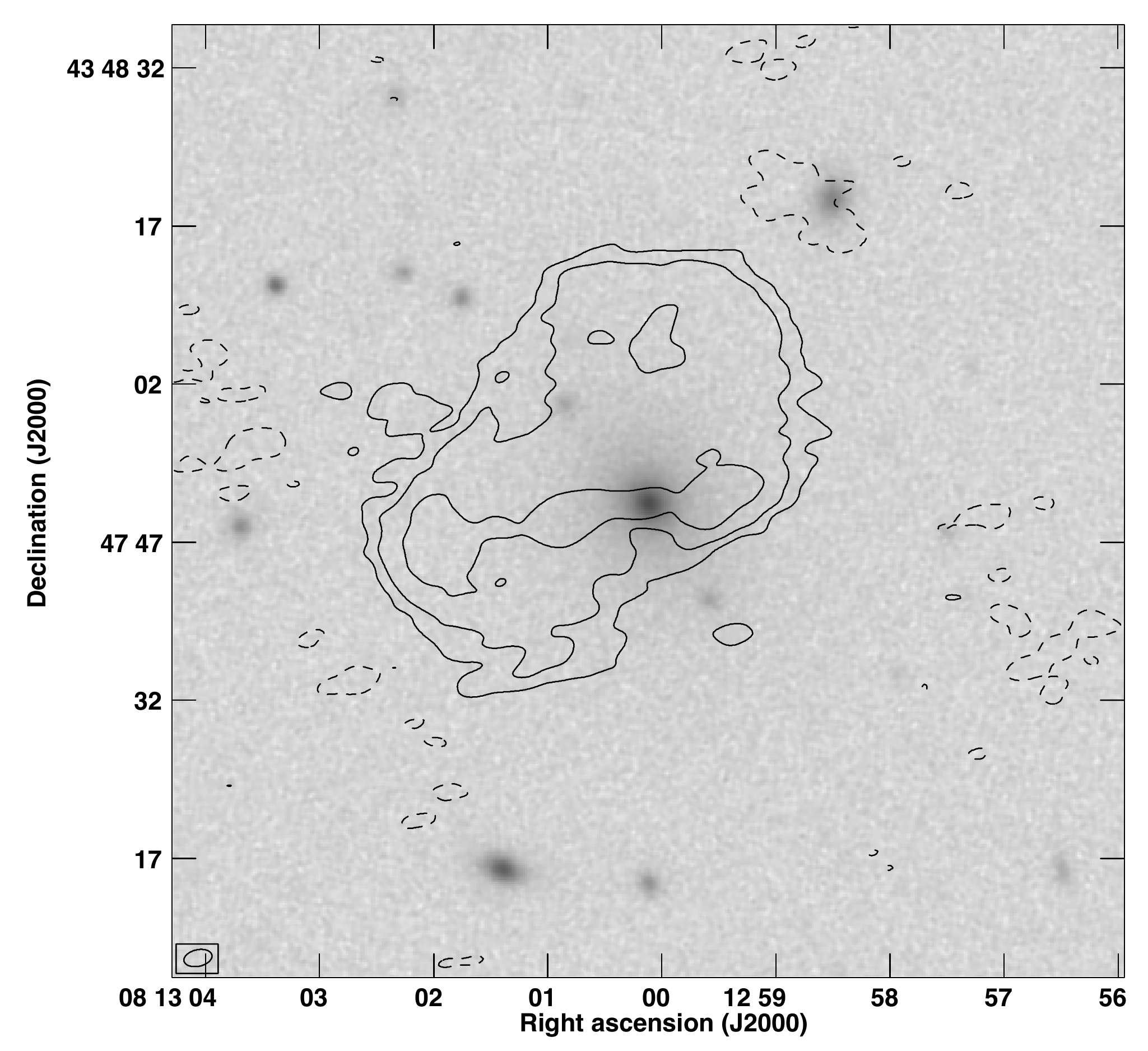}
\caption[J0813+4347 (L)]{J0813+4347. (left) VLA image at L band, (right) VLA image overlaid on red SDSS image. Lowest contour = 0.125~mJy/beam, peak = 12.0~mJy/beam.  \label{fig:J0813+4347}}
\end{figure}

\noindent J0813+4347 (Figure~\ref{fig:J0813+4347}).  Our higher resolution map reveals a central triple structure (two lobes straddling a core) which is itself embedded in a much larger diffuse emission region. 

\begin{figure}[ht] 
\includegraphics[width=0.45\columnwidth]{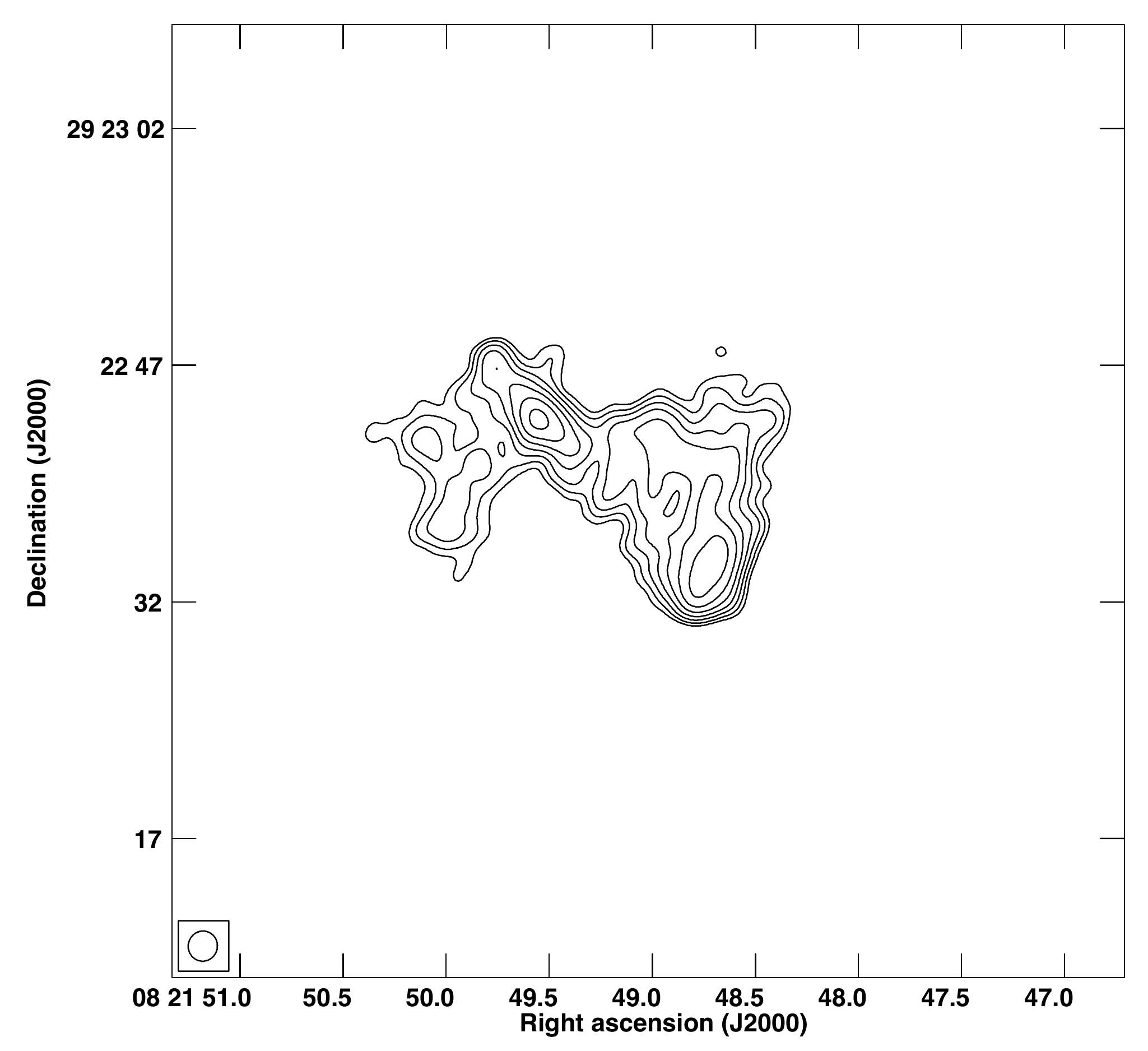}
\includegraphics[width=0.45\columnwidth]{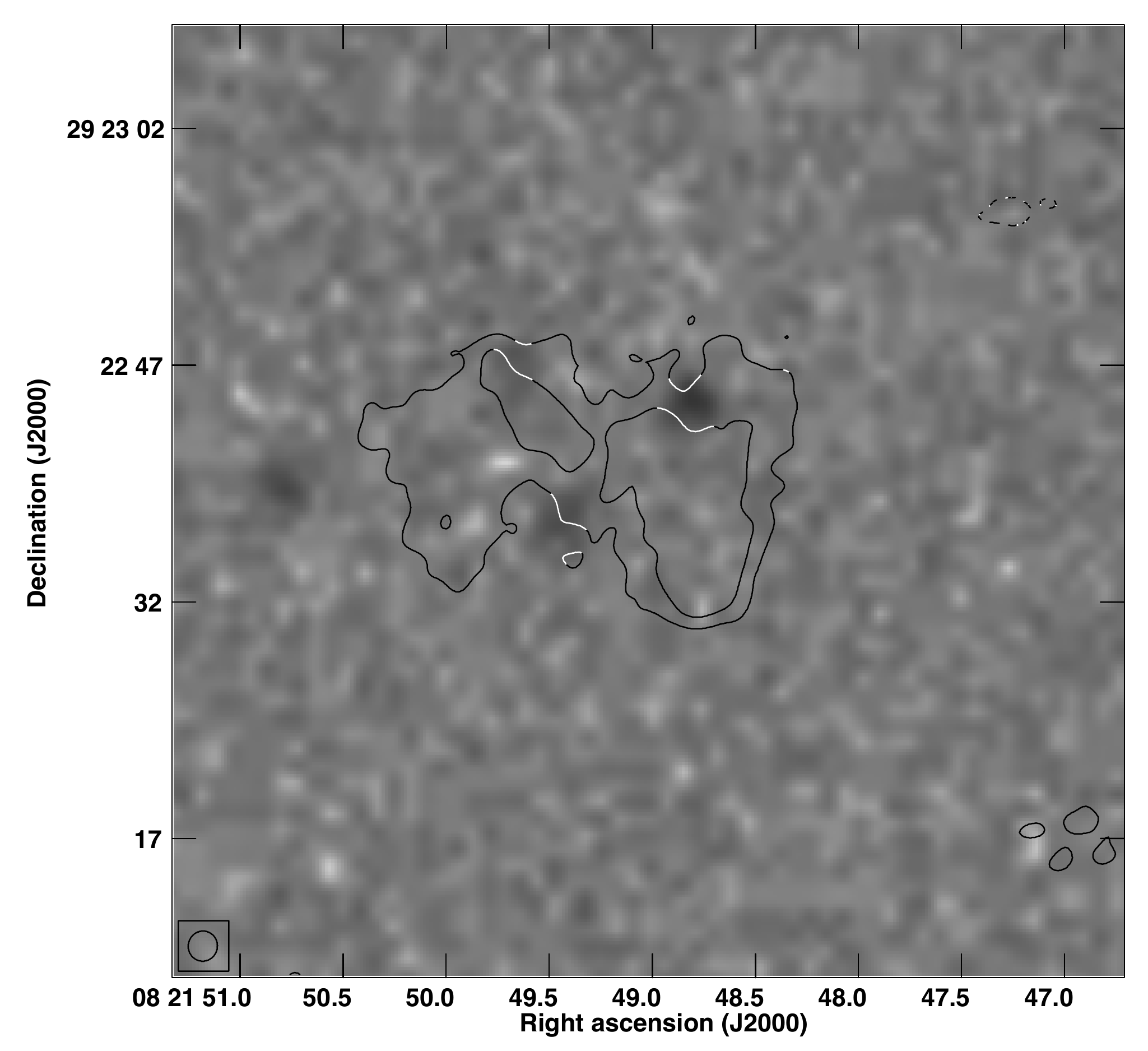}
\caption[J0821+2922 (L)]{J0821+2922.  (left) VLA image at L band, (right) VLA image overlaid on red SDSS image. Lowest contour = 0.2~mJy/beam, peak  = 2.89~mJy/beam.  \label{fig:J0821+2922}}
\end{figure}

\noindent J0821+2922 (Figure~\ref{fig:J0821+2922}). The map shows a compact core at the location of a faint galaxy straddled by two hotspots. The more compact NE hotspot is connected to it by a jet. The SW hotspot is offset to the west from the core-jet axis. The hotspots are accompanied by regions of diffuse emission that extend in opposite directions.

\begin{figure}[ht] 
\includegraphics[width=0.45\columnwidth]{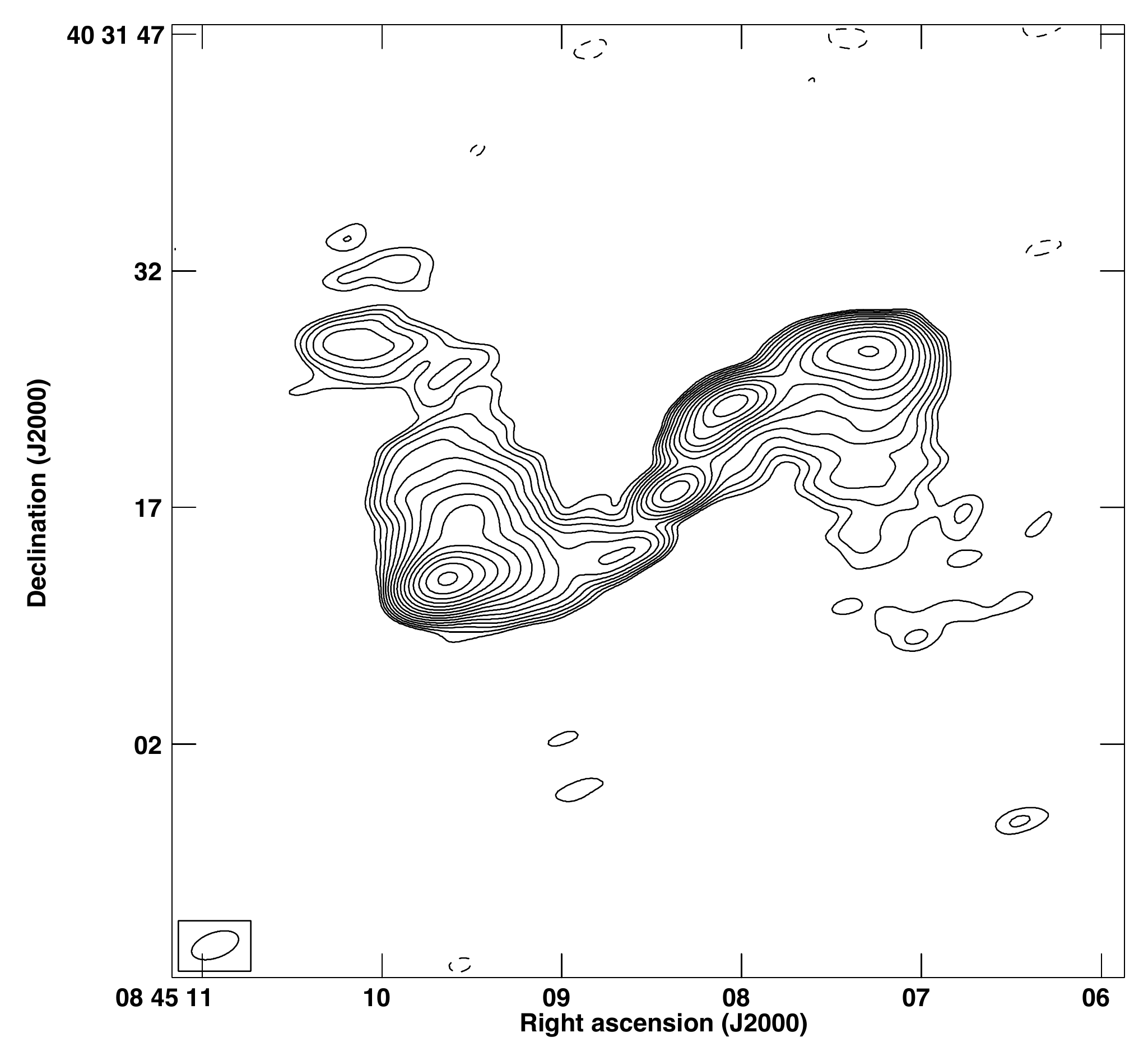}
\includegraphics[width=0.45\columnwidth]{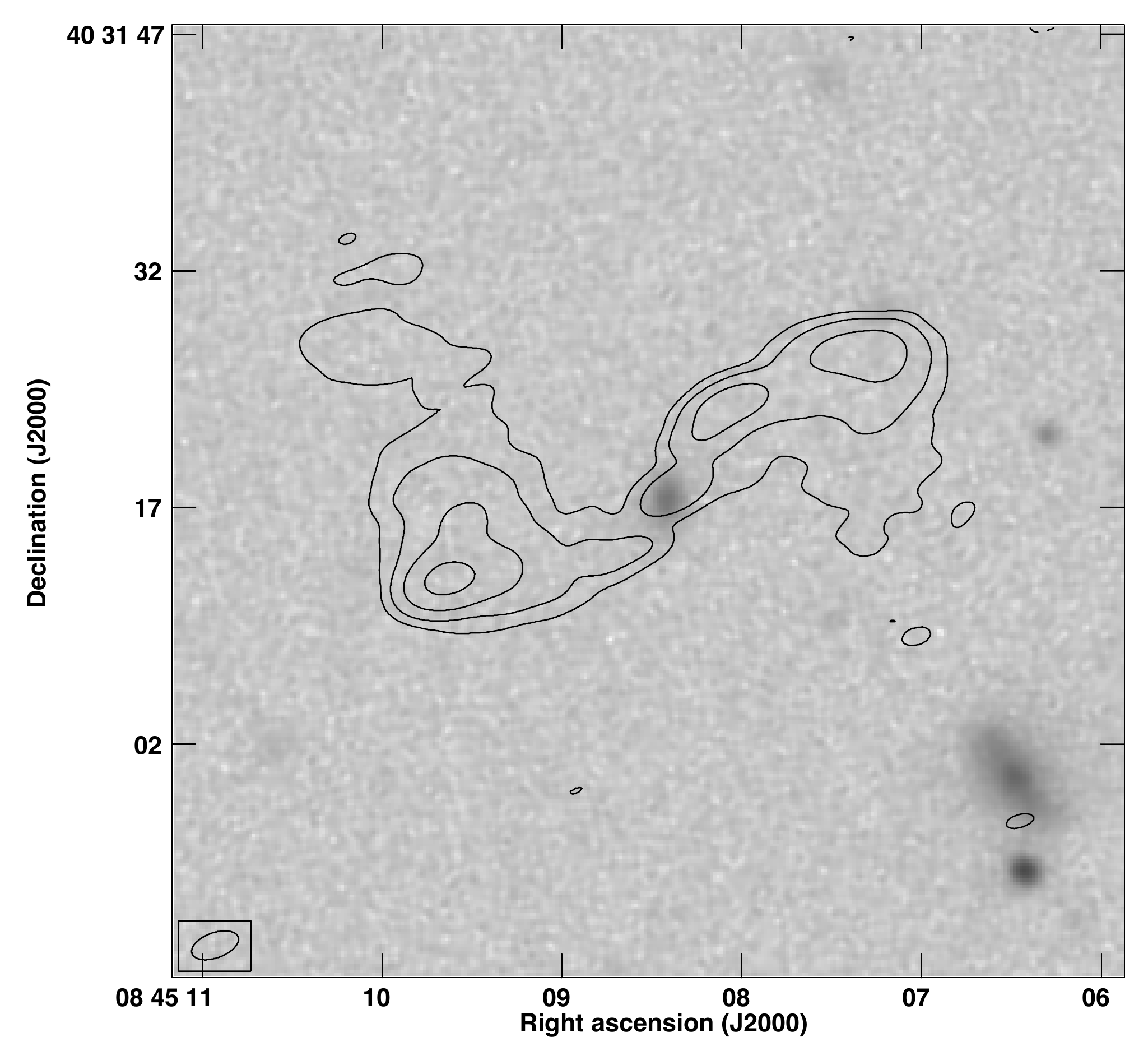}
\caption[J0845+4031 (L)]{J0845+4031. (left) VLA image at L band, (right) VLA image overlaid on red SDSS image. Lowest contour = 0.15~mJy/beam, peak  = 20.8~mJy/beam. \label{fig:J0845+4031}}
\end{figure}

\noindent J0845+4031 (Figure~\ref{fig:J0845+4031}). The map reveals a very inversion-symmetric structure. There is an inner pair of emission peaks along an axis after which the respective outer (edge-brightened) lobes bend in opposite directions. Both outer lobes are associated with trailing faint emission extended again in opposite directions.  We regard this source as a prime example of an AGN with rotating jets.

\begin{figure}[ht] 
\includegraphics[width=0.45\columnwidth]{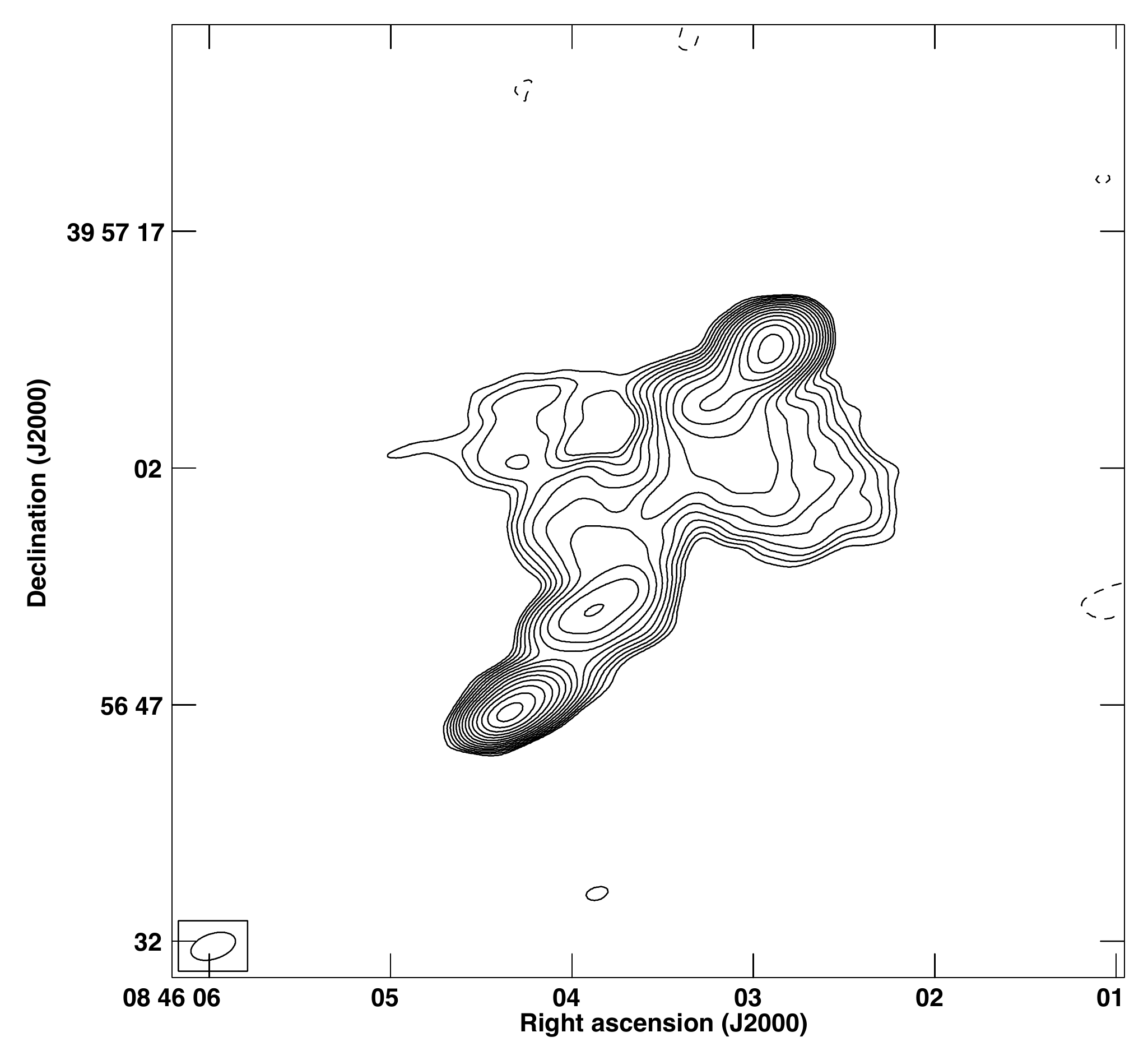}
\includegraphics[width=0.45\columnwidth]{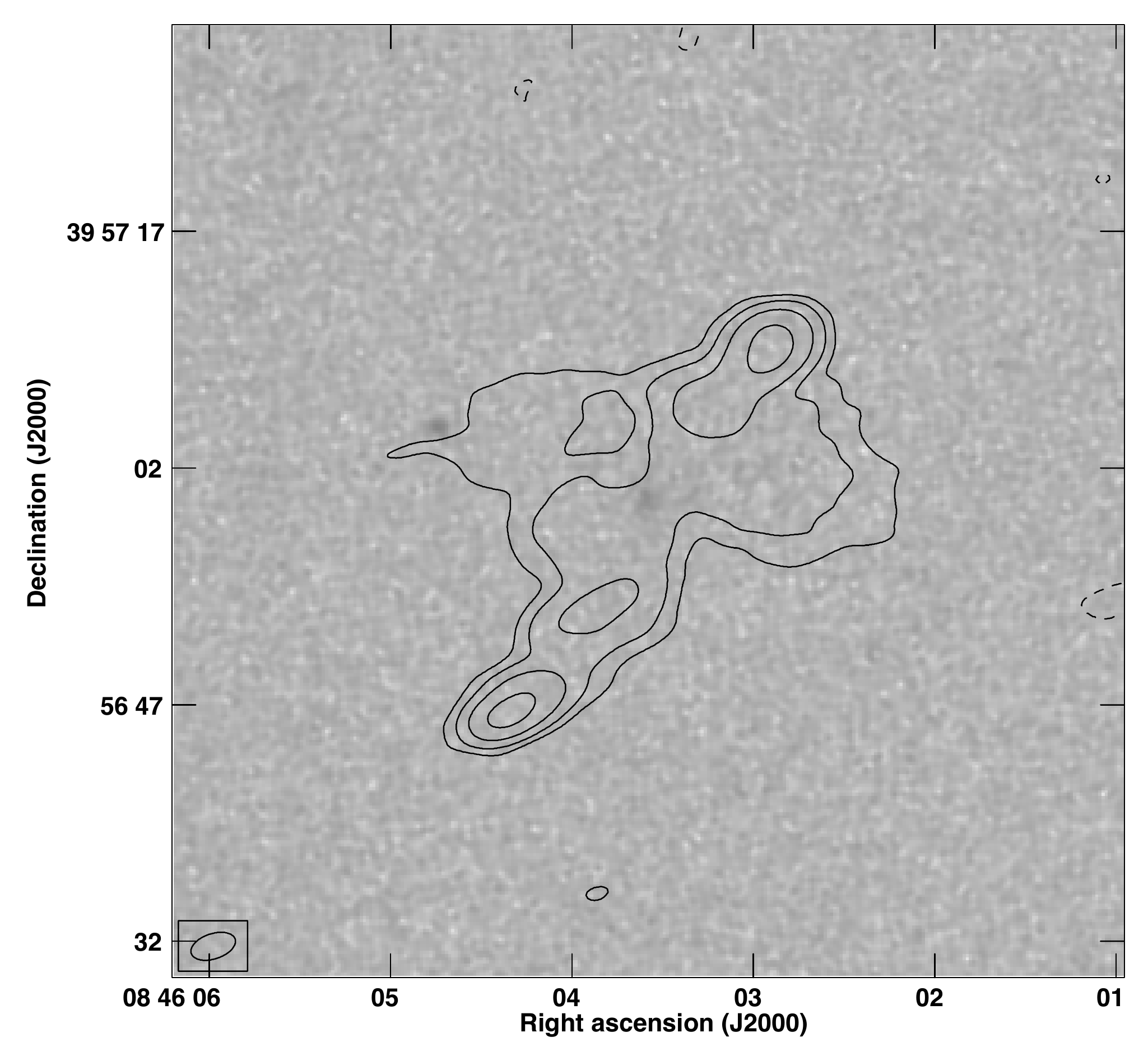}
\caption[J0846+3956 (L)]{J0846+3956. (left) VLA image at L band, (right) VLA image overlaid on red SDSS image.  Lowest contour  = 0.2~mJy/beam, peak  = 21.4~mJy/beam.  \label{fig:J0846+3956}}
\end{figure}

 \noindent J0846+3956 (Figure~\ref{fig:J0846+3956}). The new map hints at a core at the location of a faint galaxy at the center. The transverse  extensions are clearly seen to be associated with the individual lobes. 

\begin{figure}[ht] 
\includegraphics[width=0.45\columnwidth]{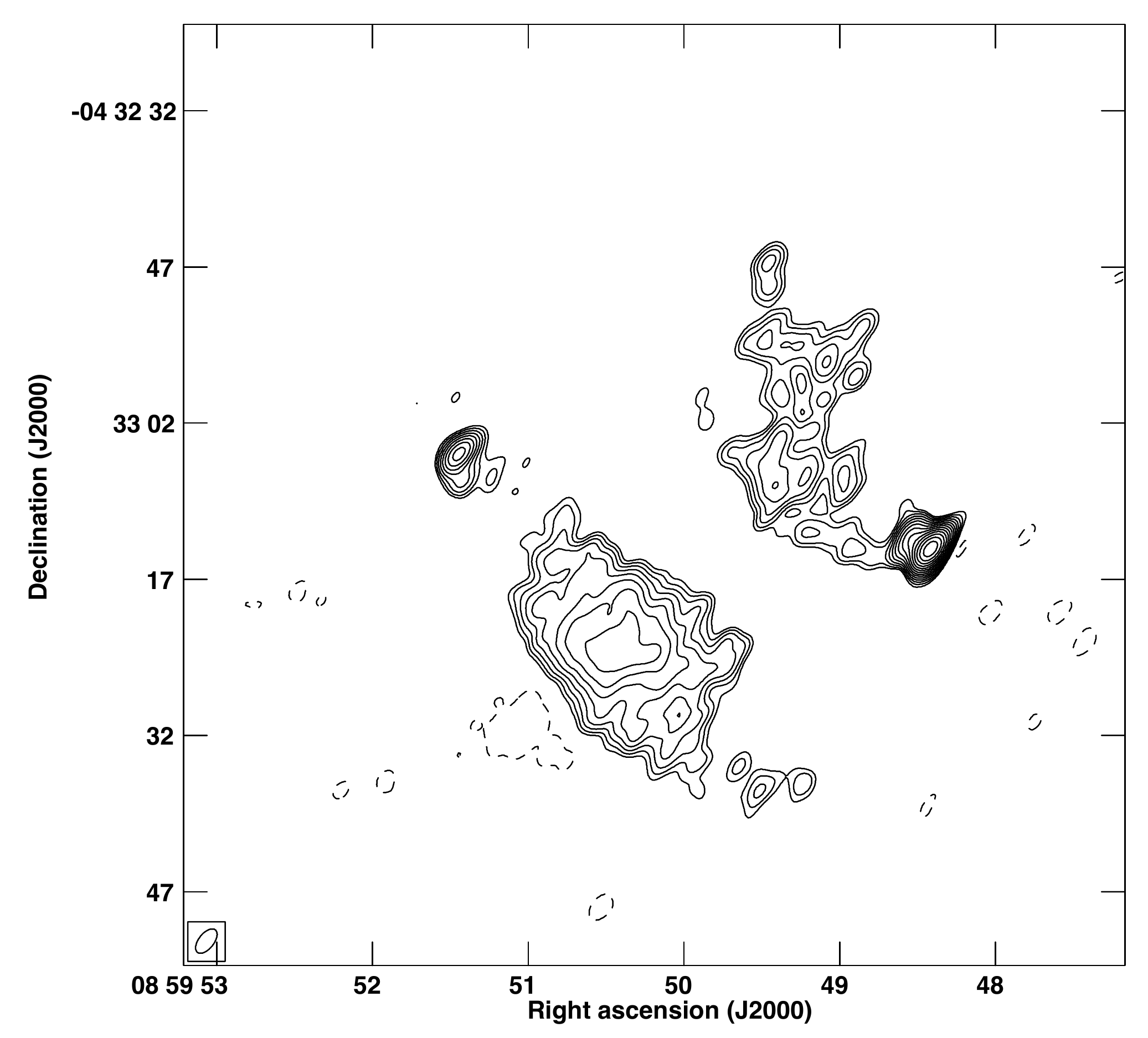}
\includegraphics[width=0.45\columnwidth]{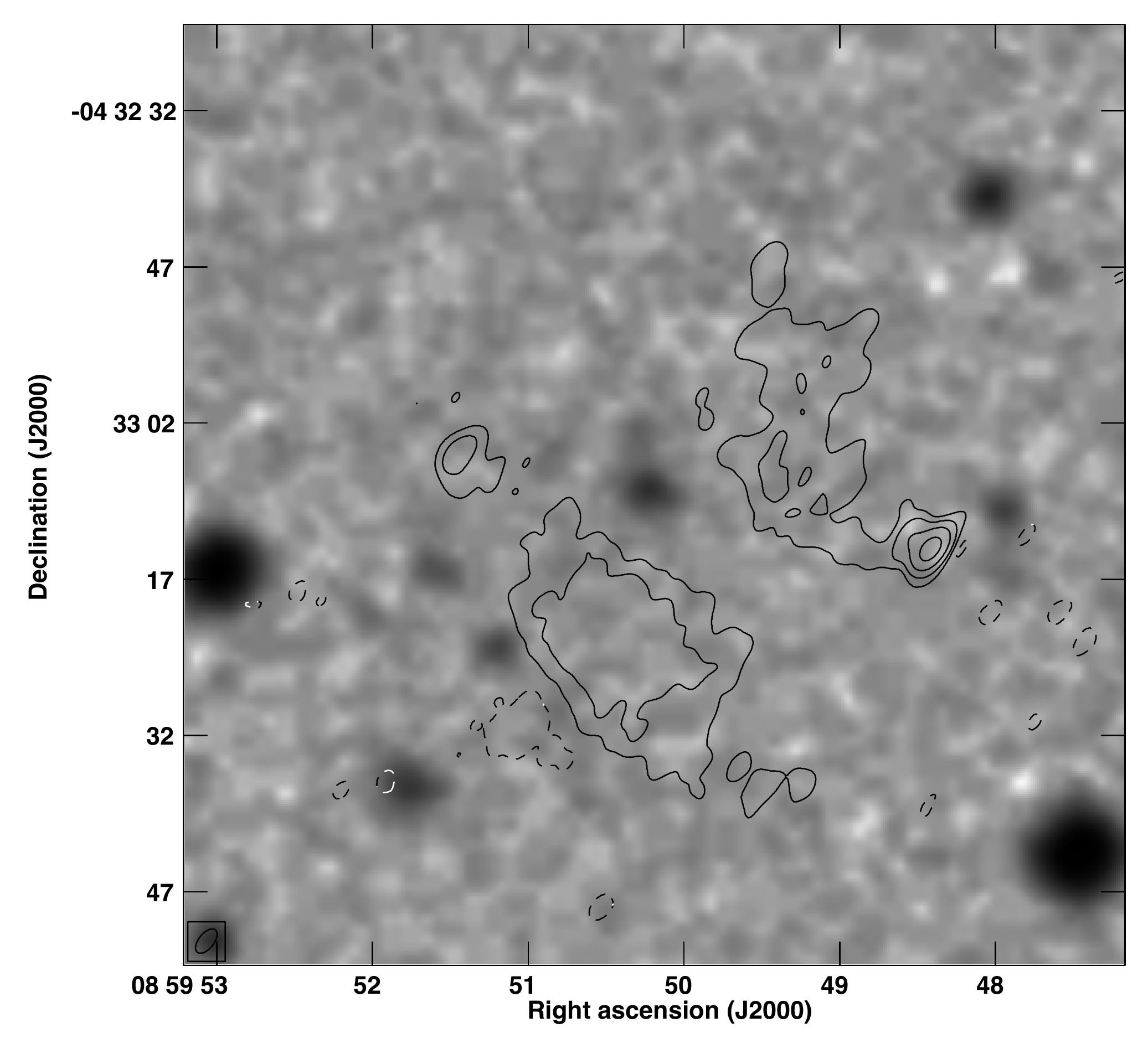}
\caption[J0859$-$0433 (L)]{J0859$-$0433.  (left) VLA image at L band, (right) VLA image overlaid on red DSS\,II image. Lowest contour = 0.2~mJy/beam, peak =  23.4~mJy/beam. \label{fig:J0859-0433}}
\end{figure}

\noindent J0859$-$0433 (Figure~\ref{fig:J0859-0433}). Our higher resolution map fails to detect a core in this source. The two hotspots are well mapped and the transverse emission to the north is clearly seen to connect to the W hotspot. The diffuse emission feature to the south is seen to have a bounded appearance with clear edges.

\begin{figure}[ht] 
\includegraphics[width=0.45\columnwidth]{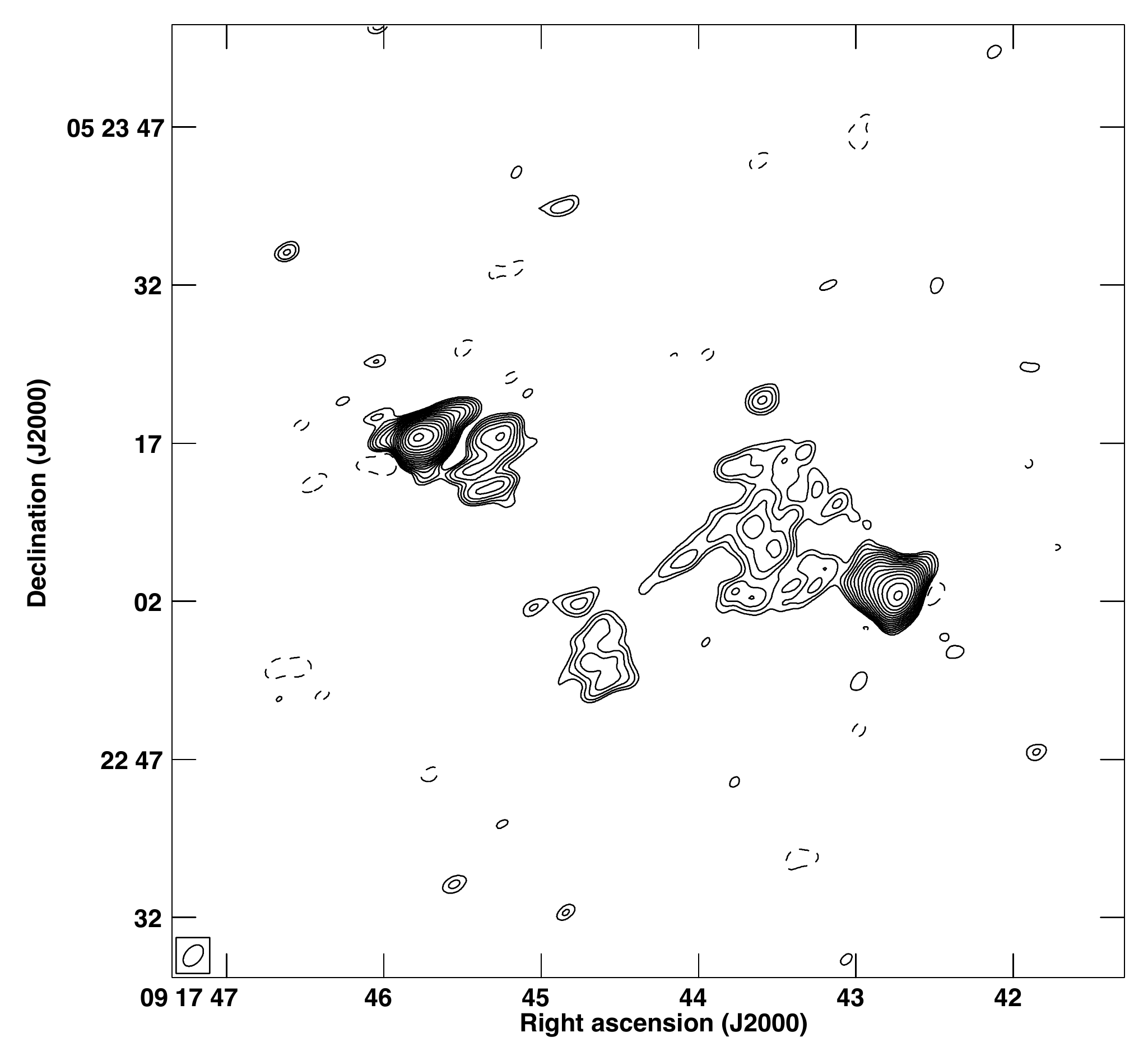}
\includegraphics[width=0.45\columnwidth]{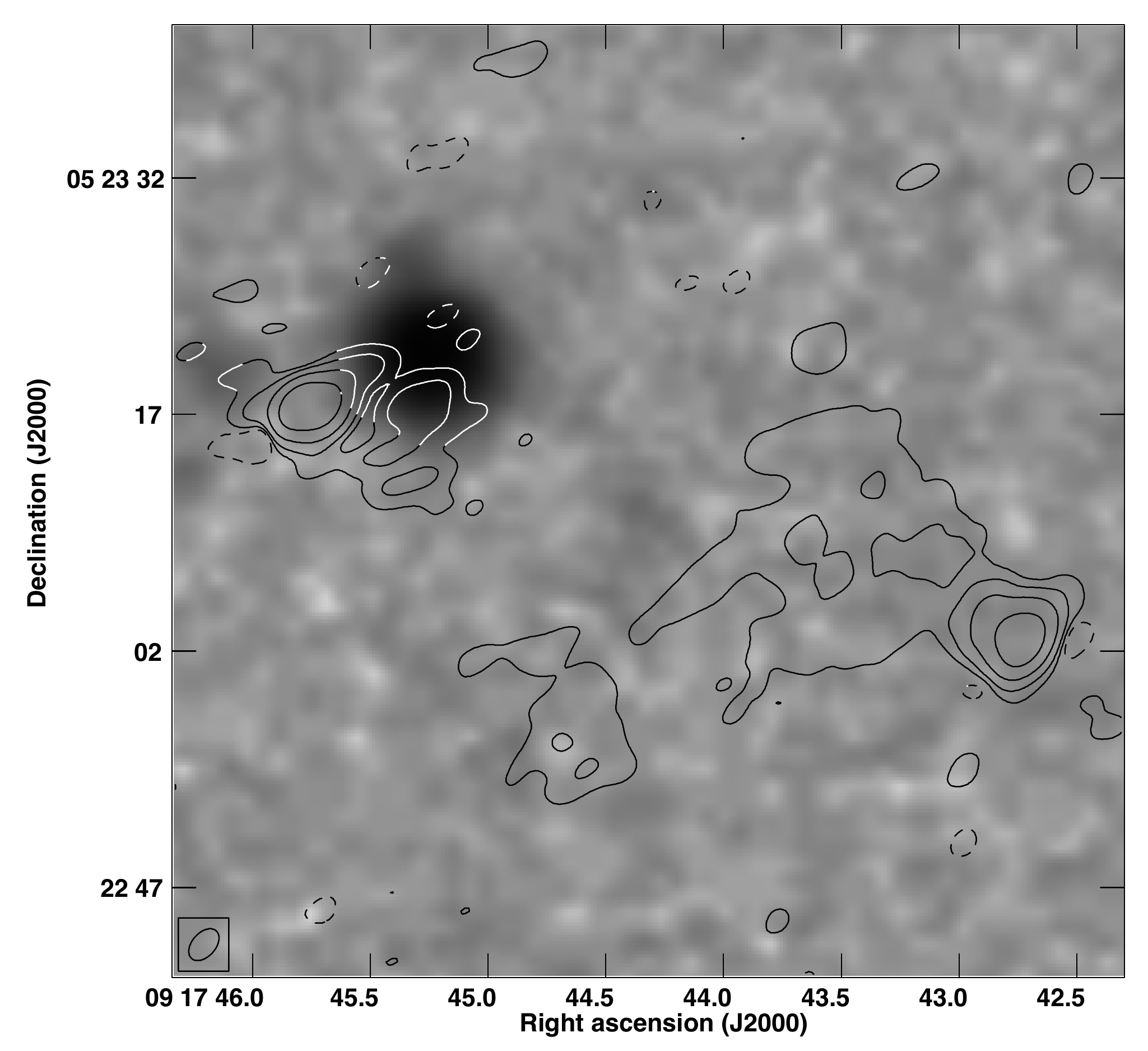}
\caption[J0917+0523 (L)]{J0917+0523. (left) VLA image at L band, (right) VLA image overlaid on red SDSS image. Lowest contour =  0.6~mJy/beam, peak = 117~mJy/beam. \label{fig:J0917+0523}}
\end{figure}

\noindent J0917+0523 (Figure~\ref{fig:J0917+0523}). This source is very similar to J0859-0433. No core is detected in our map. The two hotspots are well mapped although the extended transverse emission is only partially imaged.

\begin{figure}[ht] 
\includegraphics[width=0.45\columnwidth]{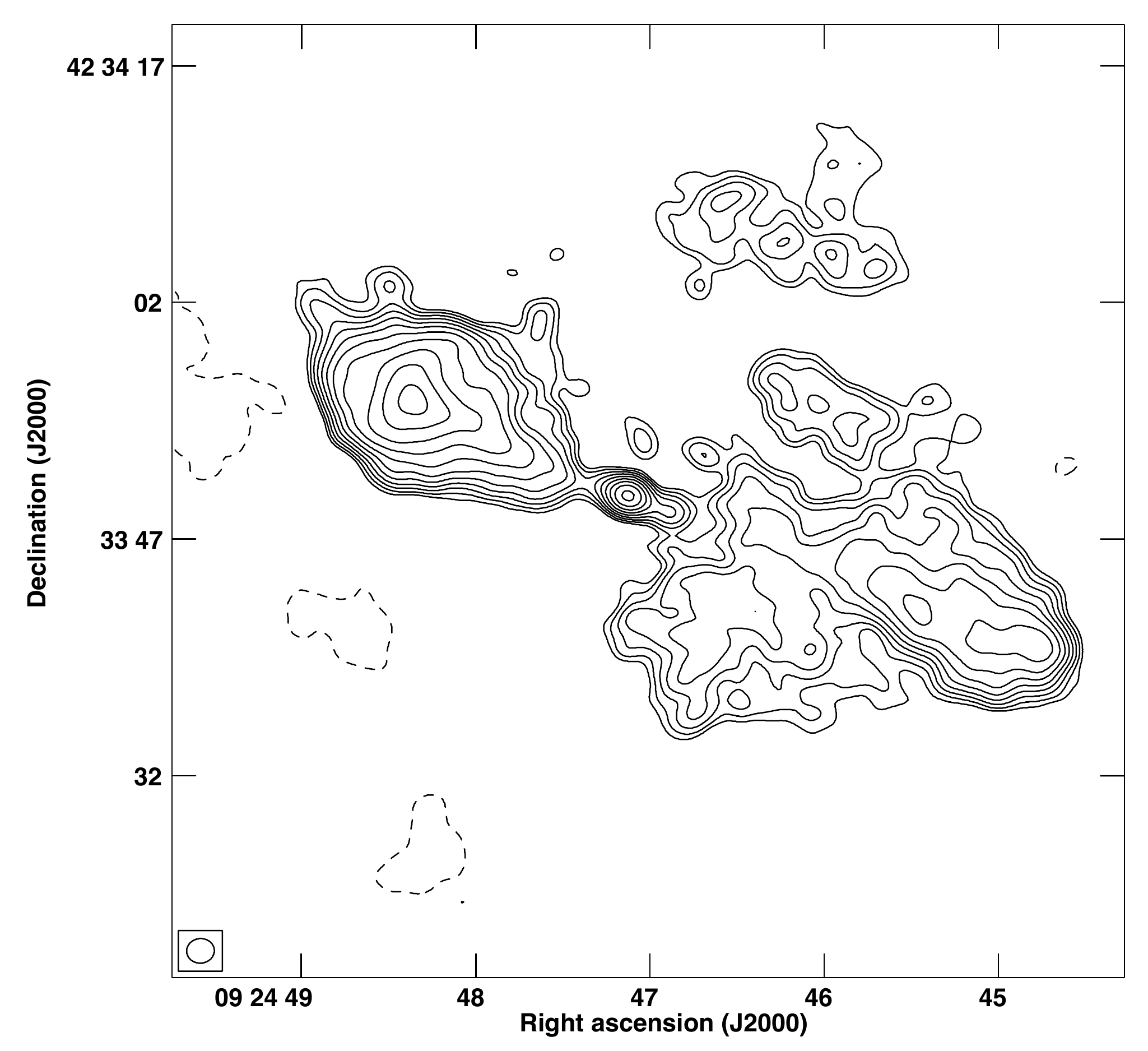}
\includegraphics[width=0.45\columnwidth]{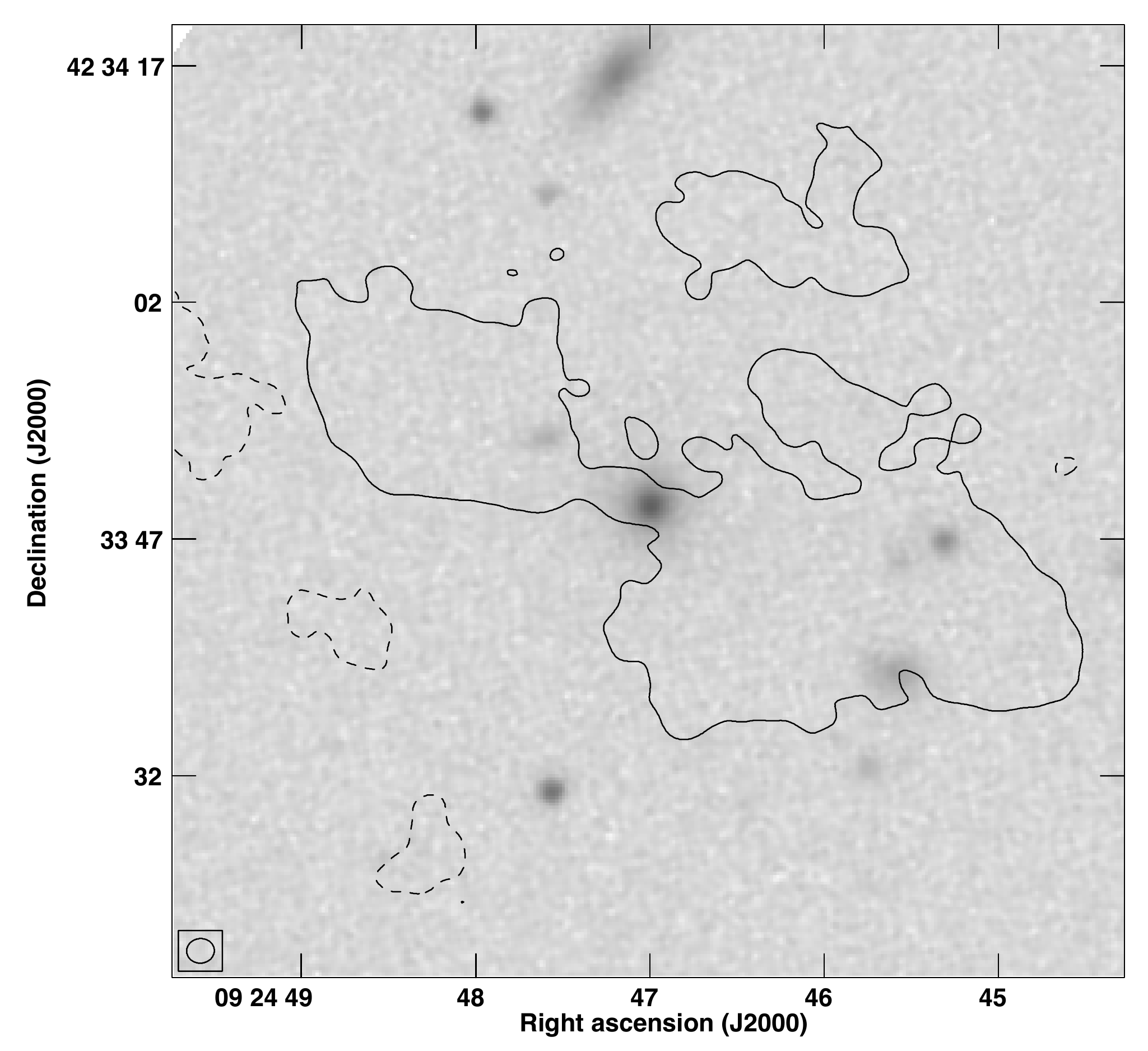}
\caption[J0924+4233 (L)]{J0924+4233. (left) VLA image at L band, (right) VLA image overlaid on red SDSS image. Lowest contour = 0.15~mJy/beam, peak  = 5.40~mJy/beam. \label{fig:J0924+4233}}
\end{figure}

\noindent J0924+4233 (Figure~\ref{fig:J0924+4233}). No distinct core is seen in the map. The bright ID is straddled to the east by a core-like feature and to the west by a short terminated jet. The structure suggests a restarted AGN activity. The diffuse lobe emission to the south is traced clearly and seen to be linked with the W lobe. The northern diffuse extension is mostly resolved out.

\begin{figure}[ht] 
\includegraphics[width=0.45\columnwidth]{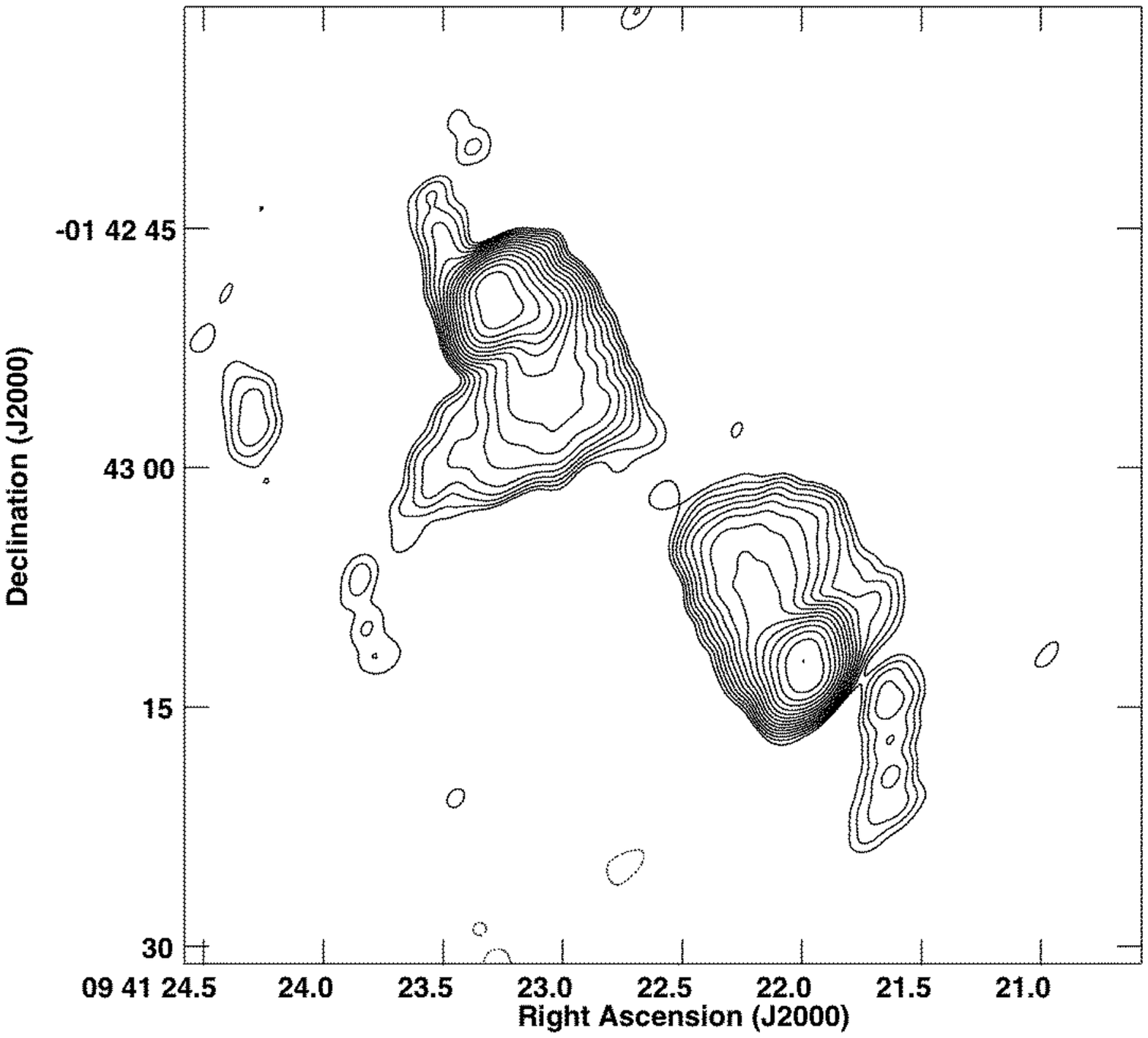}
\includegraphics[width=0.45\columnwidth]{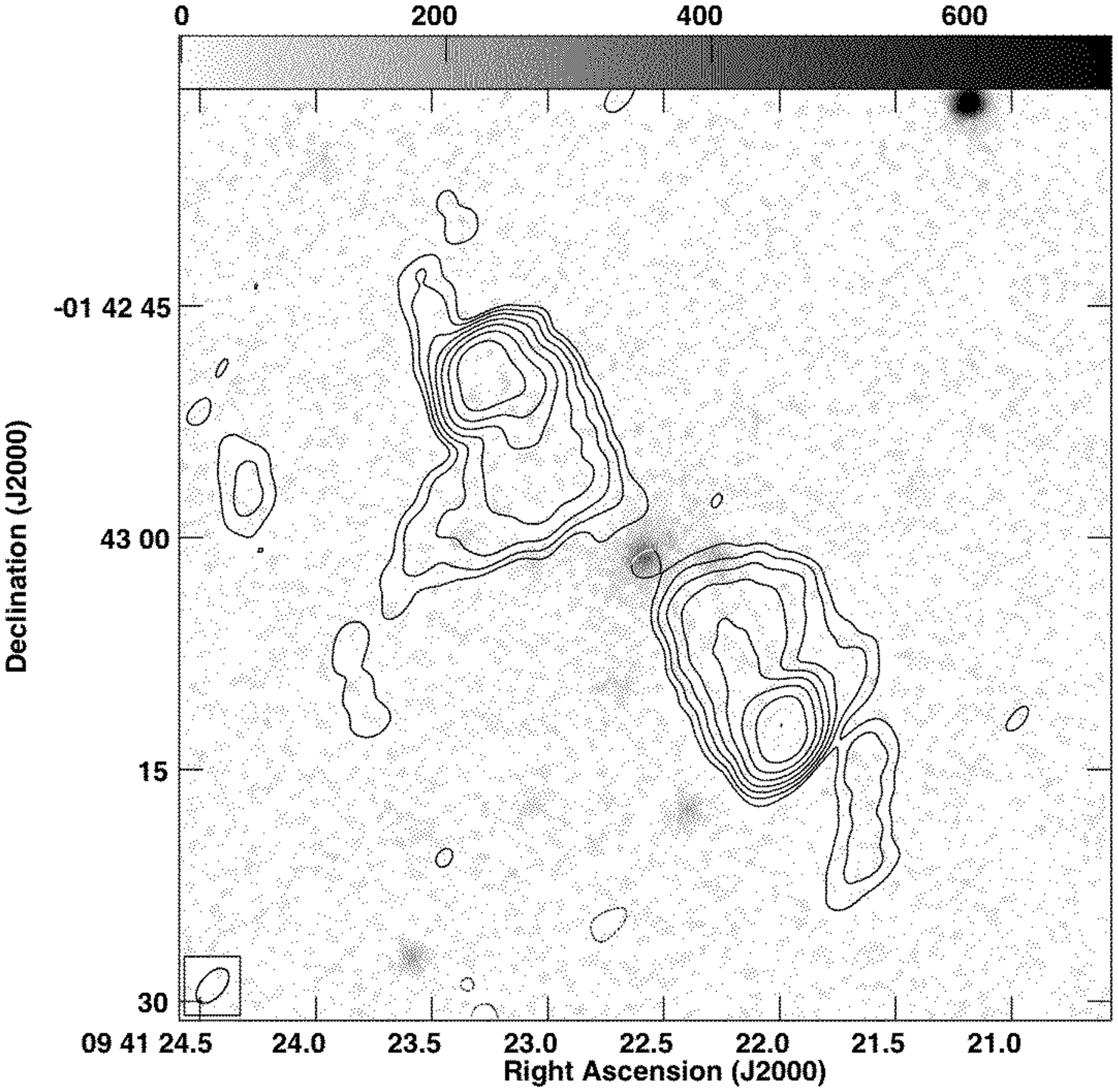}
\caption[J0941$-$0143 (L)]{J0941$-$0143. (left) VLA image at L band, (right) VLA image overlaid on red SDSS image.  Lowest contour = 0.60~mJy/beam, peak  = 73.1~mJy/beam. \label{fig:J0941-0143}}
\end{figure}

\noindent J0941$-$0143 (Figure~\ref{fig:J0941-0143}). We detect a weak core at the location of a galaxy ID. The source structure is revealed to be inversion-symmetric about the ID. The prominent SE extension to the northern lobe is only partially imaged. 

\begin{figure}[ht] 
\includegraphics[width=0.45\columnwidth]{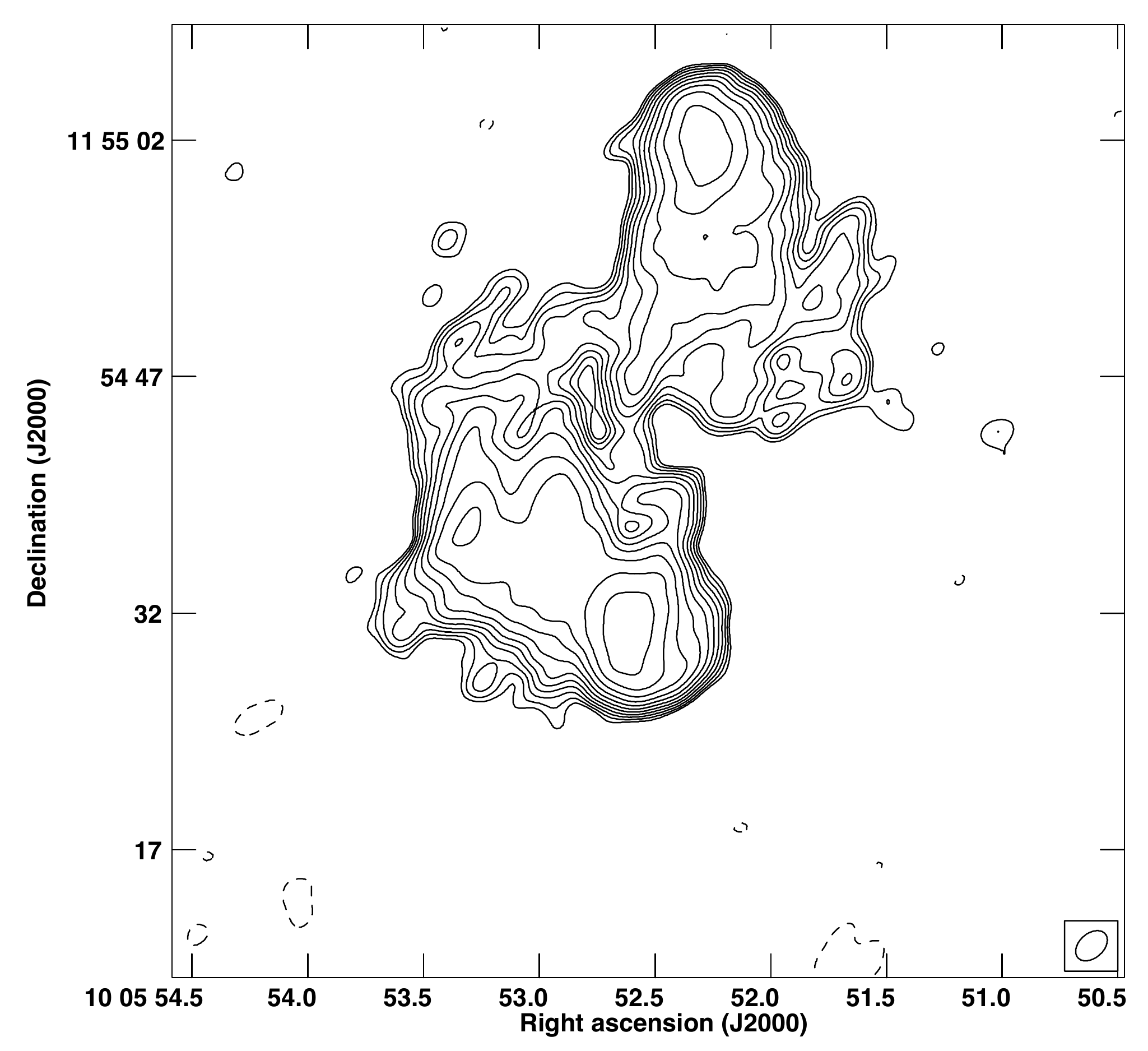}
\includegraphics[width=0.45\columnwidth]{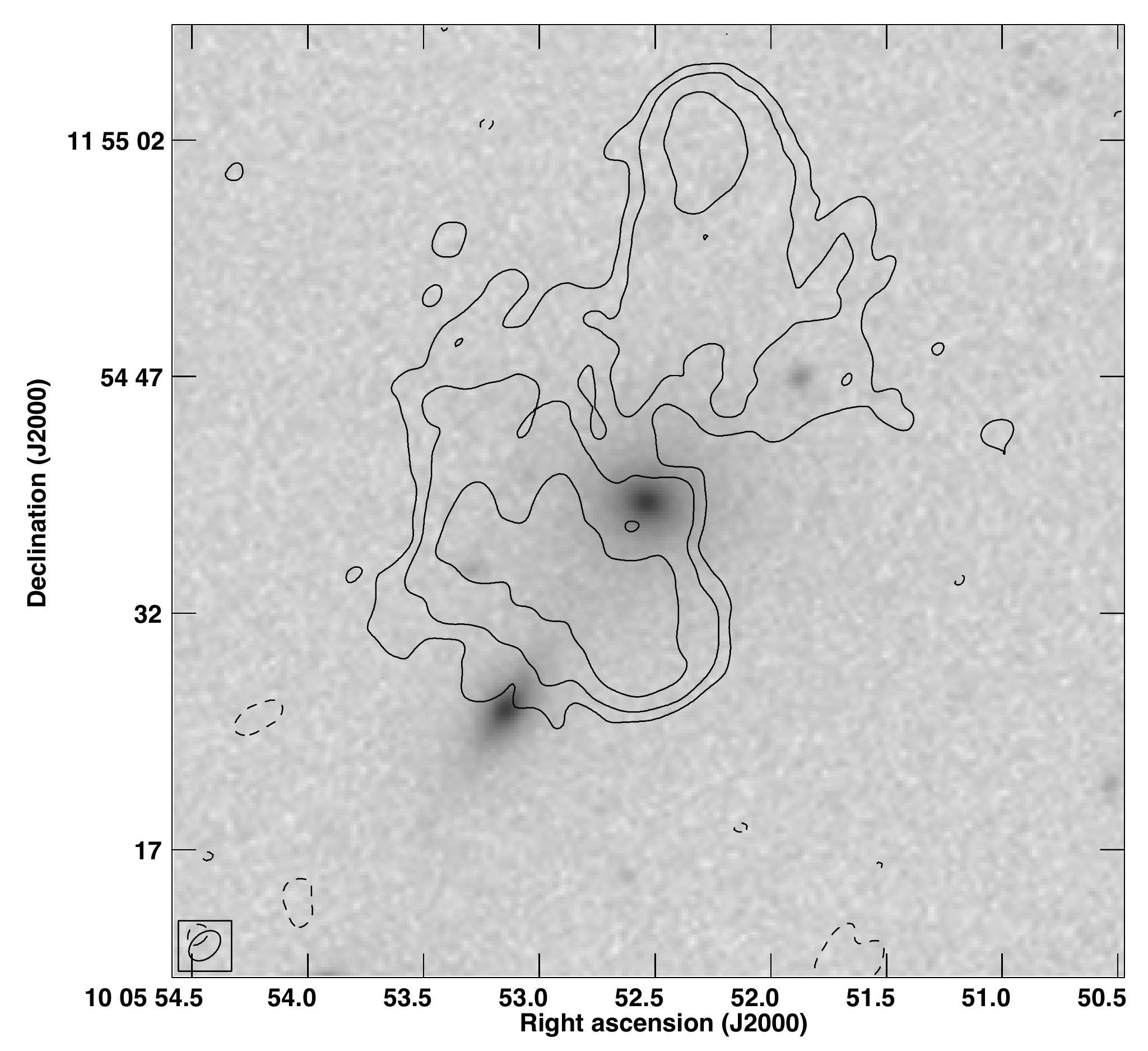}
\caption[J1005+1154 (L)]{J1005+1154.  (left) VLA image at L band, (right) VLA image overlaid on red SDSS image. Lowest contour = 0.09~mJy/beam, peak  = 3.77~mJy/beam.  \label{fig:J1005+1154}}
\end{figure}

\noindent J1005+1154 (Figure~\ref{fig:J1005+1154}).  The map detects compact emission at the location of an optical object. It makes the source highly asymmetric in extent. The northern lobe reaches most of the way to the core along the radio axis before appearing to change direction to the west well ahead of the core. The offset emission shows a well-bounded inner edge. 

\begin{figure}[ht] 
\includegraphics[width=0.45\columnwidth]{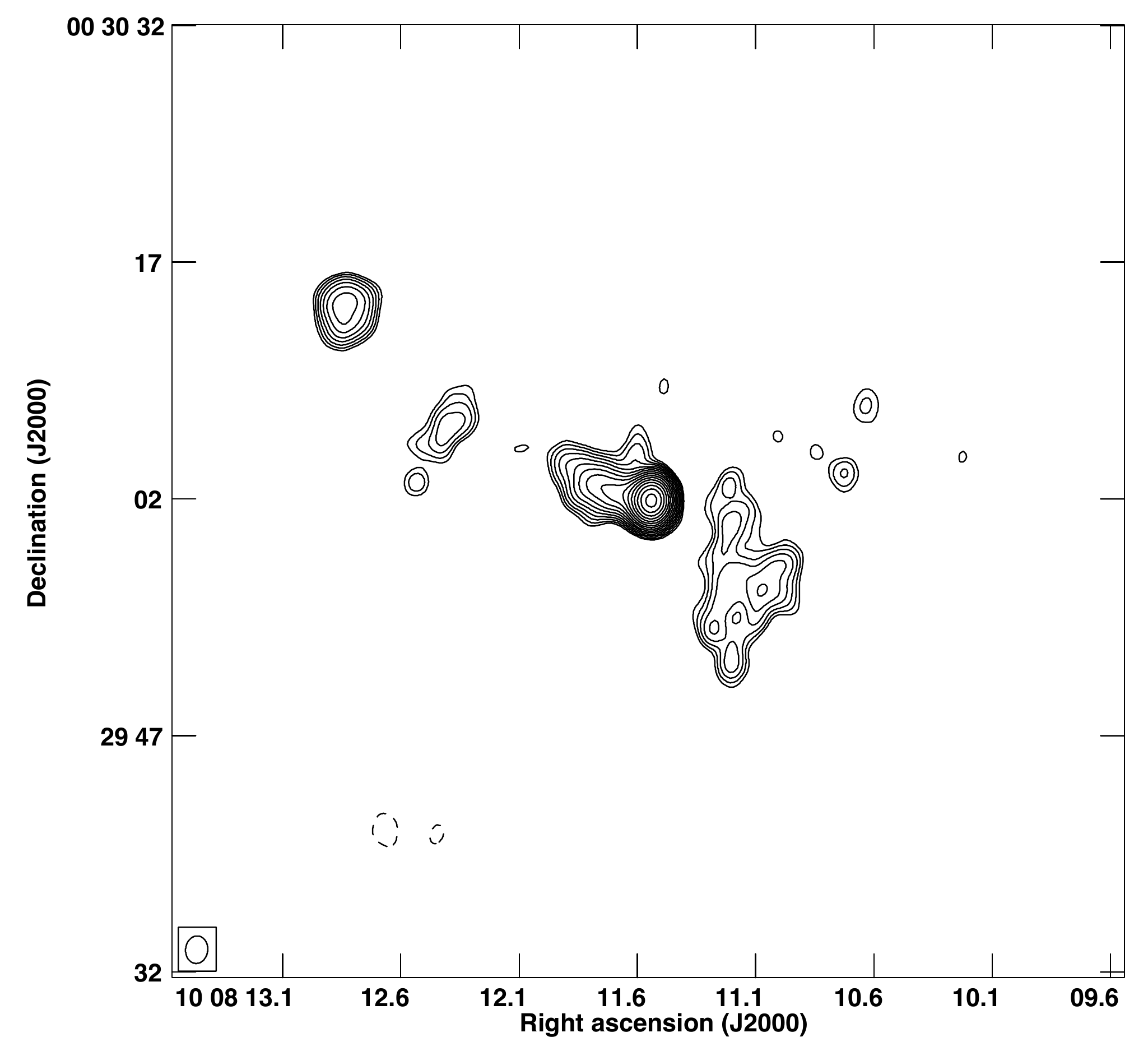}
\includegraphics[width=0.45\columnwidth]{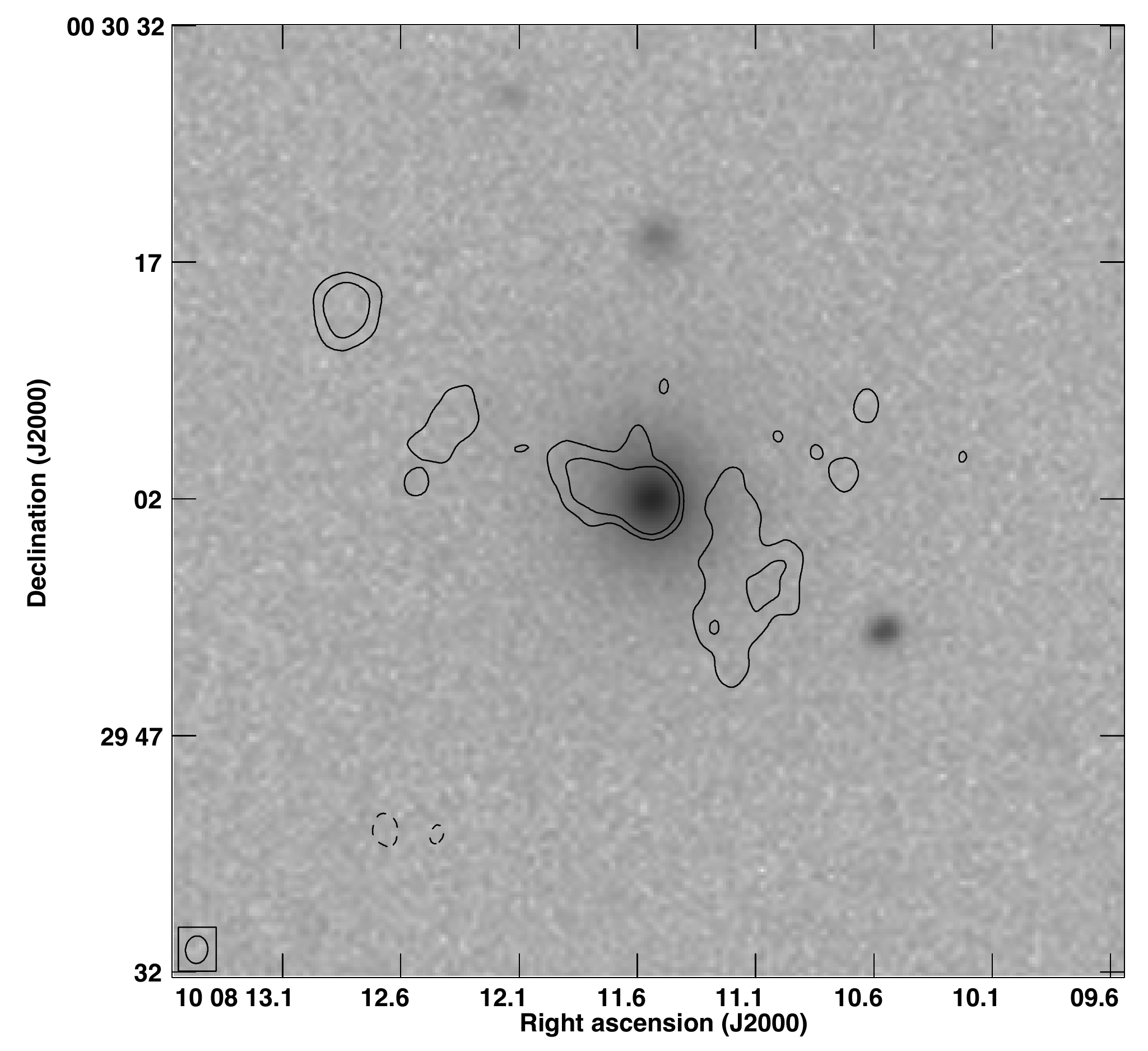}
\caption[J1008+0030 (L)]{J1008+0030.  (left) VLA image at L band, (right) VLA image overlaid on red SDSS image. Lowest contour = 0.3~mJy/beam, peak = 63.5~mJy/beam. \label{fig:J1008+0030}}
\end{figure}

\noindent J1008+0030 (Figure~\ref{fig:J1008+0030}). The halo-like diffuse emission in which a central radio source is embedded (as seen in the FIRST map) is completely resolved out. Instead the map reveals a bright core from which a short bright jet is seen extending toward the NE hotspot.

\begin{figure}[ht] 
\includegraphics[width=0.45\columnwidth]{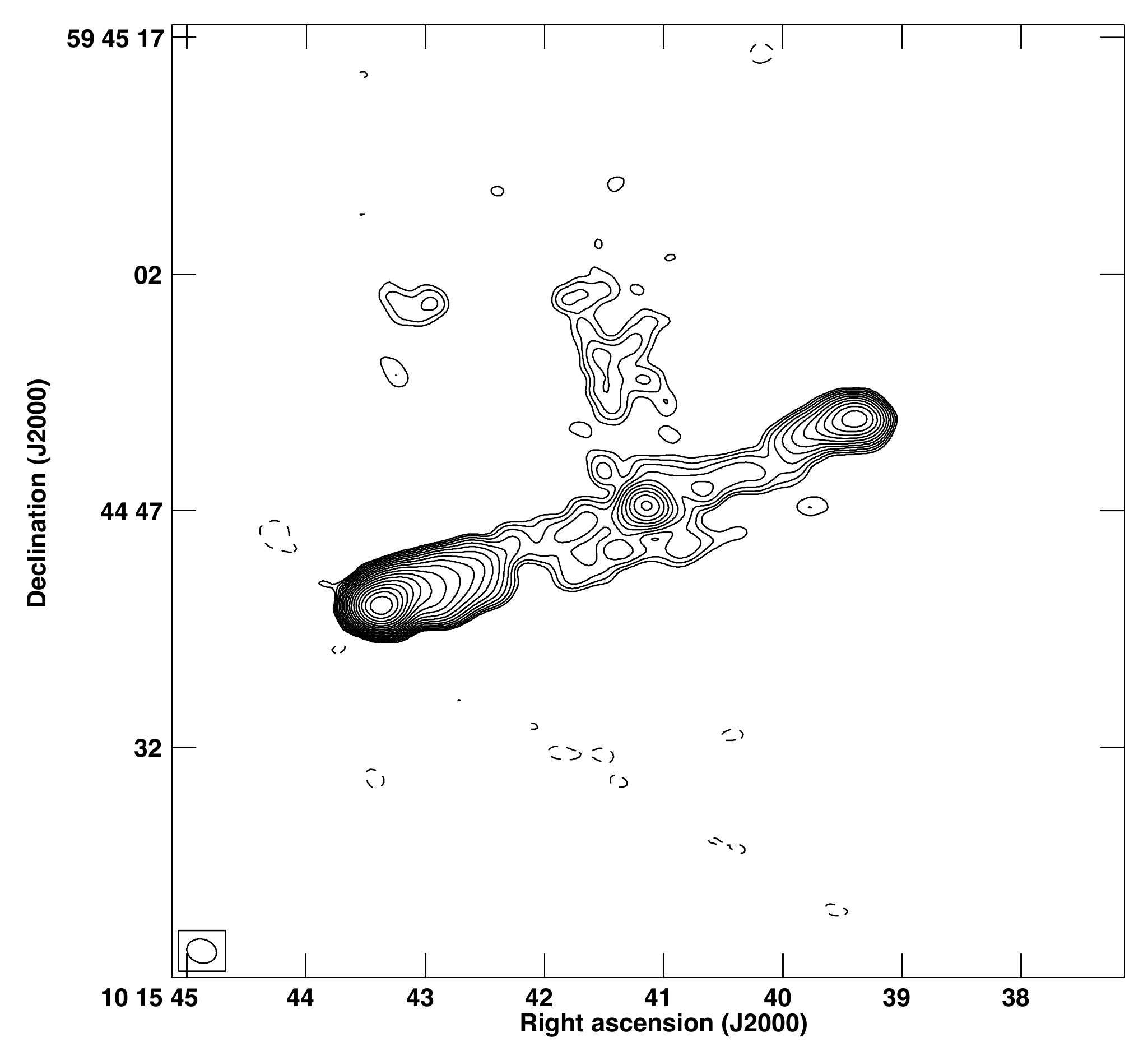}
\includegraphics[width=0.45\columnwidth]{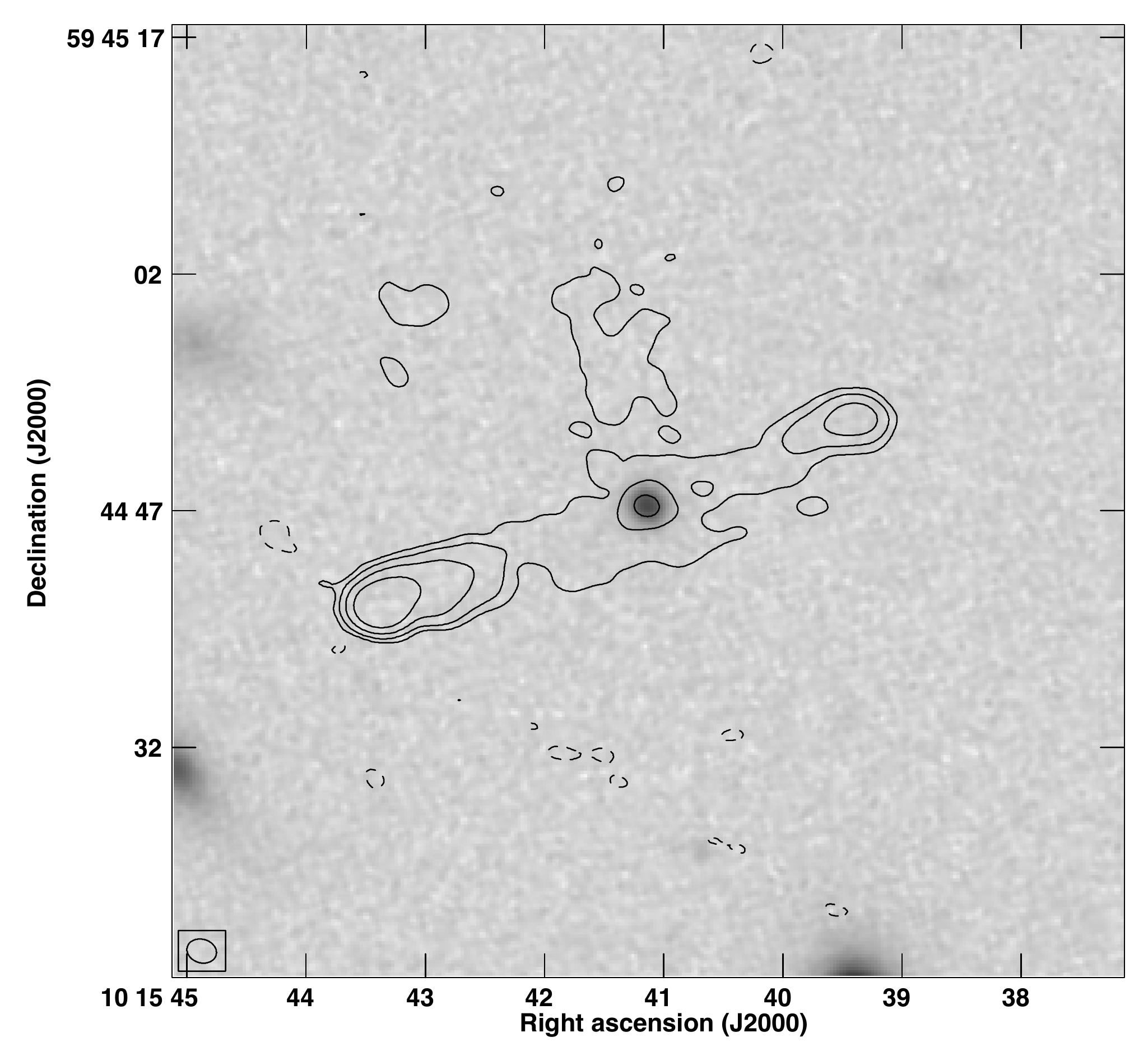}
\caption[J1015+5944 (L)]{J1015+5944. (left) VLA image at L band, (right) VLA image overlaid on red SDSS image. Lowest contour = 0.2~mJy/beam, peak  = 98.1~mJy/beam. \label{fig:J1015+5944}}
\end{figure}

\noindent J1015+5944 (Figure~\ref{fig:J1015+5944}). The source is revealed to be a narrow edge-brightened radio galaxy. The northern diffuse extension is detected as a narrow feature with a sharp inner edge. 

\begin{figure}[ht] 
\includegraphics[width=0.45\columnwidth]{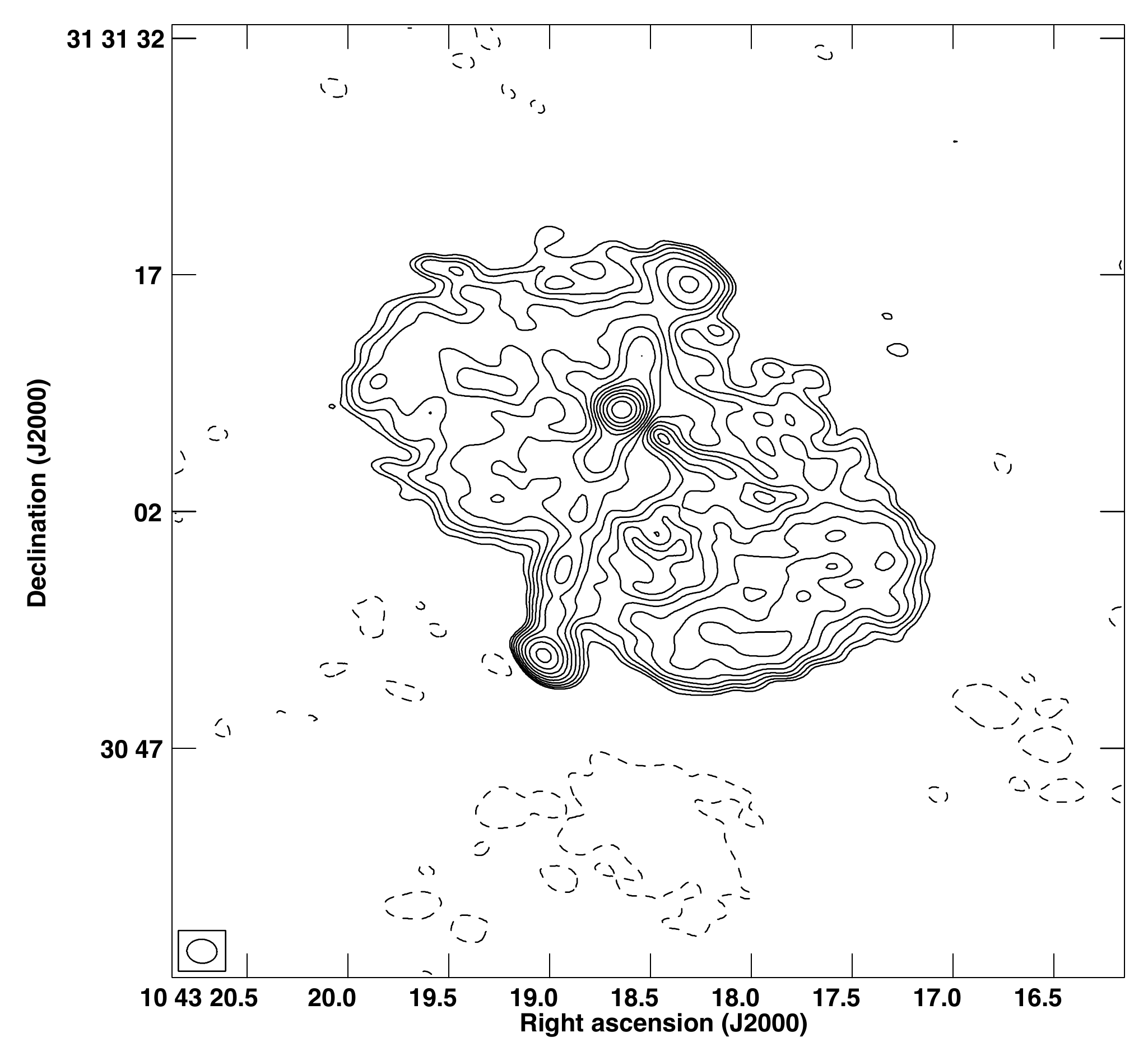}
\includegraphics[width=0.45\columnwidth]{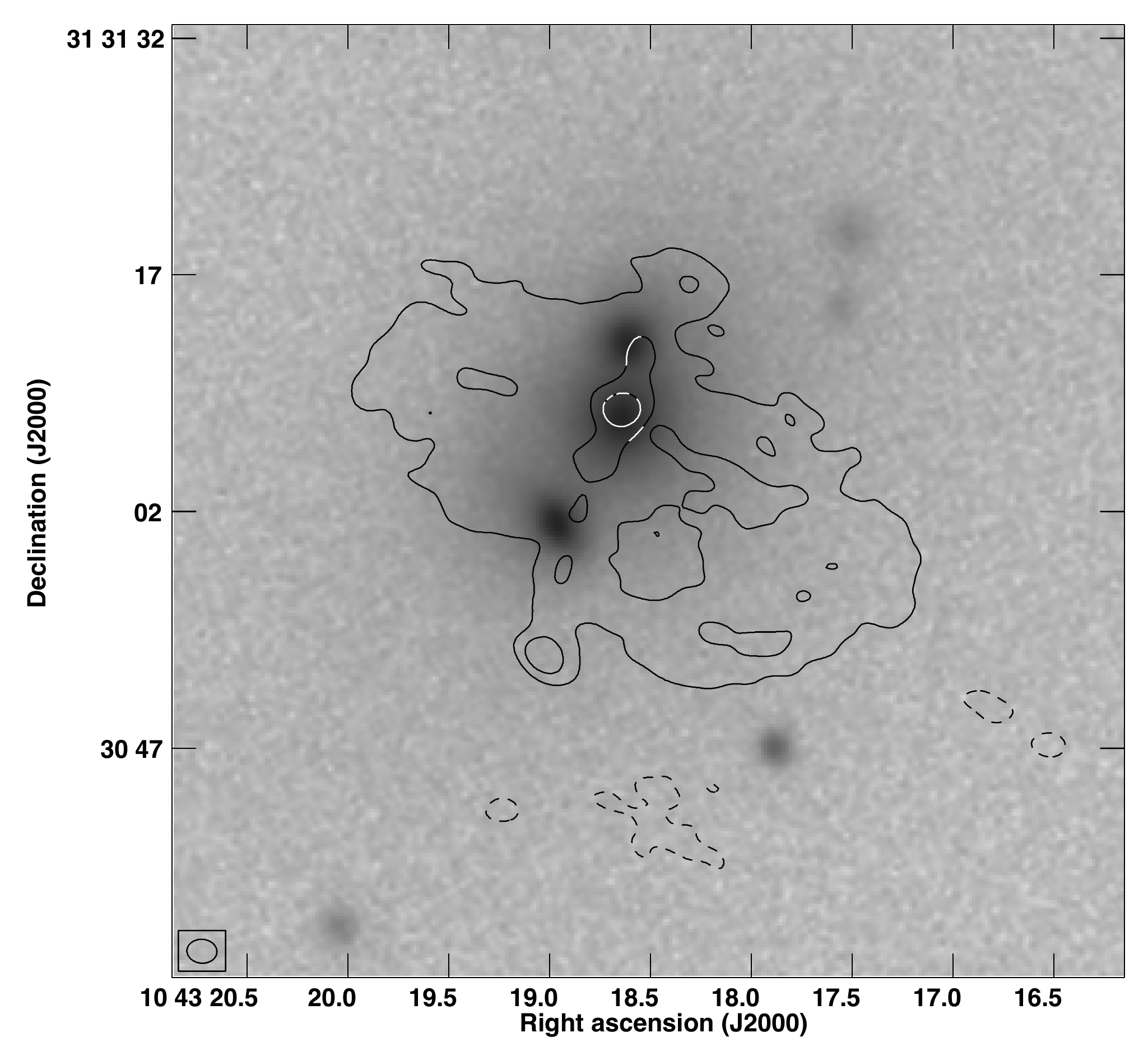}\\
\includegraphics[width=0.45\columnwidth]{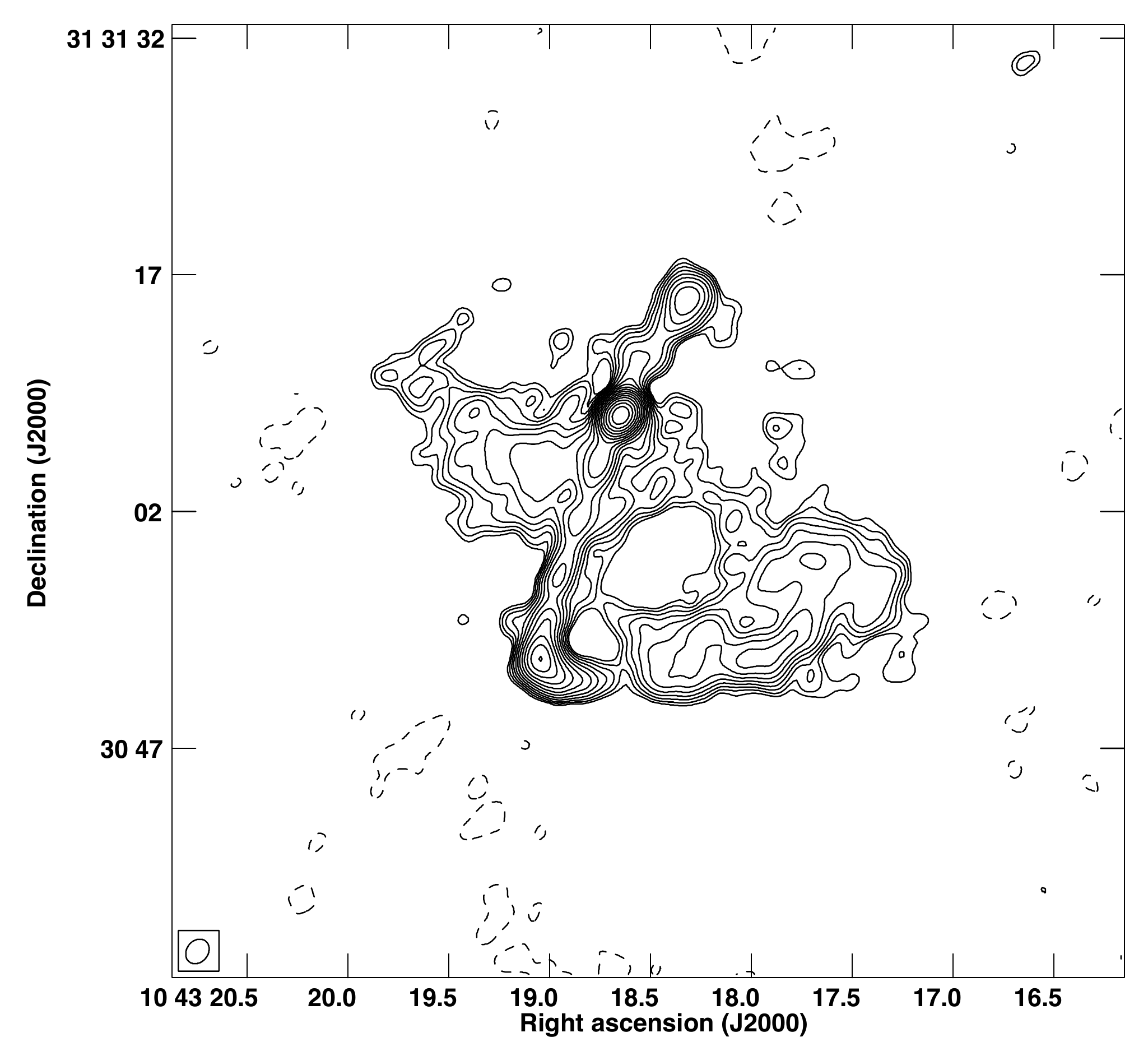}
\includegraphics[width=0.45\columnwidth]{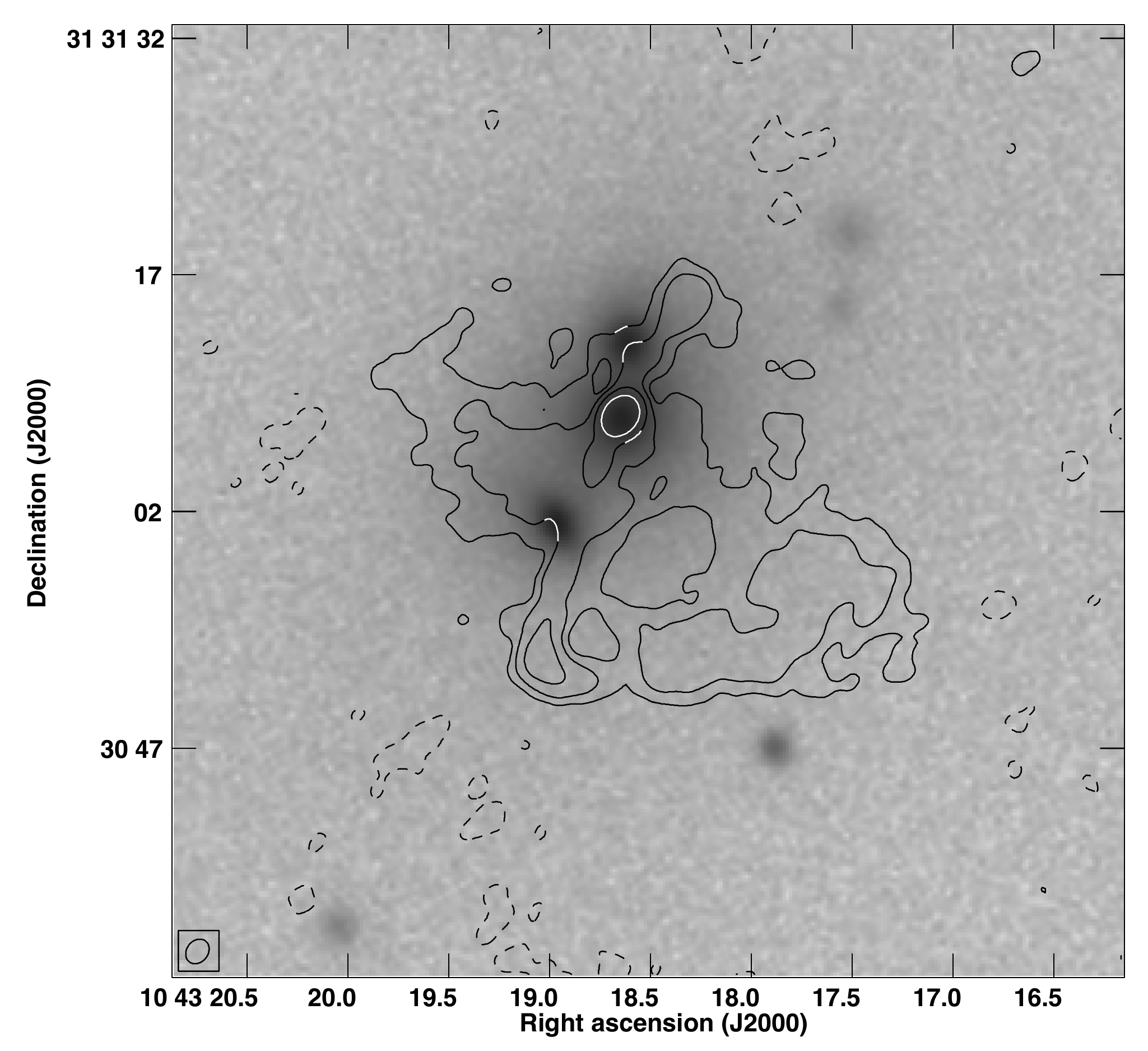}\\
\caption[J1043+3131 (L \& C)]{J1043+3131.  (top) (left) VLA image at L band and (right) VLA image overlaid on red SDSS image. Lowest contour = 0.4~mJy/beam, peak  = 34.0~mJy/beam. (bottom) (left) VLA image at C band and (right) VLA image overlaid on red SDSS image. Lowest contour = 0.08~mJy/beam, peak  = 44.6~mJy/beam.  \label{fig:J1043+3131}}
\end{figure}

\noindent J1043+3131 (Figure~\ref{fig:J1043+3131}). The source structure is well imaged in the new maps, which show a compact core connected by a pair of narrow straight jets to two compact hotspots at the extreme ends of the source. The core coincides with a centrally located and brighter of three galaxies that lie in a line along the source axis. Our map reveals a well-bounded wide feature oriented orthogonal to the source axis. An interesting, narrow jet-like feature is seen to the west associated with the core along the axis of this broad diffuse feature.

\begin{figure}[ht] 
\includegraphics[width=0.45\columnwidth]{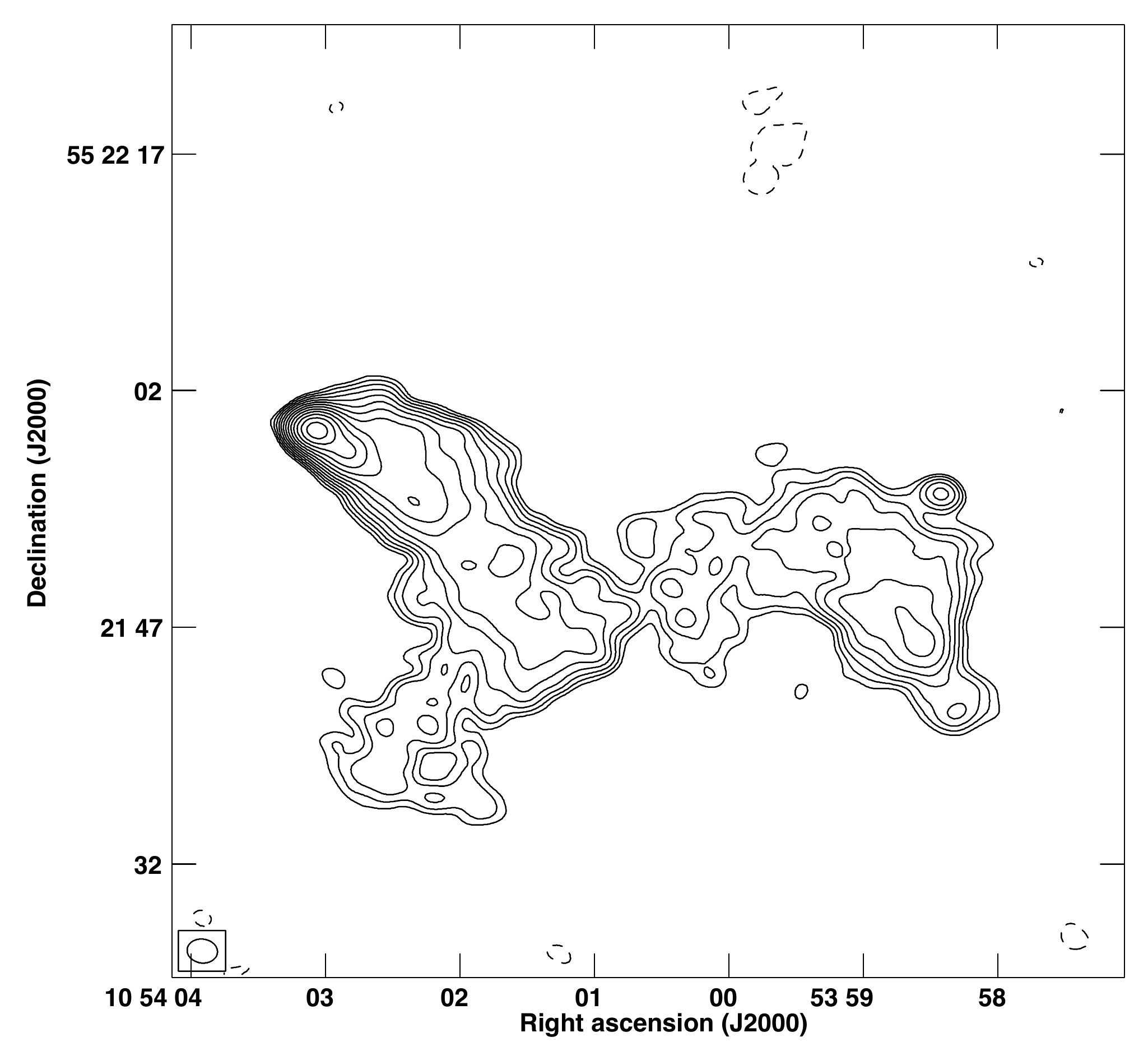}
\includegraphics[width=0.45\columnwidth]{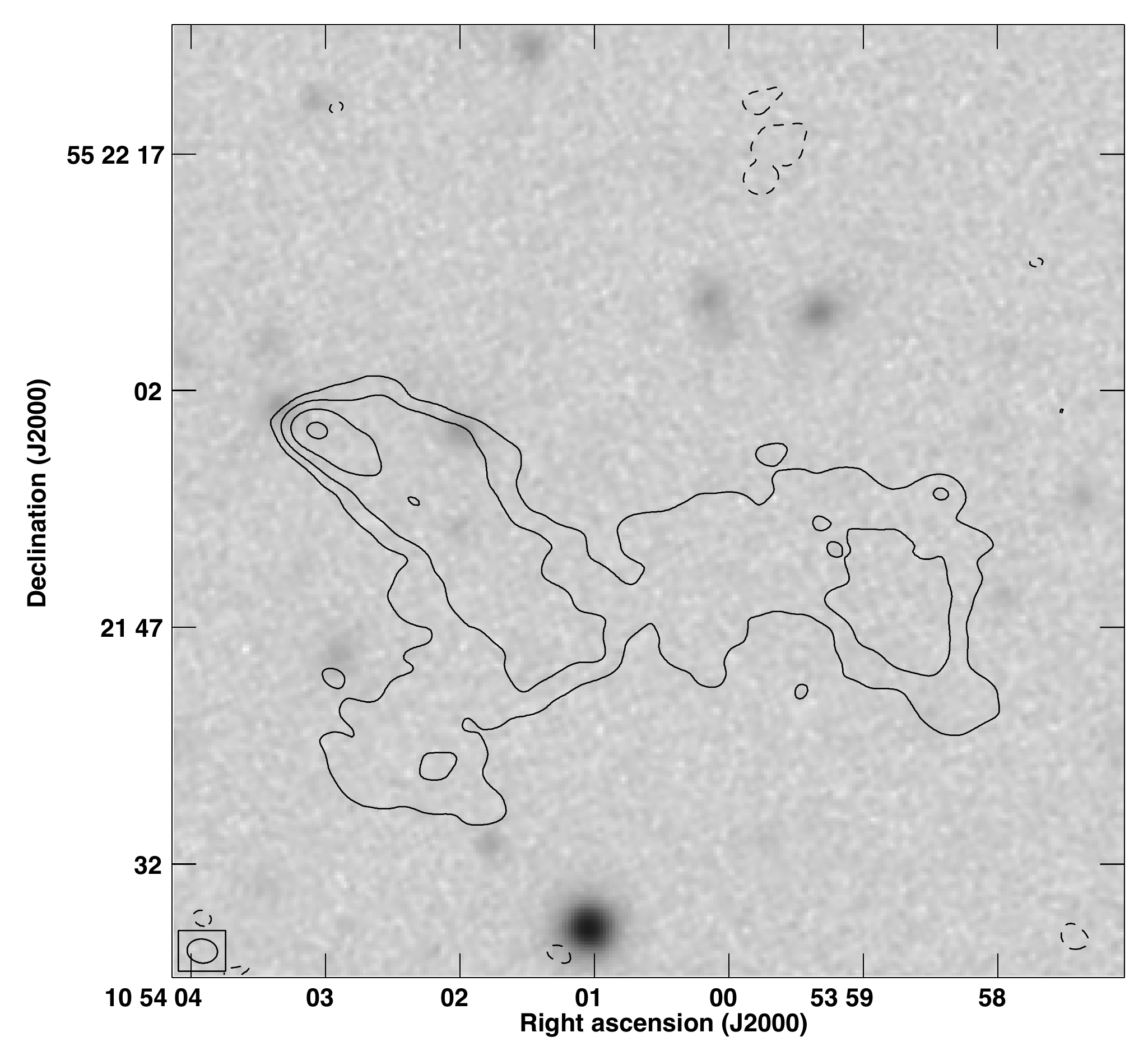}
\caption[J1054+5521 (L)]{J1054+5521. (left) VLA image at L band, (right) VLA image overlaid on red SDSS image.  Lowest contour = 0.175~mJy/beam, peak  = 14.1~mJy/beam. \label{fig:J1054+5521}}
\end{figure}

\noindent J1054+5521 (Figure~\ref{fig:J1054+5521}). The map does not reveal a compact core although we note a weak source between the lobes at the location of a very faint object. The edge-brightened lobes extend along axes that are offset from each other. The off axis emission connected to the well-bounded eastern lobe shows a sharp inner edge.

\begin{figure}[ht] 
\includegraphics[width=0.45\columnwidth]{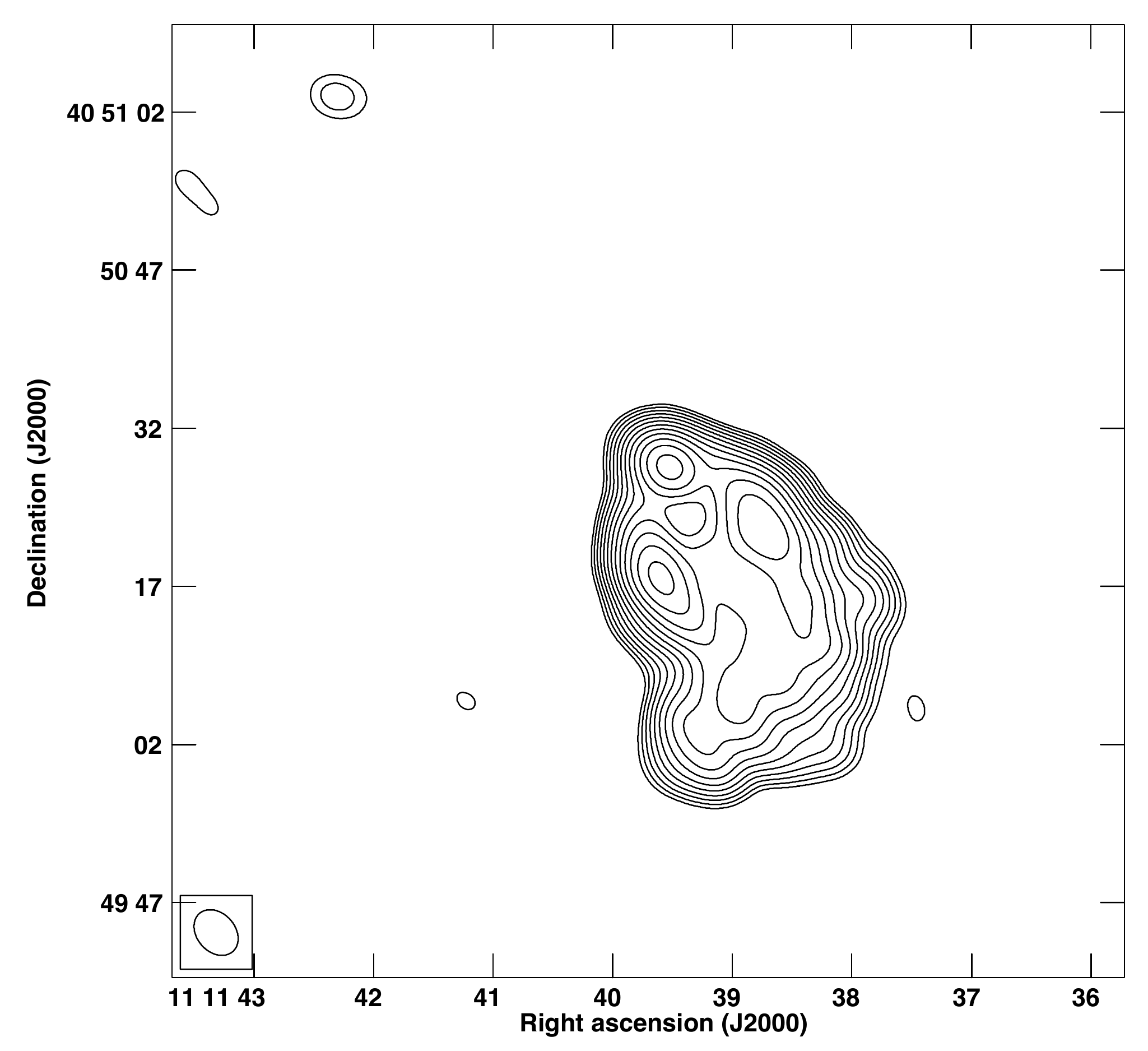}
\includegraphics[width=0.45\columnwidth]{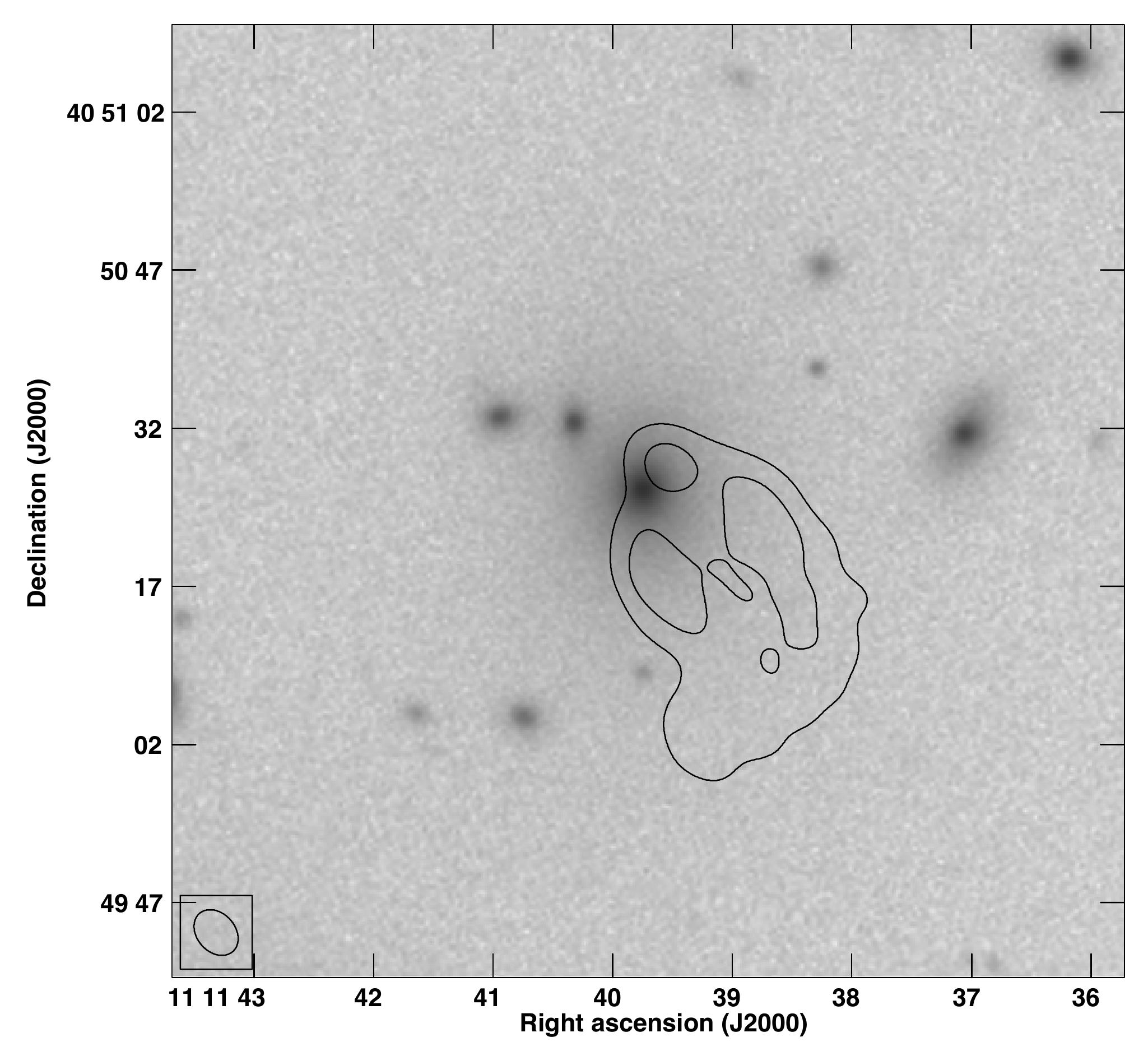}\\
\includegraphics[width=0.45\columnwidth]{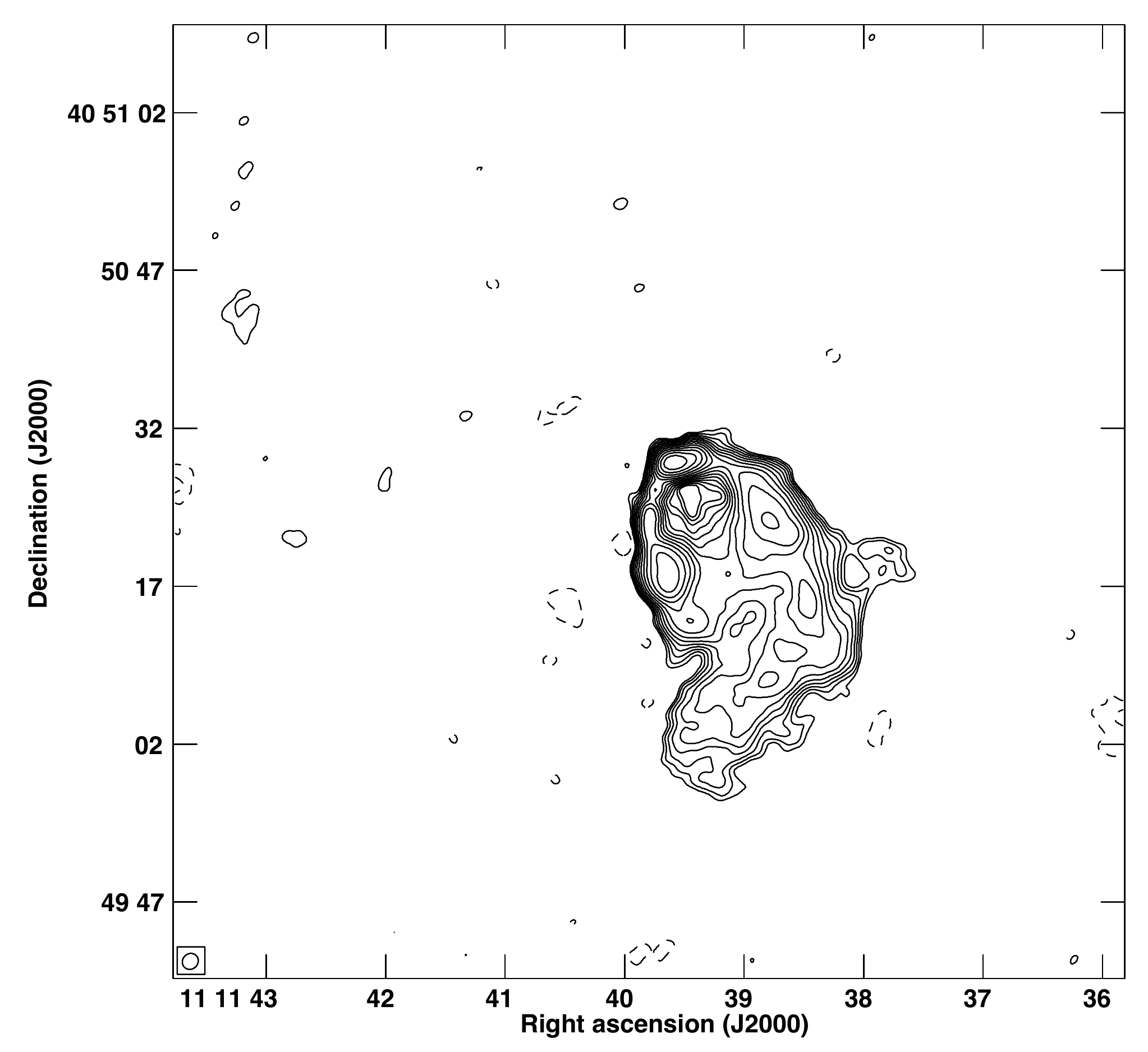}
\includegraphics[width=0.45\columnwidth]{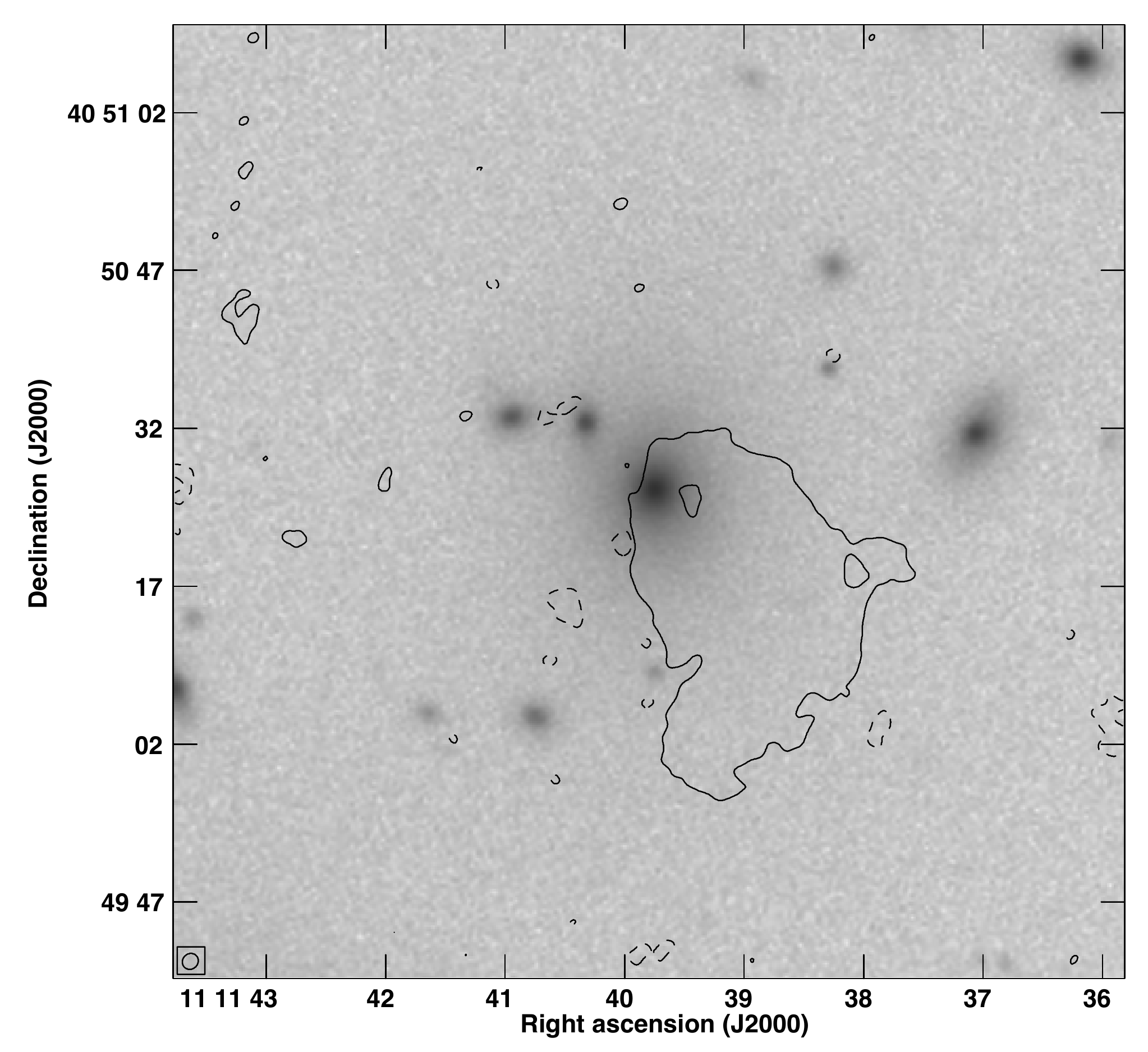}\\
\caption[J1111+4050 (L \& C)]{J1111+4050.  (top) (left) VLA image at L band and (right) VLA image overlaid on red SDSS image. Lowest contour = 1.0~mJy/beam, peak = 73.3~mJy/beam. (bottom) (left) VLA image at C band and (right) VLA image overlaid on red SDSS image. Lowest contour = 0.125~mJy/beam, peak = 11.3~mJy/beam. \label{fig:J1111+4050}}
\end{figure}

\noindent J1111+4050 (Figure~\ref{fig:J1111+4050}).  The source is clearly seen to be a radio galaxy with bent jets, a likely wide-angle tail source. 

\begin{figure}[ht] 
\includegraphics[width=0.45\columnwidth]{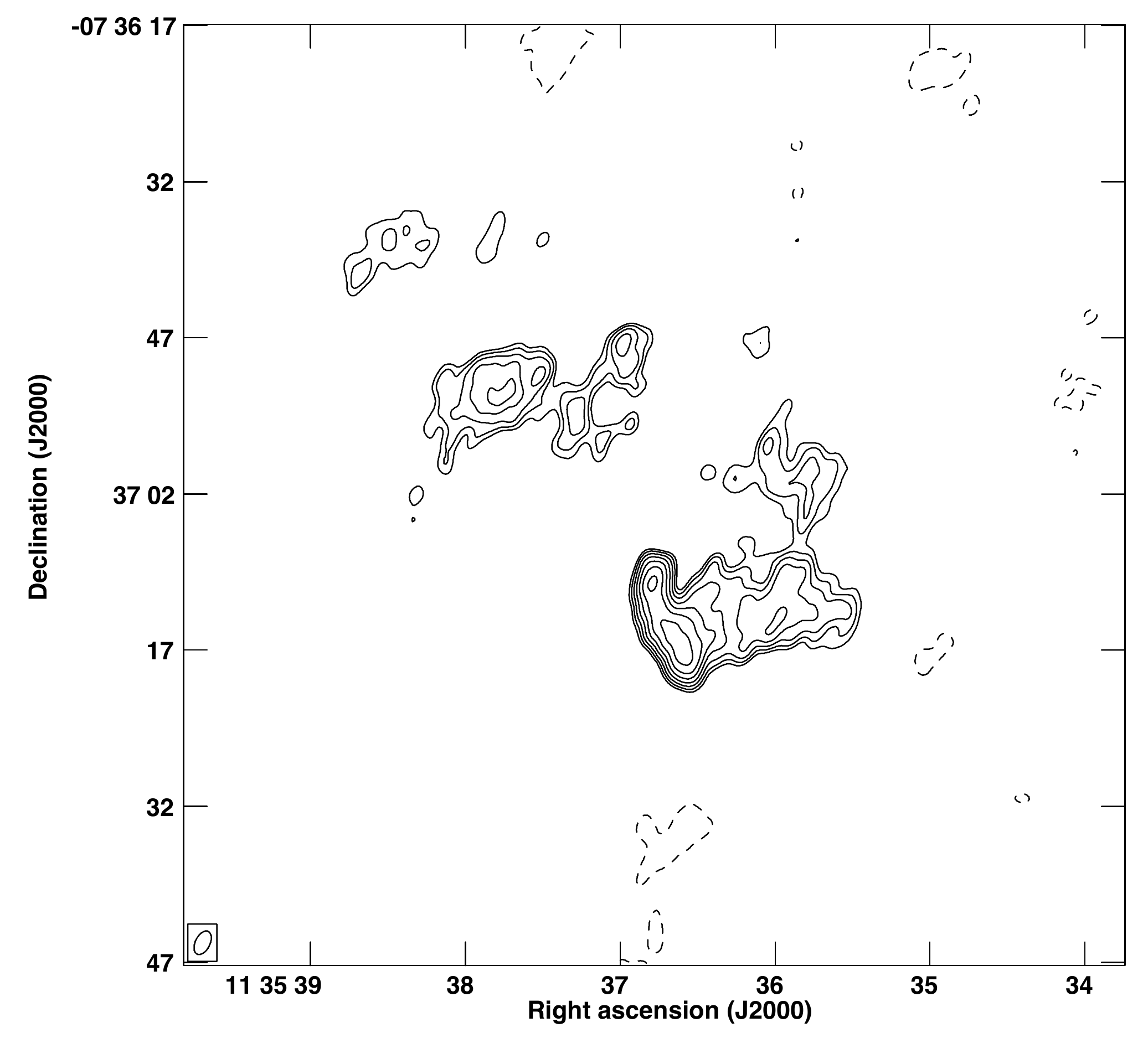}
\includegraphics[width=0.45\columnwidth]{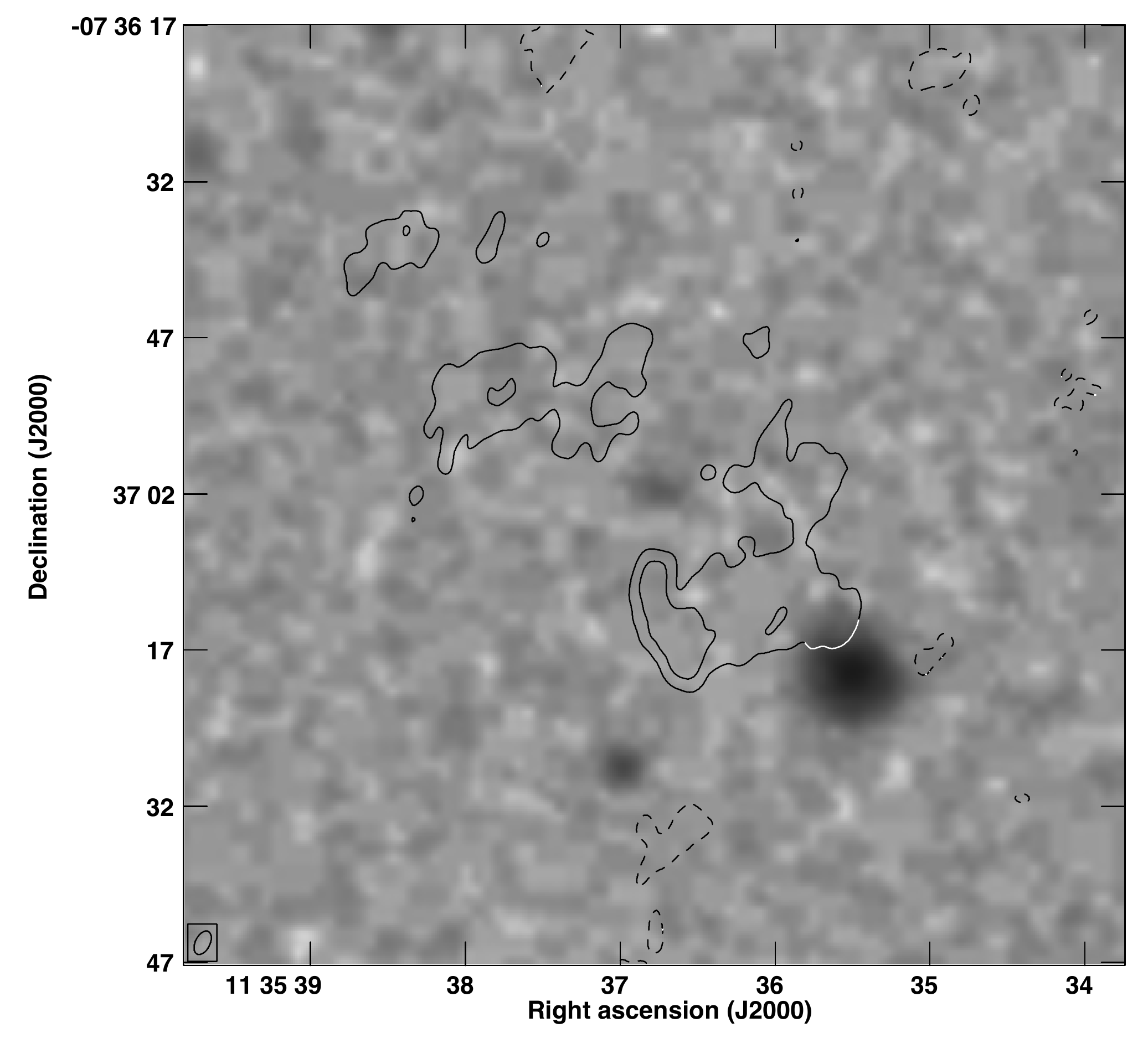}
\caption[J1135$-$0737 (L)]{J1135$-$0737. (left) VLA image at L band, (right) VLA image overlaid on red DSS\,II image. Lowest contour = 0.2~mJy/beam, peak = 1.99~mJy/beam. \label{fig:J1135-0737}}
\end{figure}

\noindent J1135$-$0737 (Figure~\ref{fig:J1135-0737}).  A weak hotspot is seen associated with the north lobe, which is accompanied by lobe emission to its east, whereas in the south lobe a narrow extended feature is seen accompanied by lobe emission to the west. No core is detected in the new map. However an optical object is seen along the axis formed by the northern hotspot and the narrow, elongated feature in the south.

\begin{figure}[ht] 
\includegraphics[width=0.45\columnwidth]{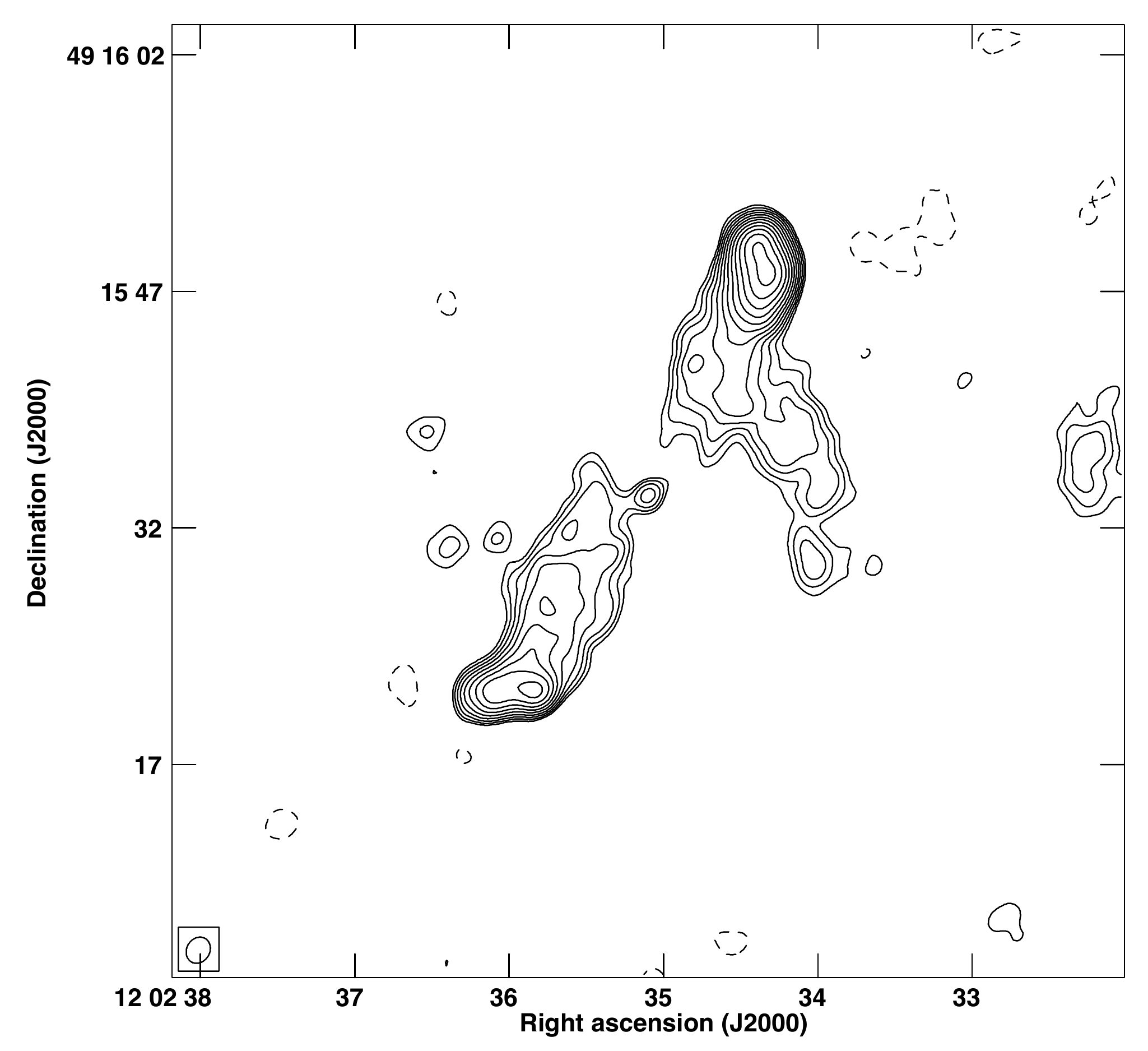}
\includegraphics[width=0.45\columnwidth]{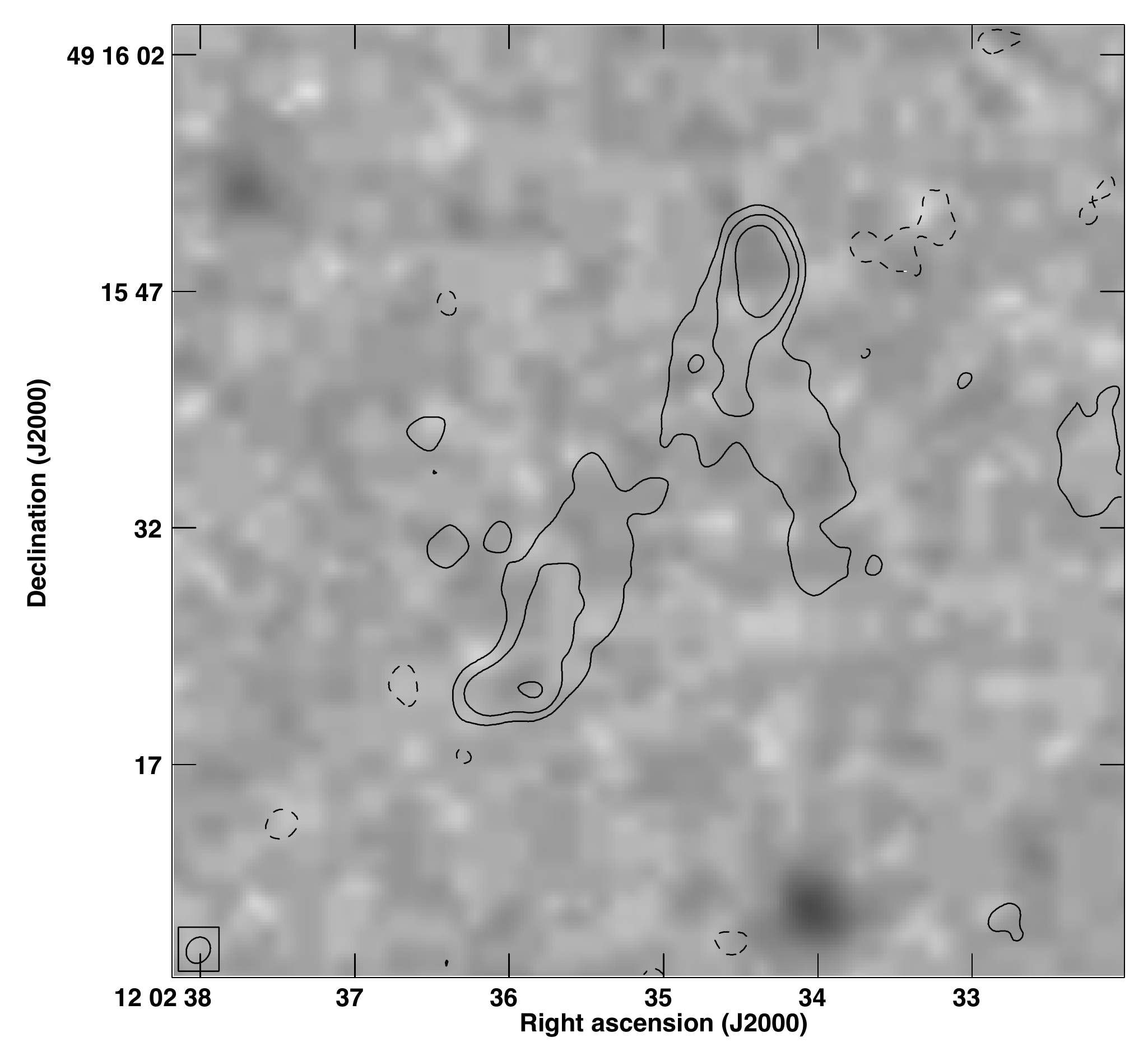}
\caption[J1202+4915 (L)]{J1202+4915. (left) VLA image at L band, (right) VLA image overlaid on red SDSS image.  Lowest contour = 0.15~mJy/beam, peak  = 7.84~mJy/beam. \label{fig:J1202+4915}}
\end{figure}

\noindent J1202+4915 (Figure~\ref{fig:J1202+4915}). The edge-brightened lobes are both revealed to have extended hotspot structures at their ends. The new map also reveals a compact radio core although no optical ID is visible. The diffuse extensions to the lobes are mostly resolved out especially that associated with the southern lobe. 

\begin{figure}[ht] 
\includegraphics[width=0.45\columnwidth]{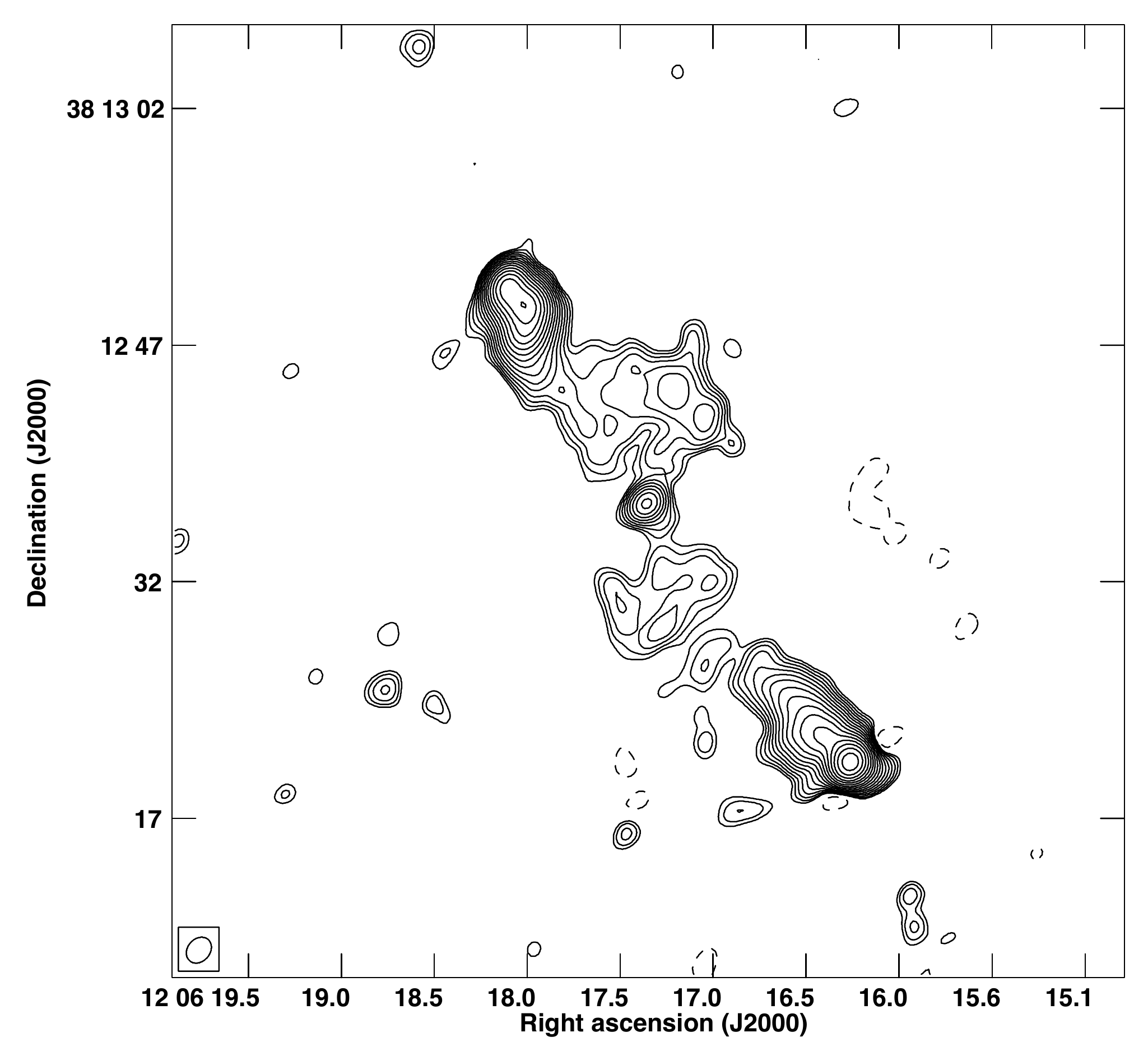}
\includegraphics[width=0.45\columnwidth]{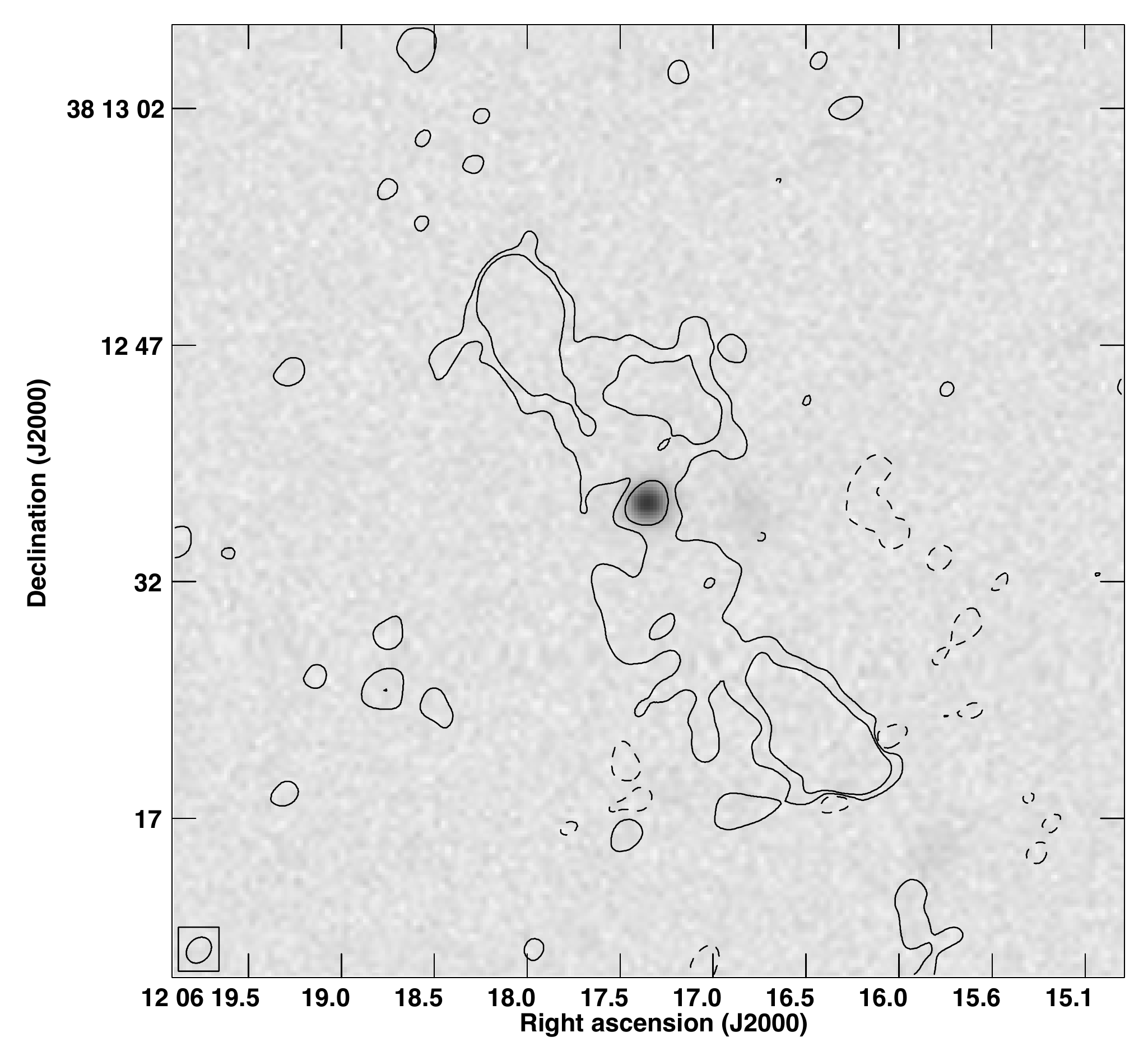}\\
\includegraphics[width=0.45\columnwidth]{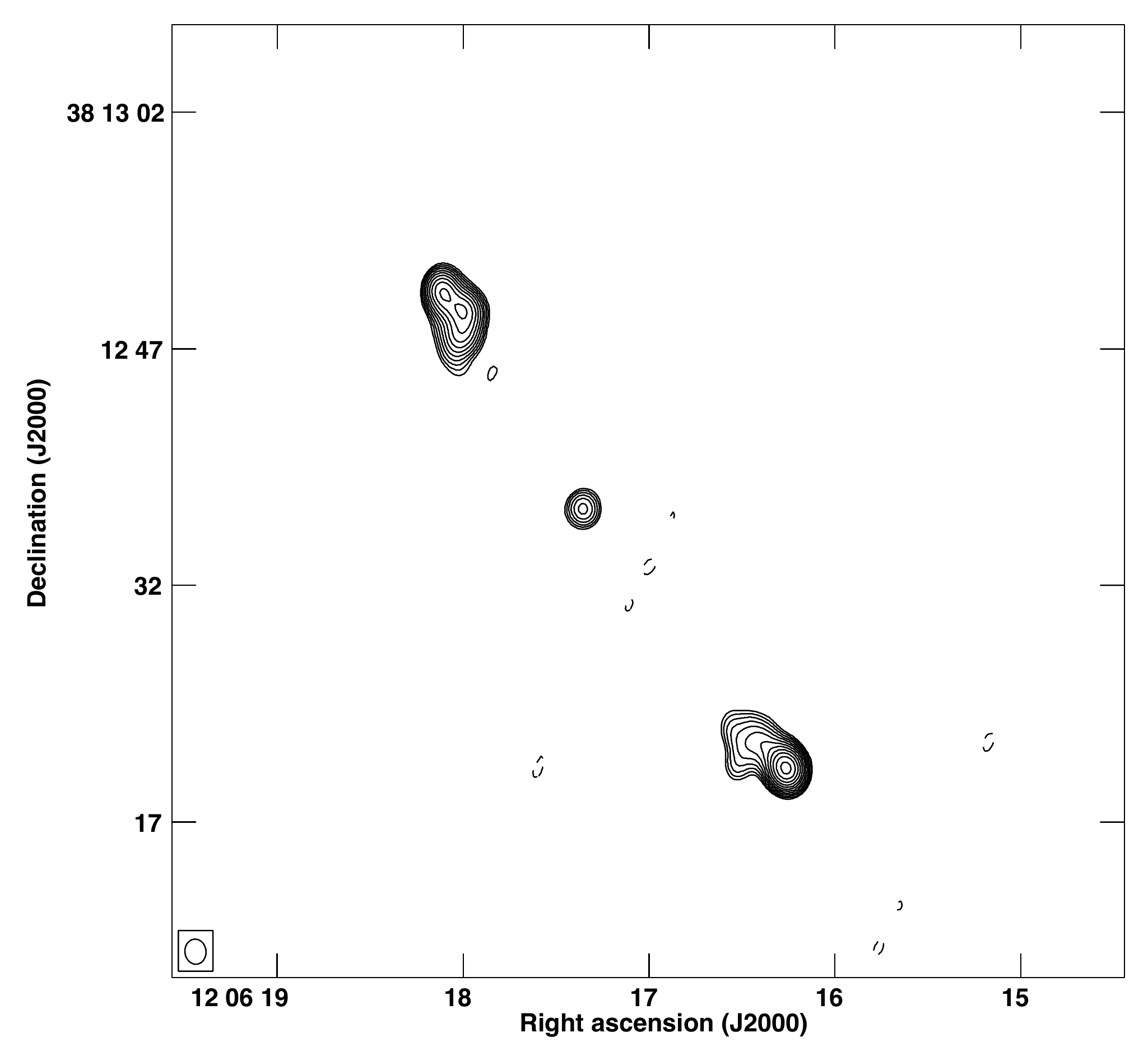}
\includegraphics[width=0.45\columnwidth]{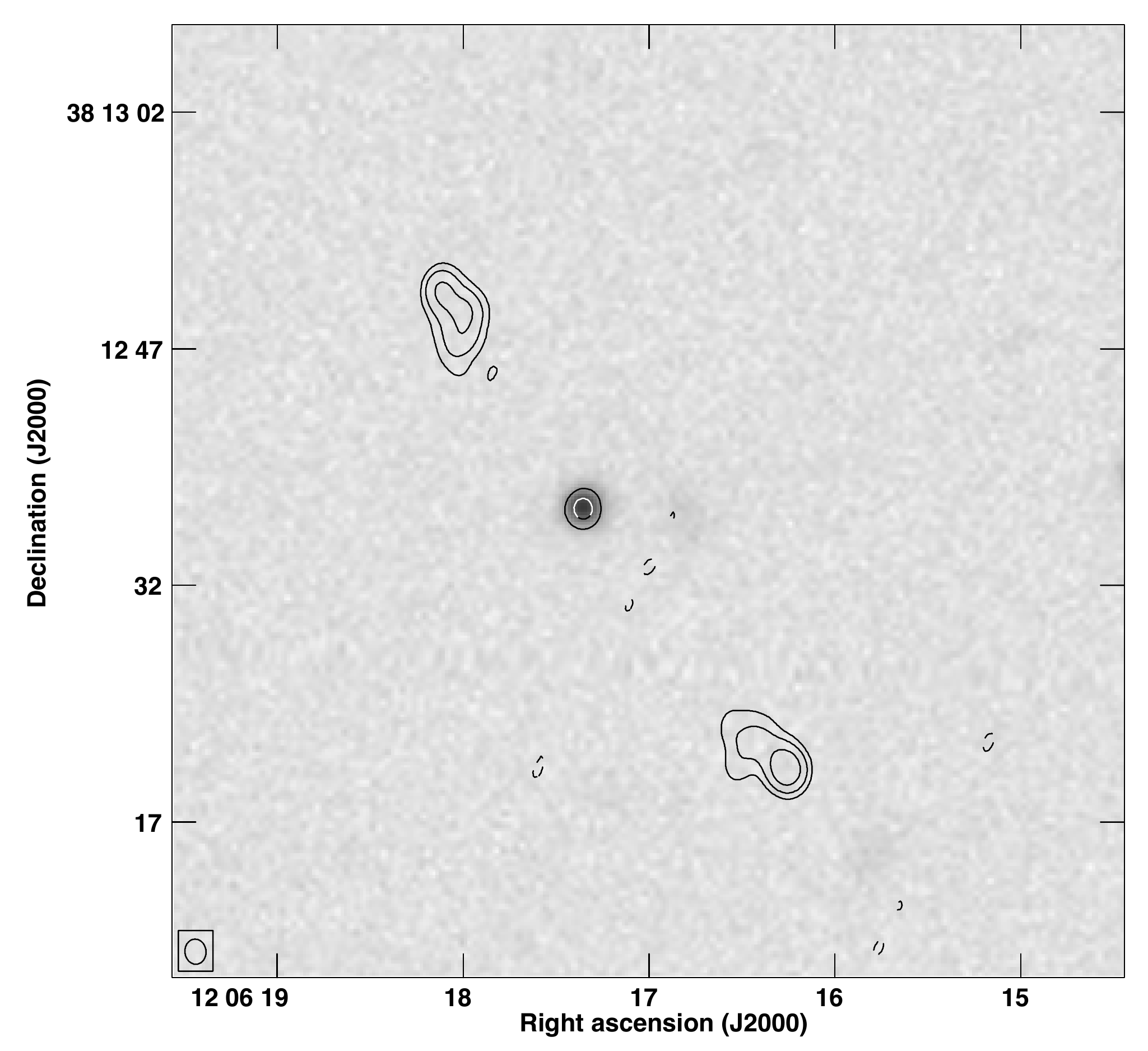}\\
\caption[J1206+3812 (L)]{J1206+3812.  (top) (left) VLA image at L band, (right) VLA image overlaid on red SDSS image. Lowest contour = 0.2~mJy/beam, peak  = 47.0~mJy/beam. (bottom) (left) VLA image at C band, (right) VLA image overlaid on red SDSS image. Lowest contour = 0.3~mJy/beam, peak  = 15.6~mJy/beam.  \label{fig:J1206+3812}}
\end{figure}

\noindent J1206+3812 (Figure~\ref{fig:J1206+3812}). The maps have revealed the hotspots in finer detail and the bridge emission leading from them along the source axis towards the compact core.  

\begin{figure}[ht] 
\includegraphics[width=0.45\columnwidth]{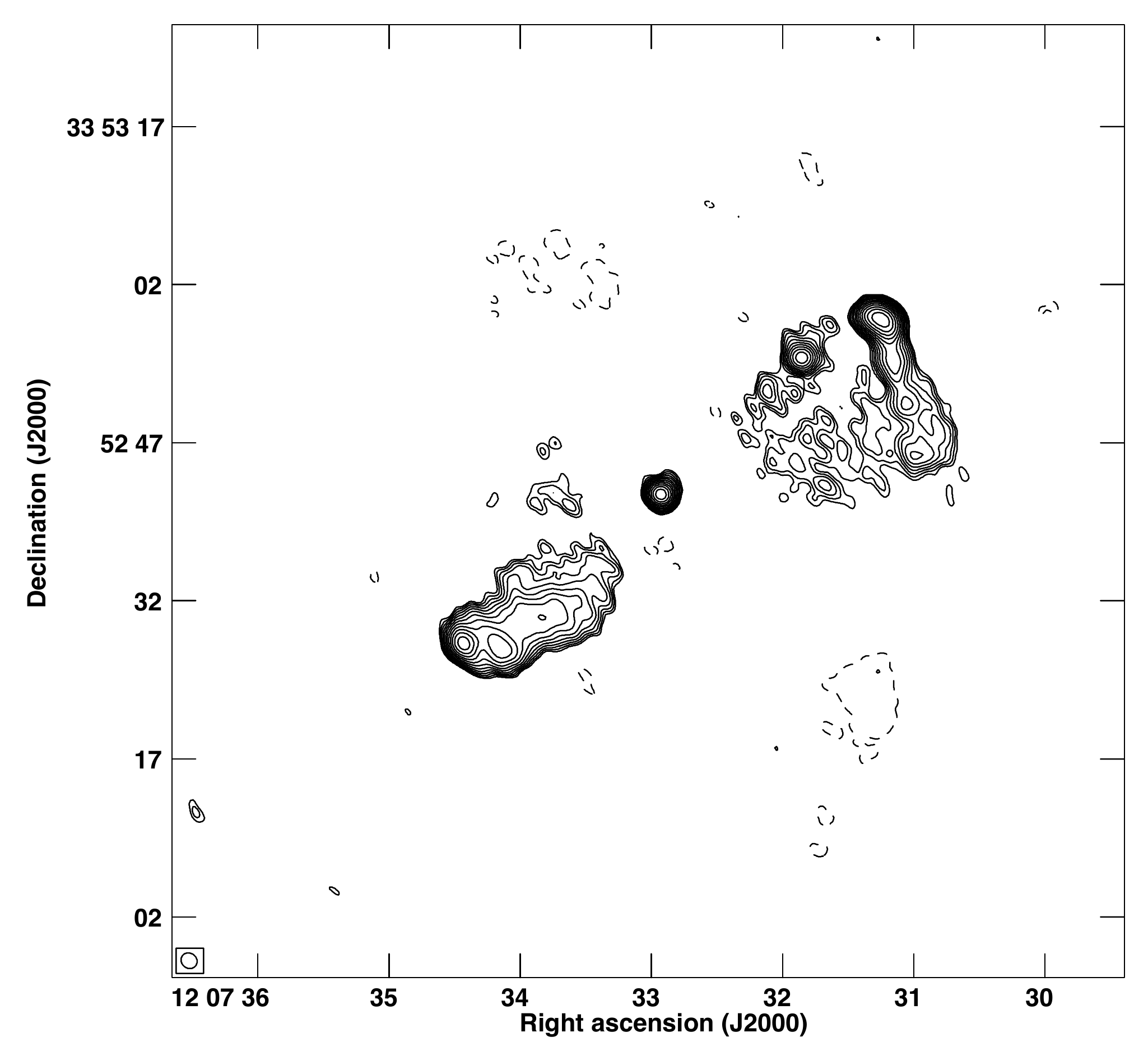}
\includegraphics[width=0.45\columnwidth]{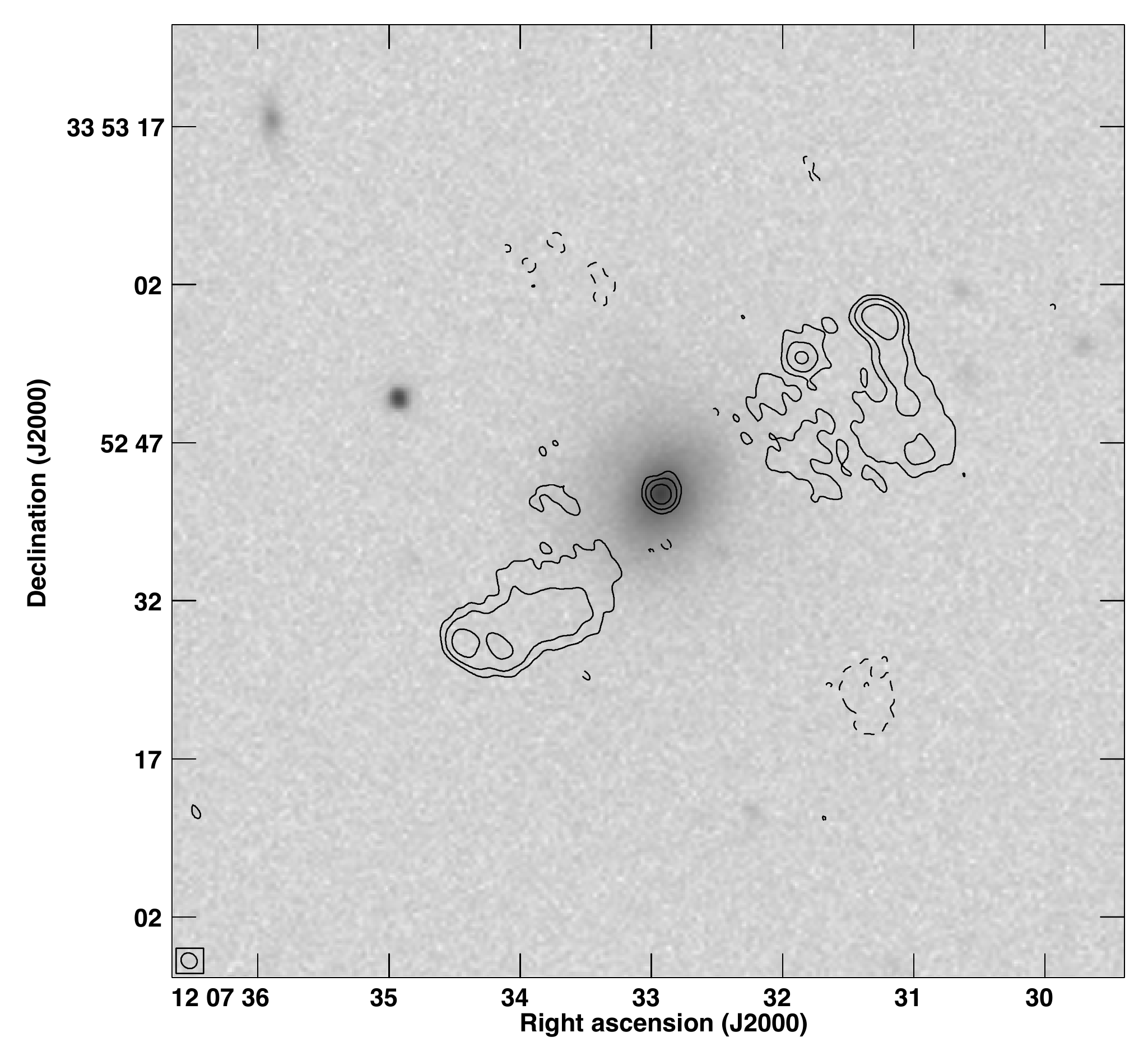}\\
\includegraphics[width=0.45\columnwidth]{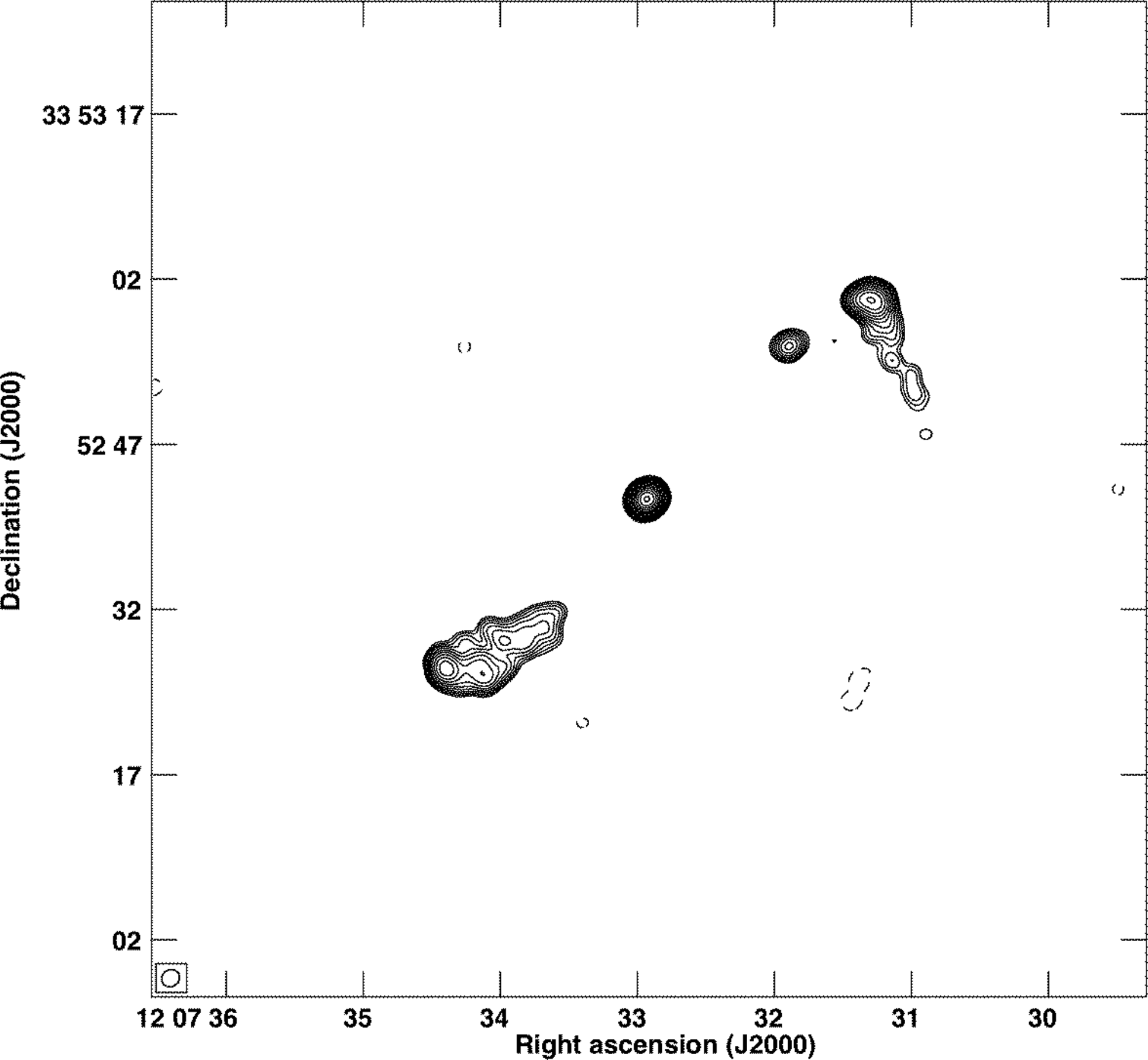}
\includegraphics[width=0.45\columnwidth]{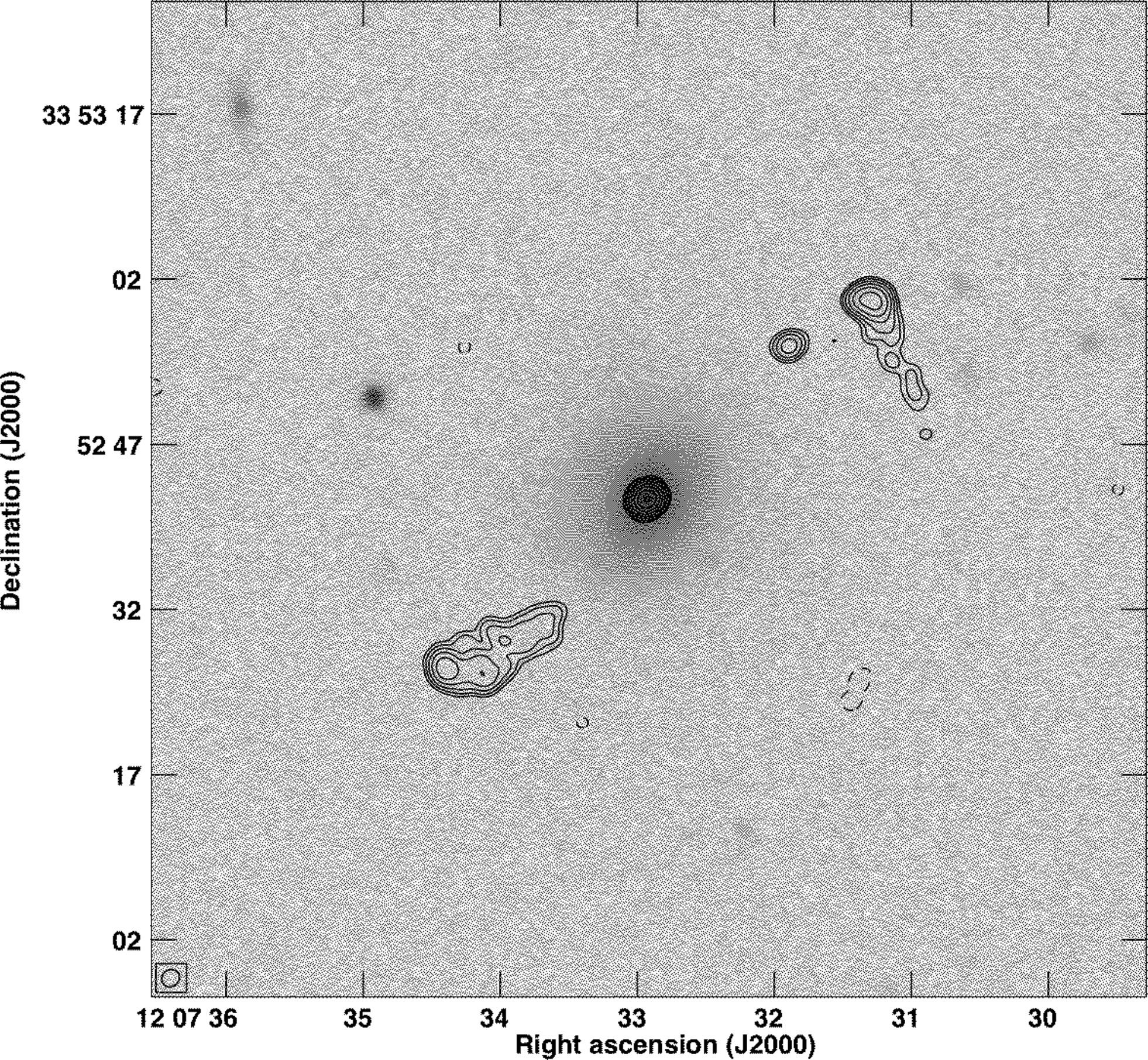}\\
\caption[J1207+3352 (L \& C)]{J1207+3352.  (top) (left) VLA image at L band and (right) VLA image overlaid on red SDSS image. Lowest contour = 0.3~mJy/beam, peak  = 25.2~mJy/beam. (bottom) (left) VLA image at C band and (right) VLA image overlaid on red SDSS image. Lowest contour = 0.1~mJy/beam, peak = 27.5~mJy/beam. \label{fig:J1207+3352}}
\end{figure}

\noindent J1207+3352 (Figure~\ref{fig:J1207+3352}). This source appears to be one where offset emission is seen originating both from inner end (of both lobes in the FIRST map) as well as outer end (of the NW lobe in the shown figure). The two recessed inner peaks in the two lobes are aligned with the core whereas the two outer peaks are separately aligned with the core. The two axes are separated by less than $10^\circ$.

\begin{figure}[ht] 
\includegraphics[width=0.45\columnwidth]{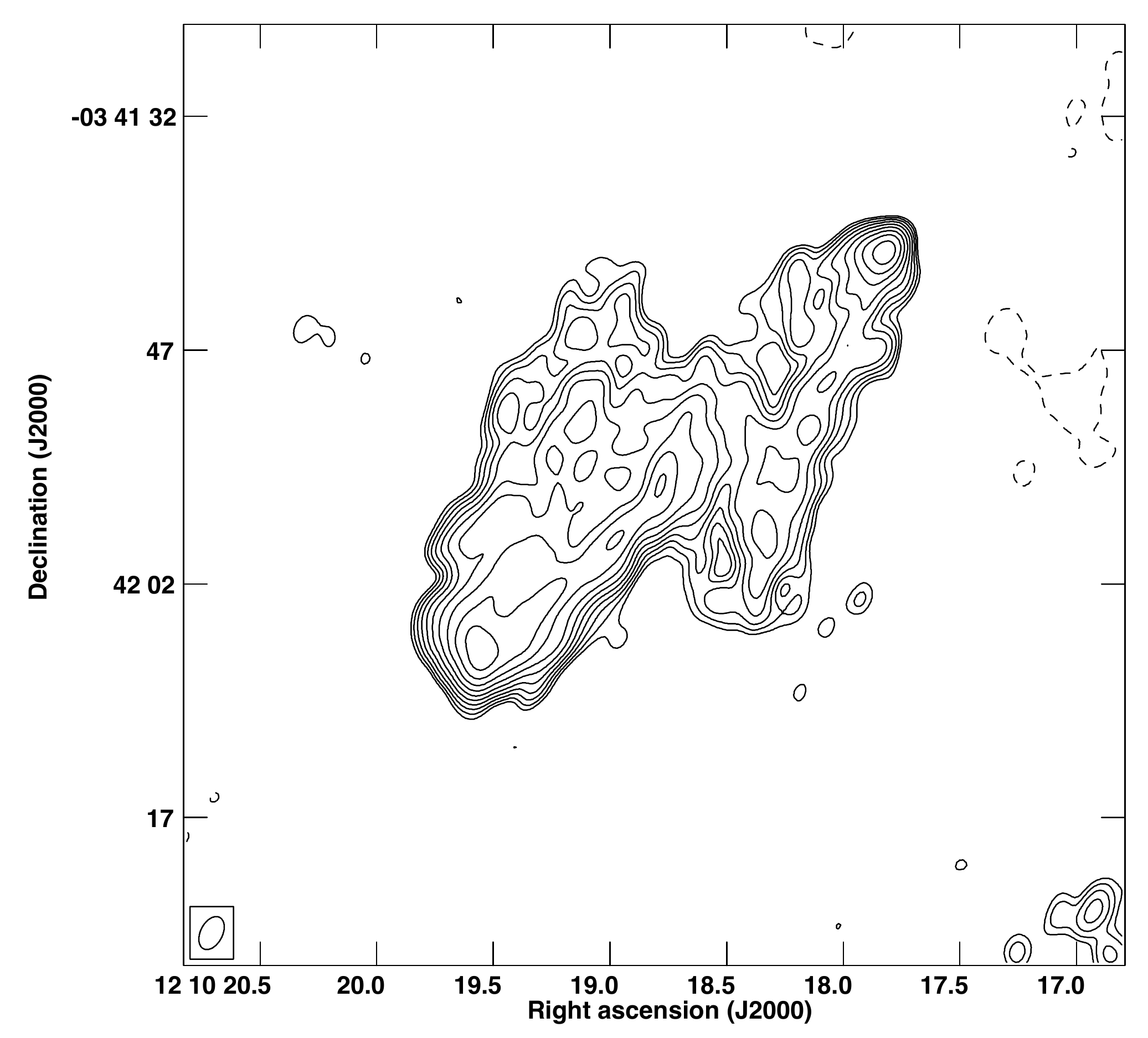}
\includegraphics[width=0.45\columnwidth]{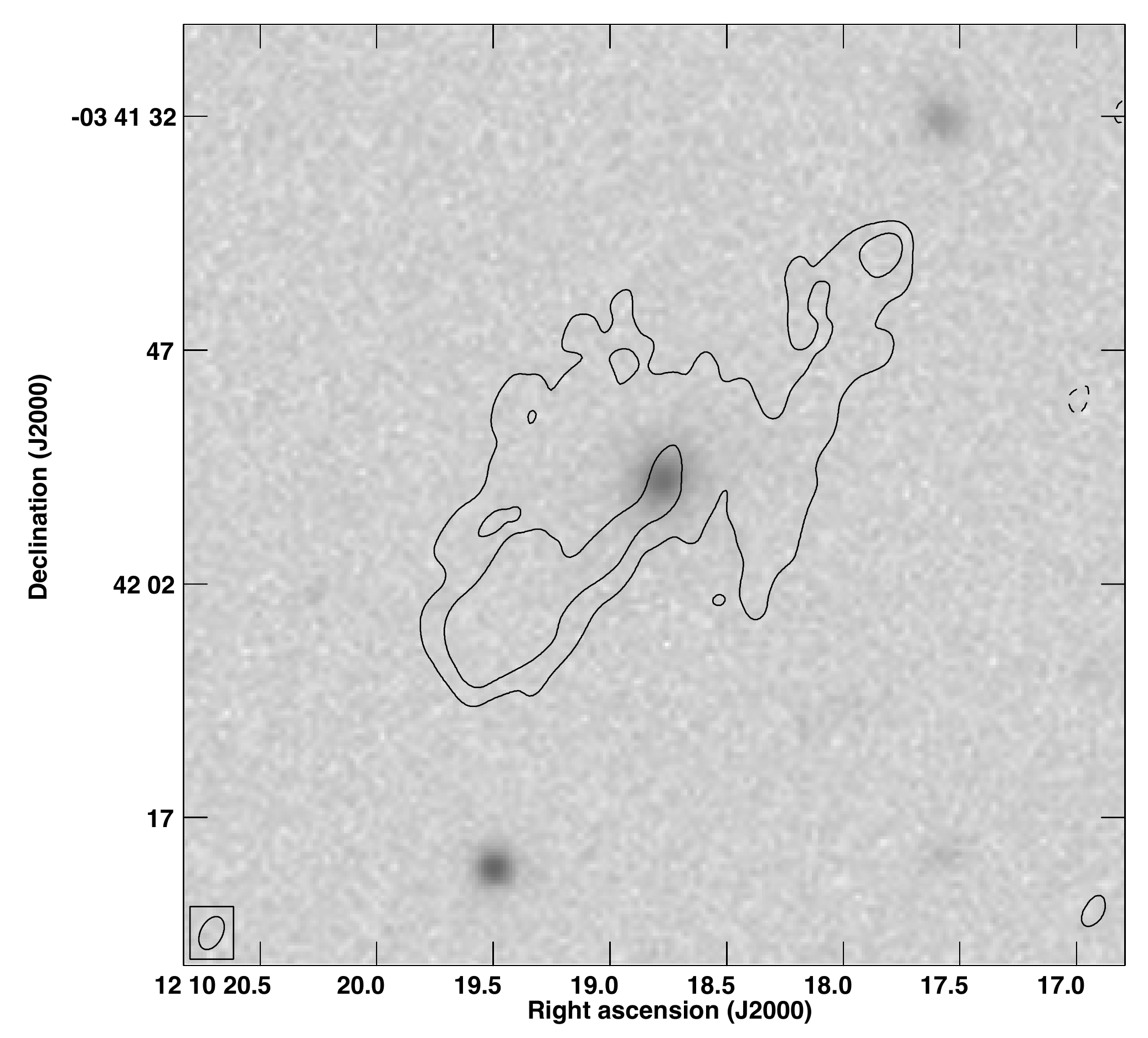}
\caption[J1210$-$0341 (L)]{J1210$-$0341.  (left) VLA image at L band, (right) VLA image overlaid on red SDSS image. Lowest contour = 0.1~mJy/beam, peak  = 2.95~mJy/beam.   \label{fig:J1210-0341}}
\end{figure}

\noindent J1210$-$0341 (Figure~\ref{fig:J1210-0341}). The new map reveals sharply bounded edge-brightened lobes connected to inversion-symmetric transverse emission features. A weak core is seen associated with the central bright galaxy connected to the SE hotspot by a jet. 

\begin{figure}[ht] 
\includegraphics[width=0.45\columnwidth]{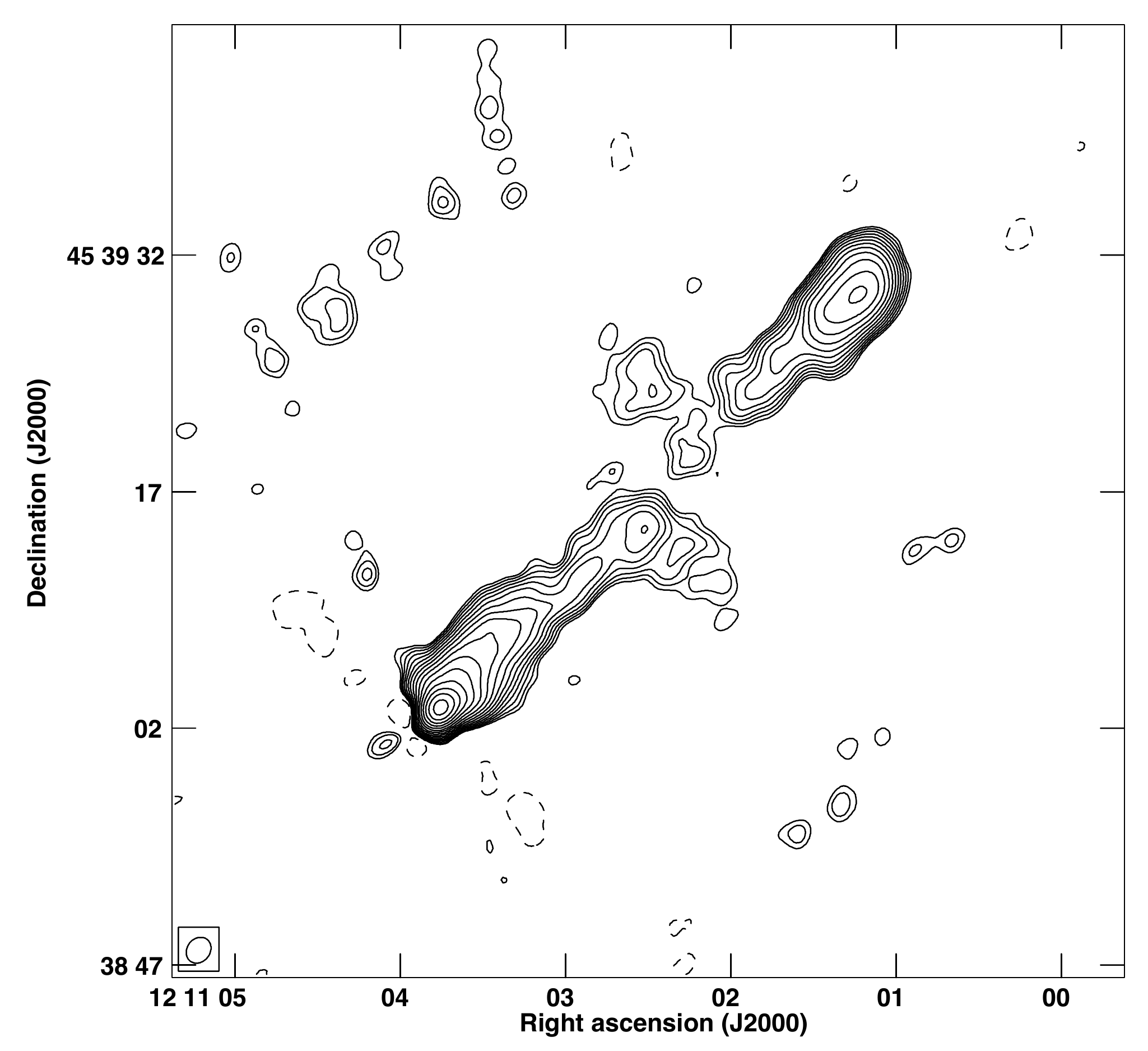}
\includegraphics[width=0.45\columnwidth]{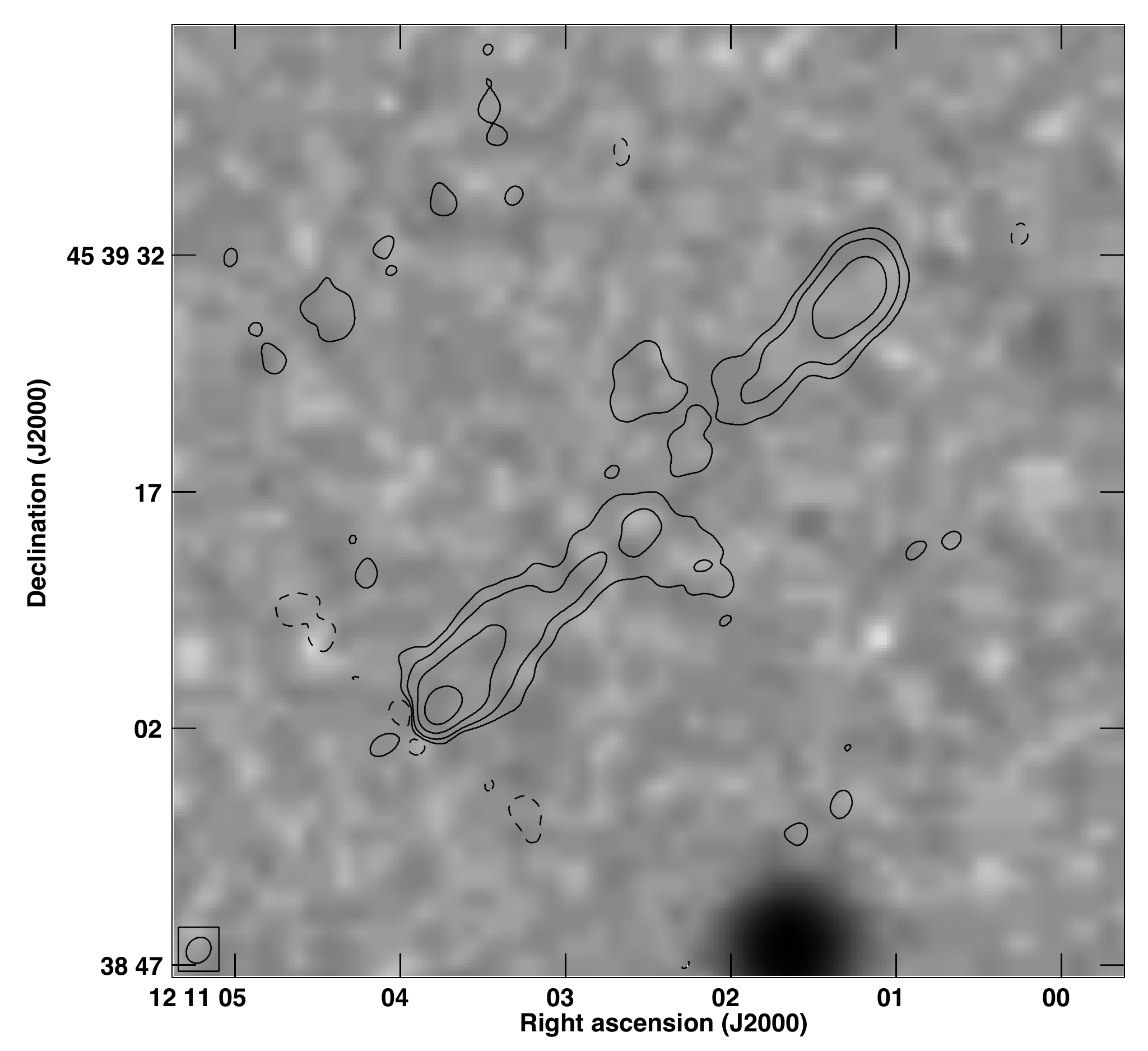}
\caption[J1211+4539 (L)]{J1211+4539. (left) VLA image at L band, (right) VLA image overlaid on red SDSS image. Lowest contour =  0.2~mJy/beam, peak = 43.6~mJy/beam. \label{fig:J1211+4539}}
\end{figure}

\noindent J1211+4539 (Figure~\ref{fig:J1211+4539}). This source is similar to J1206+3812 (Figure \ref{fig:J1206+3812}). The two lobes have high axial ratios, which have short orthogonal extensions close to the center. No core is detected in the new map. 

\begin{figure}[ht] 
\includegraphics[width=0.45\columnwidth]{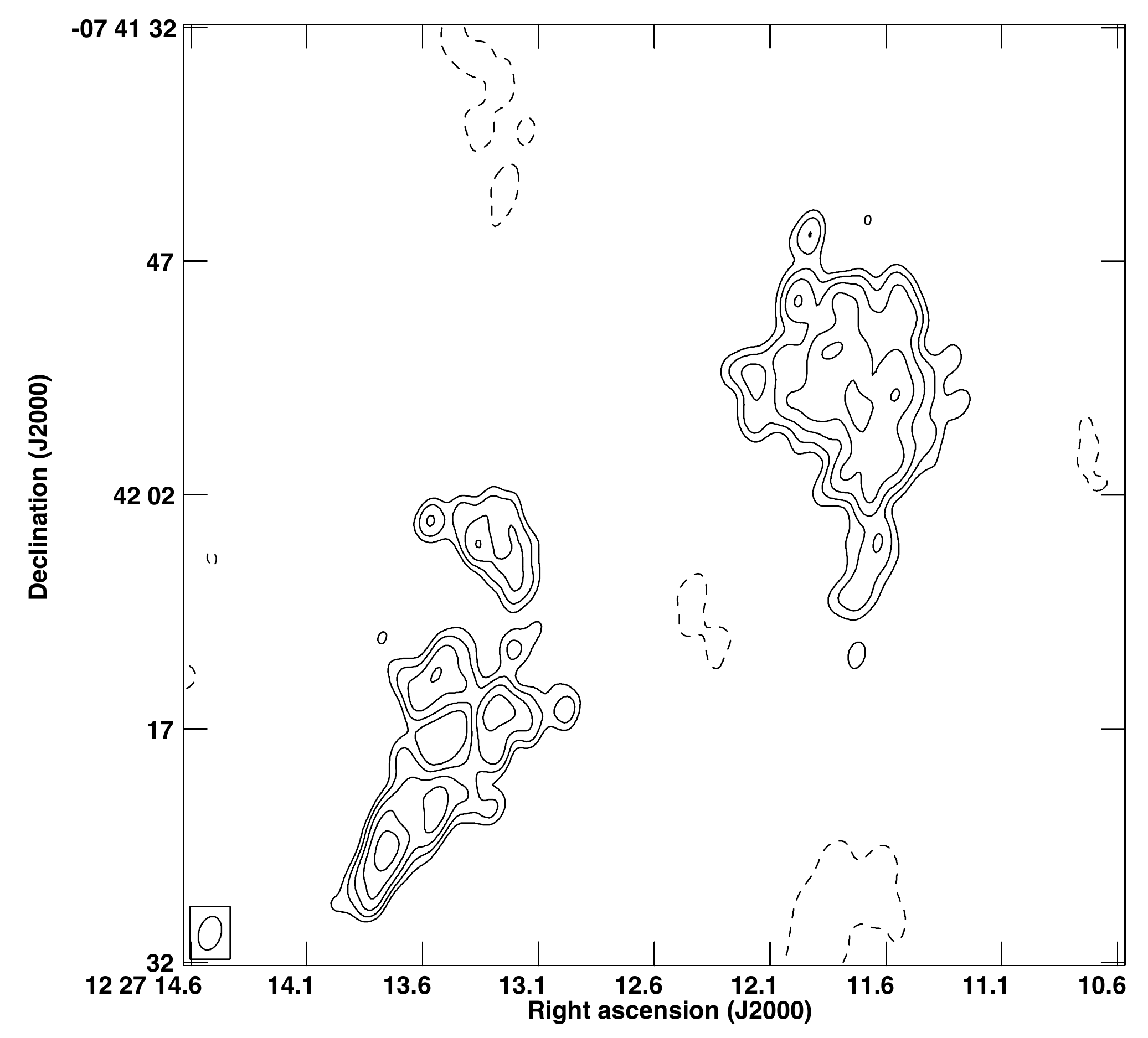}
\includegraphics[width=0.45\columnwidth]{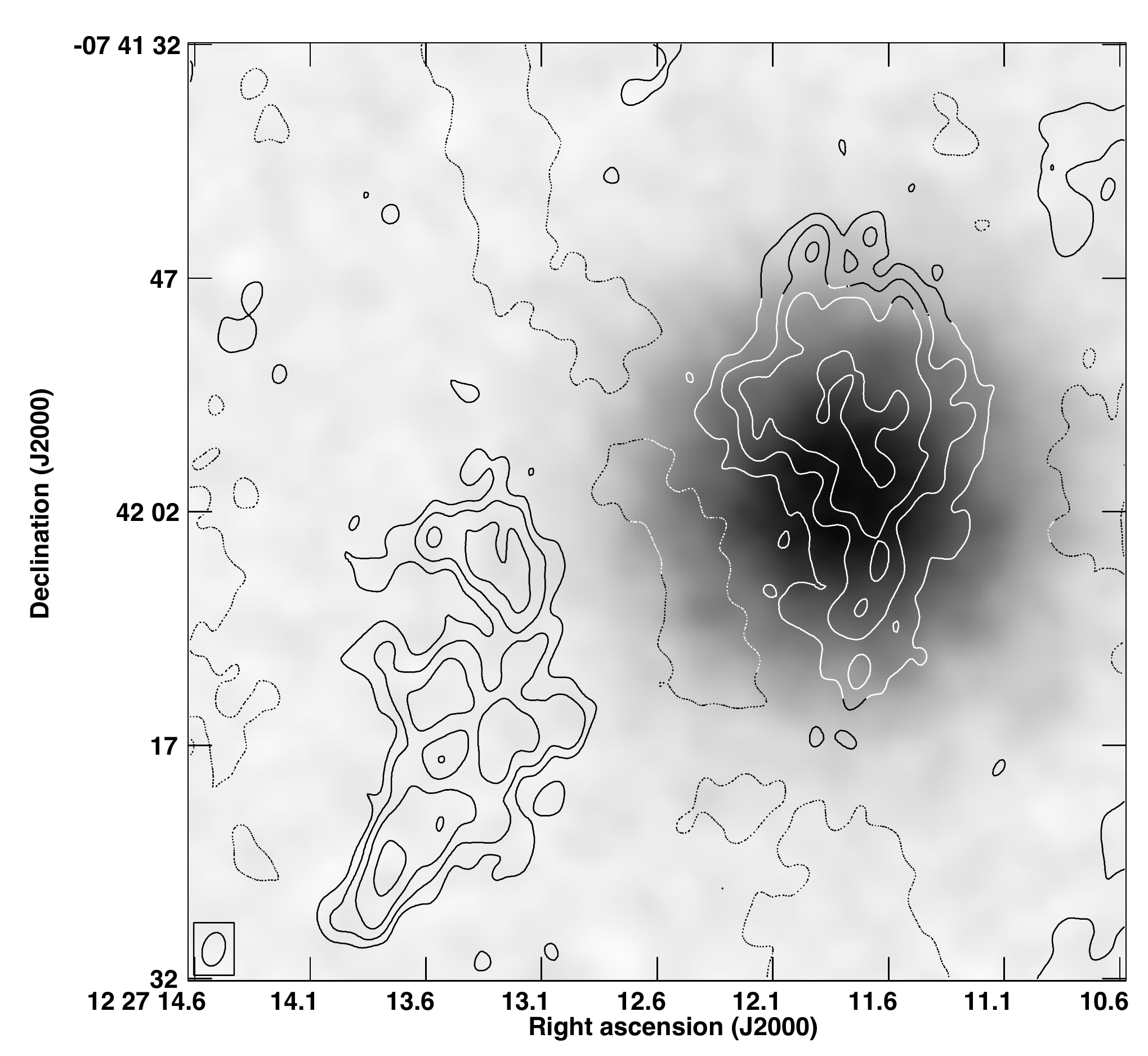}
\caption[J1227$-$0742 (L)]{J1227$-$0742. (left) VLA image at L band, (right) VLA image overlaid on red SDSS image. Lowest contour = 0.18~mJy/beam, peak  = 0.86~mJy/beam. \label{fig:J1227-0742}}
\end{figure}

\noindent J1227$-$0742 (Figure~\ref{fig:J1227-0742}). The map misses much of the extended flux seen in the FIRST image. While a hotspot is detected at the SE lobe end no core is detected nor is a hotspot seen in the western lobe. The map shows a transverse emission feature associated with the SE lobe.

\begin{figure}[ht] 
\includegraphics[width=0.45\columnwidth]{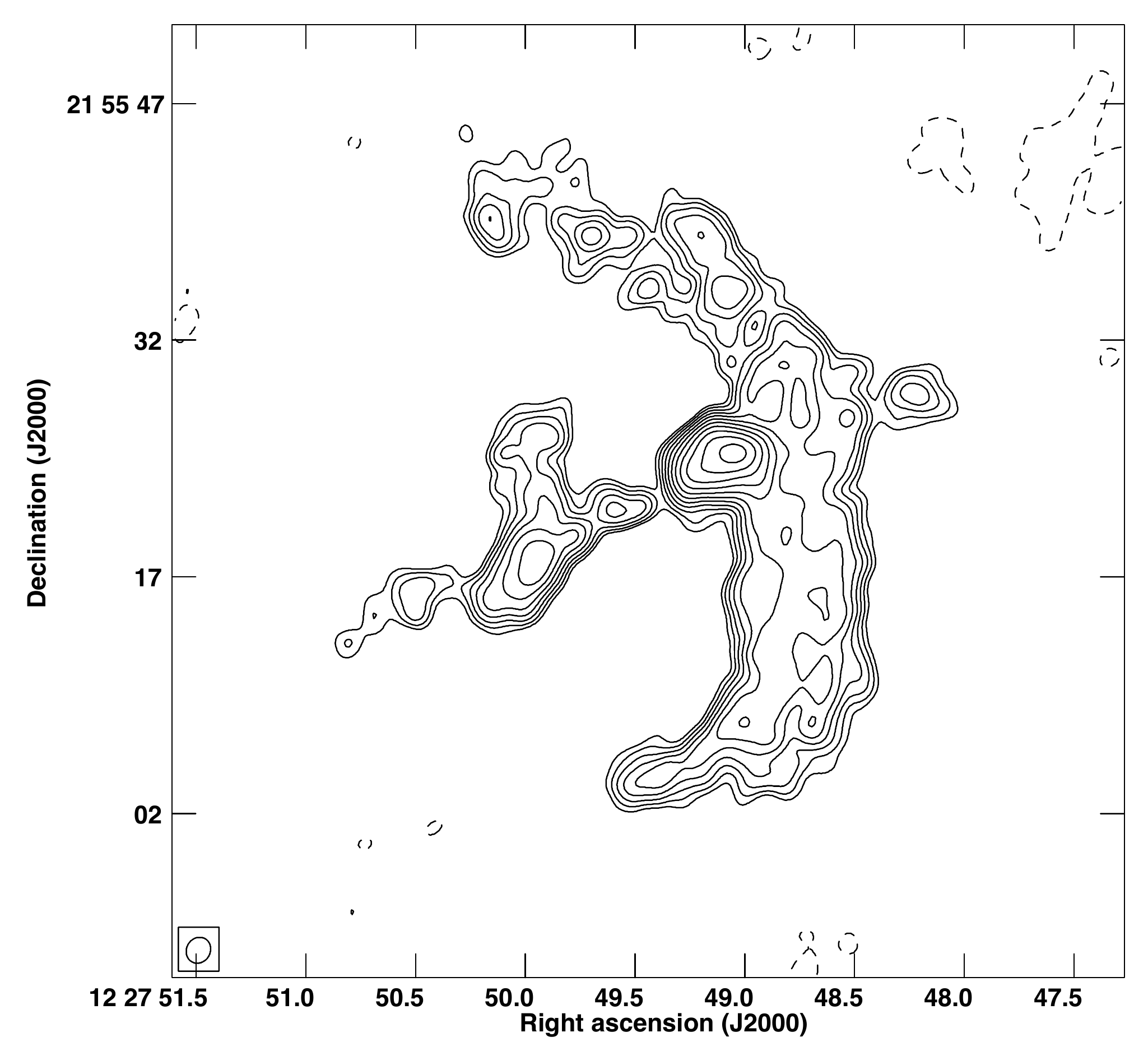}
\includegraphics[width=0.45\columnwidth]{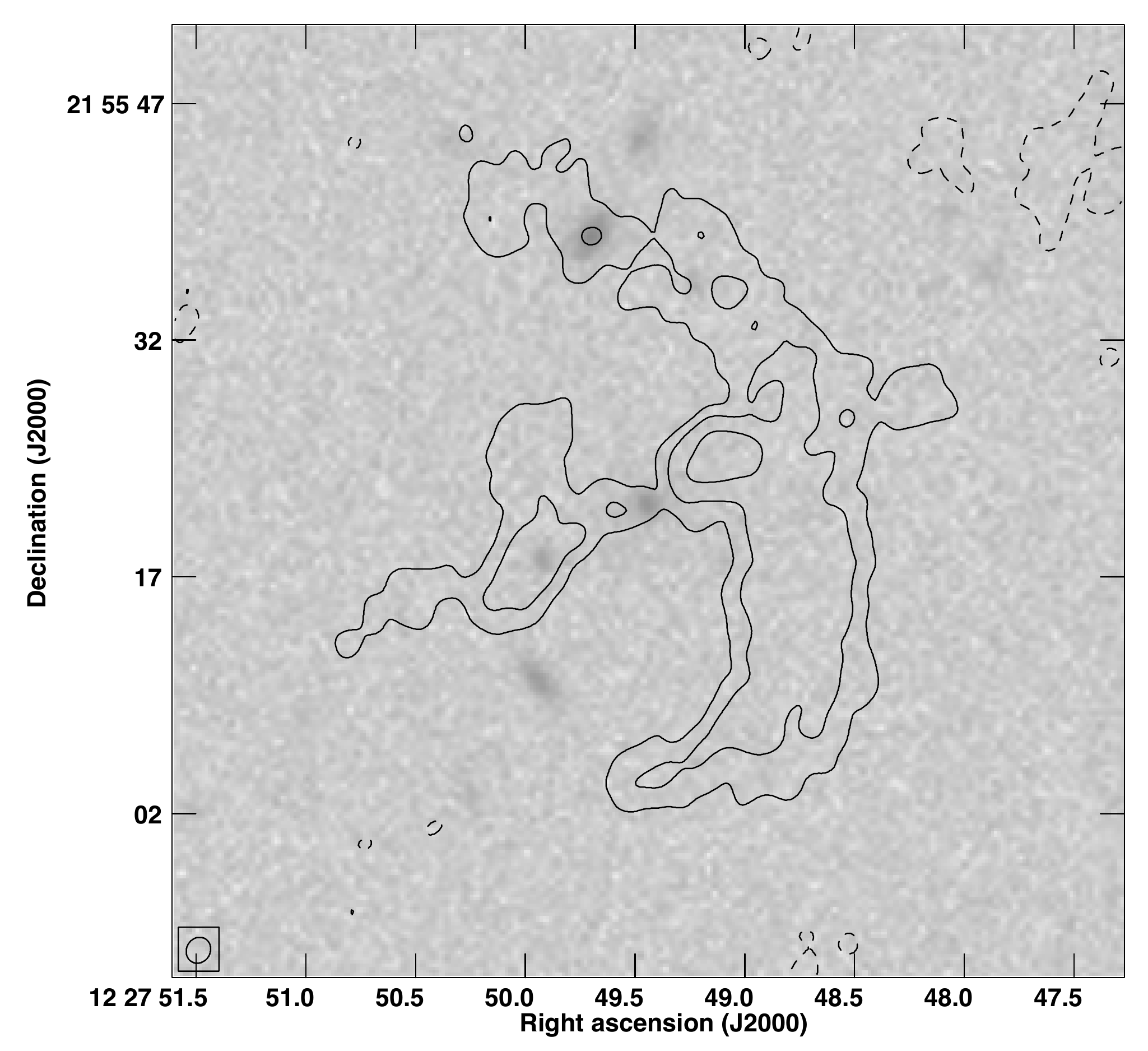}
\caption[J1227+2155 (L)]{J1227+2155. (left) VLA image at L band, (right) VLA image overlaid on red SDSS image.  Lowest contour = 0.1~mJy/beam, peak  = 3.74~mJy/beam. \label{fig:J1227+2155}}
\end{figure}

\noindent J1227+2155 (Figure~\ref{fig:J1227+2155}). A prominent feature revealed in the map is a narrow, extended arc of emission running orthogonal to another narrow
and elongated emission with prominent emission peaks, a likely radio galaxy. At its center a radio core is detected at the location of the optical object. The physical association of the curved extended feature with the radio galaxy is unclear.

\begin{figure}[ht] 
\includegraphics[width=0.45\columnwidth]{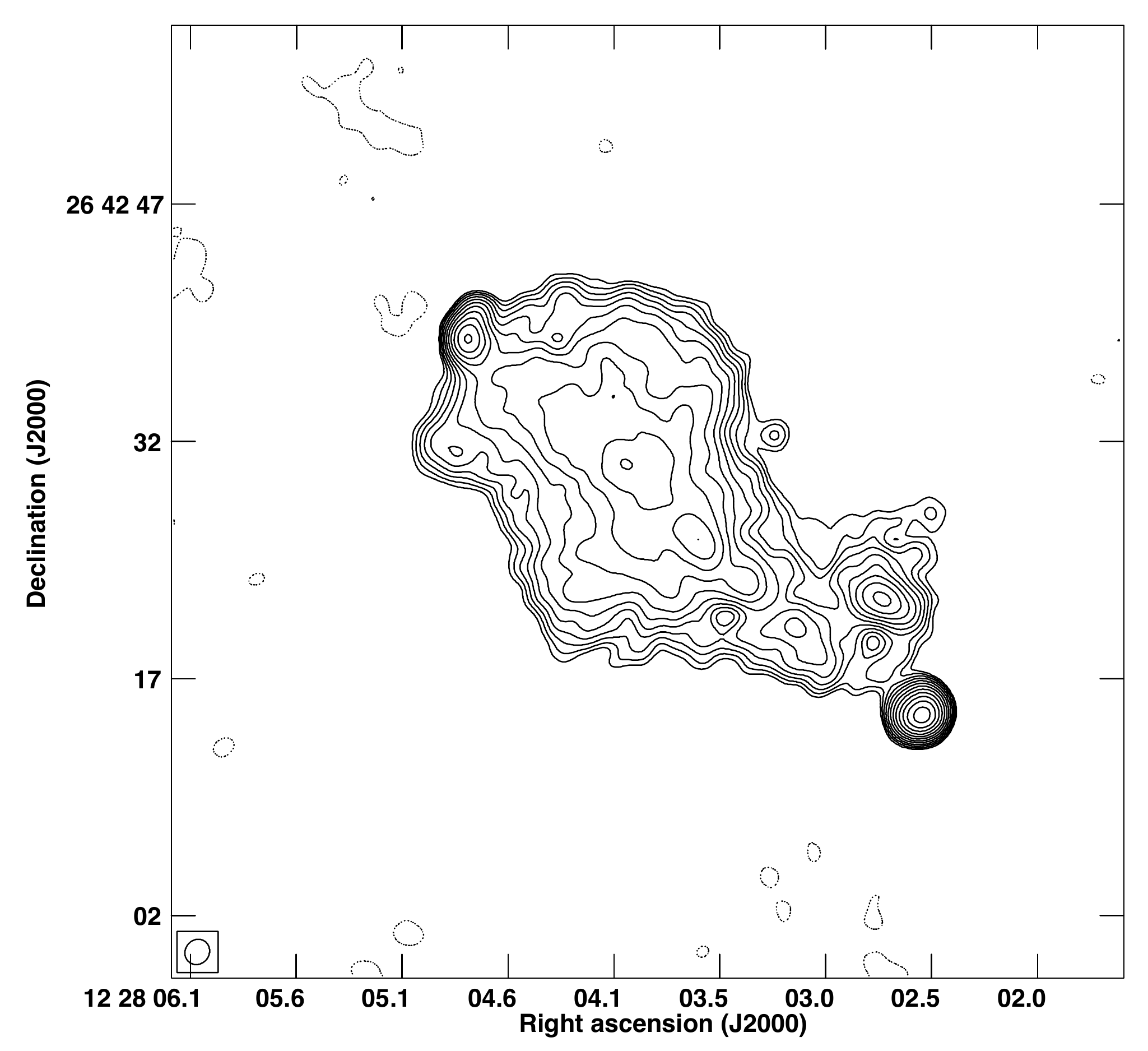}
\includegraphics[width=0.45\columnwidth]{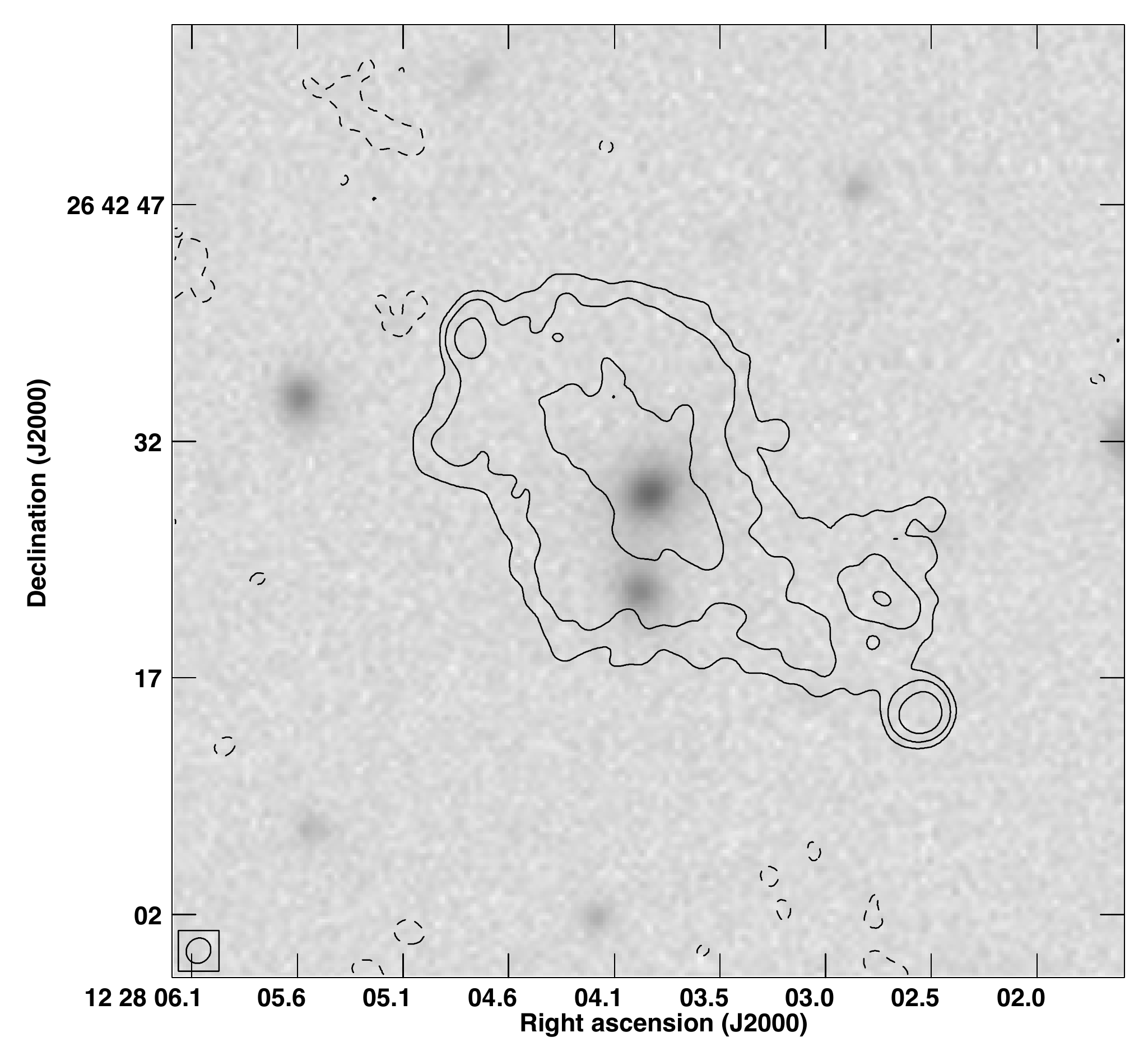}
\caption[J1228+2642 (L)]{J1228+2642. (left) VLA image at L band, (right) VLA image overlaid on red SDSS image. Lowest contour = 0.08~mJy/beam, peak  = 4.37~mJy/beam. \label{fig:J1228+2642}}
\end{figure}

\noindent J1228+2642 (Figure~\ref{fig:J1228+2642}). Our map reveals distinct hotspots at the extremities of the source. Although a distinct core is not seen a local emission peak in the center coincides with a bright elliptical galaxy. The lobes are well-bounded but are not distinct and instead form a continuous bridge between the two hotspots. 

\begin{figure}[ht] 
\includegraphics[width=0.45\columnwidth]{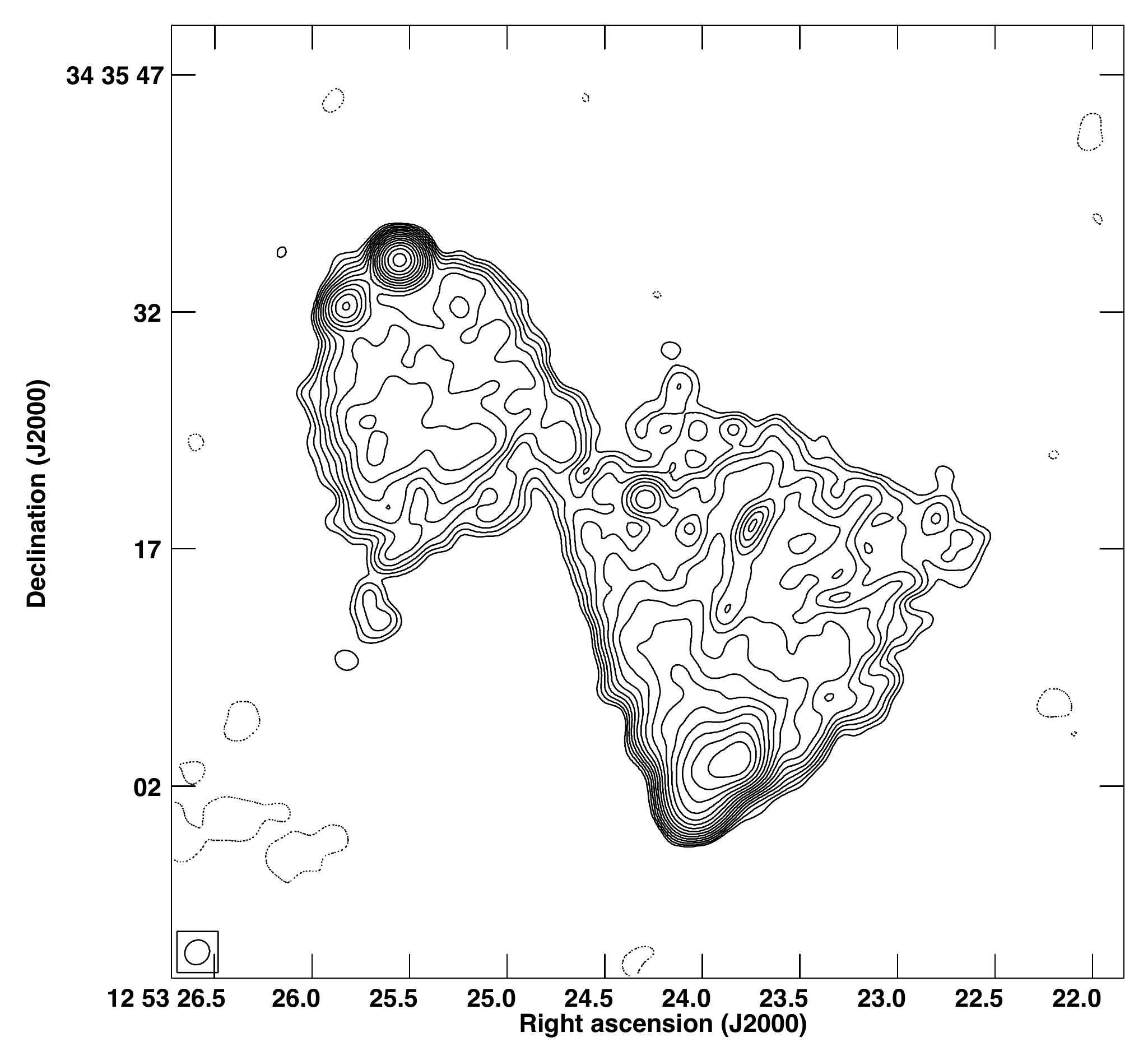}
\includegraphics[width=0.45\columnwidth]{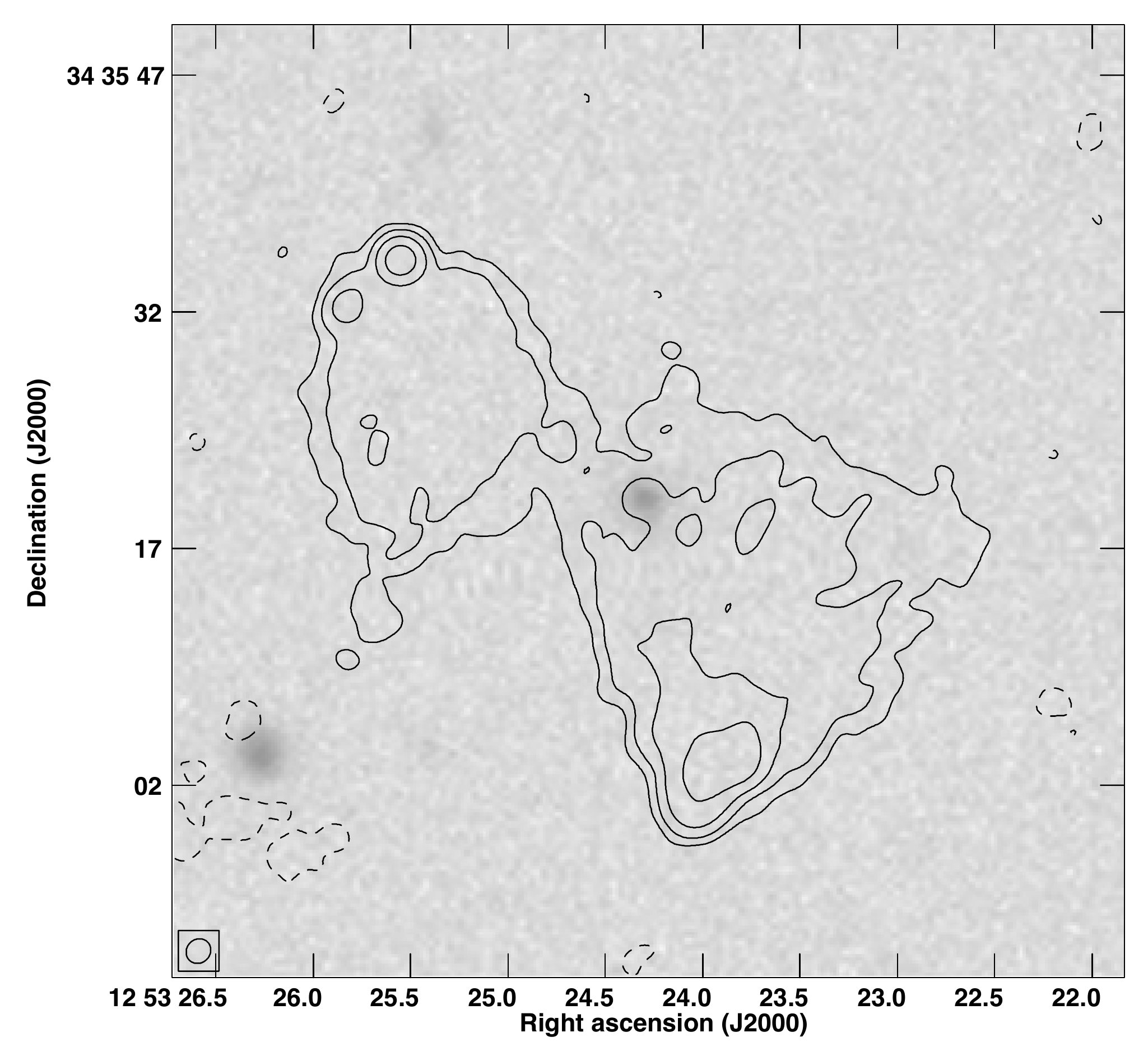}
\caption[J1253+3435 (L)]{J1253+3435.  (left) VLA image at L band, (right) VLA image overlaid on red SDSS image. Lowest contour = 0.09~mJy/beam, peak  = 15.8~mJy/beam. \label{fig:J1253+3435}}
\end{figure}

\noindent J1253+3435 (Figure~\ref{fig:J1253+3435}). A distinct core and clear hotspots are revealed for this source. The NE lobe has a pair of distinct hotspots at its extremity where as the SW lobe has an extended emission peak at the lobe end indicative of a pair of hotspots. The source is skewed, with both hotspot pairs displaced to the east with respect to the core. The lobes show sharp boundaries. 

\begin{figure}[ht] 
\includegraphics[width=0.45\columnwidth]{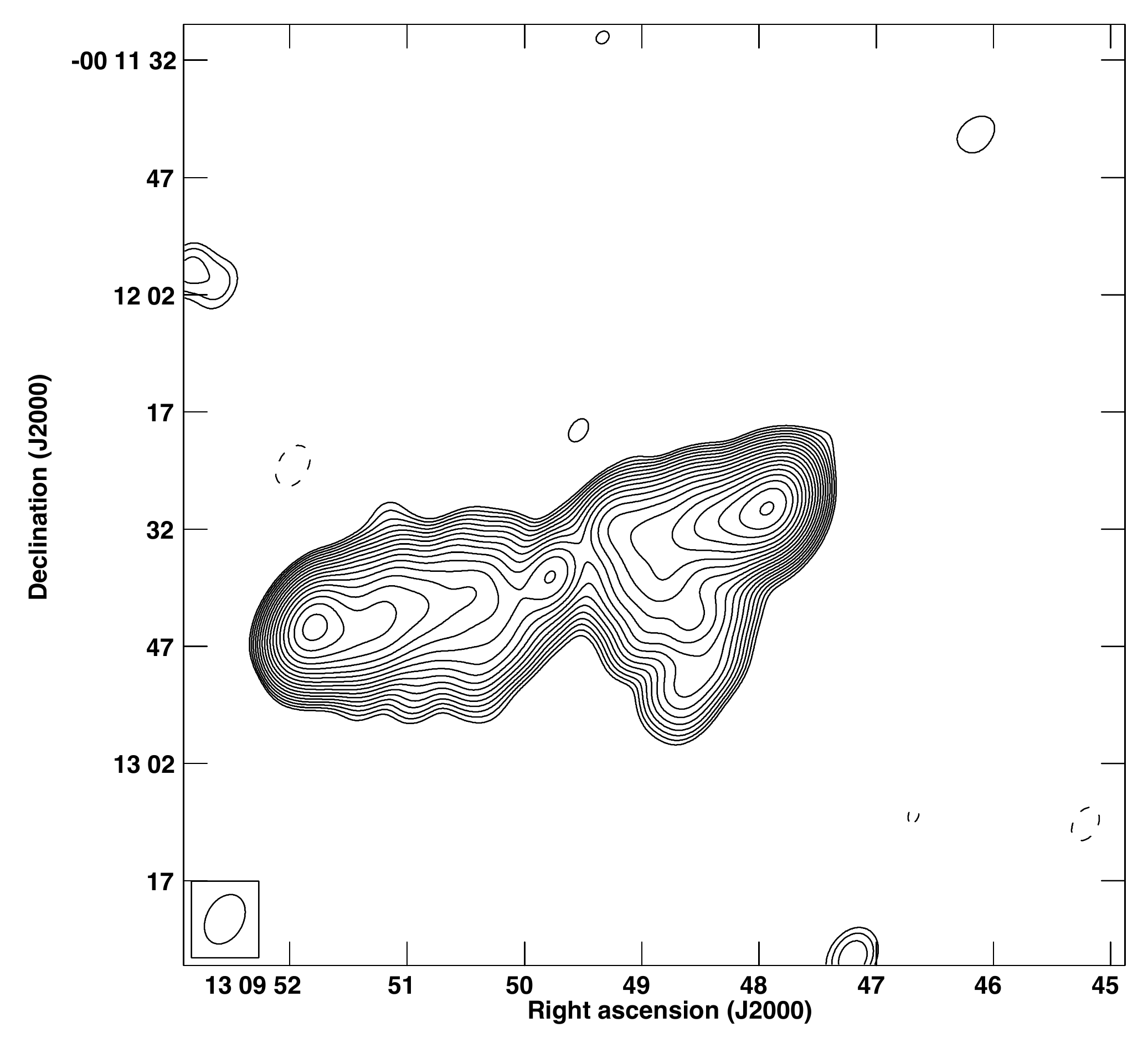}
\includegraphics[width=0.45\columnwidth]{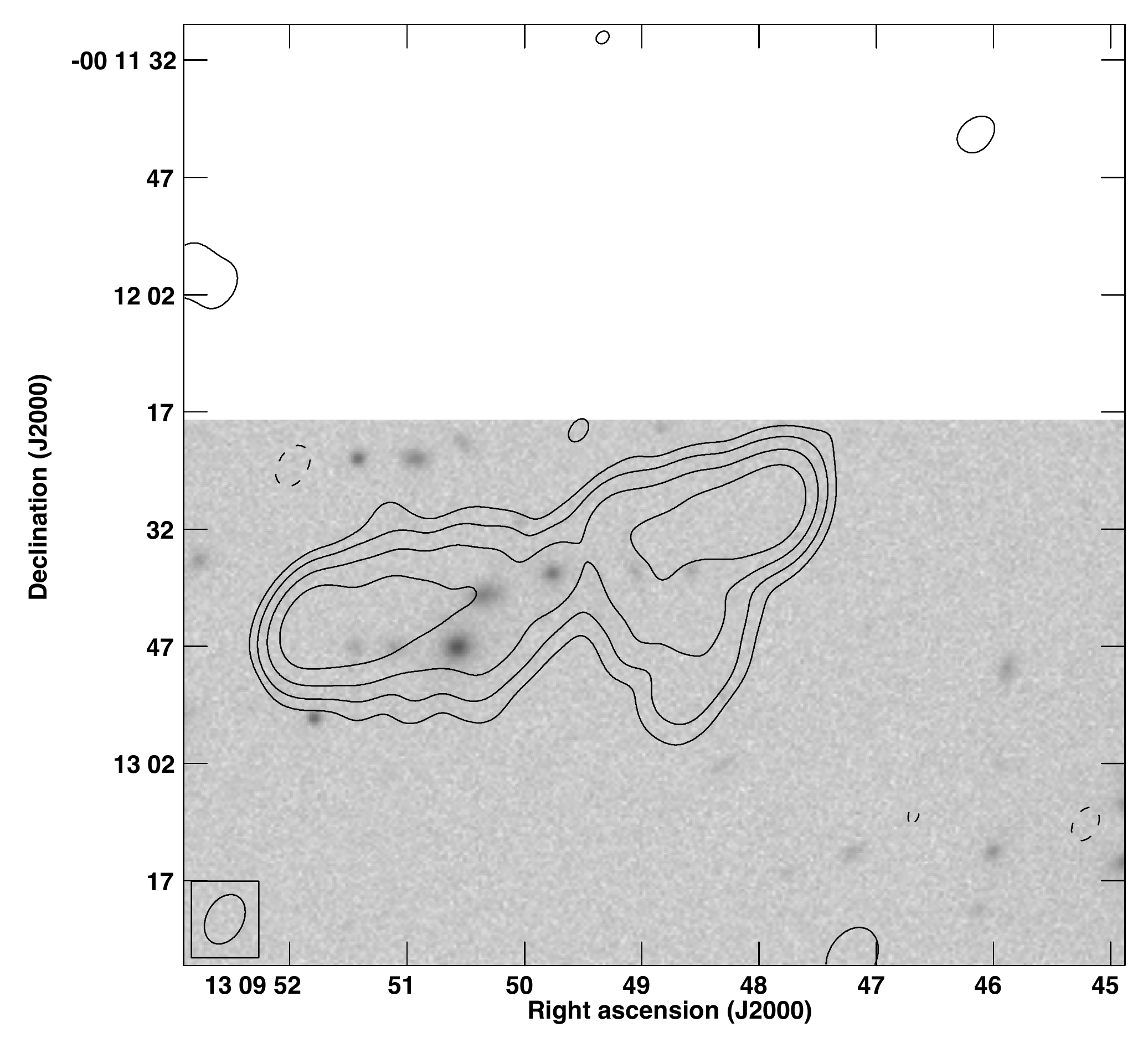}
\caption[J1309$-$0012 (C)]{J1309$-$0012. (left) VLA image at C band, (right) VLA image overlaid on red SDSS image. Lowest contour = 0.3~mJy/beam, peak  = 113~mJy/beam. \label{fig:J1309-0012}}
\end{figure}

\noindent J1309$-$0012 (Figure~\ref{fig:J1309-0012}). As seen in the FIRST image, transverse emission is seen associated only with the western lobe; this has a sharp inner edge that is almost orthogonal to the source axis.
 
\begin{figure}[ht] 
\includegraphics[width=0.45\columnwidth]{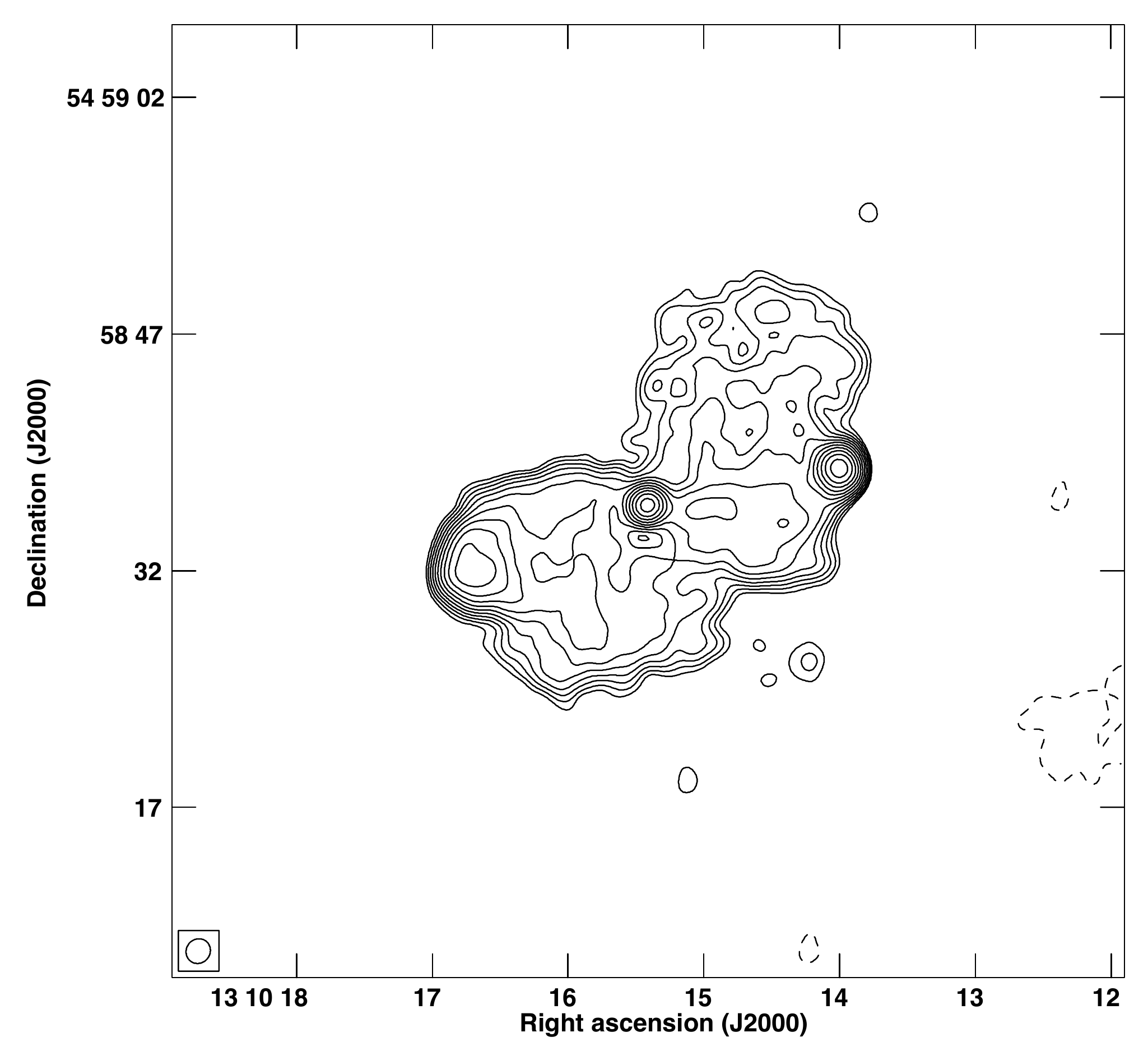}
\includegraphics[width=0.45\columnwidth]{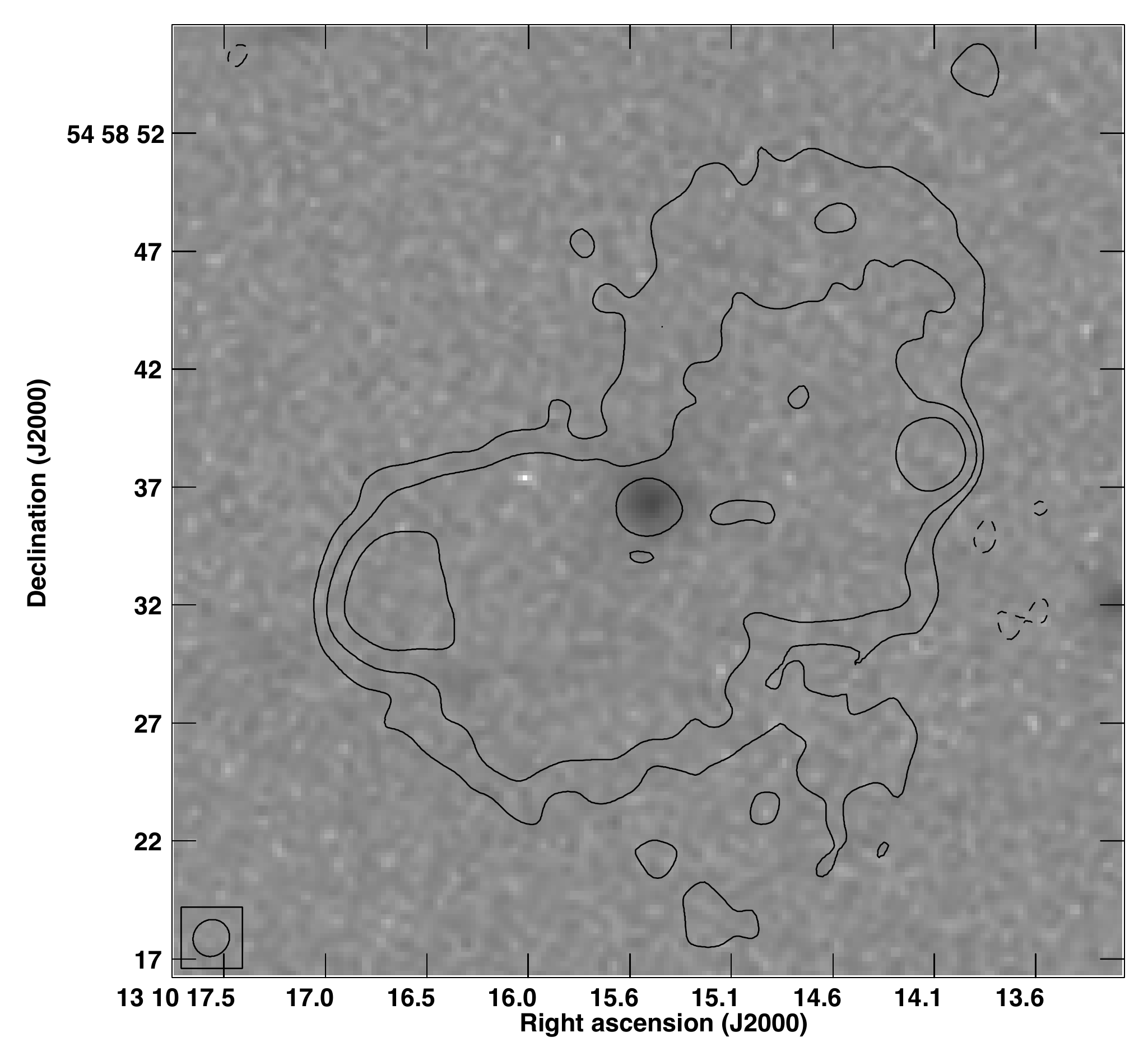}
\caption[J1310+5458 (L)]{J1310+5458.  (left) VLA image at L band, (right) VLA image overlaid on red SDSS image.  Lowest contour = 0.2~mJy/beam, peak  = 11.4~mJy/beam. \label{fig:J1310+5458}}
\end{figure}

\noindent J1310+5458 (Figure~\ref{fig:J1310+5458}). A strong core and hotspots are seen in our high-resolution map. The source has prominent lobes with low axial ratio. 

\begin{figure}[ht] 
\includegraphics[width=0.45\columnwidth]{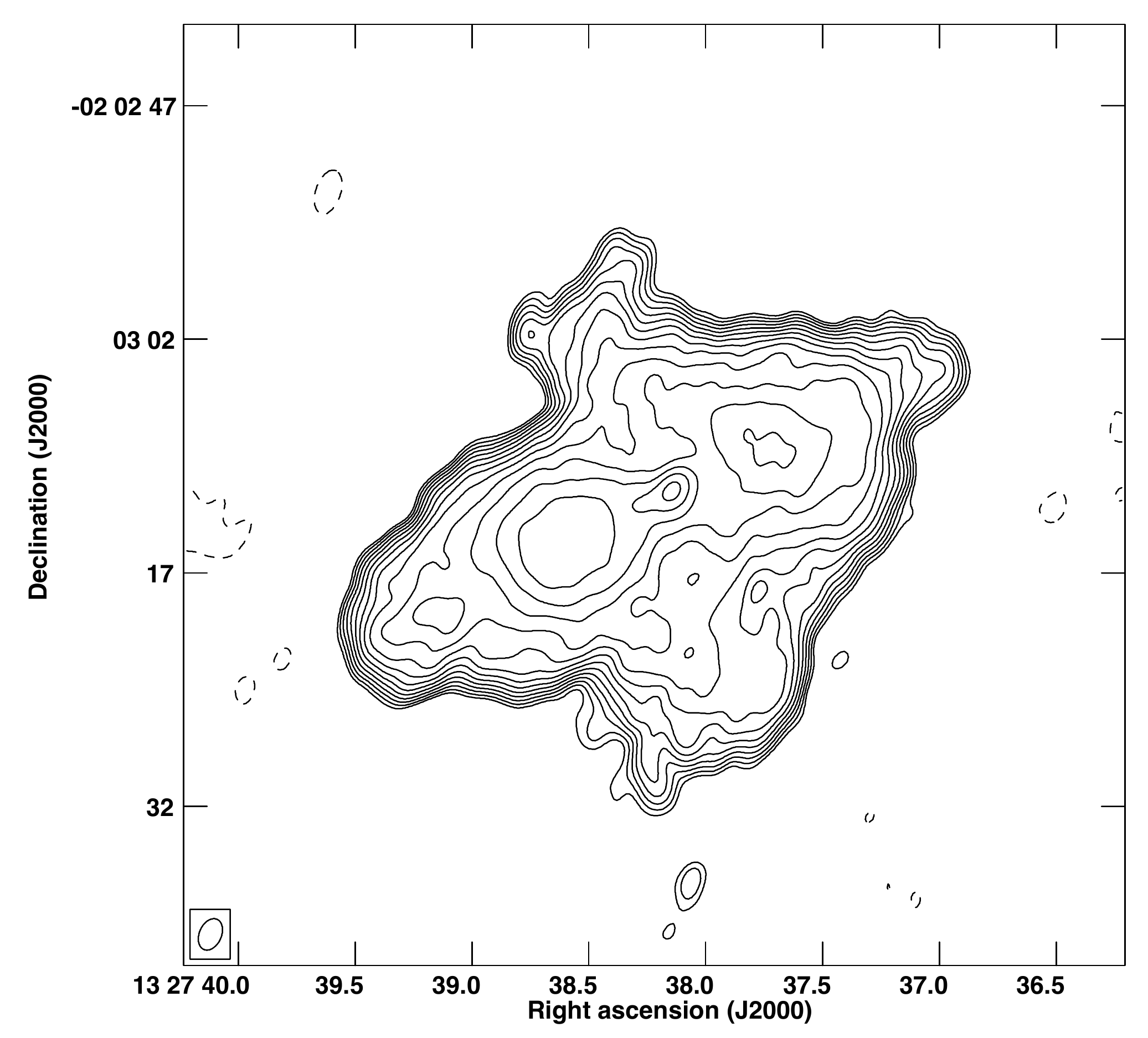}
\includegraphics[width=0.45\columnwidth]{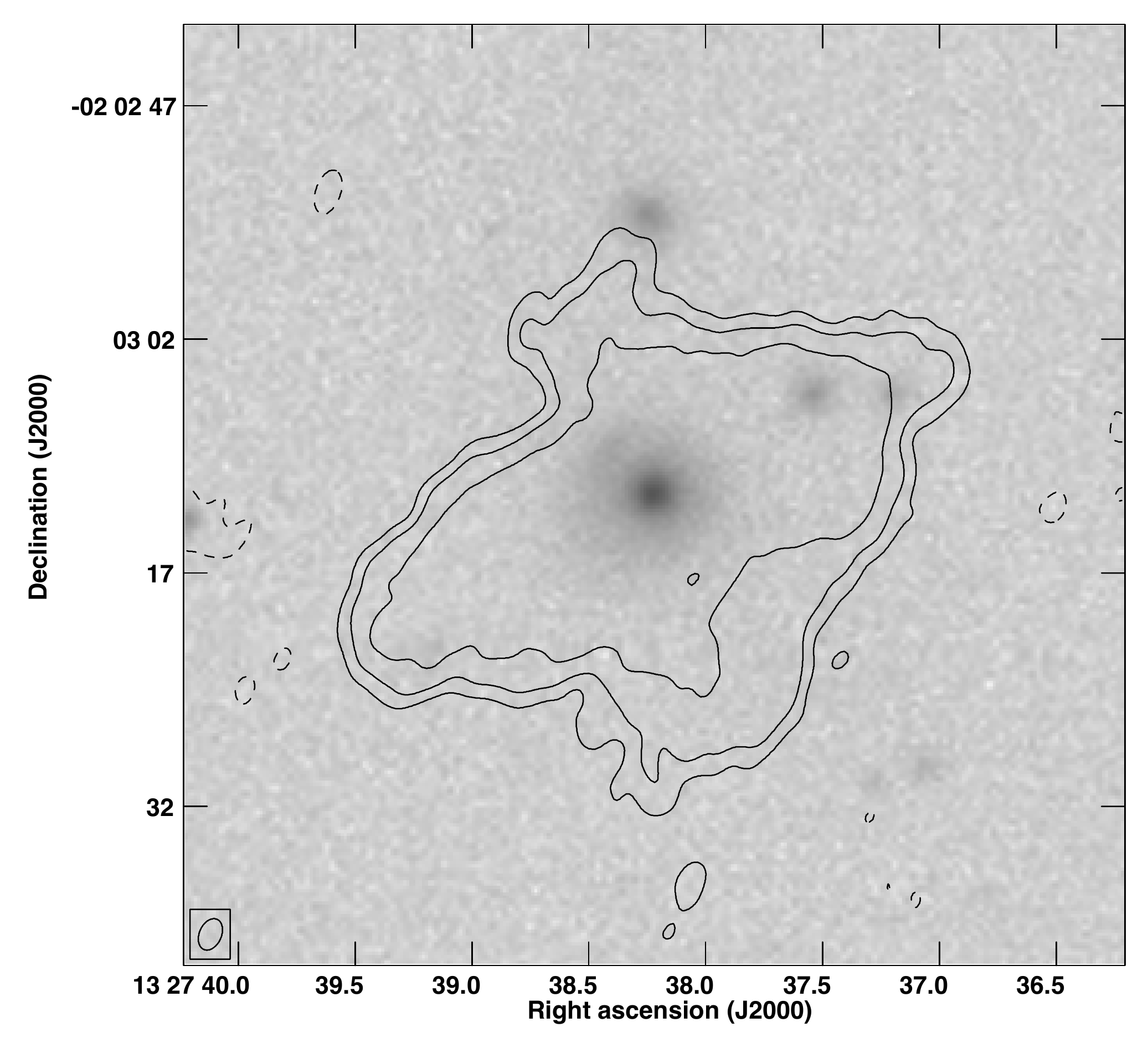}
\caption[J1327$-$0203 (L)]{J1327$-$0203. (left) VLA image at L band, (right) VLA image overlaid on red SDSS image. Lowest contour = 0.2~mJy/beam, peak  = 20.9~mJy/beam. \label{fig:J1327-0203}}
\end{figure}

\noindent J1327$-$0203 (Figure~\ref{fig:J1327-0203}).  Our new map reveals a central core. The map shows prominent, well-bounded lobes that are devoid of compact hotspots. The transverse, central protrusions are also seen to be well bounded.

\begin{figure}[ht] 
\includegraphics[width=0.45\columnwidth]{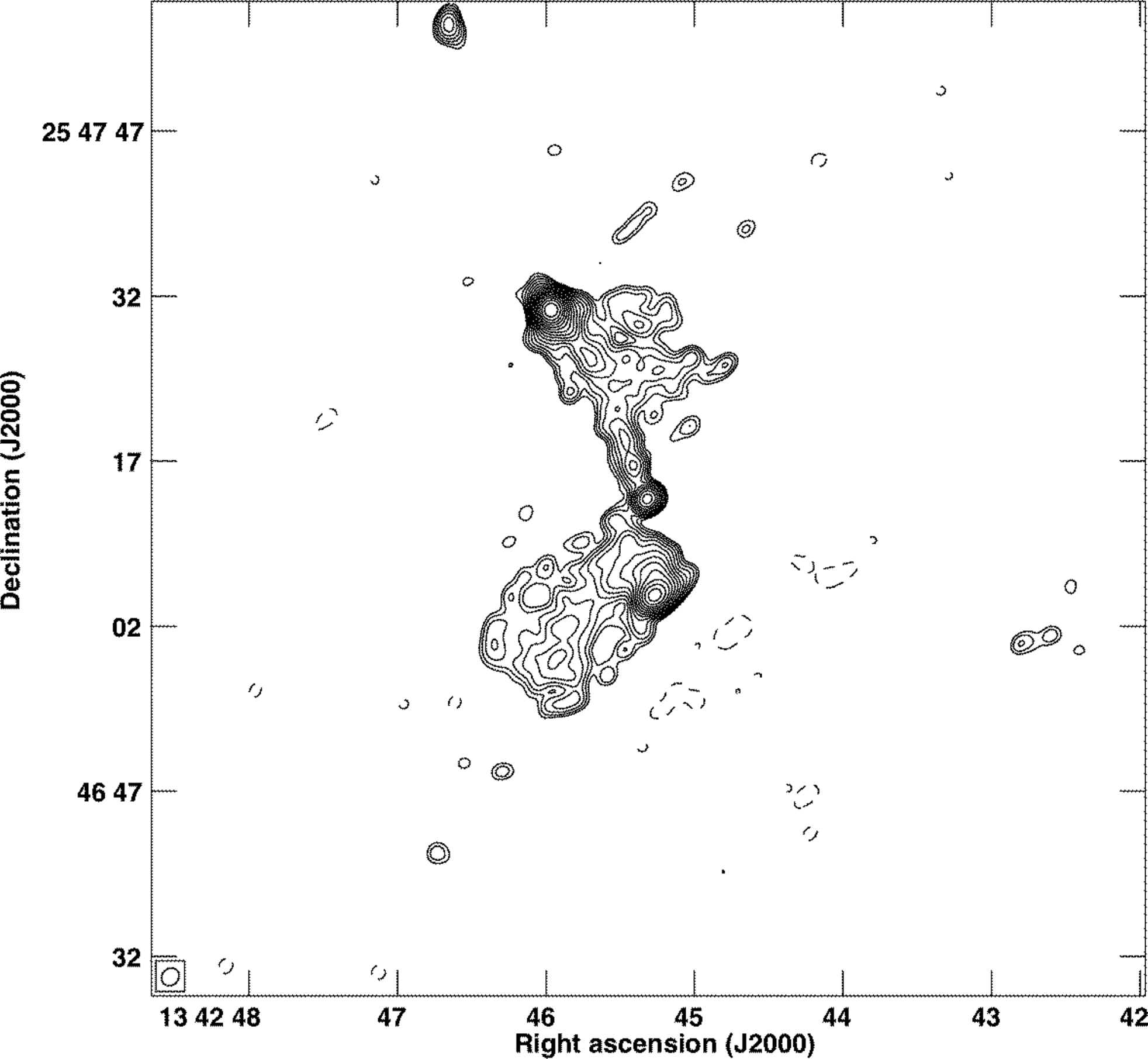}
\includegraphics[width=0.45\columnwidth]{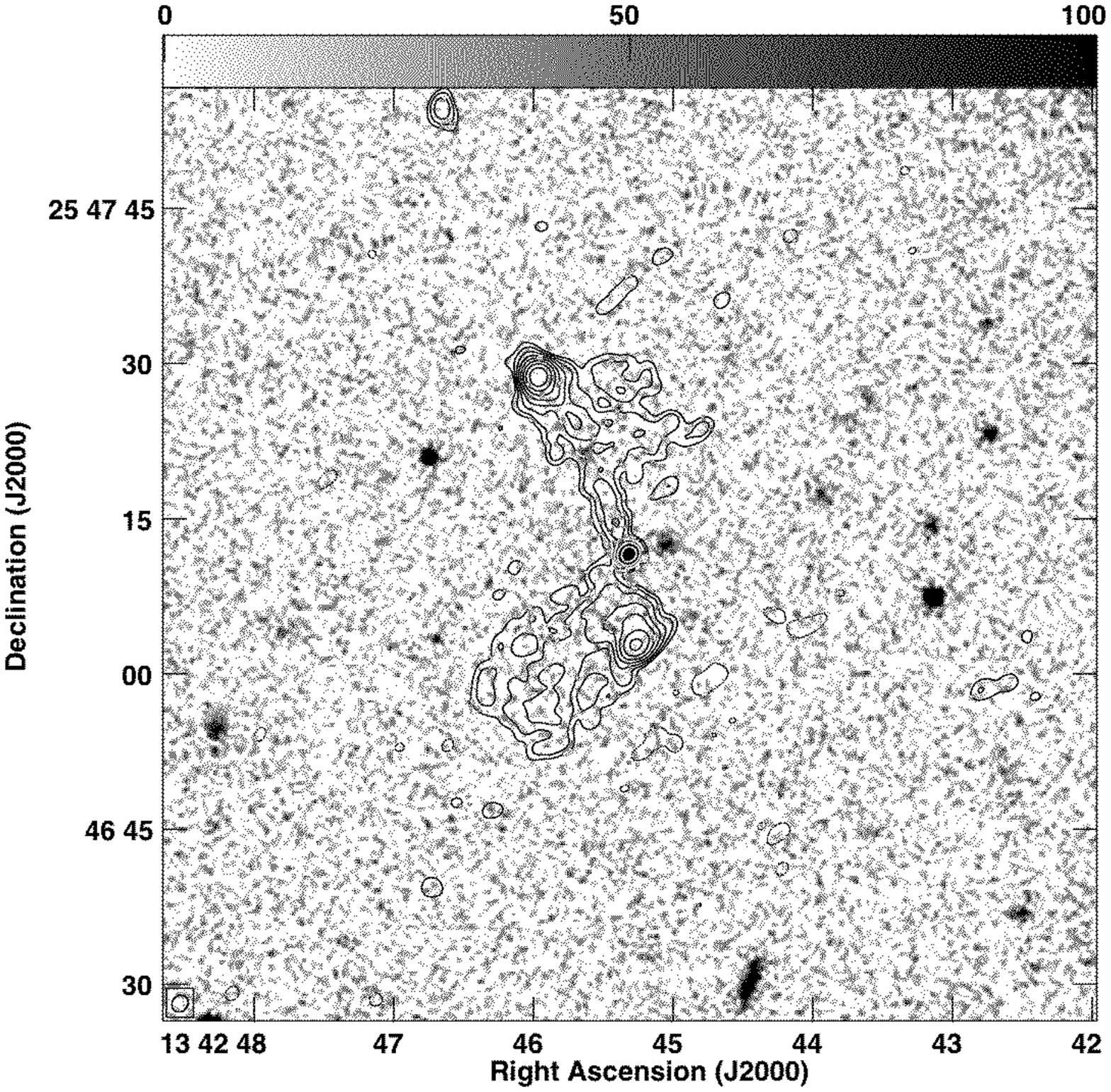}
\caption[J1342+2547 (L)]{J1342+2547. (left) VLA image at L band, (right) VLA image overlaid on red SDSS image. Lowest contour = 0.3~mJy/beam, peak = 72.5~mJy/beam.  \label{fig:J1342+2547}}
\end{figure}

\noindent J1342+2547 (Figure~\ref{fig:J1342+2547}). The source shows several signs of projection effects: strong core, the asymmetry in separation of the hotspots from the core, and the presence of a narrow jet on the side of the more distant hotspot. The southern lobe is traced as a distinct, broad feature that is extended at an angle of nearly $120^\circ$ from the source axis.

\begin{figure}[ht] 
\includegraphics[width=0.45\columnwidth]{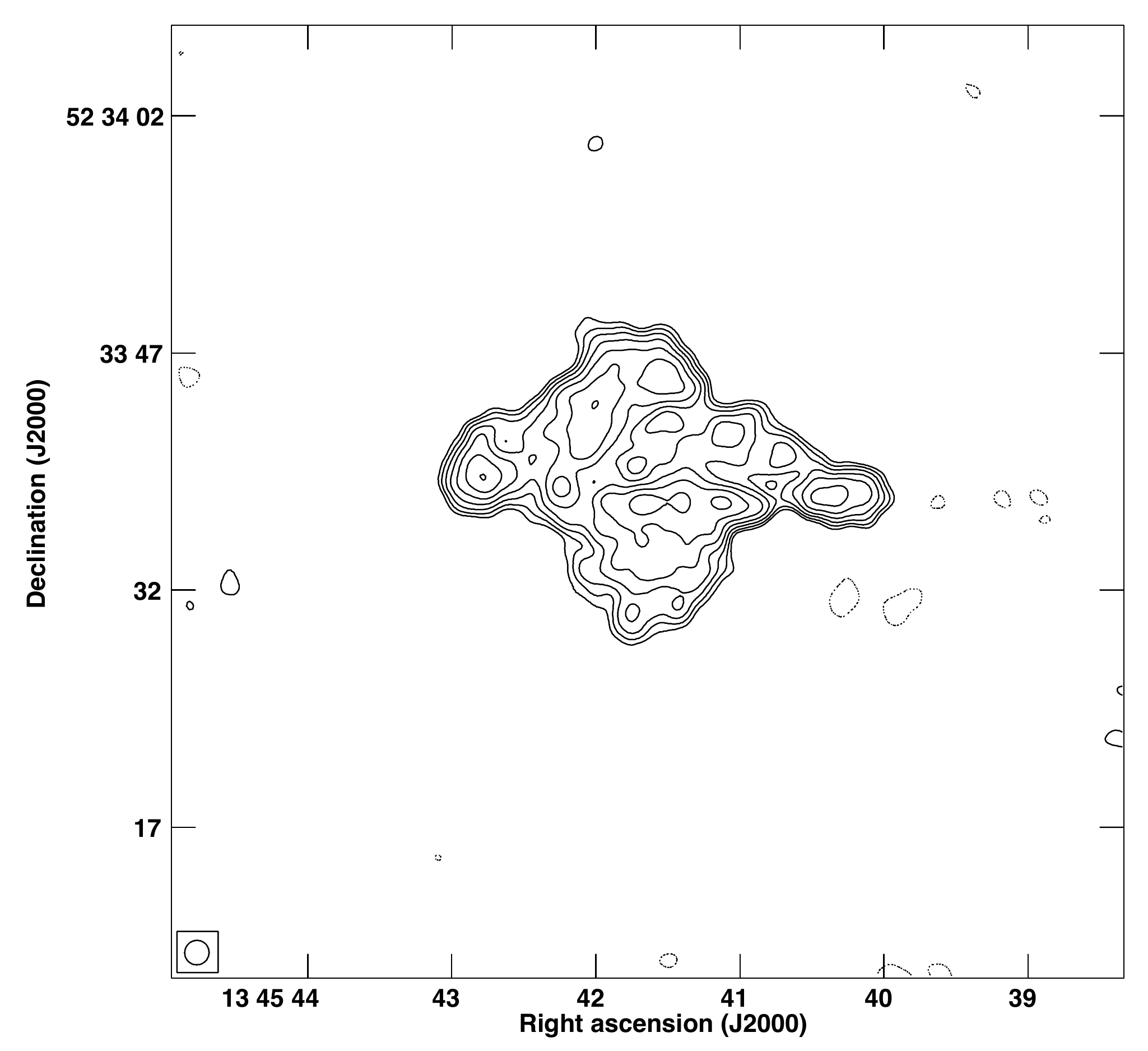}
\includegraphics[width=0.45\columnwidth]{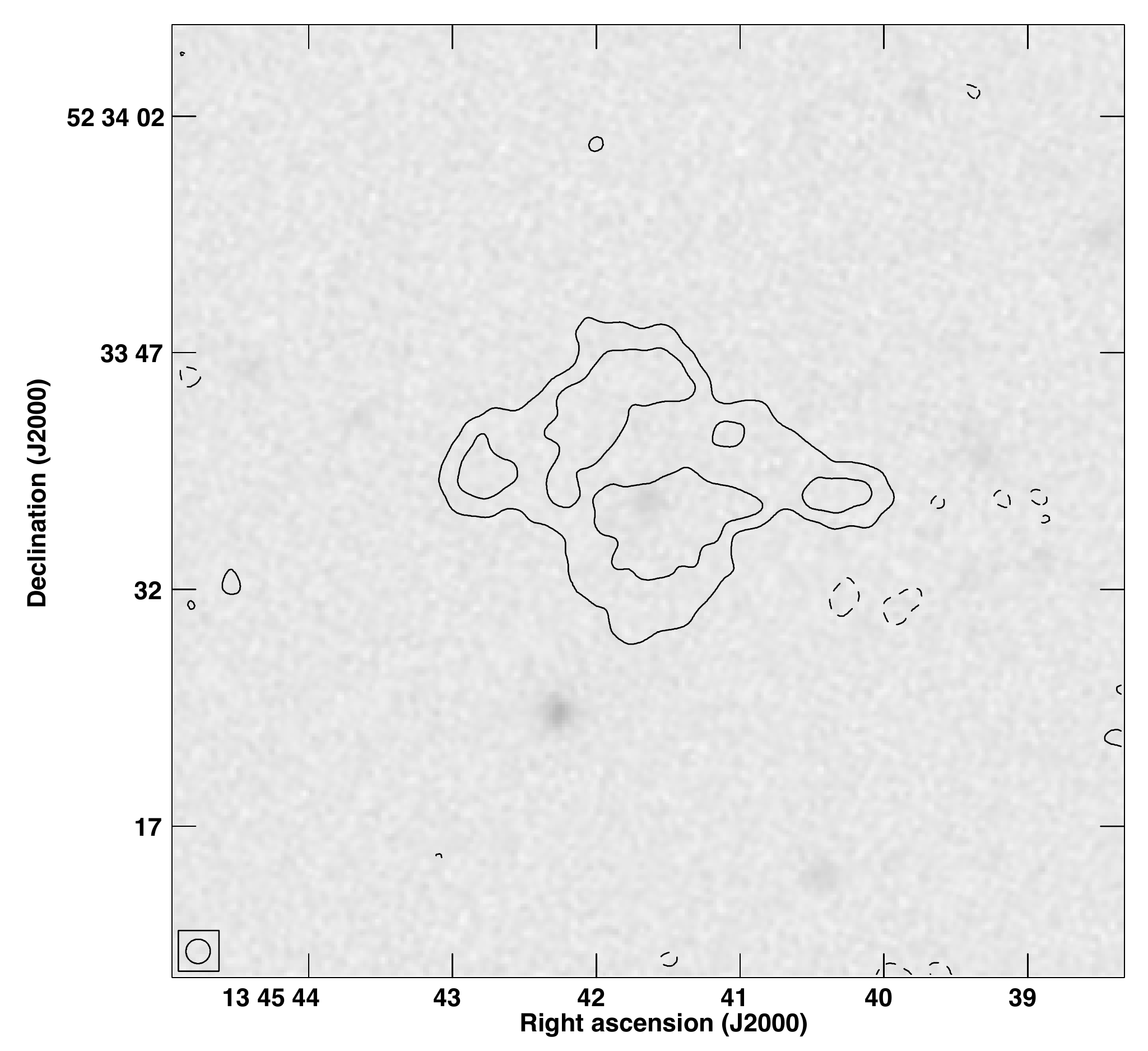}
\caption[J1345+5233 (L)]{J1345+5233. (left) VLA image at L band, (right) VLA image overlaid on red SDSS image. Lowest contour = 0.09~mJy/beam, peak  = 0.92~mJy/beam. \label{fig:J1345+5233}}
\end{figure}

\noindent J1345+5233 (Figure~\ref{fig:J1345+5233}).  The source is revealed to have two edge-brightened lobes . A weak core may be detected at the location of the central faint object.

\begin{figure}[ht] 
\includegraphics[width=0.45\columnwidth]{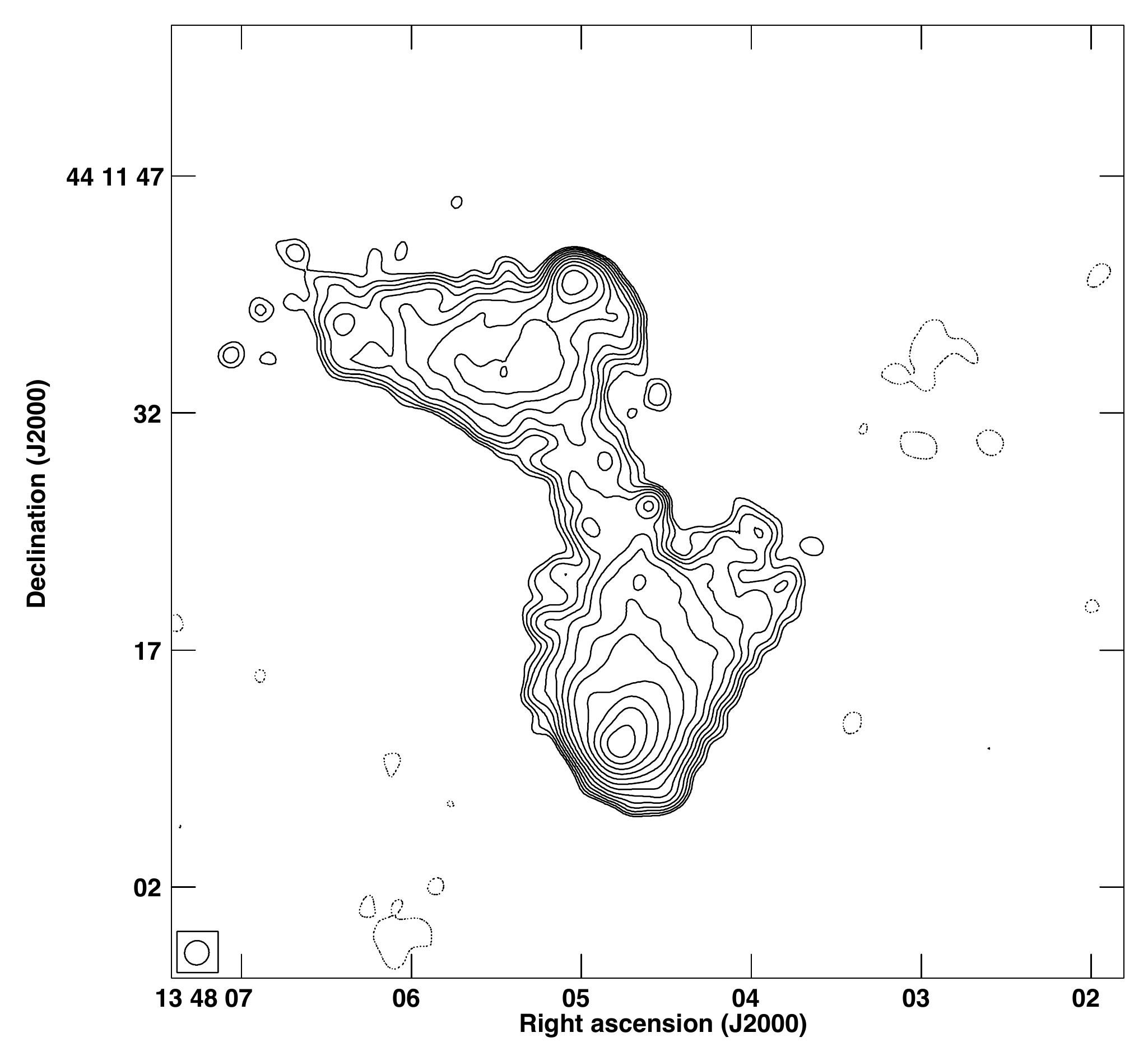}
\includegraphics[width=0.45\columnwidth]{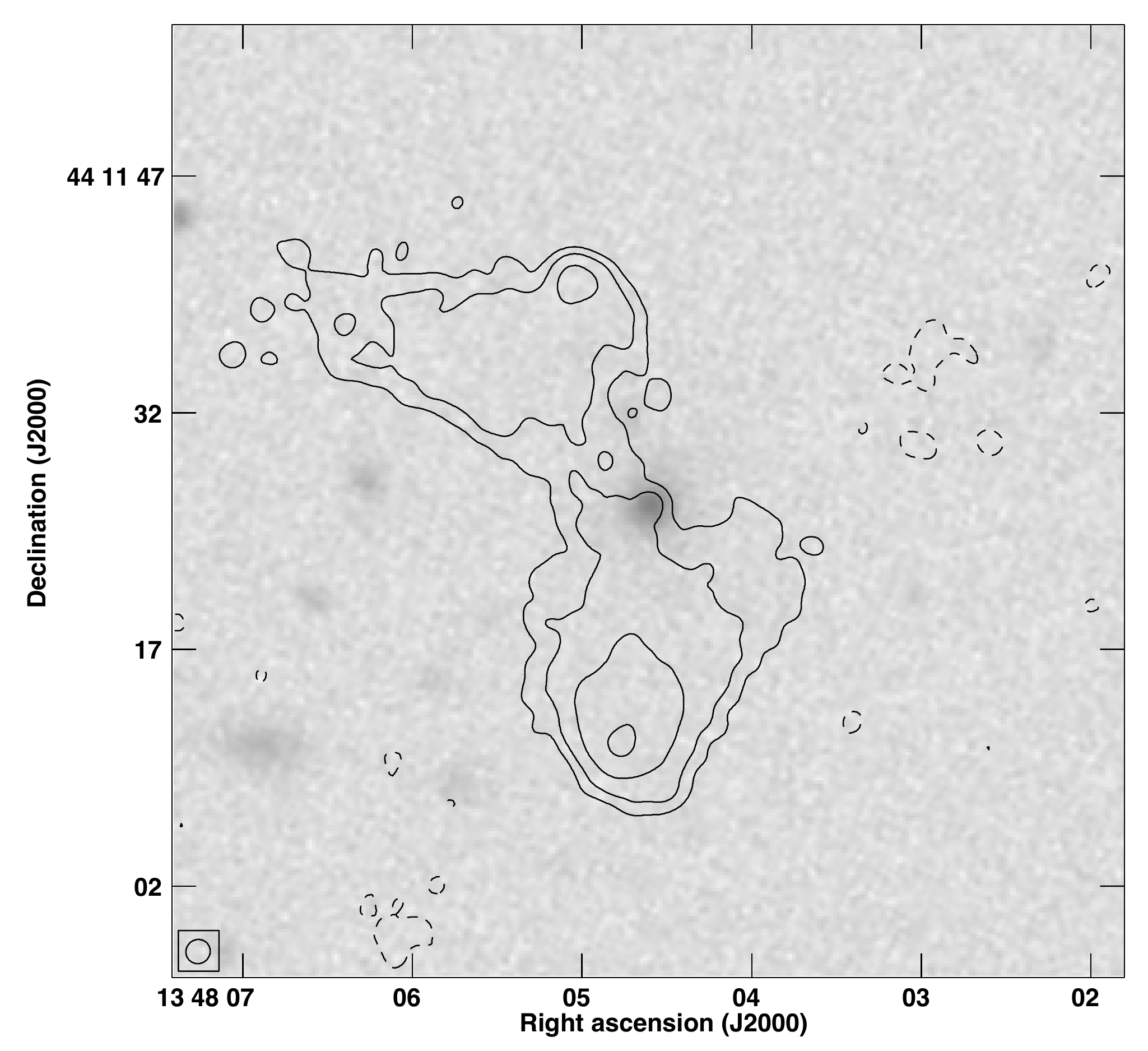}
\caption[J1348+4411 (L)]{J1348+4411.  (left) VLA image at L band, (right) VLA image overlaid on red SDSS image. Lowest contour = 0.07~mJy/beam, peak = 5.96~mJy/beam.  \label{fig:J1348+4411}}
\end{figure}

\noindent J1348+4411 (Figure~\ref{fig:J1348+4411}). The map reveals a non-collinear, edge-brightened source, where the two hotspots lie along axes that make an angle of nearly $30^\circ$ to each other. The presumed jet to the north appears to bend by nearly $40^\circ$ to the east at a location about halfway to the northern hotspot. The lobes steer away from these respective axes with the northern lobe extended away by more than $90^\circ$. 

\begin{figure}[ht] 
\includegraphics[width=0.45\columnwidth]{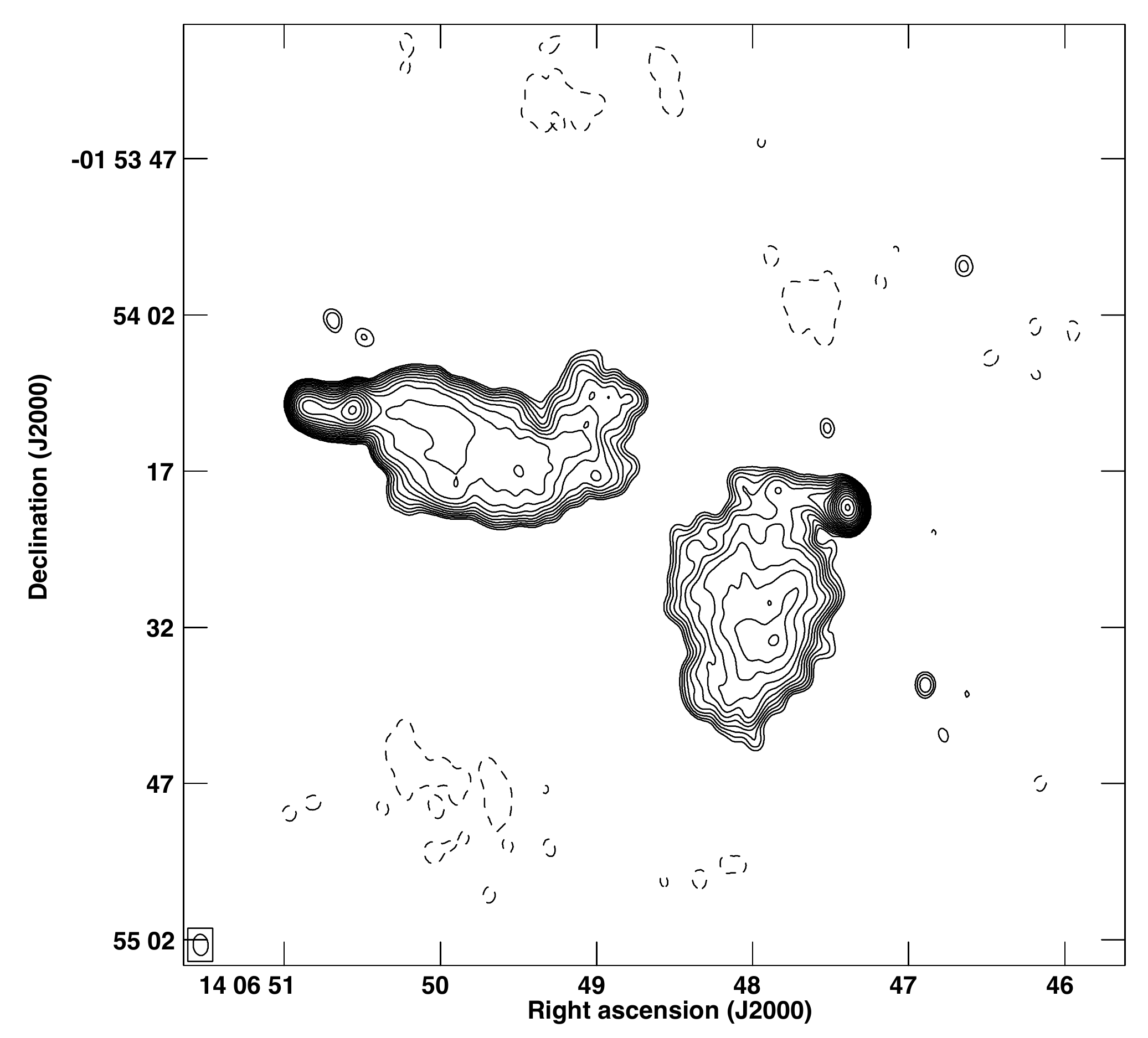}
\includegraphics[width=0.45\columnwidth]{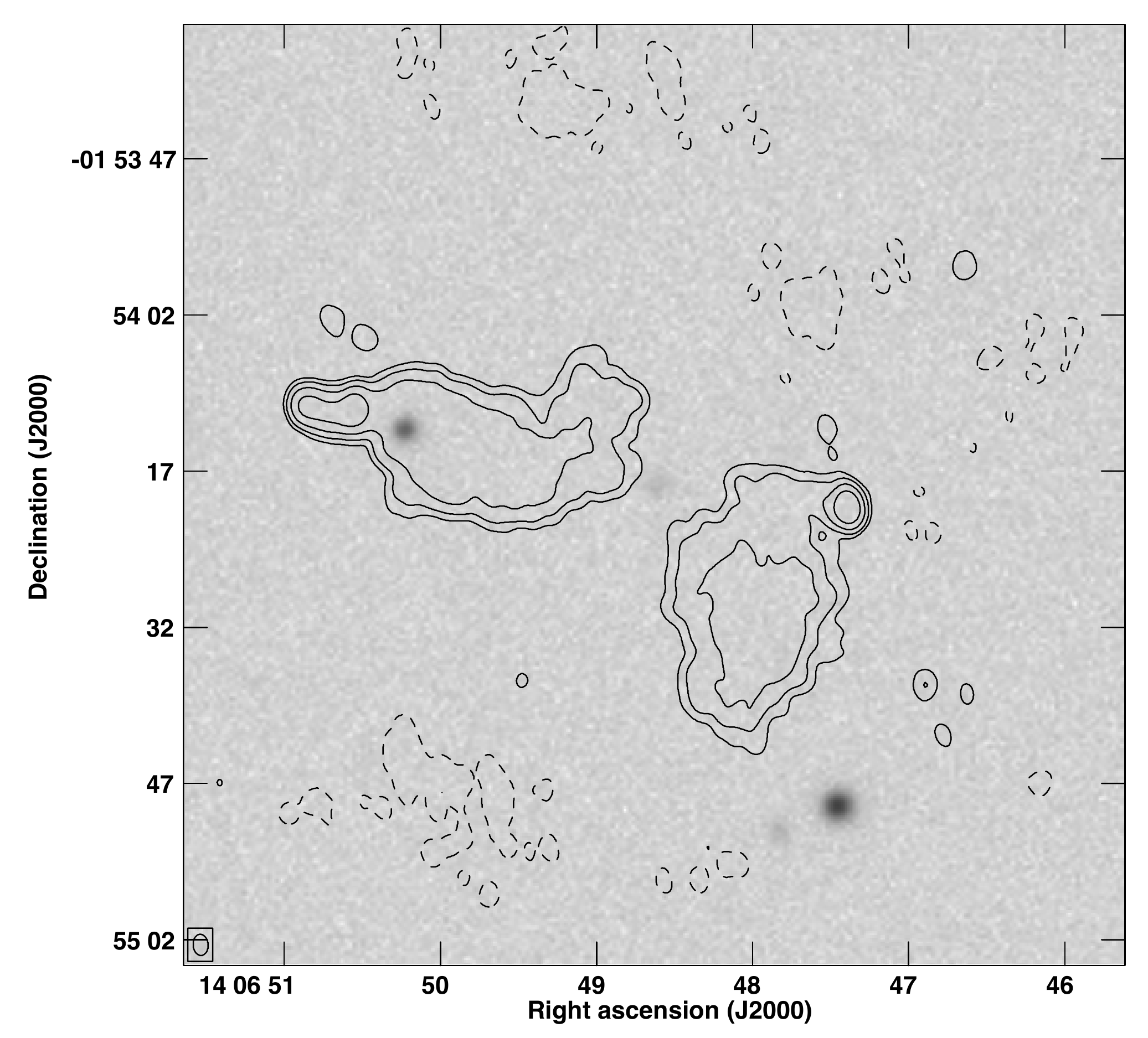}
\caption[J1406$-$0154 (L)]{J1406$-$0154. (left) VLA image at L band, (right) VLA image overlaid on red SDSS image. Lowest contour = 0.3~mJy/beam, peak = 57.6~mJy/beam. \label{fig:J1406-0154}}
\end{figure}

\noindent J1406$-$0154. (Figure~\ref{fig:J1406-0154}).  Bright hotspots are seen at the extremities of the lobes. No core is detected although there is a faint object close to the radio axis between the two lobes. Both lobes are well confined including the two transverse emission regions which are also separated by a significant gap in emission.

\begin{figure}[ht] 
\includegraphics[width=0.45\columnwidth]{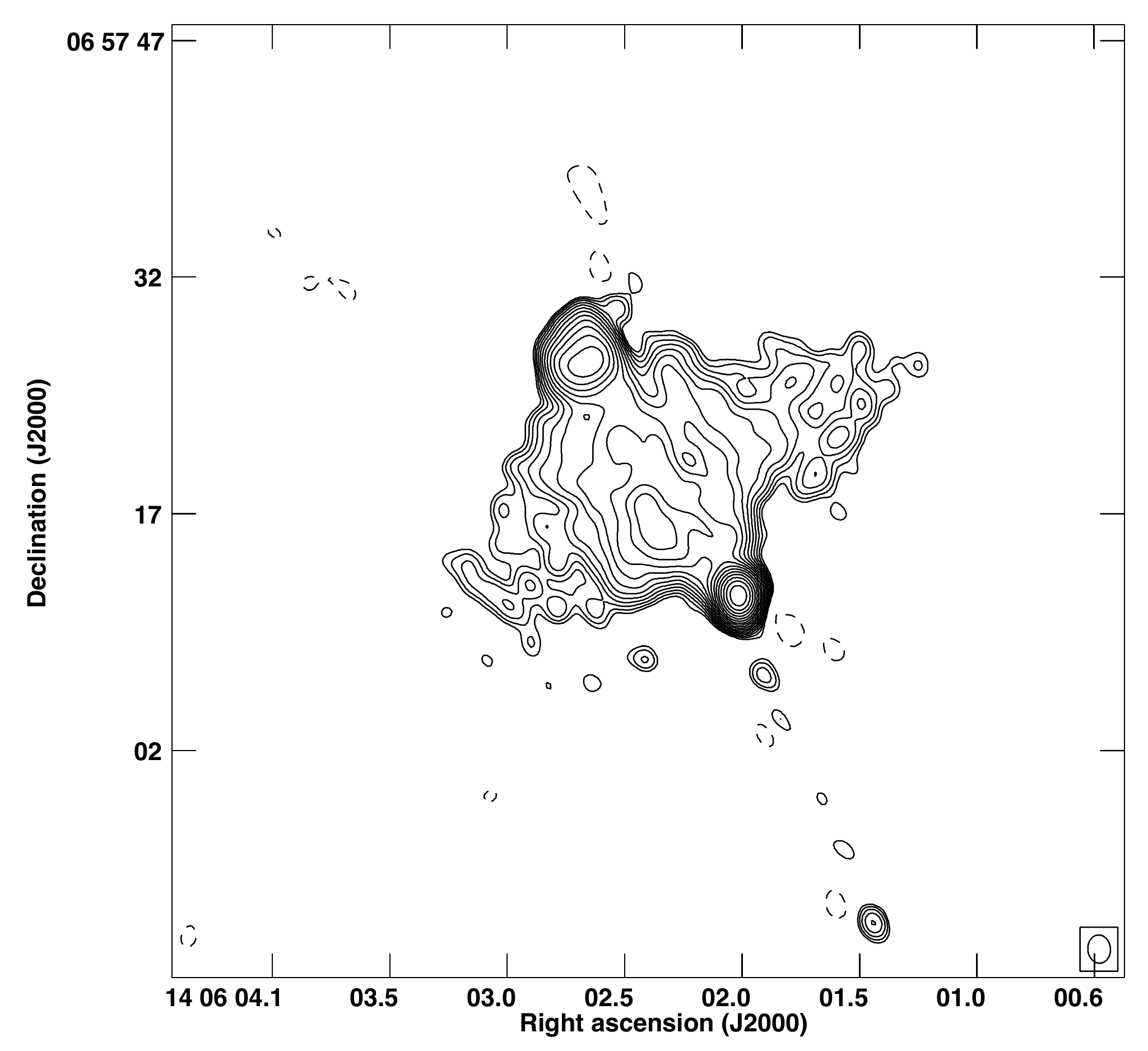}
\includegraphics[width=0.45\columnwidth]{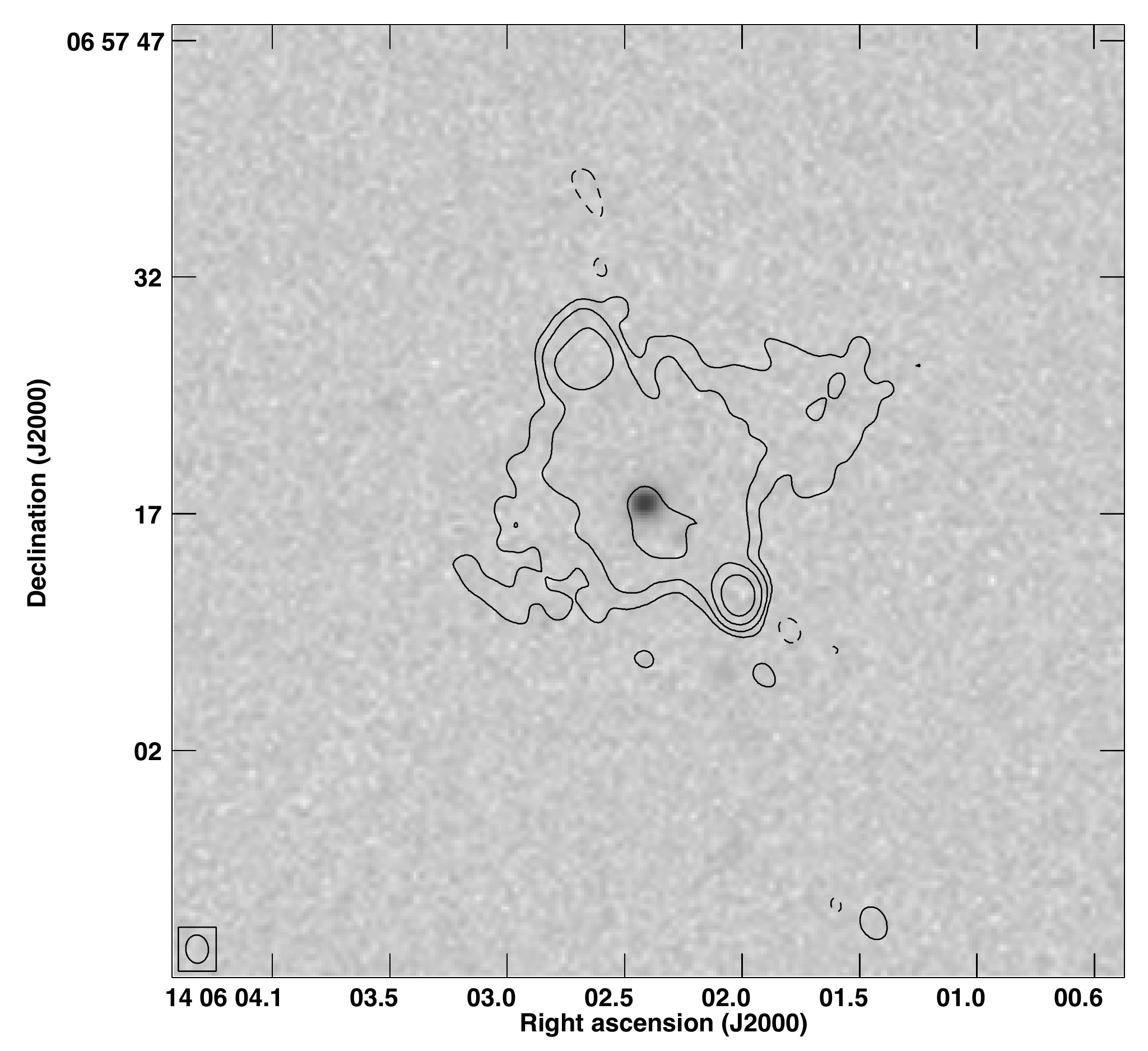}
\caption[J1406+0657 (L)]{J1406+0657.  (left) VLA image at L band, (right) VLA image overlaid on red SDSS image. Lowest contour = 0.2~mJy/beam, peak = 71.5~mJy/beam. \label{fig:J1406+0657}}
\end{figure}

\noindent J1406+0657 (Figure~\ref{fig:J1406+0657}). Our new map reveals two bright hotspots at the lobe ends. Lobes are not seen as distinct components and instead a tapering broad swathe of emission, which is centrally located, is seen across the source axis. A weak core may be present at the location of a galaxy close to the source axis. 

\begin{figure}[ht] 
\includegraphics[width=0.45\columnwidth]{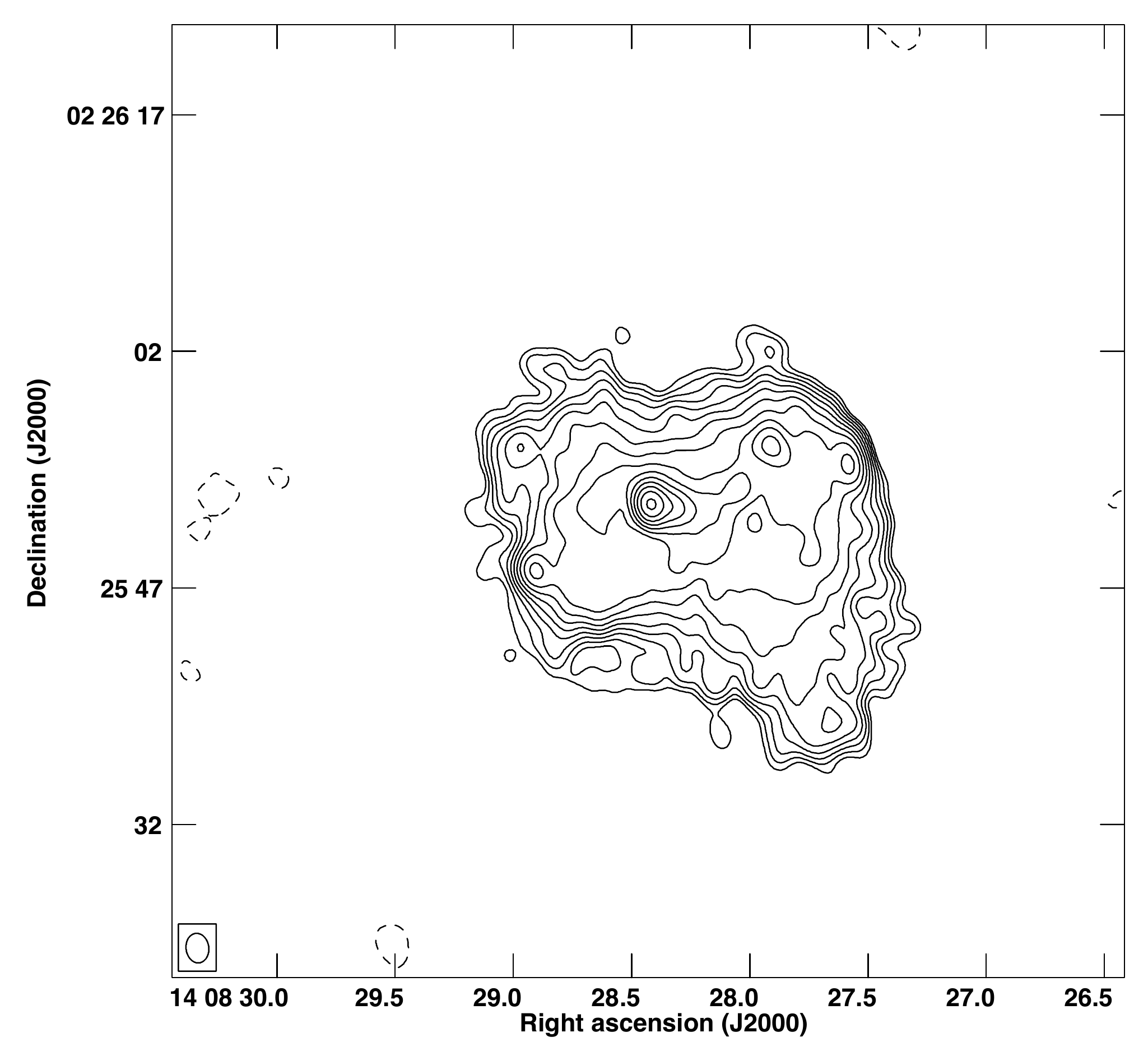}
\includegraphics[width=0.45\columnwidth]{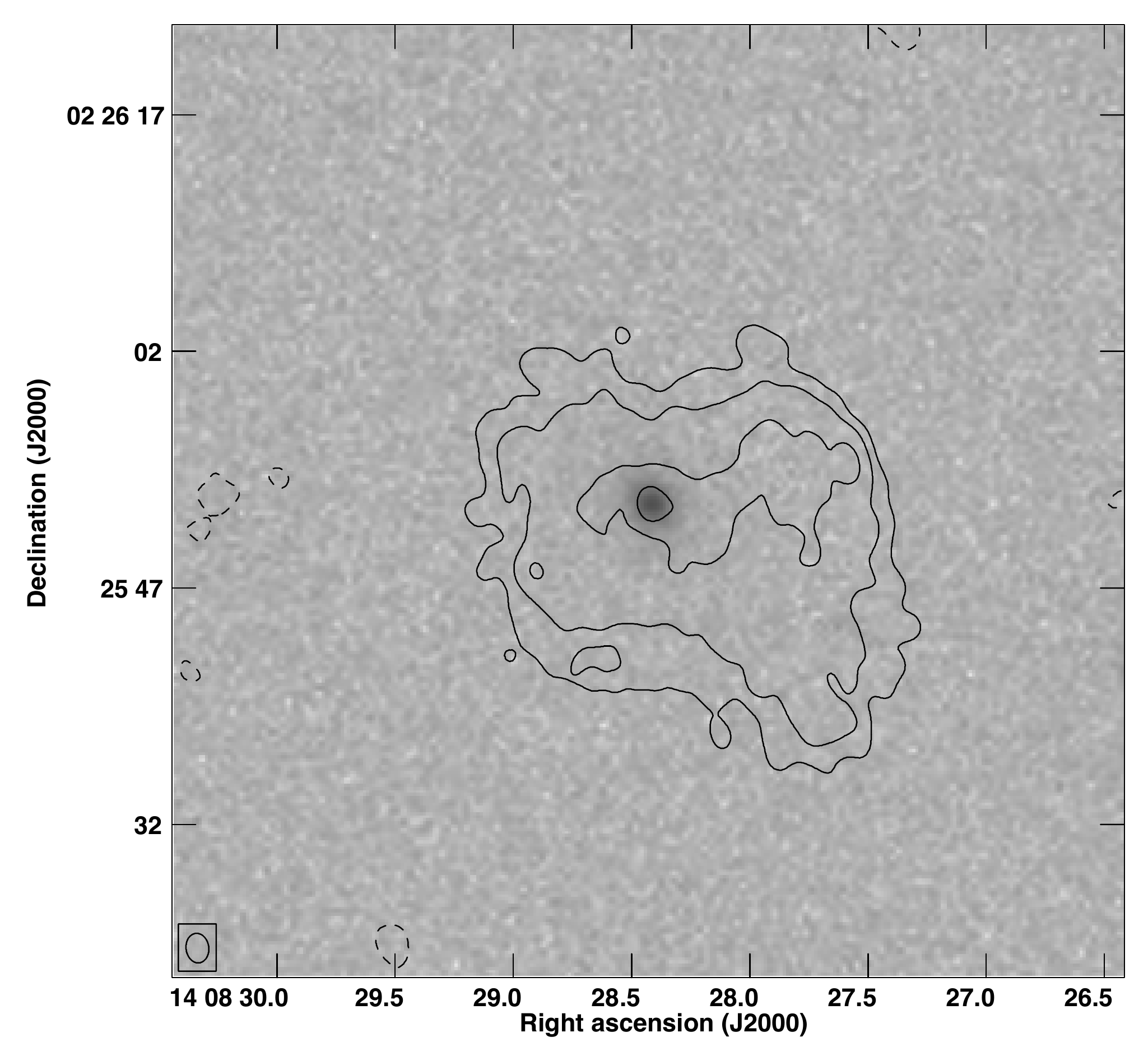}
\caption[J1408+0225 (L)]{J1408+0225. (left) VLA image at L band, (right) VLA image overlaid on red SDSS image. Lowest contour = 0.1~mJy/beam, peak = 14.0~mJy/beam. \label{fig:J1408+0225}}
\end{figure}

\noindent J1408+0225 (Figure~\ref{fig:J1408+0225}).  The almost featureless FIRST source is revealed to have a complex structure in our high-resolution map. Several compact sources and emission peaks are seen embedded within a large emission region. A prominent elongated core is seen at the location of the galaxy at the center. The elongation is along the direction to the hotspot at the northeast. Interestingly the core is located on an axis formed by another pair of hotspots at an angle of $60^\circ$ Higher resolution observations are required to understand the complex structure.

\begin{figure}[ht] 
\includegraphics[width=0.45\columnwidth]{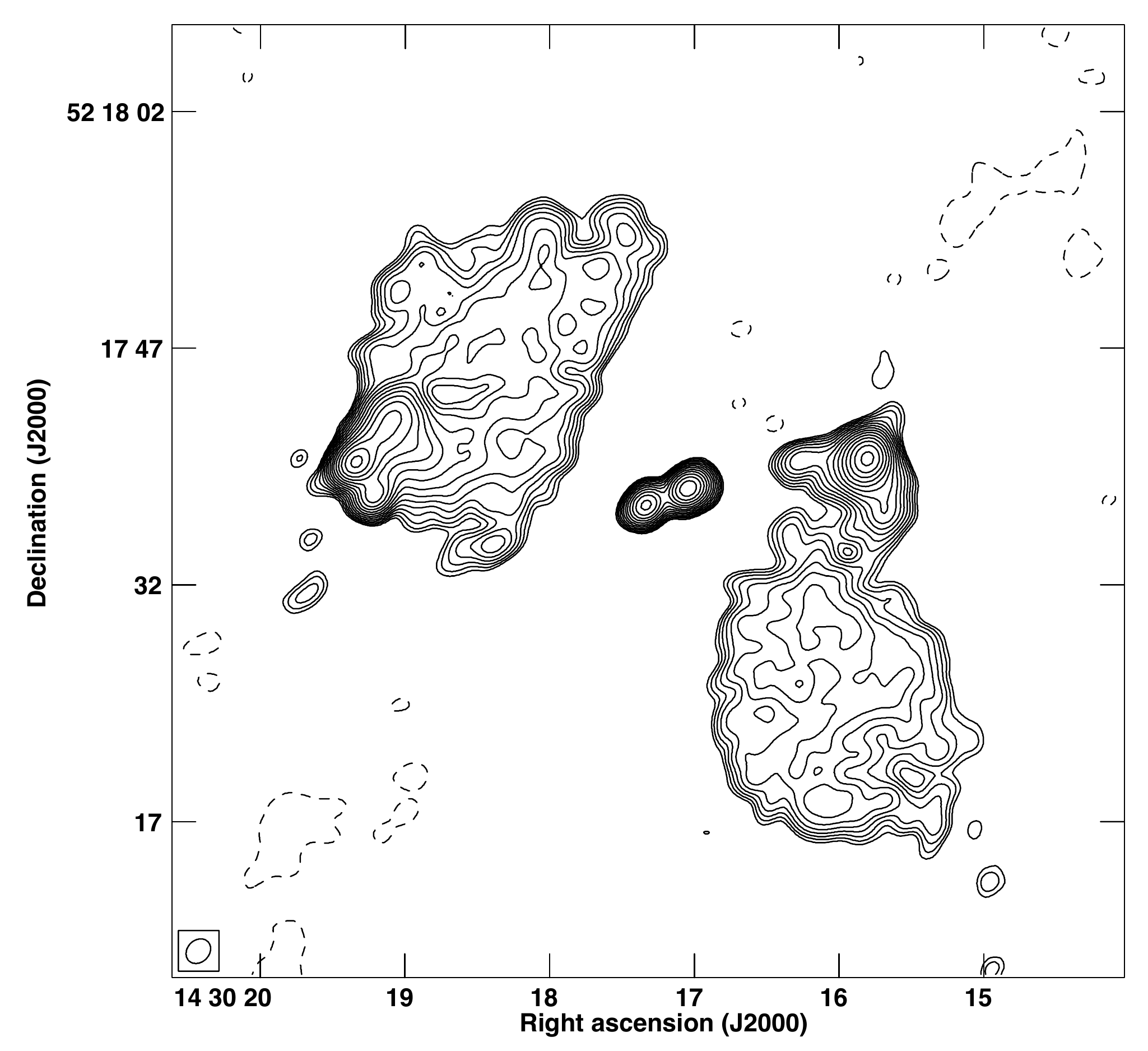}
\includegraphics[width=0.45\columnwidth]{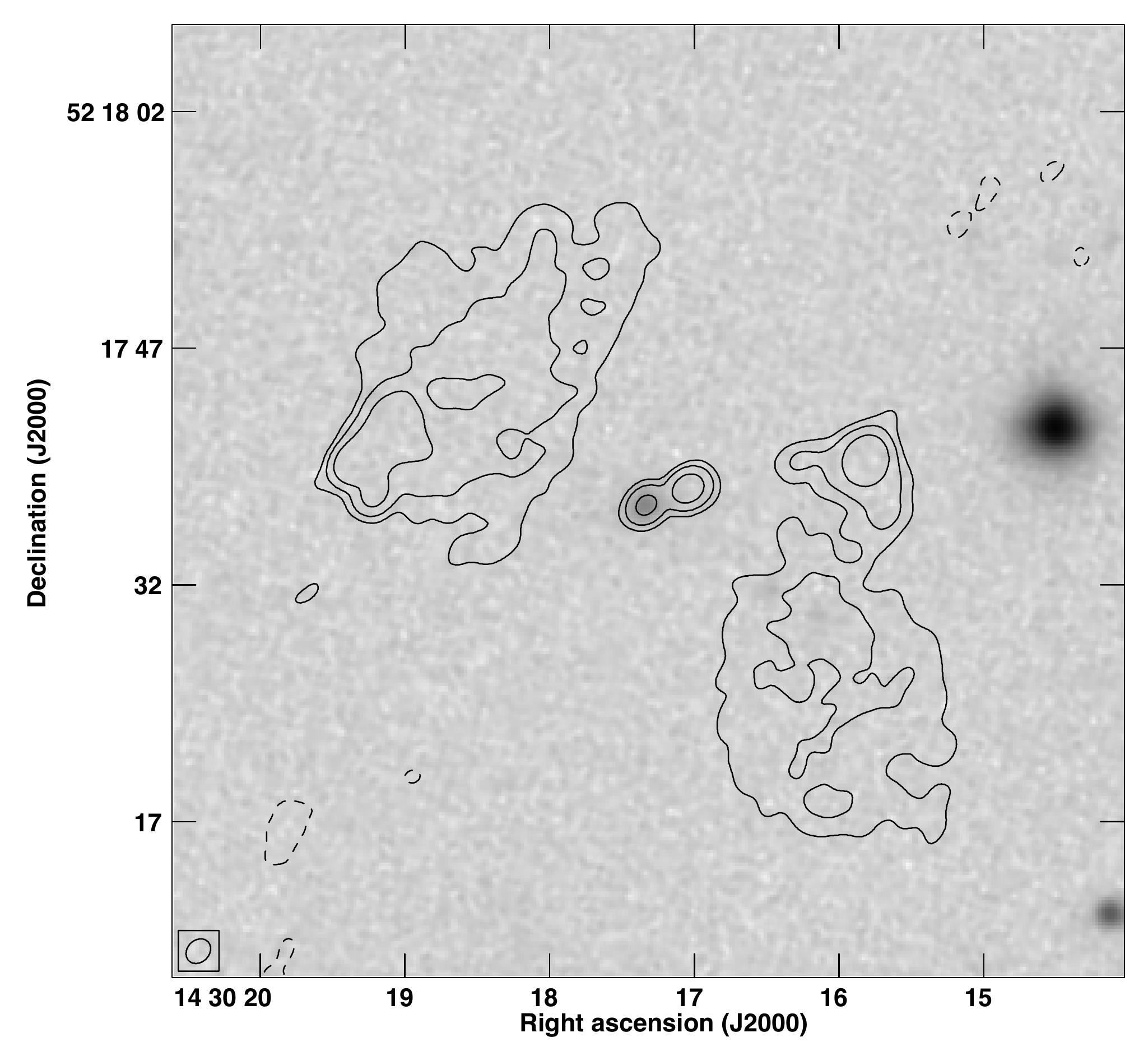}
\caption[J1430+5217 (L)]{J1430+5217.  (left) VLA image at L band, (right) VLA image overlaid on red SDSS image. Lowest contour = 0.2~mJy/beam, peak  = 40.2~mJy/beam. \label{fig:J1430+5217}}
\end{figure}

\noindent J1430+5217 (Figure~\ref{fig:J1430+5217}).  Our high-resolution map has resolved the central emission region into a pair of peaks one of which (to the SE) coincides with a galaxy. The second peak, which is slightly less compact, could be part of a jet to the west. Further to the west along the axis of the twin peaks lies a compact hotspot which also shows a partial collimated feature along the axis towards the pair of peaks. The bright region at the leading end of the eastern lobe may be showing a complex hotspot. This hotspot region does not lie on the source axis formed by the core, jet and the western hotspot. However, the eastern lobe shows an elongated feature along the axis offset from the rest of the lobe. Both lobes deflect away nearly orthogonally from the source axis.  

\begin{figure}[ht] 
\includegraphics[width=0.45\columnwidth]{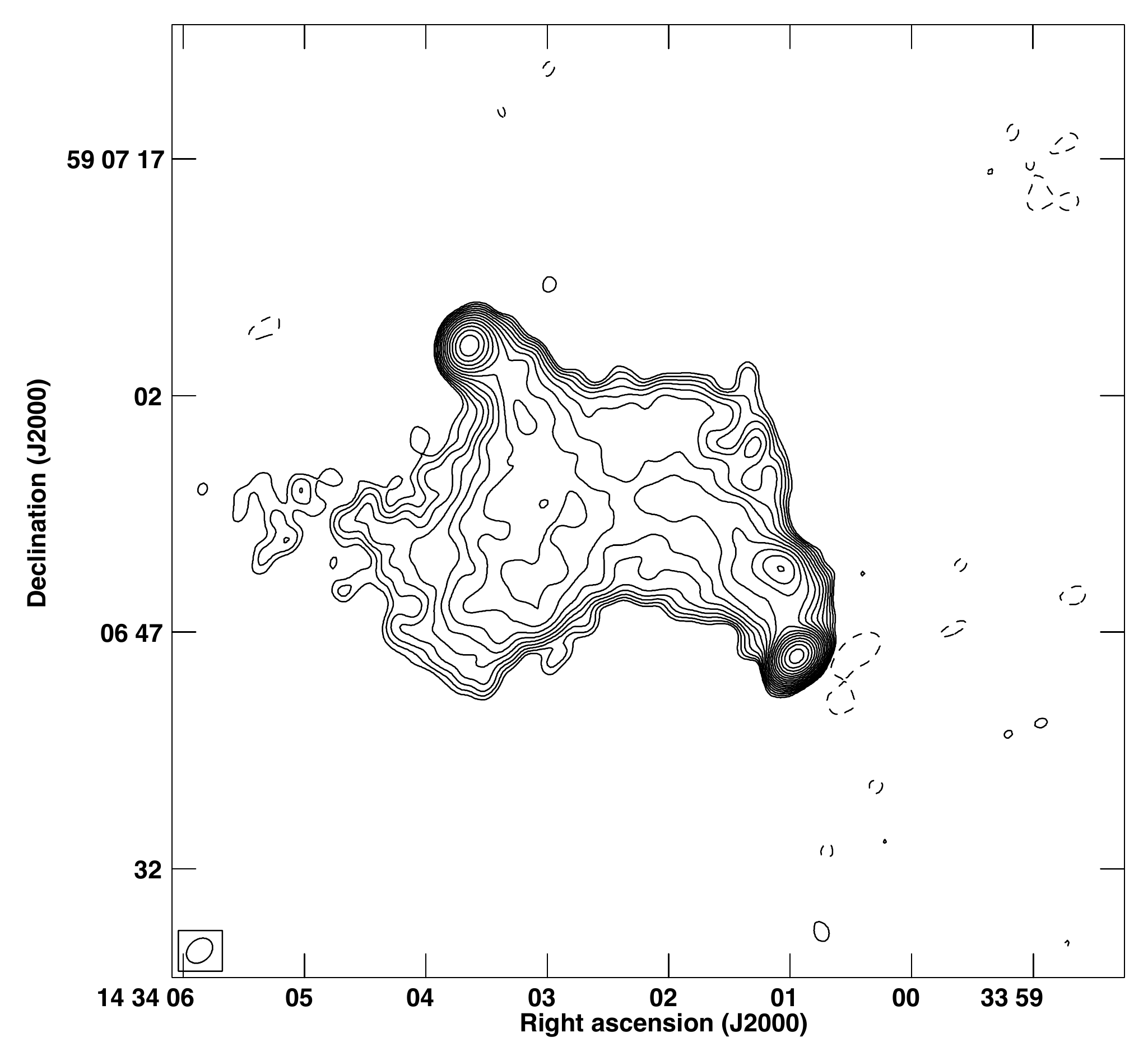}
\includegraphics[width=0.45\columnwidth]{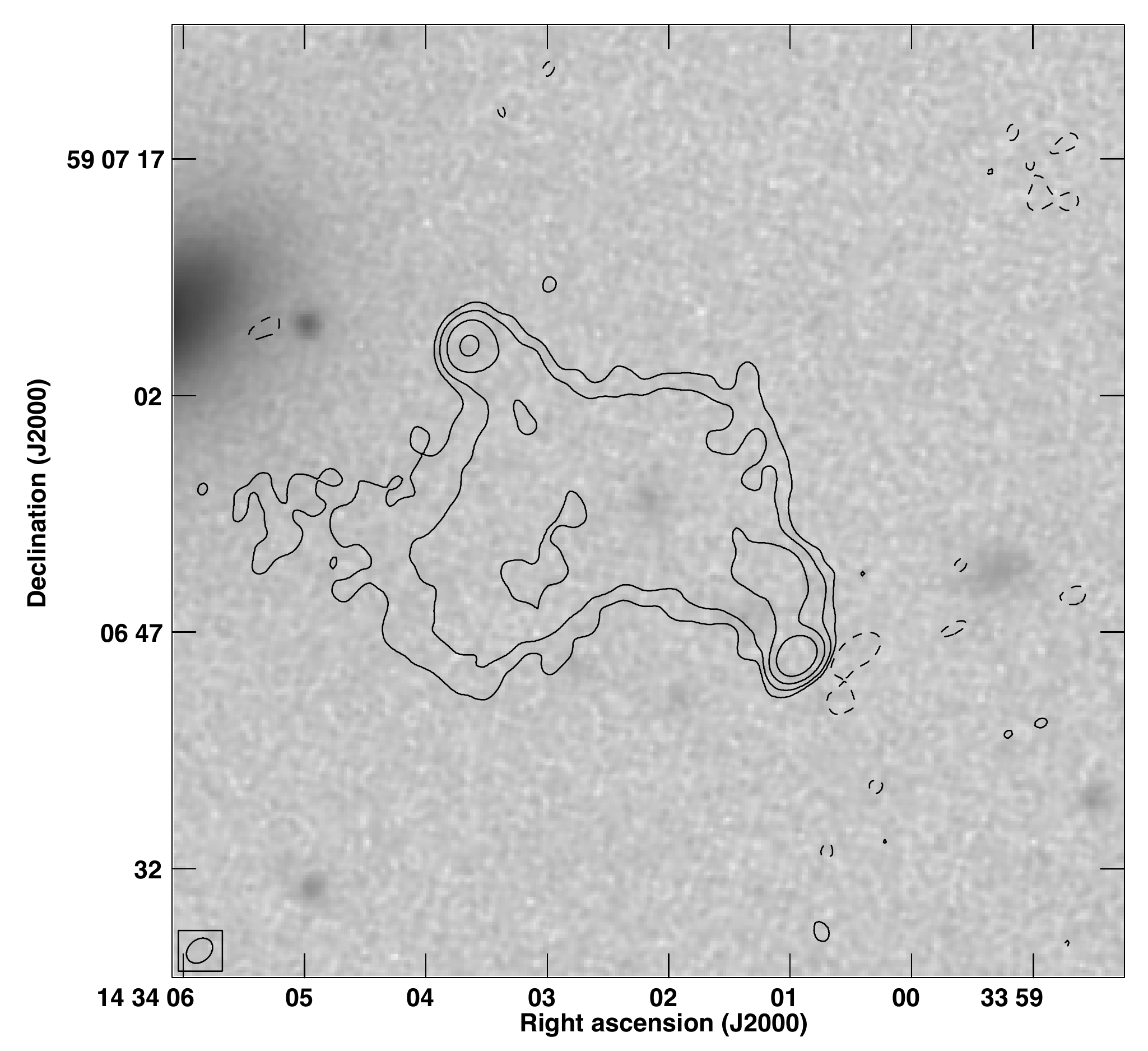}
\caption[J1434+5906 (L)]{J1434+5906.  (left) VLA image at L band, (right) VLA image overlaid on red SDSS image. Lowest contour = 0.15~mJy/beam, peak = 47.3~mJy/beam. \label{fig:J1434+5906}}
\end{figure}

\noindent J1434+5906 (Figure~\ref{fig:J1434+5906}).  There are two bright and compact hotspots at the extremities along with a swathe of broad emission orthogonal to the source axis mostly seen on one side. A weak core may be present at the location of the faint object in the center.

\begin{figure}[ht] 
\includegraphics[width=0.45\columnwidth]{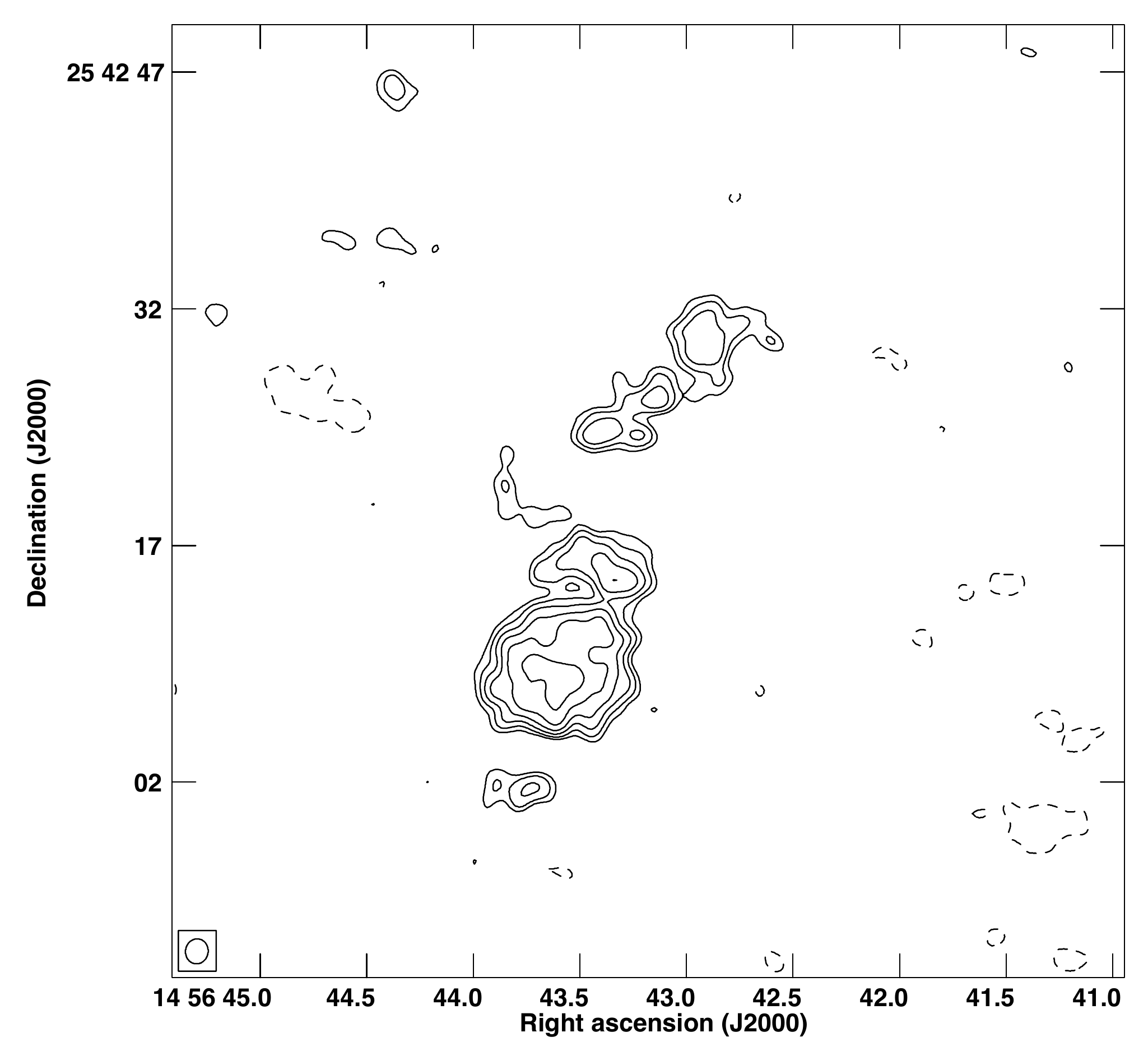}
\includegraphics[width=0.45\columnwidth]{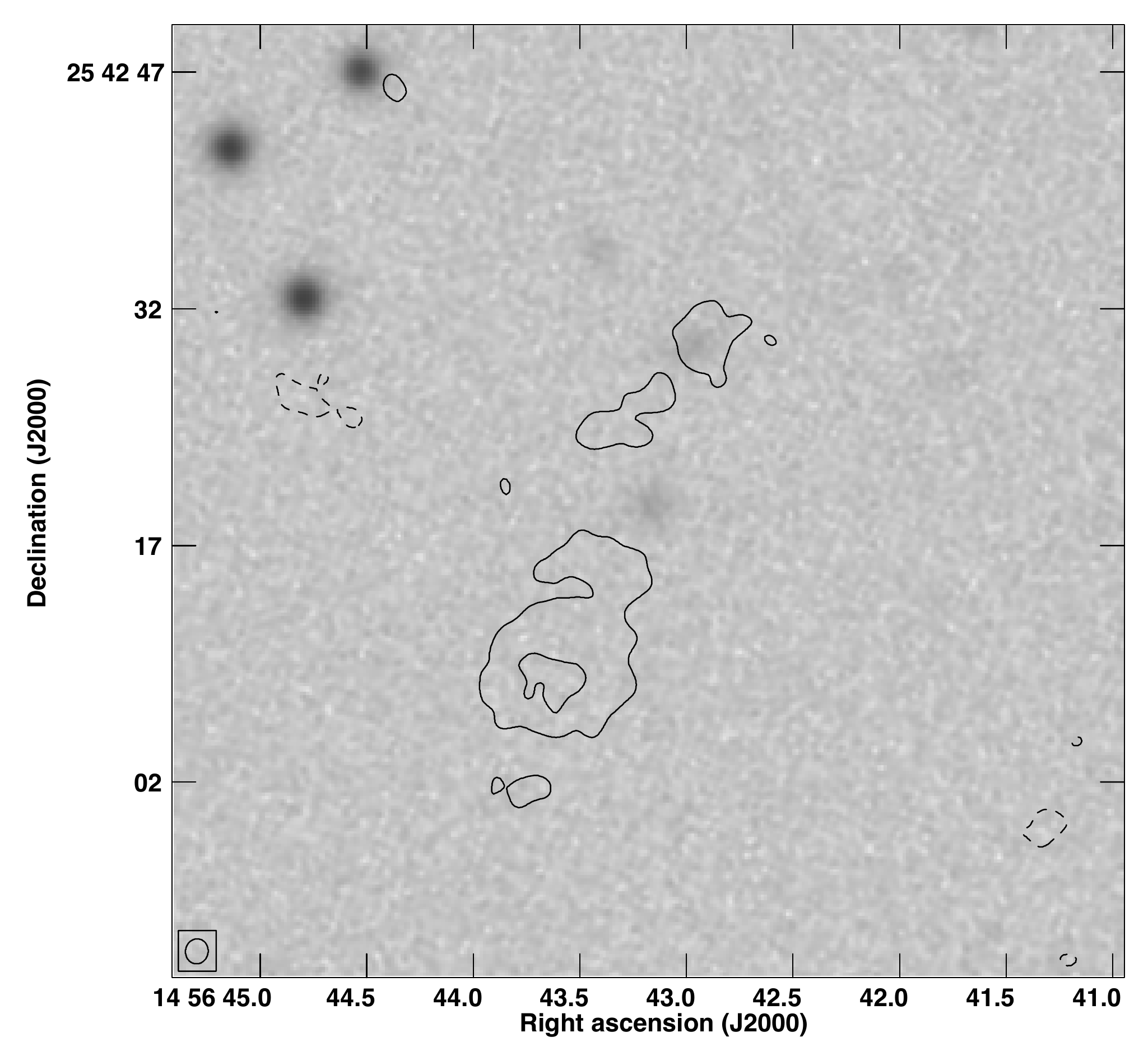}
\caption[J1456+2542 (L)]{J1456+2542. (left) VLA image at L band, (right) VLA image overlaid on red SDSS image. Lowest contour = 0.11~mJy/beam, peak  = 1.16~mJy/beam.  \label{fig:J1456+2542}}
\end{figure}

\noindent J1456+2542 (Figure~\ref{fig:J1456+2542}). The map shows an edge-brightened morphology for the source. Neither of the lobes however is found to have a compact hotspot nor is a core or jet seen, suggesting a relic nature for the source. The extended diffuse emission region is resolved out in the new map.

\begin{figure}[ht] 
\includegraphics[width=0.45\columnwidth]{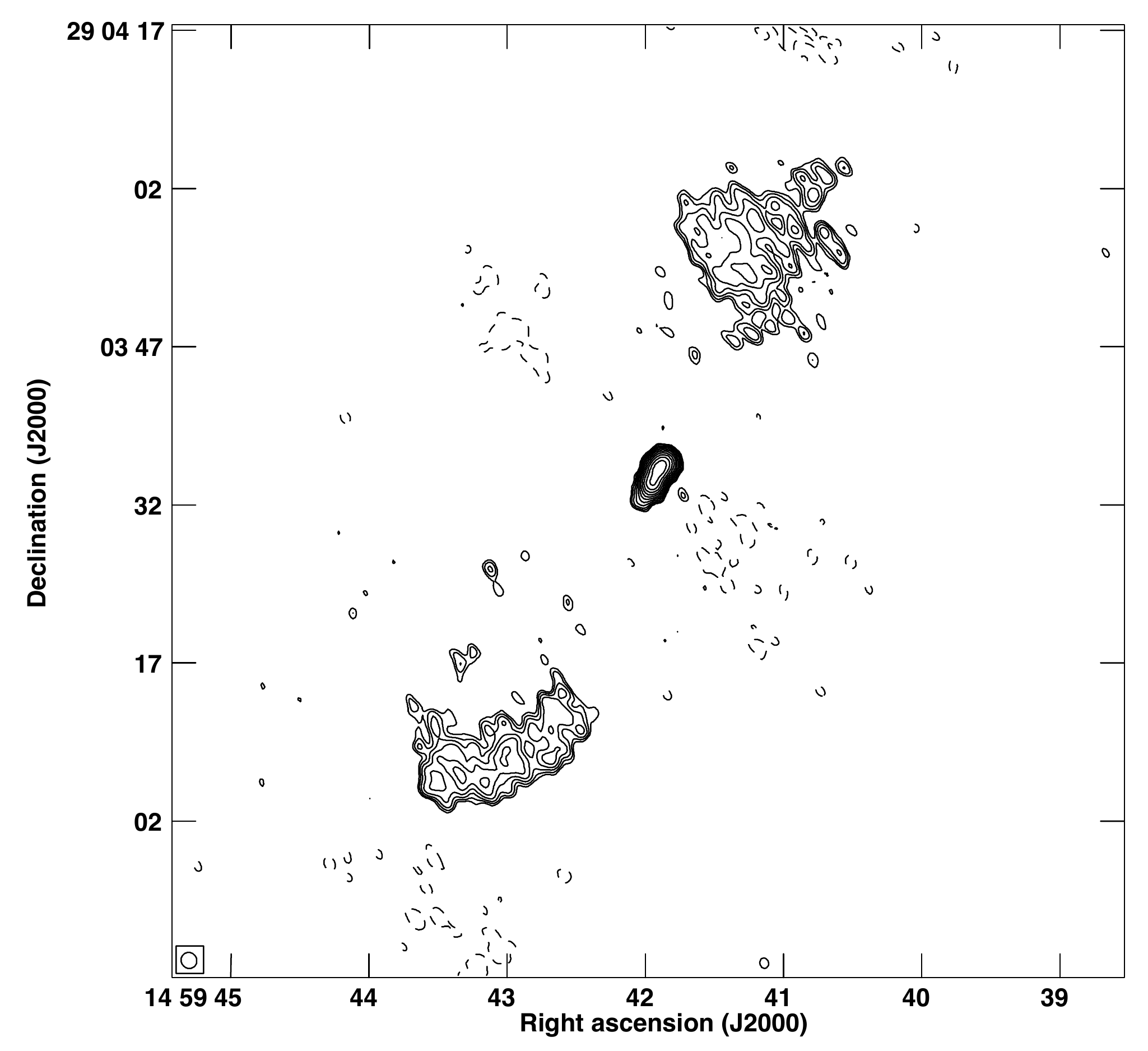}
\includegraphics[width=0.45\columnwidth]{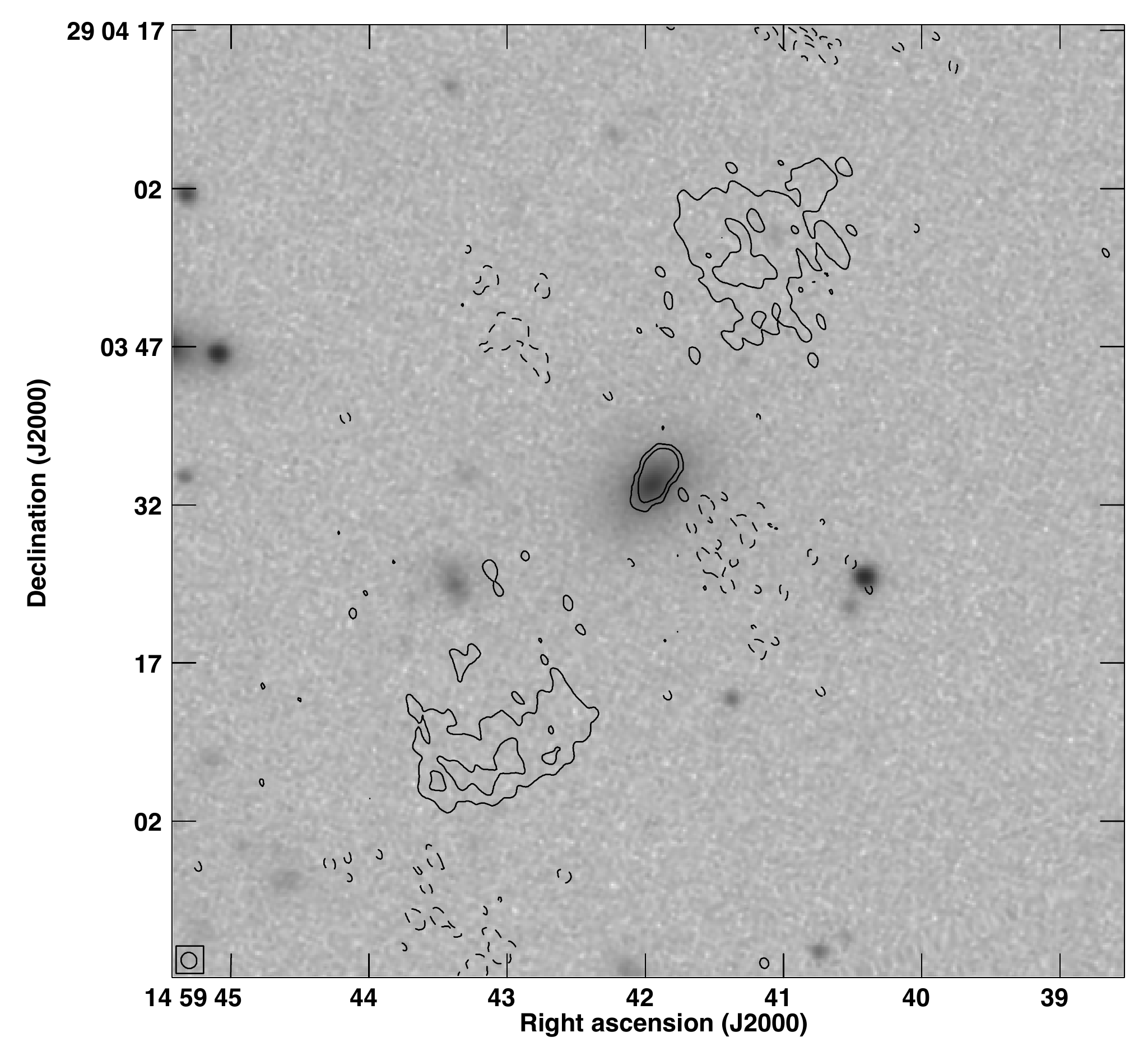}\\
\includegraphics[width=0.45\columnwidth]{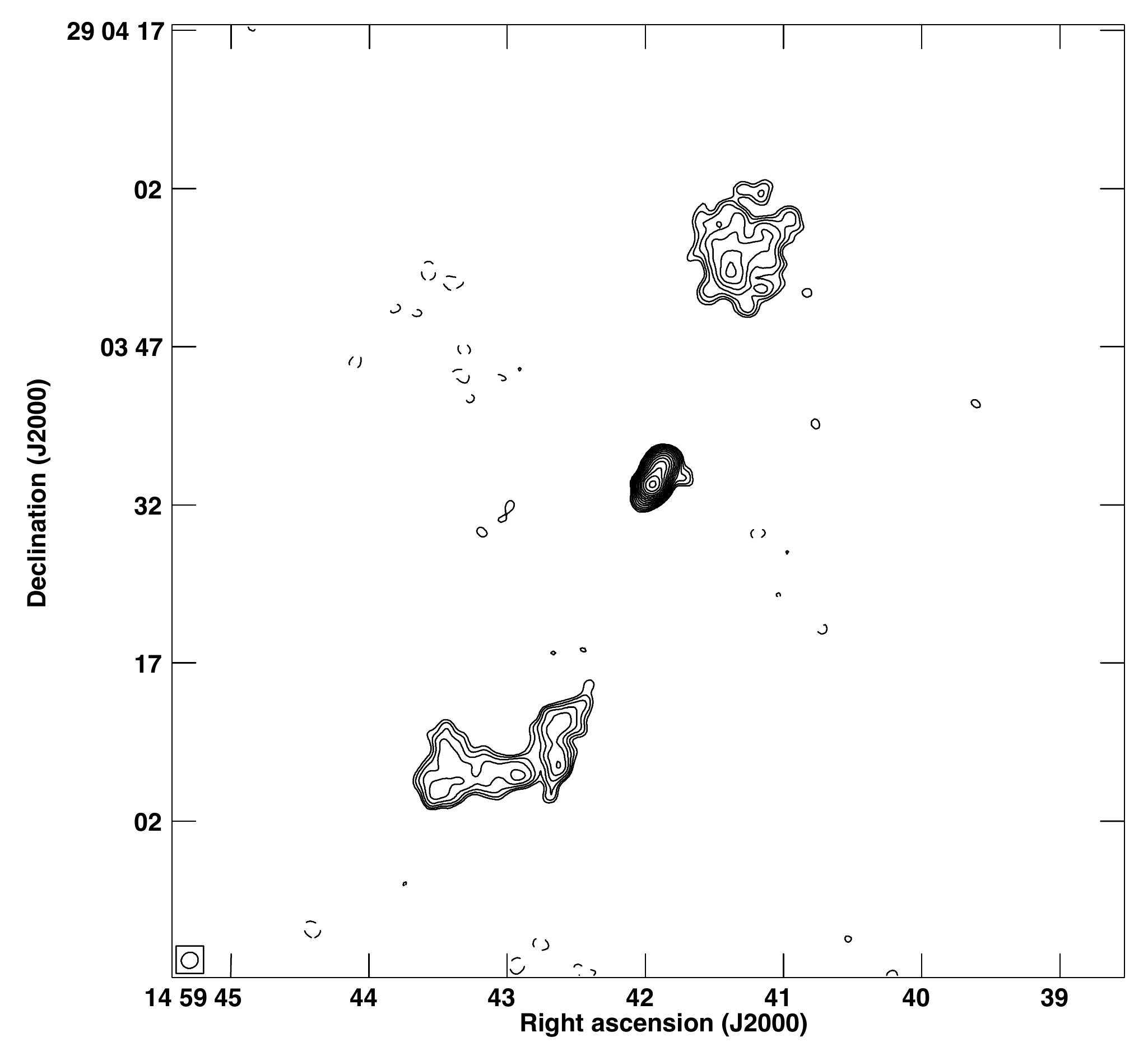}
\includegraphics[width=0.45\columnwidth]{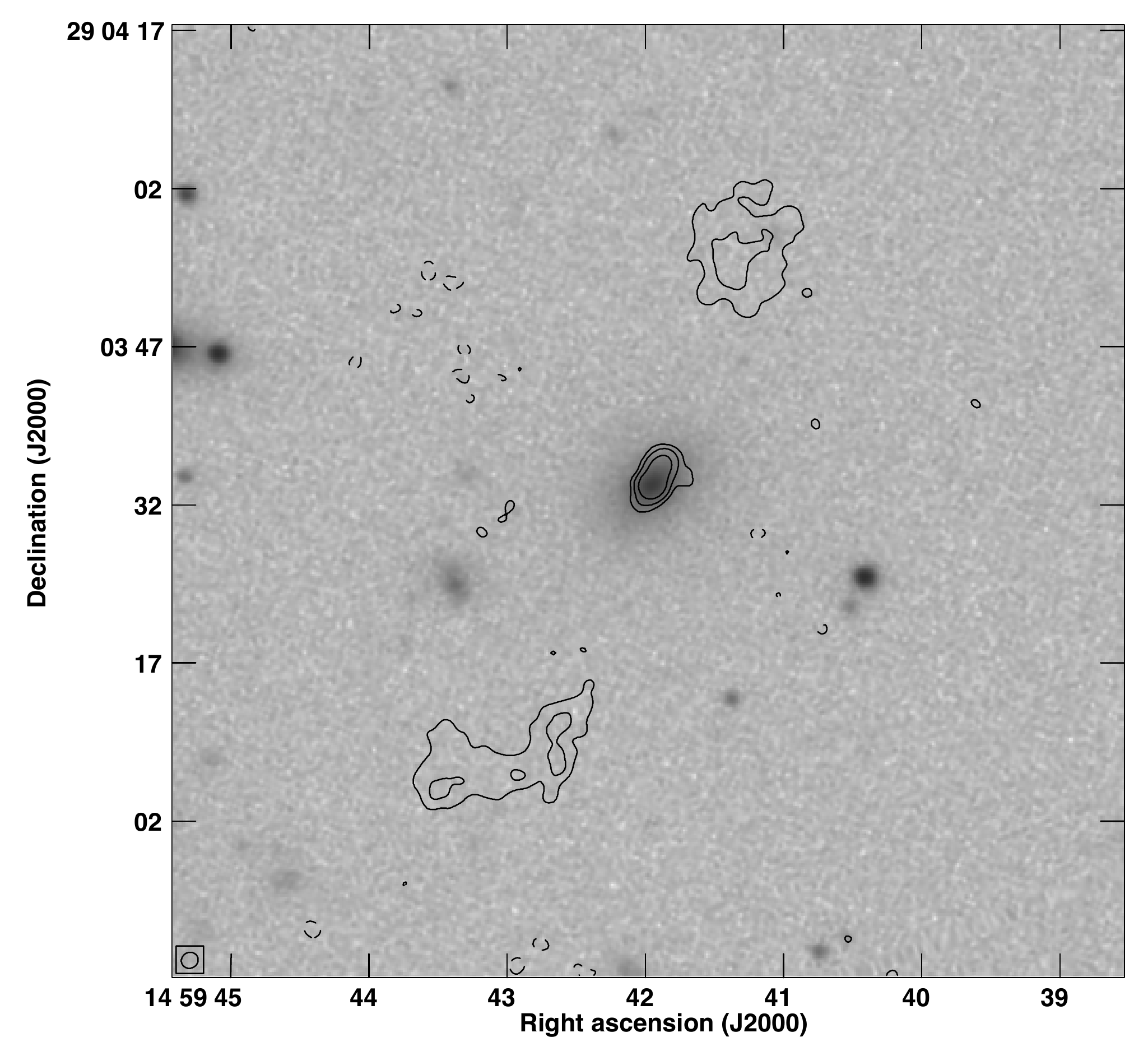}\\
\caption[J1459+2903 (L \& C)]{J1459+2903.  (top) (left) VLA image at L band and (right) VLA image overlaid on red SDSS image. Lowest contour = 0.2~mJy/beam, peak  = 8.67~mJy/beam. (bottom) (left) VLA image at C band and (right) VLA image overlaid on red SDSS image. Lowest contour = 0.1~mJy/beam, peak  = 10.1~mJy/beam. \label{fig:J1459+2903}}
\end{figure}

\noindent J1459+2903 (Figure~\ref{fig:J1459+2903}). See Figure~\ref{fig:J1459+2903}. The core emission is elongated with jet-like extensions on either side of a bright optical ID. Although edge-brightened no compact hotspots are seen. Much of the extended emission is resolved out in our map.

\begin{figure}[ht] 
\includegraphics[width=0.45\columnwidth]{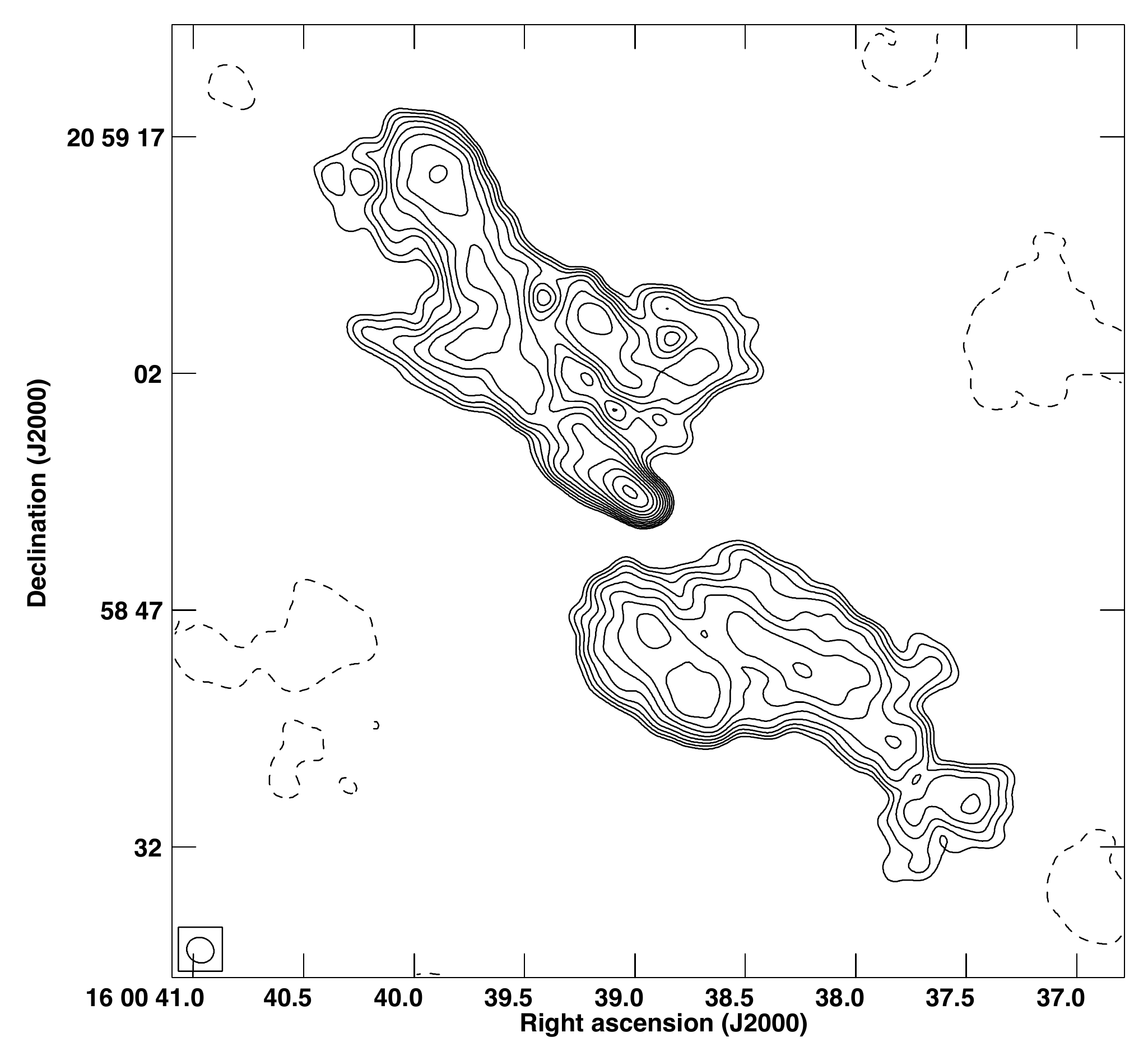}
\includegraphics[width=0.45\columnwidth]{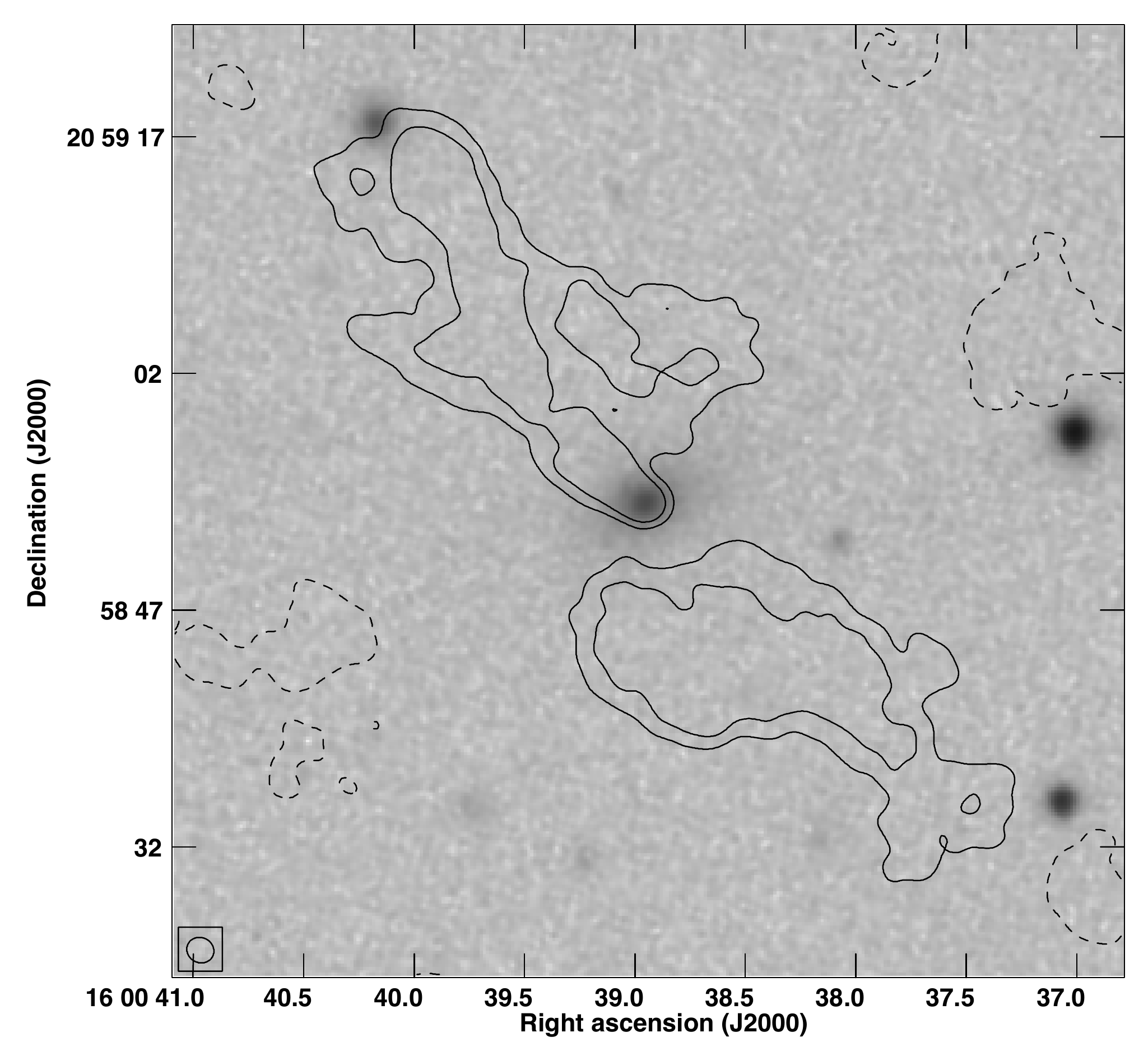}
\caption[J1600+2058 (L)]{J1600+2058. (left) VLA image at L band, (right) VLA image overlaid on red SDSS image. Lowest contour = 0.2~mJy/beam, peak = 6.89~mJy/beam. \label{fig:J1600+2058}}
\end{figure}

\noindent J1600+2058 (Figure~\ref{fig:J1600+2058}). Although the source is edge brightened the lobes are found to have only relatively weak and recessed emission peaks. Both lobes are narrow with multiple emission peaks. Both off axis emission regions are sharply bounded and there is a clear emission gap between the southern lobe and the core.

\begin{figure}[ht] 
\includegraphics[width=0.45\columnwidth]{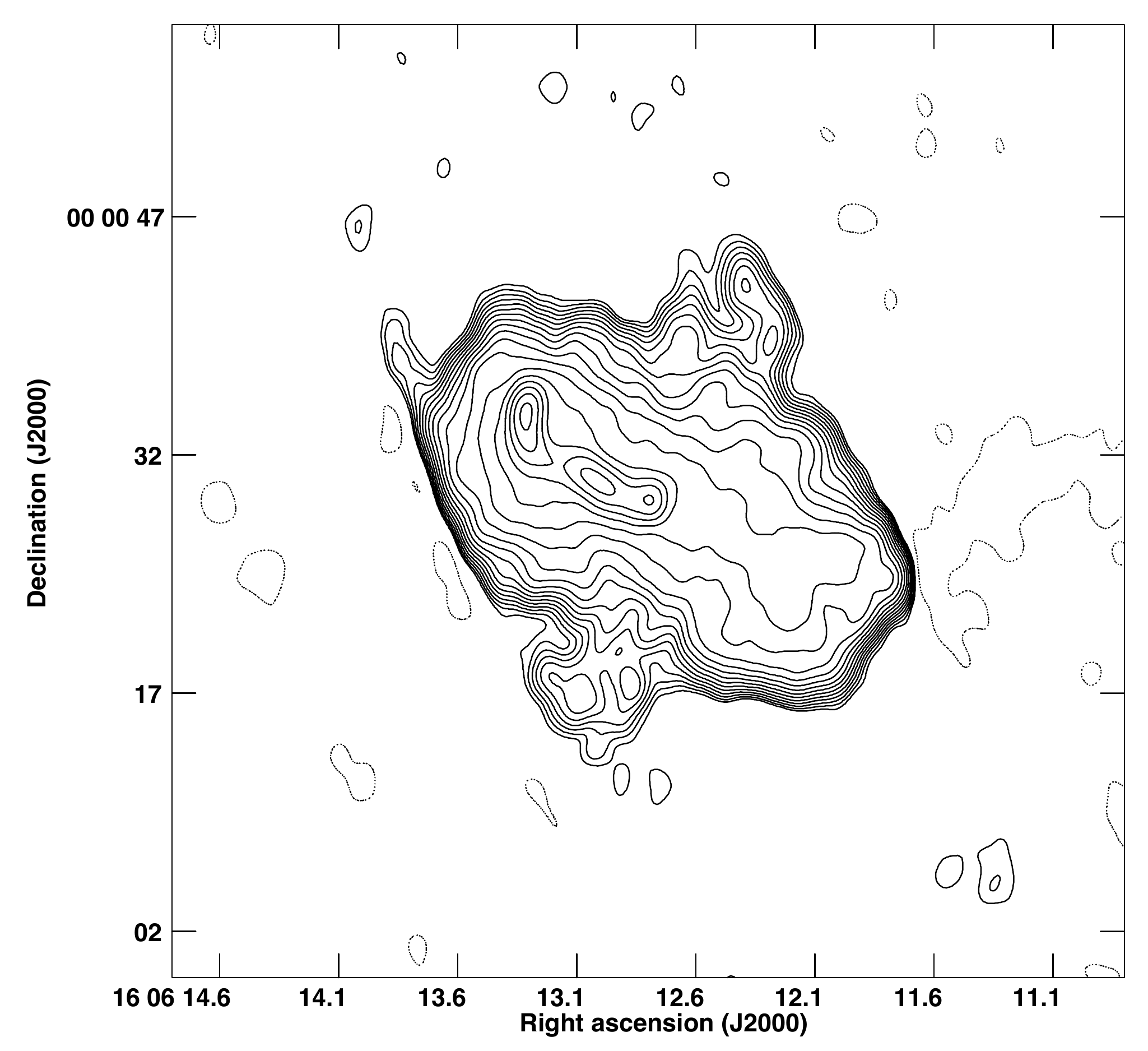}
\includegraphics[width=0.45\columnwidth]{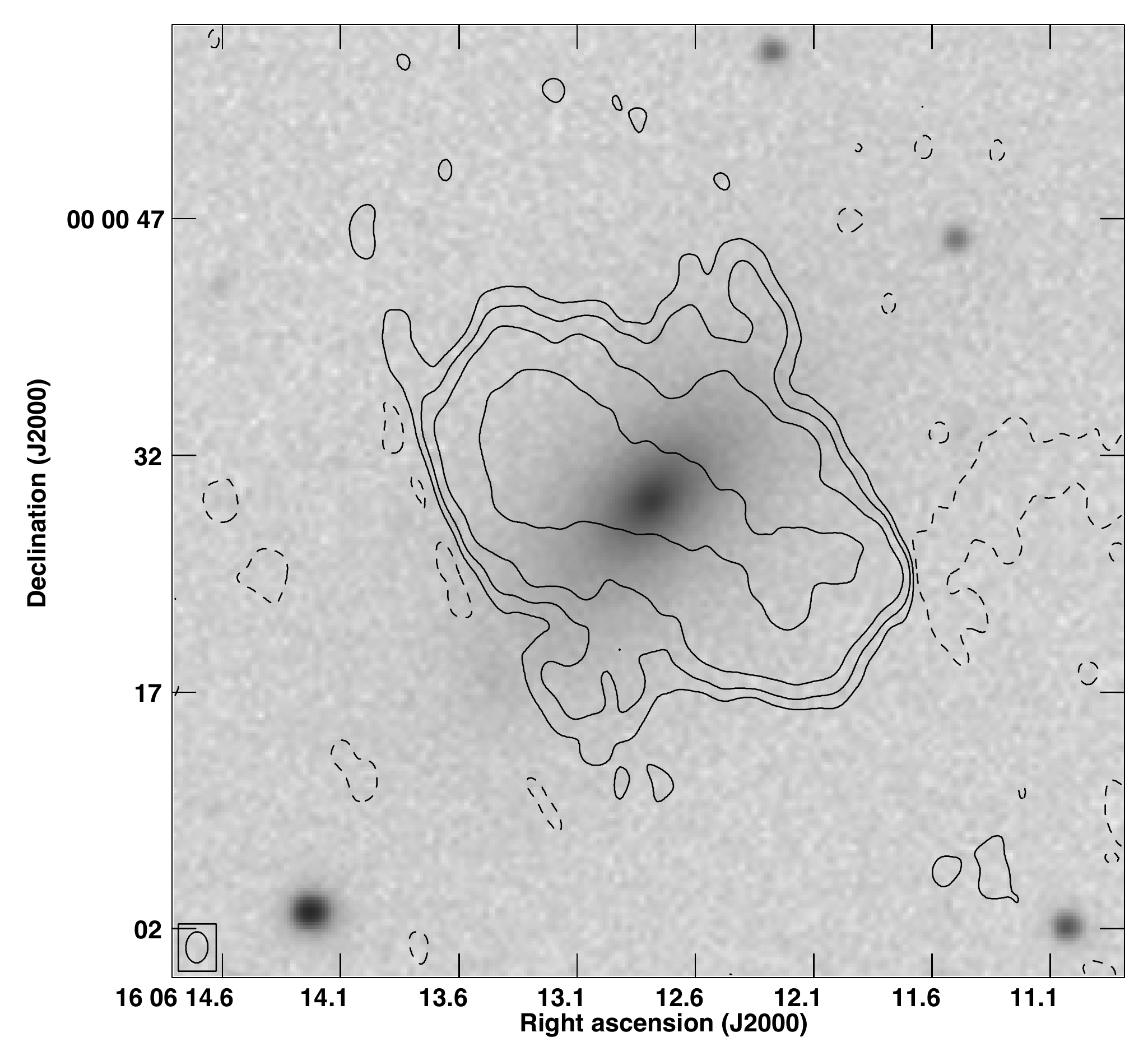}
\caption[J1606+0000 (L)]{J1606+0000. (left) VLA image at L band, (right) VLA image overlaid on red SDSS image. Lowest contour = 0.2~mJy/beam, peak = 82.0~mJy/beam.  \label{fig:J1606+0000}}
\end{figure}

\noindent J1606+0000 (Figure~\ref{fig:J1606+0000}). Our new map fails to detect any of the extended emission which is distributed skewed with respect to the main source in the FIRST map (see \cite{HK2010} for a detailed study of this source). Our map reveals a largely well confined source (also seen previously) except along a direction (through the core) that is nearly orthogonal with respect to the source axis. At these locations there are two protrusions on either side that may form the base of the elongated feature seen in the FIRST map. 

\begin{figure}[ht] 
\includegraphics[width=0.45\columnwidth]{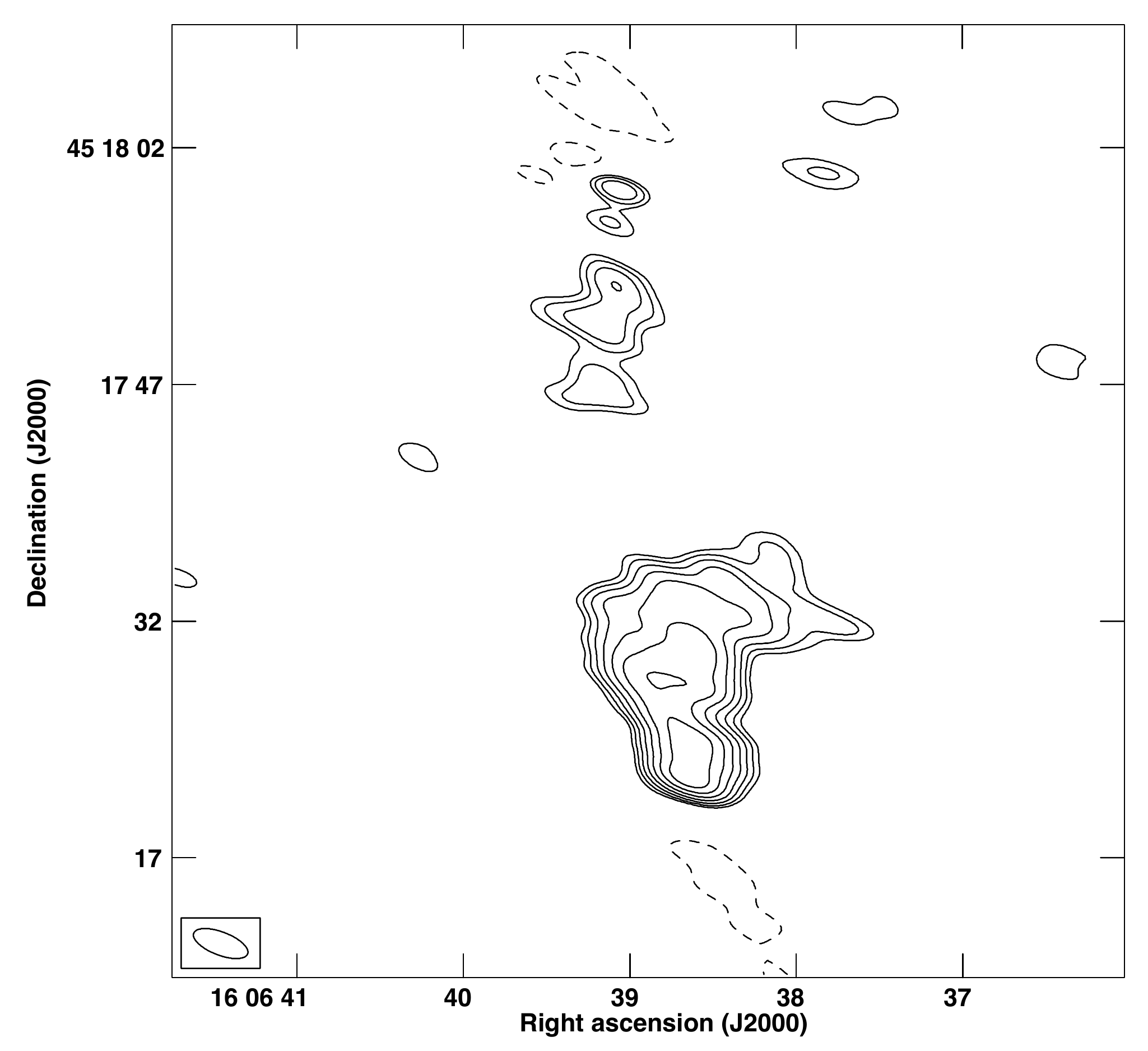}
\includegraphics[width=0.45\columnwidth]{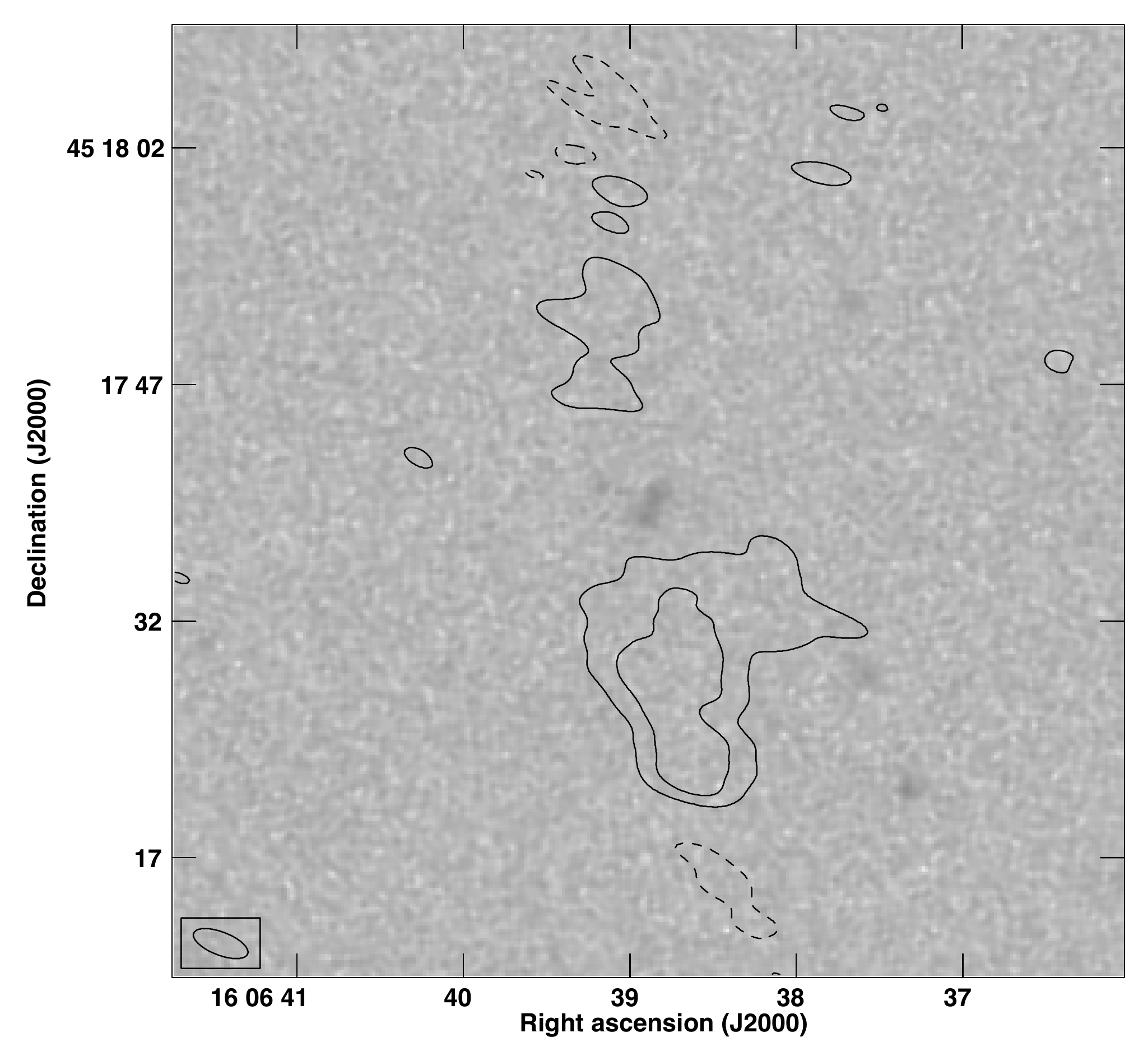}
\caption[J1606+4517 (L)]{J1606+4517. (top)  (left) VLA image at L band, (right) VLA image overlaid on red SDSS image. Lowest contour = 0.3~mJy/beam, peak  = 3.21~mJy/beam. \label{fig:J1606+4517}}
\end{figure}

\noindent J1606+4517 (Figure~\ref{fig:J1606+4517}). The edge brightened lobes are found to have compact hotspots at their extremities. The transverse inner extension to the southern lobe has a sharp inner edge. Much of the extended emission seen in the FIRST map is resolved out. 

\begin{figure}[ht] 
\includegraphics[width=0.45\columnwidth]{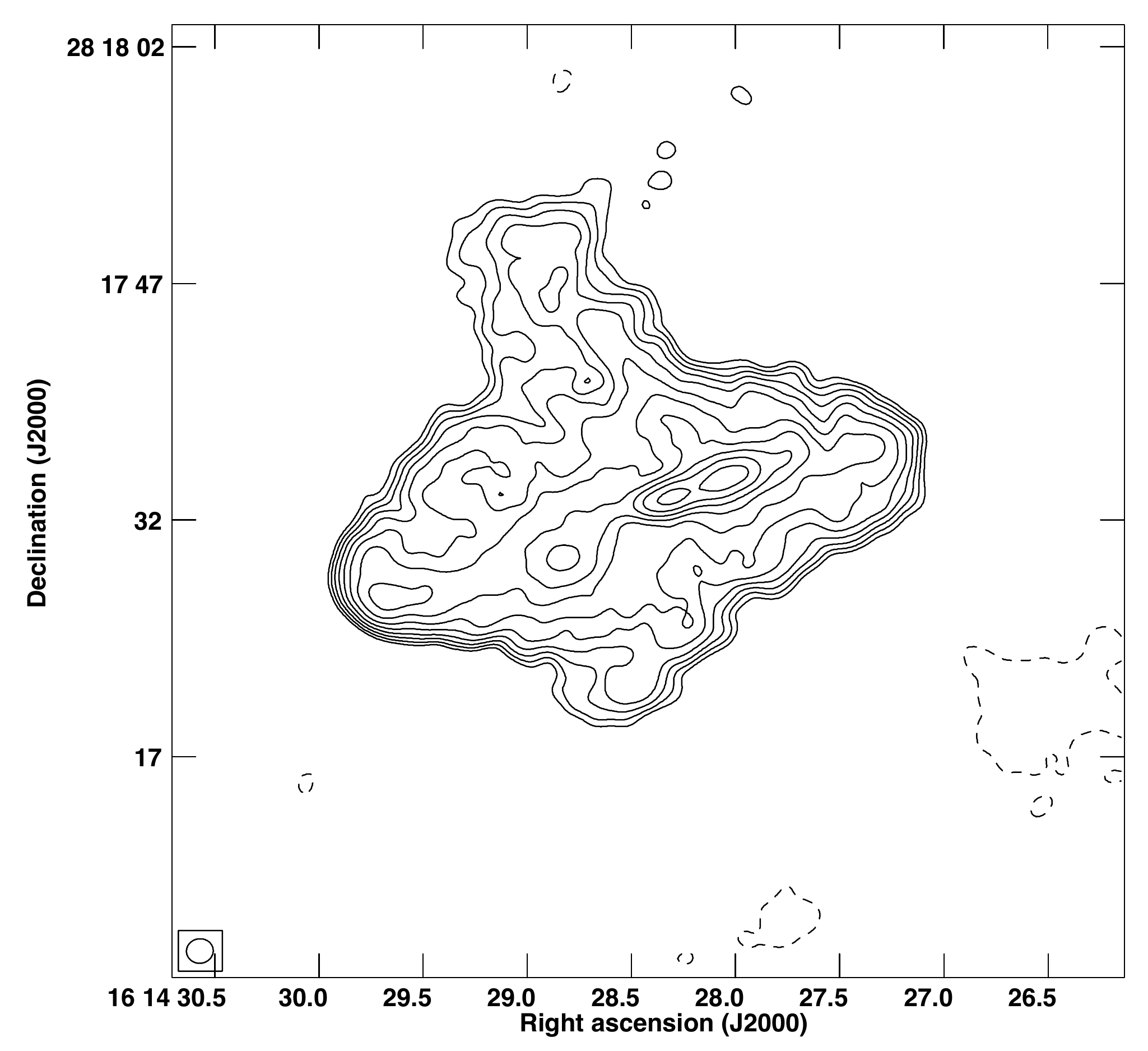}
\includegraphics[width=0.45\columnwidth]{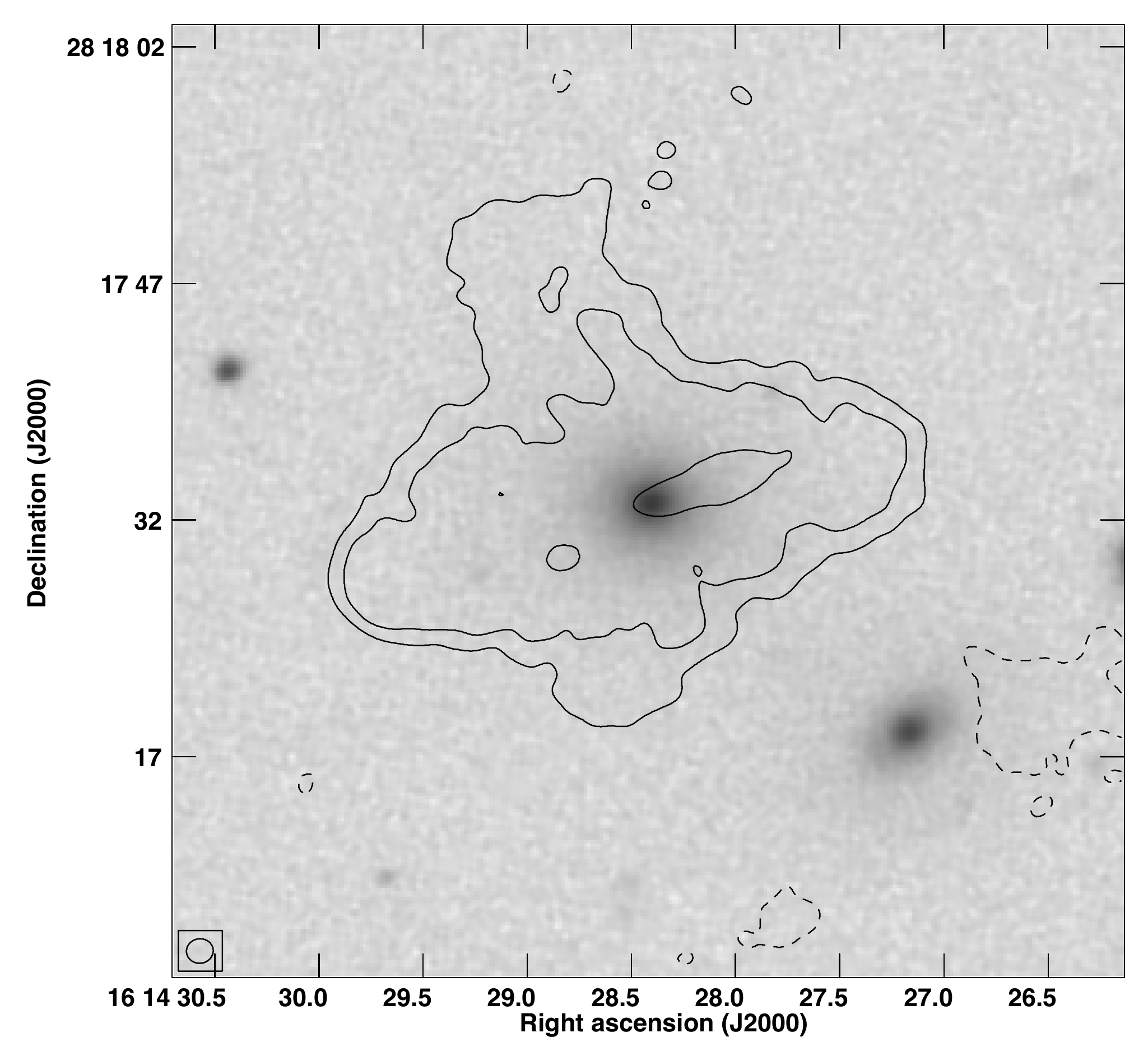}
\caption[J1614+2817 (L)]{J1614+2817.  (left) VLA image at L band, (right) VLA image overlaid on red SDSS image. Lowest contour = 0.2~mJy/beam, peak  = 8.02~mJy/beam. \label{fig:J1614+2817}}
\end{figure}

\noindent J1614+2817 (Figure~\ref{fig:J1614+2817}). A narrow channel is seen to run through the length of the source, ending in weak and bounded emission peaks at the leading ends. The source is well-bounded with two transverse extensions with only the northern extension orthogonal and centered on the core.

\begin{figure}[ht] 
\includegraphics[width=0.45\columnwidth]{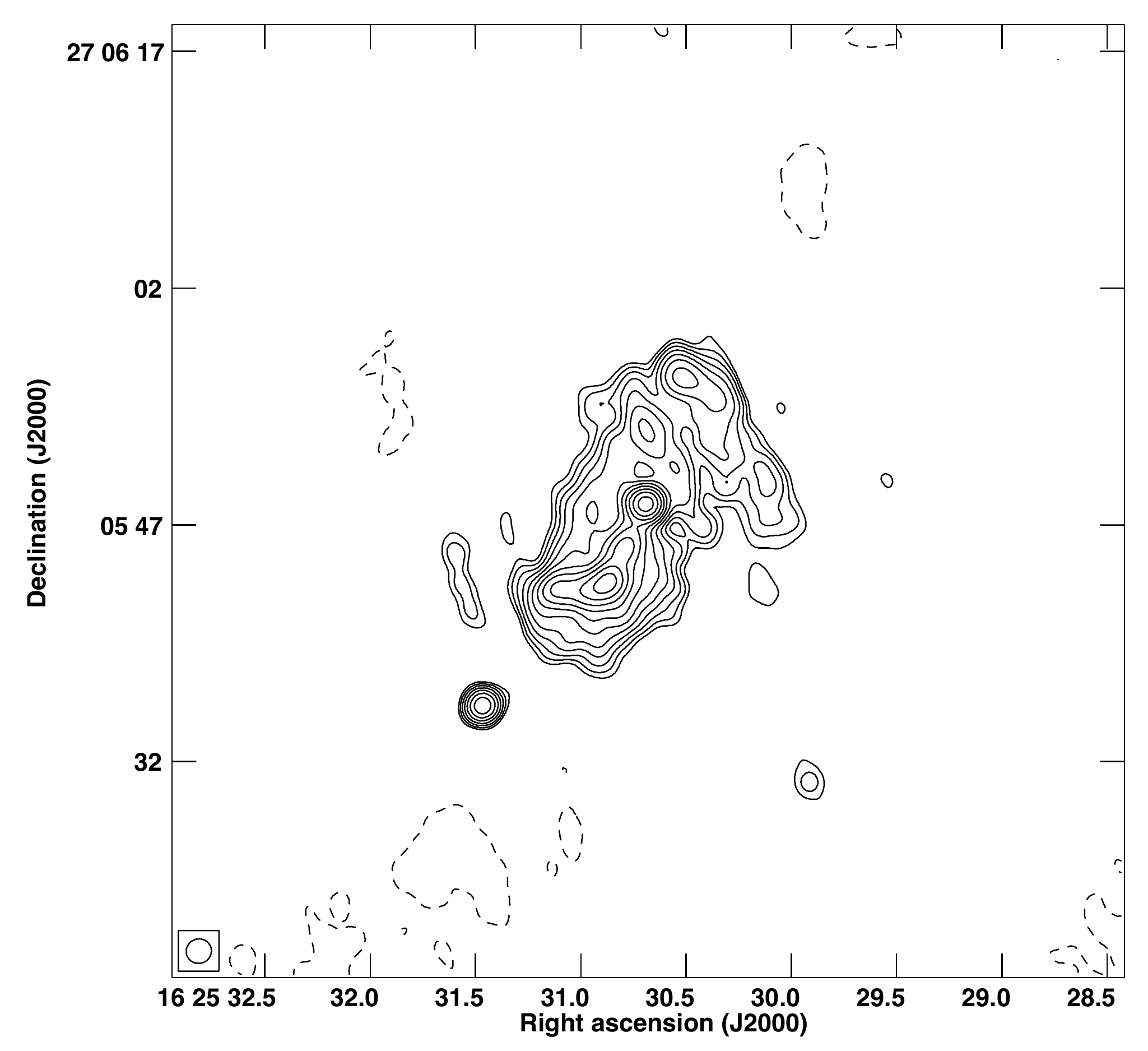}
\includegraphics[width=0.45\columnwidth]{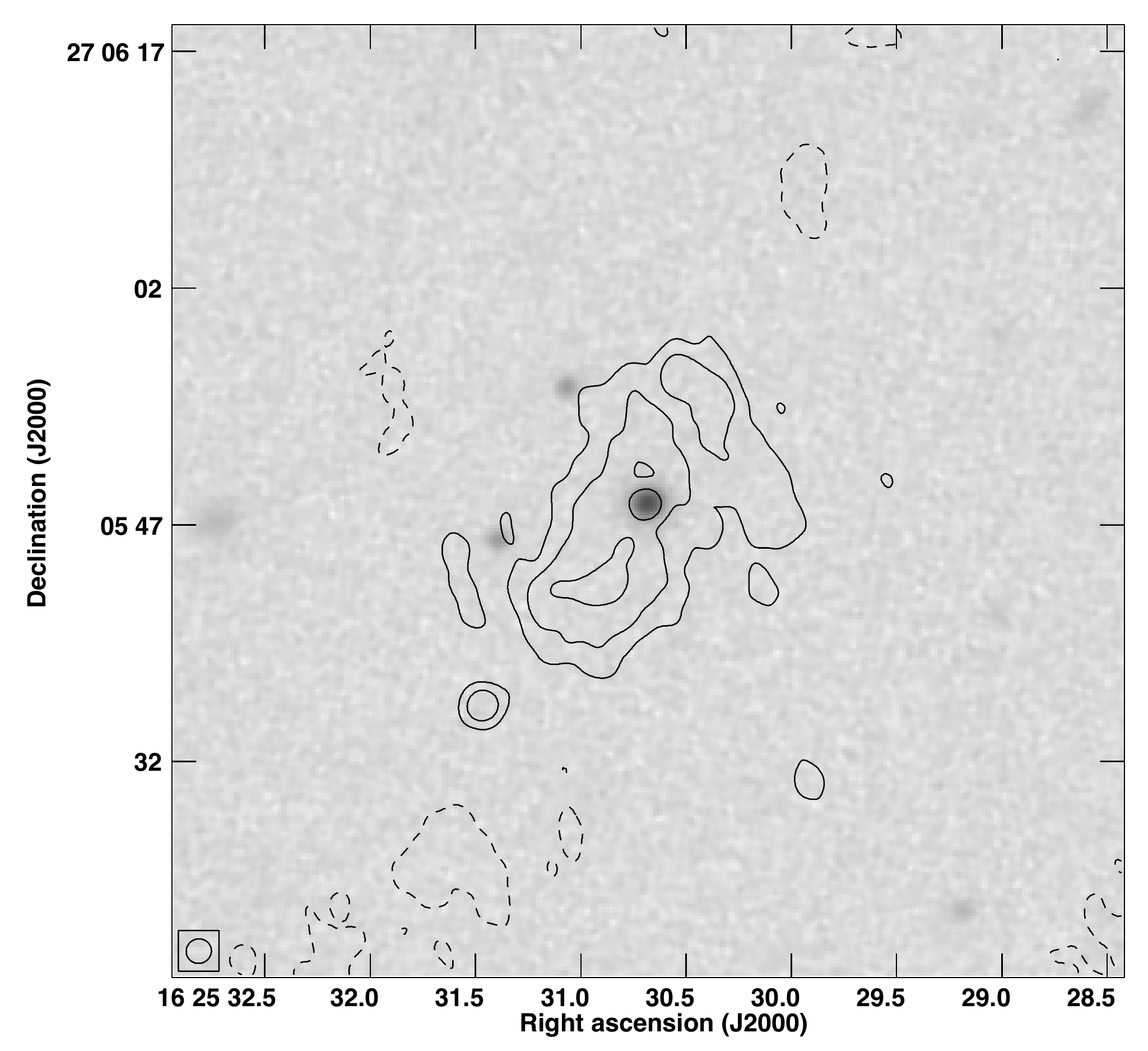}
\caption[J1625+2705 (L)]{J1625+2705. (left) VLA image at L band, (right) VLA image overlaid on red SDSS image. Lowest contour = 0.7~mJy/beam, peak = 28.0~mJy/beam.  \label{fig:J1625+2705}}
\end{figure}

\noindent J1625+2705 (Figure~\ref{fig:J1625+2705}). The quite featureless FIRST source is found to have a wealth of structure in our map. Two compact hotspots are seen. The SE hotspot is isolated appearing almost as a background source. The bright core, jet pointing to the farther and stronger SE hotspot reflect the effects of projection in the broad line AGN. A compact isolated hotspot is seen at the SE end along the axis.  

\begin{figure}[ht] 
\includegraphics[width=0.45\columnwidth]{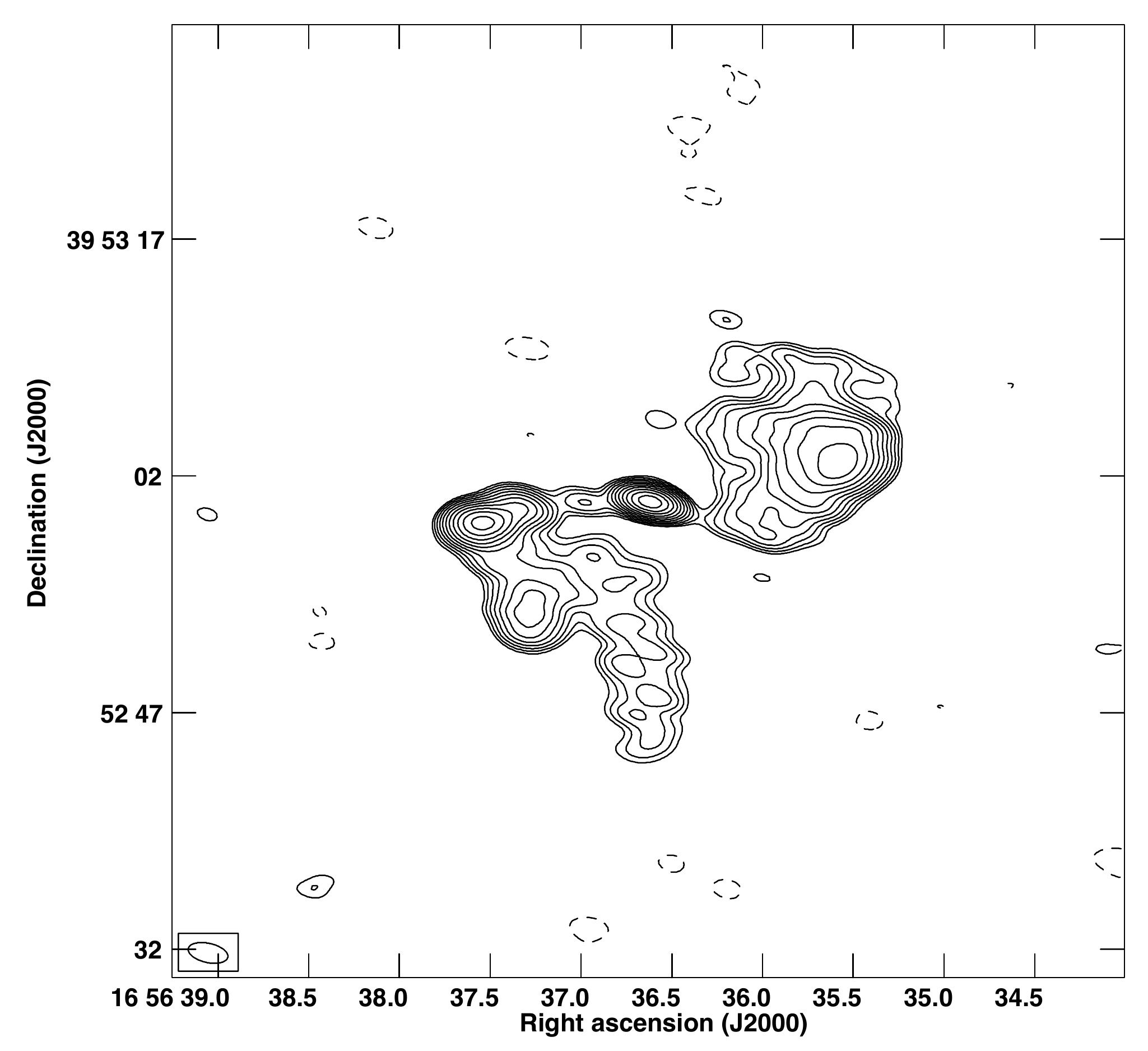}
\includegraphics[width=0.45\columnwidth]{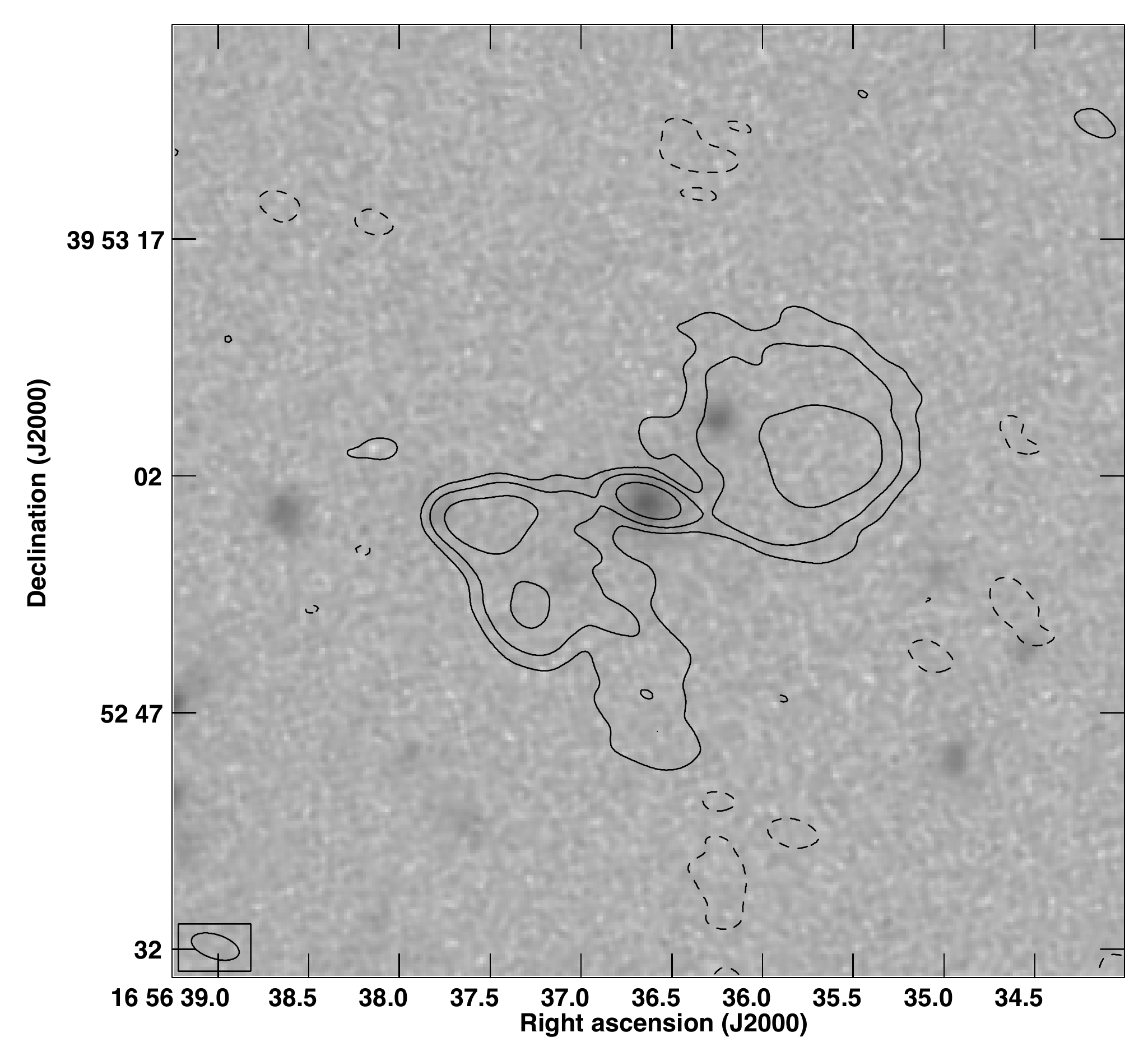}
\caption[J1656+3952 (L)]{J1656+3952.  (left) VLA image at L band, (right) VLA image overlaid on red SDSS image. Lowest contour = 0.2~mJy/beam, peak = 8.06~mJy/beam. \label{fig:J1656+3952}}
\end{figure}

\noindent J1656+3952 (Figure~\ref{fig:J1656+3952}). A classic inversion symmetric structure is revealed for this source. A bright core is seen straddled by two edge-brightened lobes to which it is connected by narrow jets. The two lobes extend away from the source axis in opposite, orthogonal directions. 

\section{DISCUSSION} \label{s:discussion}

Although a persistent structure associated with extended radio galaxies is that of two lobes straddling an elliptical galaxy (often also associated with a radio core), deviations from the radio axis are common. These deviations are most often associated with the diffuse lobes although jets are also found to show sharp bends.  The sample of 52 sources observed with the VLA and which forms the data presented herein is a subset of the 100 low axial-ratio FIRST radio sources compiled by \citet{C2007} that were chosen to be candidates for XRG morphologies \citep{C2009}, displaying recognized signs of X-shapes.  The sample has also been observed in the optical and spectra and redshifts have been obtained for most  \citep{L2010}. The redshift range for this subsample is 0.05--0.8. The new VLA observations were aimed at obtaining structural details. Our analysis of follow-up high resolution observations of these sources, selected on the basis of their low axial ratios, provides an excellent opportunity to (1) examine the types of different structural distortions in radio galaxies, (2) examine their occurrence rates and, most importantly, (3) determine the fraction that might be classified as a genuine X-shaped radio source. The observed properties of the sources are given in Table~\ref{tab:Observed} and derived properties in Table~\ref{tab:Derived}.

\subsection{Classifying Radio Lobe Distortions} \label{s:classfiication}

Although several sources appear, in the imaging of the new VLA data, to have well-resolved structures, in a significant number of new high resolution images the diffuse large-scale radio emission, which is seen in the FIRST images presented by \citep{C2007}, is completely resolved out (e.g. J0143$-$0119). Among the well-imaged sources the new observations reveal structural distortions where lobe deviations appear to occur at `strategic' locations in the radio galaxy, either from the outer lobe ends (with or without hotspots) and extending away from the radio axis at large angles or from the inner lobe ends closer to the host galaxy. In each case the off-axis emission appears connected and a continuation of one of the two individual radio lobes. The deviations in the two lobes are mostly inversion symmetric (about the center or host galaxy or radio core). In a small fraction of sources the off-axis emission in the central part is in the form of a common band of emission running through the host galaxy or radio core.  A few radio galaxies have jets seen to be bent into an ``integral'' (gradual ``S'') sign shape.

These off-axis deviations to radio galaxy structures, which otherwise exhibit basic twin-lobe morphologies, represent influences that need to be understood. Of course, effects of projection also enter and need to be considered in understanding the origin of the distortions. In an early study, \citet{L1984} noted and categorized the different kinds of distortions found in the well-studied 3CR radio galaxy sample and attempted to relate them with thermal halo gas associated with host galaxies at the centers as well as with any previous beam-activity episodes experienced by the AGN. However, lobe deviations seen originating from the outer ends of radio galaxies, far from the host galaxy influence, may have different underlying physical mechanisms at work. In the case of radio sources with a single broad swathe of emission at the centers that appear as a central band of emission a mechanism independent of the ongoing activity has also been proposed (e.g. \cite{M2002}).

\subsection{The Nature of Low Axial Ratio Sources} \label{s:nature}

Examination of the higher resolution images in conjunction with low resolution images allows us to gain insights into factors that may be causing deviations from the simple classical twin-lobe structure (thereby rendering a low axial ratio to the sources at low resolution) that is expected in the beam model.  Our advantage in this exercise is that we have a fairly large sample of radio galaxies all of which were selected in a uniform manner from a single survey and all of which have now been imaged at higher resolution with a similar setup (in frequency and array) allowing us to also note the occurrence rate of different types of structures responsible for the low axial ratios. We have chosen three categories for classifying the structures of the 52 radio sources: sources where the non-collinear structures are caused by features originating at the inner ends of lobes, outer ends of lobes and the third category which includes the rest where neither condition holds.

Among the 52 low-axial-ratio sources, there are 25 sources where the deviating structures originate at the inner ends of lobes and eight where the origin appears to be at the outer ends of lobes (see Table~\ref{tab:Observed}). The remaining 19 have structures that remain unclassified in this respect. In nearly all of the 33 sources with off-axis deviations connected to the individual lobes (whether at the inner ends or outer ends) the distortions are in opposite directions except in one case, J1434+5906 (with lobe extension at the inner end), where the deviant emission is mostly on one side. The sample sources are predominantly edge-brightened FR-II type and only seven source morphologies are either of FR-I type or blends of multiple radio sources or remain unclear.

Among the 19 radio sources that fall in neither category there are some for which the data do not allow for classifying their low-axial ratio structures (e.g., J0001$-$0033, J0049+0059, J0143$-$0119, J0145$-$0159, J0813+4347, J1227$-$0742, J1227+2155, \& J1228+2642). In these cases we will need to pursue further, higher, resolution imaging.  

Among the 19 sources  there are 11 sources where there is a possibility of an independent transverse feature centered on the host (J0144$-$0830, J1008+0030, J1015+5944, J1043+3131, J1327$-$0203, J1345+5233, J1406+0657, J1408+0225, J1606+0000, J1614+2817, \& J1625+2705). In all these sources there is extended transverse emission seen either in the FIRST image (but not imaged in our high resolution maps) or in our maps which cannot be traced to either of the two lobes in the radio galaxy.  Given the lack of association of this extended emission with the radio galaxy components (whether individual lobes or hotspots as seen in the two groups discussed above) we consider these eleven sources as potential candidates for ``genuine'' X-shaped radio galaxies, although better imaging would help clarify the nature of the extended emission in these sources further.

\subsection{Characteristics of Sources Identified with Inner-end and Outer-end Lobe Deviations} \label{s:inner outer}

Having classified the different types of deviations that are revealed when low-axial ratio sources are imaged at higher resolution, we attempt to characterize the properties of sources in the two groups, one with sources having inner-end deviations and the other with sources having outer-end deviations  (see Table~\ref{tab:Observed} for the two groups of sources listed separately).
For this we have used different measures and associations like the projected physical separation (where possible) between the locations of deflections in each lobe, the fractional extents of the deflections, the presence or absence of a radio core, fractional core flux, whether the sources have FR-I, FR-II or hybrid-type structures and the presence of broad emission lines in the optical spectra (where available).

Among the 25 sources where the deviations occur at the inner ends of lobes we have measured structural parameters (from radio maps) for 19. For the remainder, the exercise was hampered by absence of a core and host galaxy or poor quality of the radio image (see Table~\ref{tab:Observed}).  The variety in radio structures lead to the following inferences:
\begin{enumerate}

\item All 25 sources with inner-end deviations (Table~\ref{tab:Observed}) have FR-II morphologies.

\item Within the limits of sensitivity more often than not the transverse deviations on the two sides are unequal in extent.

\item Emission gaps between the location where the transverse deviation occurs and the cores are common. Fourteen out of 19 show distinct inner edges to the deviations which are also separated by recognizable or significant gaps (see Table~\ref{tab:Observed} for the sources that show sharp edges). The physical extents of the gaps are at most 60~kpc from the center of the host galaxy. 

\item In sources with clear hotspots on the two sides, we detect no discernible correlation or anti-correlation between presence of stronger hotspot on one side and size of the  gap or extent of the off-axis emission on that side. 

\item In five cases  (0211$-$0920, 0702+5002, 0859$-$0433, 0941$-$0143, \& 1054+5521) the transverse deviations extend to as much or more than their respective lobe extents, at least on one side. In three cases  (J0702+5002, J0859$-$0433, \& 0941$-$0143) both transverse deviations have fractional extents exceeding unity. It may be noted that our measurements of the wing extents is mostly based on the lower resolution FIRST survey images.

\end{enumerate}

The fact that the inner-end deviations are not collinear and centered on the host and are instead separated by clear gaps implies that they may not be representing visible lobes created in a previous activity epoch or even channels left behind in a previous epoch and which are now made visible by new lobe plasma that has flowed into them. 

We have also measured the position angles of the major axes of the host elliptical galaxies for a few of the sources in this group. 
The purpose was to examine if sources with inner-end deviations adhered to the same tendency of the radio axis being closer to the host major axis as shown by XRGs \citep{C2002, S2009} and the subset of 3CR radio galaxies with central distortions to the lobes \citep{S2009}. With the prevalence of central lobe distortions in radio galaxy samples and the adherence to the same correlation in radio-optical axes as shown by the much longer winged XRGs the latter authors had suggested a generic physical mechanism like deflection of backflows by thermal gaseous halos \citep{L1984}  rather than jet axis flips as the mechanism that may be causing the commonly seen small-extent central distortions as well as the more extreme ones seen in XRGs. For large angle flips in jet axis to be the responsible mechanism it would require axis flips to be commonly occurring and would need the jet axis to flip from minor axis to the major axis (see discussion of the contending XRG models by \cite{S2009}).

Measurements were possible for sources whose hosts were bright; values have been noted only for those that were clearly non-circular in appearance. For a total of eight sources it was possible to measure the major axis position angle (we used the ELLIPSE task in IRAF). For each of the sources we also measured the position angle of the radio axis. The axes were defined by the line joining the core and the hotspots in each of the lobes; where the core was not seen the axis was defined to be the line between the (likely) host galaxy and hotspots. For this group of galaxies the radio axes are within 20$\degr$ of their respective host major axes in six out of eight sources. This correspondence between host major axis and radio axis is consistent with that found previously for a subsample of 3CR radio galaxies with off-axis and inversion symmetric lobe distortions \citep{S2009} suggesting the possibility of a generic mechanism like backflow deflection underlying commonly seen off-axis lobe distortions as well as possibly also in the more extreme XRGs. 

For the smaller sample of eight sources having outer-end deviations, all of which have FR-II morphologies, this exercise was possible only for two sources and in both cases the radio axes are within 30$\degr$ of the host major axes. 

Three out of eight outer-end deviation sources (J0845+4031, J1253+3435, \& J1430+5217) show structures in their lobes that may be attributed to a drift or rotation in axis. In J0845+4031 each lobe has corresponding emission peaks and trailing emission in opposing directions that form a clear ``S''  with the inner peaks forming an axis with the core. In J1253+3435 the edge-brightened lobes have twin or extended hotspots at the lobe ends accompanied by sharp-edged oppositely-extending lobe emission, whereas in J1430+5217 there are several emission peaks and collimated extensions on either side indicating axis change besides the nearly transverse twin-hotspots at the lobe ends and oppositely extended trailing lobe emission. In the ``neither'' category of sources no source shows signs of axis rotation. Interestingly only one of the 25 inner-end deviation sources (J1207+3352) shows circumstantial evidence of axis rotation. 

Two of the ``inner-end'' deviation sources, J0924+4233 and J1459+2903, show clear signs that the AGN beam activity has in the past ceased and restarted; their structures display an inner-double source embedded within a large pair of outer lobes that are devoid of compact hotspots.

\subsection{Physical Implications} \label{s:implications}

Whether the transverse structures originate at the inner ends or outer ends of lobes can have quite different physical significance. 
Transverse structures originating at the inner ends of lobes are also seen in several known XRGs. Given the observational characteristics displayed by XRGs (described in Section~\ref{s:intro}; \cite{S2009}) the axis-flip model would require the axis to mostly flip from host minor axis direction to near host major-axis direction, would require the minor mergers responsible for axis flips or drifts to also displace the galaxy by several tens of kiloparsecs and for the relic emission to remain visible when often there is a deficit of relic radio galaxies, for the relic lobes to always have edge-darkened (FR-I) morphology and for the active main lobes to mostly have FR-II morphology. Where as the prevalence of inner-end distortions to radio galaxies (although less pronounced in lateral extent than in the more extreme wings in XRGs), the continuum of properties related to orientations of radio axis and optical axis \citep{S2009} between the two populations of radio galaxies, the presence of X-ray halos with major axis close to host major axis \citep{K2005, HK2010}, the mostly FR-II morphology of the main lobes, the clear connection seen between the off axis distortions and individual lobes, the separation between the off axis features and the often sharp inner edges to the off axis emission all are more simply explained via models that suggest backflowing lobe synchrotron plasma getting deflected in the thermal halos associated with the host galaxies \citep{C2002, S2009, HK2011}. In the case of off-axis emission connected to outer ends of lobes the wings might represent relic synchrotron plasma deposited in outer lobe regions as the jets drifted in position angle, perhaps in a precessing beam. 

Neither of the two physical mechanisms can be supported at present via evidence other than circumstantial but simulations have been carried out for the former mechanism (\cite{C2002, HK2011}) as well as for reproducing the inner-lobe deviations via radio axis precession where effects of projection and light-travel time differences play a major role \citep{G2011}. It is nevertheless important to identify such sources that may be used as test-beds for mechanisms such as backflow deflection and radio axis rotation that are fundamental to understanding the AGN central engine and stability of its black hole spin axis.

\section{SUMMARY AND CONCLUSIONS} \label{s:summary}

We have analyzed 1.4  and 5~GHz archival VLA continuum data on a sample of 52 FIRST radio sources selected on the basis of low-axial ratio radio structures. Our primary results are:
\begin{enumerate} 

\item The exercise has allowed examination of features that contribute to off-axis emission in radio sources that is not expected to naturally arise in the standard beam model for radio galaxies.

\item Our higher resolution imaging has aided in characterizing low-axial radio sources into ones where the off-axis emission is traced to individual radio lobes and ones where it instead appears as a common swathe of emission through the center and across the source axis.

\item A large fraction of the sample (60\%) constitutes sources where the off-axis emission is traced to individual lobes.

\item Eleven sources (20\% of our sample) have been identified as potentially genuine X-shaped radio galaxy candidates. Although the parent sample from which the 52-source subsample (used here) has been drawn is itself drawn from those FIRST fields that had ``sufficient dynamic range in the images to be able to see extended low surface brightness wings''  \citep{C2007} we cannot discount sources with even fainter extended wings that FIRST survey may have missed. 

\end{enumerate}
 
The implications of these results for the predicted gravitational wave background are discussed in \cite{PaperII}.

\section{ACKNOWLEDGMENTS}

The National Radio Astronomy Observatory is a facility of the National Science Foundation, operated under cooperative agreement by Associated Universities, Inc. 

Funding for the SDSS and SDSS-II has been provided by the Alfred P.\ Sloan Foundation, the Participating Institutions, the National Science Foundation, the U.\ S.\ Department of Energy, the National Aeronautics and Space Administration, the Japanese Monbukagakusho, the Max Planck Society, and the Higher Education Funding Council for England. The SDSS Web Site is http://www.sdss.org/.

D.\ H.\ R.\ gratefully acknowledges the support of the William R. Kenan, Jr.\ Charitable Trust.

We thank Anand Sahay who assisted in the derivation of host galaxy position angles. The anonymous referee is thanked for helpful comments that led to an improved presentation.

\bigskip

\noindent {\em Facilities:}: Historical VLA (data archives, project codes: AB808, AC406, AC4450, AC572, AC818, AD100, AF91, AF918, AG143B, AJ250, AL252, AM67, AM222, AM364, AO80, AO80D, AP326, AR123, VC35).

\clearpage


\input{table1}
\input{table2}

\end{document}

%% file: table1.tex
								
\begin{deluxetable}{lccccccccc}		
\tablecaption{Observed Source Properties \& Classifications \label{tab:Observed}}
\tablecolumns{10}

\tablehead{ \colhead{Name} & Figure &\colhead{Type\tablenotemark{a}} &\colhead{$z$\tablenotemark{b}}	 & \colhead{$S_{core}$\tablenotemark{c}} & \colhead{$S_{tot}$\tablenotemark{c}} &	\colhead{$S_{tot,FIRST}$\tablenotemark{d}} & \colhead{$\theta_{max}$\tablenotemark{e}} &  Inner Edges? & Large-Scale Emission  \\
\colhead{(J2000)} & Number & \colhead{} & \colhead{} & \colhead{(mJy)} & \colhead{(mJy)} & \colhead{(mJy)}  & \colhead{(arcsec)} && Resolved Out?\tablenotemark{f}} 

\startdata												
\multicolumn{10}{c}{} \\ 
\multicolumn{10}{c}{Sources with ``Bends From Inner Ends''} \\ 
&&&&&&&&&\\ \tableline
J0045+0021	& 2 &	\dots&	\dots	&	\dots &	432	&	509   &	31  & yes? &  \dots  \\
J0113+0106	& 4 &G&	0.281	&	1.1(L) ID?	&	221	&	391	 &	170   & yes &  \dots  \\
J0211$-$0920	& 8 &	\dots&	\dots		&	$<0.21$ 	&	69	&	180	 & 	56  & yes & yes \\
J0702+5002	& 9 &G&	0.0946	&	6.0	&	282	&	334	 & 	59  & yes & \dots \\
J0846+3956	& 13 &	\dots&	\dots		&	$<2.$	&	170	&	197	 & 	37  & ? &  \dots  \\
J0859$-$0433	& 14 &G&	0.356	&	\dots	&	123	&	237	  & 	52  & yes &  \dots \\
J0917+0523	& 15 &G&	0.591	&	\dots	&	508	&	612	 & 	56  & ? &  \dots  \\
J0924+4233	& 16 &G&	0.2274	&	2.7	&	169	&	292	 & 	56  & yes &  \dots  \\
J0941$-$0143	& 17 &G&	0.382	&	$<1.4$	&	790	&	830	 & 	36  & yes &  \dots  \\
J1005+1154	& 18&G&	0.1656	&	0.84	&	149	&	205	 & 	42  & yes &  \dots \\
J1054+5521\tablenotemark{g}	& 22 &	\dots&	\dots		&	\dots	&	132	&	222	 & 	50  & \dots &  \dots  \\
J1202+4915	& 25 &	\dots&	\dots		&	0.51	&	72	&	104	 & 	38  & yes &  \dots  \\
J1206+3812	& 26 &BG&	0.838	&	3.5(L), 1.9(C)	&	217	&	247 	 & 	42  & yes &  \dots  \\
J1207+3352\tablenotemark{g}		& 27&G&	0.0788	&	18.(L), 27.(C)	&	263	&	490	 & 	56  & \dots, dual morphology & \dots\\
J1210$-$0341	& 28&G&  0.178, 0.26	& 0.16	&	93	&	151	 & 	42  & yes but no gap & \dots \\
J1211+4539	& 29&	\dots&	\dots		&	\dots	&	209	&	232	 & 	44  & yes &  \dots  \\
J1227$-$0742\tablenotemark{g}		& 30&	\dots&	\dots		&	\dots	&	33	&	168	 & 	55  & \dots & \dots  \\
J1309$-$0012	& 34&G&	0.419	&	14.(C)	&	663	& 1637 & 78  & yes? & \dots  \\
J1310+5458	& 35&G&	0.356	&	11.	&	190	&	226	 & 	29  & no & \dots  \\
J1406$-$0154	& 40&G&	0.641	&	$<0.3$	&	772	&	1143	 & 	58  & yes &  \dots  \\
J1434+5906\tablenotemark{g}		& 44&G&	0.538	&	$<1.8$	&	255	&	315	 & 	33  & \dots &  \dots \\
J1456+2542\tablenotemark{g}		& 45&G&	0.536	&	$<0.06$	&	12	&	 36	 & 	35  & \dots &  \dots \\
J1459+2903\tablenotemark{g}		& 46&G&	0.1460	&	7.1(L), 10.(C)	&	59	&	367 	& 	70  & \dots & yes\\
J1600+2058	& 47&G&	0.174	&	6.7	&	158	&	523	 & 	58  & yes & yes \\
J1606+4517	& 49&G&	0.556	&	\dots	&	41	&	116	 & 	49  & yes & yes \\ \tableline
&&&&&&&&&\\
\multicolumn{10}{c}{Sources with ``Bends From Outer Ends''} \\  
&&&&&&&&& \\ \tableline
J0821+2922	& 11&G&	0.246	&	$<1.6$	&	62	&	117	 & 	24  & &  \dots \\
J0845+4031	& 12&G&	0.429	&	3.0	&	124	&	164	 & 	40  & &  \dots \\	
J1135$-$0737	& 24&G&	0.602	&	\dots	&	44	&	106	 & 	62  & & yes \\	
J1253+3435	& 33&G&	0.358	&	1.4	&	244	&	362	 & 	44  & &  \dots \\	
J1342+2547	& 37&BG&	0.585	&	8.7	&	274	&	365	 & 	33  & &  \dots \\	
J1348+4411	& 39&G&	0.267	&	0.62	&	96	&	158	 & 	36  & &  \dots \\	
J1430+5217	& 43&BG&	0.3671	&	10.	&	469	&	671	 & 	38  & &  \dots \\ 
J1656+3952	& 52&	\dots&	\dots	 	&	8.1	&	83	&	147	 & 	30  & &  \dots \\ \tableline
&&&&&&&&& \\ 
\multicolumn{10}{c}{Other Sources}  \\ 
&&&&&&&&&\\ \tableline
J0001$-$0033	& 1&G&	0.2469	&	2.4	&	32	&	73	 &	28  & & yes \\
J0049+0059	& 3&G&	0.3044	&	0.71	&	82	&	155	 & 	46  & &  \dots  \\
J0143$-$0119	& 5&G&	0.520	&	43.(L), 51.(C)	&	493	&	823	 & 	43  & &  \dots \\
J0144$-$0830	& 6&G&	0.181	&	$<0.5$	&	24	&	47	 & 	34  & &  \dots \\
J0145$-$0159	& 7&G&	0.1264	&	$<0.5$	&	25{\bf ??}	&	272	 & 	46  & & yes\\
J0813+4347	& 10&G&	0.1282	&	12.	&	288	&	333	 & 	51  & &  \dots \\
J1008+0030	& 19&G&	0.0977	&	63.	&	107	&	484	 & 	36  & & yes\\
J1015+5944	& 20&BG&	0.5271	&	4.9	&	203	&	221	 & 	38  & &  \dots \\
J1043+3131	& 21&G&	0.0357	&	34.(L), 36.(C)	&	280	&	749	 & 	40   & & yes\\
J1111+4050	& 23&G&	0.0737	&	53.(L), 11.(C)	&	740	&	819	 & 	41  & & \dots  \\
J1227+2155	& 31&	\dots&	\dots		&	$<0.3$	&	70	&	199	 & 	44  & & yes \\
J1228+2642	& 32&G&	0.201	&	$<0.4$	&	126	&	185	 & 	42  & &  \dots \\
J1327$-$0203	& 36&G&	0.183	&	21.	&	1085	&	1179	 & 	44  & &  \dots \\
J1345+5233	& 38&	\dots&	\dots		&	$<0.92$	&	33	&	71	 & 	29  & & yes\\
J1406+0657	& 41&BG&	0.550	&	$8.6$	&	294	&	339	 & 	24  & &  \dots  \\
J1408+0225	& 42&	\dots&	\dots		&	14.	&	163  &	244	 & 	26  & &  \dots \\
J1606+0000	& 48&G&	0.059	&	54.	&	1889	&	2385	 & 	38  & &  \dots \\
J1614+2817	& 50&G&	0.1069	&	7.4	&	271	&	424	 & 	39  & &  \dots \\
J1625+2705	& 51&BG&	0.5259	&	28.	&	255	&	531	 & 	29  & & yes\\

\enddata

\tablenotetext{a}{G = narrow line radio galaxy, BG = broad line radio galaxy.}
\tablenotetext{b}{All redshifts from \cite{C2007} or \cite{C2009}.}	
\tablenotetext{c}{This work.}	
\tablenotetext{d}{From \cite{C2007}.}	
\tablenotetext{e}{\,Largest angular size in arcseconds.}	
\tablenotetext{f}{Visibility less than 0.5.}
\tablenotetext{g}{ For these sources absence of core or ID and poor radio maps disallowed examining radio structural properties.}

\end{deluxetable}

%% file: table2.tex
\begin{deluxetable}{ccc}		
\tablecaption{Derived Source Properties \label{tab:Derived}}
\tablecolumns{3}

\tablehead{ \colhead{Name} & \colhead{$L_{proj}$\tablenotemark{a}} & \colhead{$f_{core}$\tablenotemark{b}} \\
\colhead{} &  \colhead{(kpc)}  & \colhead{(percent)} } 

\startdata												

\multicolumn{3}{c}{} \\ 
\multicolumn{3}{c}{Sources with ``Bends From Inner Ends''} \\ 
&&\\ \tableline
J0045+0021	&	\dots	&	 \dots \\
J0113+0106	&	950	&		0.28	\\
J0211$-$0920	&	\dots	&		0.12	\\
J0702+5002	&	110	&		1.8	\\
J0846+3956	&	\dots	&		$<1$	\\
J0859$-$0433	&	370	&		    \dots     \\
J0917+0523	&	650	&		\dots 	\\
J0924+4233	&	250	&		0.92	 \\
J0941$-$0143	&	280	&		$<0.2$	\\
J1005+1154	&	140	&		0.41	\\
J1054+5521	&	\dots	&		\dots	\\
J1202+4915	&	\dots	&	0.49	\\
J1206+3812	&	700	&	1.4	\\
J1207+3352	&	88	&	3.7	\\
J1210$-$0341	&  	150	&		0.11	\\
J1211+4539	&        \dots    	&	\dots	\\
J1227$-$0742	&       \dots     	&	\dots	\\
J1309$-$0012	&       650  	&	\dots \\
J1310+5458	&       210 	&	4.9	\\
J1406$-$0154	&        \dots    	&	$<0.03$	\\
J1434+5906	&       350  	&	$<0.6$	\\
J1456+2542	&       370  	&	$<0.2$	\\
J1459+2903	&      \dots     	&	1.9	\\
J1600+2058	&       200  	&	1.3	\\
J1606+4517	&       540  	&	\dots	\\ \tableline
&&\\
\multicolumn{3}{c}{Sources with ``Bends From Outer Ends''} \\ 
&&\\ \tableline
J0821+2922	&       \dots     	&	$<1.4$ 	\\
J0845+4031	&       340  	&	1.8	\\	
J1135$-$0737	&       740  	&	\dots	\\	
J1253+3435	&       \dots     	&	0.39	\\	
J1342+2547	&       390  	&	2.4	\\	
J1348+4411	&       190  	&	0.39	\\	
J1430+5217	&       280  	&	1.5	\\ 
J1656+3952	&      \dots      	&	5.5	\\ \tableline
&&\\
\multicolumn{3}{c}{Other Sources} \\ 
&&\\ \tableline
J0001$-$0033	&       140  	&	3.3	\\
J0049+0059	&       280  	&	0.46	\\
J0143$-$0119	&       \dots     	&	5.2	\\
J0144$-$0830	&      120   	&	$<1$	\\
J0145$-$0159	&      120   	&	$<0.2$	\\
J0813+4347	&      130   	&	3.6	\\
J1008+0030	&      71  	&	13.	\\
J1015+5944	&      400   	&	2.2	\\
J1043+3131	&      28  	&	4.5	\\
J1111+4050	&      60  	&	6.5	\\
J1227+2155	&       \dots &	$<0.2$	\\
J1228+2642	&      170   	&	$<0.2$	\\
J1327$-$0203	&      160   	&	1.8	\\
J1345+5233	&       \dots     	&  $<1$	\\
J1406+0657	&      270   	&	$<3$	\\
J1408+0225	&      \dots      	&	6.3	\\
J1606+0000	&      44  	&	2.3	\\
J1614+2817	&      83  	&	1.7	\\
J1625+2705	&      300   	&	5.3	\\ \tableline
\enddata	
\tablenotetext{a}{Projected size in kiloparsecs.}	
\tablenotetext{b}{Core fraction in percent, L band.}

\end{deluxetable}